\documentclass[graybox]{svmult}

\usepackage{amsmath}
\usepackage{amssymb}

\usepackage{IEEEtrantools}

\usepackage{mathptmx}       
\usepackage{helvet}         
\usepackage{courier}        
\usepackage{type1cm}        
\usepackage{makeidx}         
\usepackage{graphicx}        
\usepackage{multicol}        
\usepackage{multirow}
\usepackage[bottom]{footmisc}
\usepackage{hyperref}        
\usepackage{soul}            
\hypersetup{colorlinks=true,urlcolor=blue}
\usepackage[square,numbers]{natbib}
\usepackage[scr=boondoxo,scrscaled=1.05]{mathalfa}
\usepackage{bm}
\makeindex             

\font\ec=ecrm0800 at 11pt
\def\th{\hbox{\ec\char'336}}
\def\edth{\hbox{\ec\char'360}}

\newcommand{\mb}{{\bar{m}}}

\newcommand{\flux}{{\mathcal{F}}}
\newcommand{\fluxEI}{\mathcal{F}_E^\mathcal{I}}
\newcommand{\fluxEH}{\mathcal{F}_E^\mathcal{H}}
\newcommand{\fluxLzI}{\mathcal{F}_{L_z}^\mathcal{I}}
\newcommand{\fluxLzH}{\mathcal{F}_{L_z}^\mathcal{H}}

\newcommand{\bpsi}{\Psi}
\newcommand{\ppsi}{\psi}
\newcommand{\hpsi}{\psi}

\newcommand{\e}{\epsilon}
\newcommand{\beq}{\begin{equation}}
\newcommand{\eeq}{\end{equation}}
\newcommand{\res}{{\cal R}}
\renewcommand{\P}{{\cal P}}
\newcommand{\R}{{\rm R}}
\renewcommand{\S}{{\rm S}}
\newcommand{\Lie}{\pounds}
\renewcommand{\r}{{\sf r}}
\newcommand{\s}{{\sf s}}

\newcommand{\p}{{0}}
\newcommand{\orbit}{{o}}
\newcommand{\ro}{r_\orbit}
\newcommand{\torbit}{t_\orbit}
\newcommand{\fo}{f_\orbit}

\newcommand{\ZM}{\psi^{\rm ZM}}
\newcommand{\CPM}{\psi^{\rm CPM}}
\newcommand{\RW}{\psi^{\rm RW}}
\newcommand{\emm}{{\mathscr m}}
\newcommand{\even}{{\rm even}}
\newcommand{\odd}{{\rm odd}}
\newcommand{\lm}{{\ell\mathscr{m}}}

\begin{document}

\title*{Black hole perturbation theory and gravitational self-force}

\author{Adam Pound \and Barry Wardell} 
\institute{Adam Pound  \at School of Mathematical Sciences and STAG Research Centre, University of Southampton, Southampton, United Kingdom, SO17 1BJ, \email{A.Pound@soton.ac.uk}
\and Barry Wardell \at School of Mathematics \& Statistics, University College Dublin, Belfield, Dublin 4, Ireland, \email{barry.wardell@ucd.ie}}

{\def\addcontentsline#1#2#3{}\maketitle}

\abstract{Much of the success of gravitational-wave astronomy rests on perturbation theory. Historically, perturbative analysis of gravitational-wave sources has largely focused on post-Newtonian theory. However, strong-field perturbation theory is essential in many cases such as the quasinormal ringdown following the merger of a binary system, tidally perturbed compact objects, and extreme-mass-ratio inspirals. In this review, motivated primarily by small-mass-ratio binaries but not limited to them, we provide an overview of essential methods in (i) black hole perturbation theory, (ii) orbital mechanics in Kerr spacetime, and (iii) gravitational self-force theory. Our treatment of black hole perturbation theory covers most common methods, including the Teukolsky and Regge-Wheeler-Zerilli equations, methods of metric reconstruction, and Lorenz-gauge formulations, presenting them in a new consistent and self-contained form. Our treatment of orbital mechanics covers quasi-Keplerian and action-angle descriptions of bound geodesics and accelerated orbits, osculating geodesics, near-identity averaging transformations, multiscale expansions, and orbital resonances. Our summary of self-force theory's foundations is brief, covering the main ideas and results of matched asymptotic expansions, local expansion methods, puncture schemes, and point particle descriptions. We conclude by combining the above methods in a multiscale expansion of the perturbative Einstein equations, leading to adiabatic and post-adiabatic evolution schemes. Our presentation is intended primarily as a reference for practitioners but includes a variety of new results. In particular, we present the first complete post-adiabatic waveform-generation framework for generic (nonresonant) orbits in Kerr.}


\setcounter{tocdepth}{3}
\begingroup
\let\cleardoublepage\clearpage
\tableofcontents
\endgroup
\markboth{}{}

\section{Introduction}

Black hole perturbation theory has a long and rich history, dating back at least as far as Regge and Wheeler's study of odd-parity perturbations of
Schwarzschild spacetime in the late 1950s \cite{Regge:1957td}. This was followed up in 1970 by Zerilli's study of even-parity perturbations
\cite{Zerilli:1970se,Zerilli:1971wd}. Soon afterwards, Vishveshwara \cite{Vishveshwara:1970zz} identified quasinormal modes in perturbations of
Schwarzschild spacetime, Press \cite{Press:1971wr} studied the associated quasinormal mode frequencies, and Chandrasekhar and Detweiler
\cite{Chandrasekhar:1975zza} numerically computed the frequencies. Teukolsky's success in deriving decoupled and separable equations for perturbations
of Kerr spacetime \cite{Teukolsky:1972my,Teukolsky:1973ha} paved the way for similar progress in the Kerr case.

The idea of a self-force has an even longer history, having been studied by Dirac in 1938 in his relativistic generalization of the Abraham–Lorentz self-force to the context of an electric charge undergoing acceleration
in flat spacetime \cite{Dirac:1938nz}. In the 1960s this was extended by DeWitt and Brehme to the curved spacetime case \cite{DeWitt:1960fc}. The
gravitational self-force acting on a point mass was studied in the late 1990s by Mino, Sasaki and Tanaka \cite{Mino:1996nk} and by Quinn and Wald
\cite{Quinn:1996am}, leading to the MiSaTaQuWa equation that is named after the authors of those first papers. Subsequent work has put gravitational
self-force theory on a very strong theoretical footing \cite{Gralla:2008fg,Pound:2009sm,Pound:2010wa} and has extended the formalism to second order
in perturbation theory \cite{Rosenthal:2006iy,Detweiler:2011tt,Pound:2012nt,Gralla:2012db}.

The last 20 years have seen increasingly intense focus on the study of gravitational self-force in perturbations of black hole spacetimes. This has
been motivated to a large extent by the European Space Agency's LISA mission, which is scheduled for launch in the 2030s \cite{LISA} and which will
observe gravitational waves in the millihertz frequency band. One of the key sources for LISA will be extreme-mass-ratio inspirals (EMRIs), binary
systems consisting of a compact solar-mass object orbiting a massive black hole. The presence of a small parameter (the mass ratio, which is expected
to be in the region of $10^{-6}$) makes black hole perturbation theory an ideal tool for the development of theoretical models of the gravitational
waveforms from EMRIs. Over the several year timescale that the LISA mission is expected to run, the smaller body in an EMRI will execute $\sim 10^4$--$10^5$ intricate orbits in the strong-field regime around the central black hole, acting as a precise probe and enabling high-precision measurements of the black hole's parameters, tests of its Kerr nature, and tests of general relativity. Radiation reaction will cause the orbit  to
significantly evolve and possibly plunge into the black hole in that time, meaning that self-force effects will be important to include in waveform models. Indeed, in order to extract the maximum
information from the observation of EMRIs by LISA it has been established that it will be necessary to incorporate information at second order in
perturbation theory by computing the second order gravitational self-force \cite{Hinderer:2008dm,Isoyama:2012bx,Burko:2013cca}. Aside from EMRIs,
gravitational self-force is also potentially highly accurate for intermediate-mass-ratio inspirals (IMRIs)~\cite{vandeMeent:2020xgc}, in which the mass ratio may be
as large as $\sim10^{-2}$. This makes black hole perturbation theory and self-force also relevant for the current generation of ground-based
gravitational wave detectors including LIGO \cite{LIGO}, Virgo \cite{Virgo} and Kagra \cite{Kagra}.

There are already numerous reviews of these topics in the literature. The classic text by Chandrasekhar~\cite{Chandrasekhar:1985kt} provides a comprehensive introduction to black hole physics, linear black hole perturbation theory, and geodesic motion in black hole spacetimes. Ref.~\cite{Sasaki:2003xr} reviews linear black hole perturbation theory with an emphasis on analytical post-Newtonian expansions of the perturbation equations. Ref.~\cite{Berti:2009kk} provides a thorough introduction to quasinormal modes of black holes. Ref.~\cite{Barack:2018yvs} offers a broad introduction to self-force calculations for non-experts, including a survey of concrete physical results through 2018. Refs.~\cite{Poisson:2011nh,Pound:2015tma} cover the foundations of self-force theory, and Ref.~\cite{Harte:2014wya} provides a complementary view of the foundations from a fully nonlinear perspective. Finally, Refs.~\cite{Barack:2009ux,Wardell:2015kea} provide detailed introductions to methods of computing the self-force.

Our aim is to complement rather than reiterate these existing reviews. We keep our description of self-force theory brief, only summarizing the key ideas and methods, and we forgo a survey of physical results. Instead, we focus on detailing the main perturbative methods required to model waveforms from small-mass-ratio binaries, leading ultimately to a multiscale expansion of the Einstein equations with a small-body source. At the same time, we keep much of the material sufficiently general to apply to other scenarios of interest.

Our aim is also not to provide detailed descriptions of the numerical approaches to solving the many equations detailed in this review. Open source codes implementing state-of-the-art numerical algorithms for solving the equations of black hole perturbation theory and self-force are available through the Black Hole Perturbation Toolkit \cite{BHPToolkit}. The Black Hole Perturbation Toolkit also acts as a repository for collating data (typically in the form of numerical tables or analytical post-Newtonian series expansions) produced by the research community.

Our discussion is divided into three distinct parts. Sections 2 and 3 briefly introduce relevant background material on perturbation theory in general relativity and the Kerr spacetime. Sections 4, 5, and 6 review three disjoint topics: black hole perturbation theory; geodesics and accelerated orbits in Kerr spacetime; and the foundations of the ``local problem'' in self-force theory. These three sections are written to be largely independent of one another, and they can be read in any order. Finally, in Section 7 we bring together all three topics in a description of black hole perturbation theory with a (skeletal) small-body source, focusing on the multiscale formulation. The multiscale expansion of the Einstein equation for generic (nonresonant) orbits in Kerr, and the post-adiabatic waveform-generation framework that comes along with it, appears here for the first time.

\section{Perturbation theory in General Relativity}

The overarching framework for our review is perturbation theory in general relativity. In self-force calculations, this is typically applied to the specific case of a small object in the spacetime of a Kerr black hole, and in much of the review we specialize to that scenario. But to allow for generality in some sections, we first consider the more generic case of smooth perturbations of an arbitrary vacuum spacetime. We assume the metric can be expanded in powers of a small parameter $\e$,
\beq
g^{\rm exact}_{\mu\nu} = g_{\mu\nu} + \e h^{(1)}_{\mu\nu} + \e^2 h^{(2)}_{\mu\nu} + O(\e^3),\label{g expansion}
\eeq
where $g_{\mu\nu}$ is a vacuum metric, and that the stress-energy can be similarly expanded as 
\beq
T_{\mu\nu} = \e T^{(1)}_{\mu\nu} + \e^2 T^{(2)}_{\mu\nu} + O(\e^3).\label{T expansion}
\eeq
For later convenience, we define the total metric perturbation $h_{\mu\nu}=\sum_{n>0}\e^n h^{(n)}_{\mu\nu}$. We also warn the reader that we will later treat $\e$ as a formal counting parameter that can be set equal to 1.

To expand the Einstein equations $G_{\mu\nu}[g+h]=8\pi T_{\mu\nu}$ in powers of $\e$, we first note that the Einstein tensor of a metric $g_{\mu\nu}+h_{\mu\nu}$ can be expanded in powers of the exact perturbation $h_{\mu\nu}$:
$G_{\mu\nu}[g+h] = G_{\mu\nu}[g]+G^{(1)}_{\mu\nu}[h] + G^{(2)}_{\mu\nu}[h,h] + O(|h|^3)$.
The quantities $G^{(n)}_{\mu\nu}$ are easily obtained from the exact Riemann tensor (see, e.g., Ch. 7.5 of Ref.~\cite{Wald:1984rg}). For a vacuum background, the first two terms are
\begin{align}
G^{(1)}_{\mu\nu}[h] &= \left(g_\mu{}^\alpha g_\nu{}^\beta-\tfrac{1}{2}g_{\mu\nu}g^{\alpha\beta}\right)R^{(1)}_{\alpha\beta},\label{Einstein1}\\
G^{(2)}_{\mu\nu}[h,h] &= \left(g_\mu{}^\alpha g_\nu{}^\beta-\tfrac{1}{2}g_{\mu\nu}g^{\alpha\beta}\right)R^{(2)}_{\alpha\beta} -\tfrac{1}{2}\left(h_{\mu\nu}g^{\alpha\beta}-g_{\mu\nu}h^{\alpha\beta}\right)R^{(1)}_{\alpha\beta},\label{Einstein2}
\end{align}
where the linear and quadratic terms in the Ricci tensor are
\begin{align}
R^{(1)}_{\alpha\beta}[h] &=  -\tfrac{1}{2}\left(\Box h_{\alpha\beta}+2R_\alpha{}^\mu{}_\beta{}^\nu h_{\mu\nu}-2\bar h_{\mu(\alpha}{}^{;\mu}{}_{\beta)}\right),\\
R^{(2)}_{\alpha\beta}[h,h] &= \tfrac{1}{4}h^{\mu\nu}{}_{;\alpha}h_{\mu\nu;\beta} + \tfrac{1}{2}h^{\mu}{}_{\beta}{}^{;\nu}\left(h_{\mu\alpha;\nu} - h_{\nu\alpha;\mu}\right) - \tfrac{1}{2}\bar h^{\mu\nu}{}_{;\nu}\left(2h_{\mu(\alpha;\beta)}-h_{\alpha\beta;\mu}\right)\nonumber\\
&\quad -\tfrac{1}{2} h^{\mu\nu}\left(2h_{\mu(\alpha;\beta)\nu} - h_{\alpha\beta;\mu\nu} - h_{\mu\nu;\alpha\beta}\right). 
\end{align}
Here we have defined the trace-reversed perturbation $\bar h_{\mu\nu}:=h_{\mu\nu}-\tfrac{1}{2}g_{\mu\nu}g^{\alpha\beta}h_{\alpha\beta}$ and the d'Alembertian $\Box:=g^{\mu\nu}\nabla_{\!\mu}\nabla_{\!\nu}$. A semicolon and $\nabla$ both denote the covariant derivative compatible with $g_{\mu\nu}$.

So, substituting the expansions~\eqref{g expansion} and \eqref{T expansion} into the Einstein equations and equating powers of $\e$, we obtain
\begin{align}
 G^{(1)}_{\mu\nu}[h^{(1)}] &= 8\pi T^{(1)}_{\mu\nu},\label{EFE1}\\
 G^{(1)}_{\mu\nu}[h^{(2)}] &= 8\pi T^{(2)}_{\mu\nu} - G^{(2)}_{\mu\nu}[h^{(1)},h^{(1)}].\label{EFE2}
\end{align}

This perturbative expansion comes with the freedom to perform gauge transformations
\begin{align}
 h^{(1)}_{\mu\nu} &\to  h^{(1)}_{\mu\nu} + \Lie_{\xi_{(1)}}g_{\mu\nu},\\
 h^{(2)}_{\mu\nu} &\to  h^{(2)}_{\mu\nu} + \Lie_{\xi_{(2)}}g_{\mu\nu} + \tfrac{1}{2}\Lie^2_{\xi_{(1)}}g_{\mu\nu} +\Lie_{\xi_{(1)}}h^{(1)}_{\mu\nu},
\end{align}
where $\Lie_{\xi}$ is a Lie derivative, and $\xi^\alpha_{(n)}$ are freely chosen vector fields. In  self-force theory, this freedom is commonly used to impose the Lorenz gauge condition,
\begin{equation}
  \label{eq:LorenzGauge}
  \nabla_\alpha\bar h^{\alpha\beta}=0,
\end{equation}
in which case the linearized Einstein tensor simplifies to
\begin{equation}
  \label{eq:LorenzField}
G^{(1)}_{\mu\nu}[h] = -\tfrac{1}{2}\left(\Box \bar h_{\mu\nu}+2R_\mu{}^\alpha{}_\nu{}^\beta\bar h_{\alpha\beta}\right).
\end{equation}

A perturbed metric will come hand in hand with a perturbed equation of motion for objects in the spacetime:
\beq\label{perturbed geodesic equation}
\frac{D^2z^\mu}{d\tau^2} = f_{(0)}^\mu + \e f_{(1)}^\mu + \e^2 f_{(2)}^\mu + O(\e^3).
\eeq
Here $z^\mu(\tau)$ is a perturbed worldline, $\tau$ is its proper time as measured in the background $g_{\mu\nu}$, $\frac{D^2 z^\mu}{d\tau^2} = \frac{dz^\nu}{d\tau}\nabla_{\nu}\frac{dz^\mu}{d\tau}:=a^\mu$ is its covariant acceleration with respect to $g_{\mu\nu}$, and $f_{(n)}^\mu$ is the $n$th-order covariant force (per unit mass) driving the acceleration. In our review, we will consider the general case including a zeroth-order force, but we will focus primarily on cases with $f^\mu_{(0)}=0$. The forces $f^\mu_{(n)}$ will arise from (parts of) the metric perturbations $h^{(n)}_{\mu\nu}$ as well as from coupling of $g_{\mu\nu}$ to the matter that creates those perturbations.

Here we have limited the treatment to first- and second-order perturbations, which are expected to be necessary and sufficient for modelling small-mass-ratio binaries. In some sections we will restrict the context to first, linearized order.

\section{Isolated, stationary black hole spacetimes}
\label{sec:bh}

In most of our review, we take the background spacetime to be that of an isolated, stationary black hole. In this section we provide an overview of the properties of these spacetimes.

\subsection{Metric}
\label{sec:metric}

The Schwarzschild spacetime is a static, spherically symmetric solution of the vacuum Einstein equations representing a non-rotating black hole with mass $M$. It has a line element given by
\begin{equation}
\label{eq:SchwMetric}
ds^2 = - f(r) dt^2 + f(r)^{-1} dr^2 + r^2 \big(d\theta^2 + \sin^2\theta d\phi^2\big),
\end{equation}
where $f(r) := 1 - \frac{2M}{r}$. The Schwarzschild spacetime may be generalized to allow the black hole to have a charge per unit mass, $Q$, resulting in the Reissner-Nordstr\"{o}m solution of the Einstein-Maxwell equations, with line element
\begin{equation}
\label{eq:RNMetric}
ds^2 = - \left( 1-\frac{2M}{r} + \frac{Q^2}{r^2}\right) dt^2 + \left( 1-\frac{2M}{r} + \frac{Q^2}{r^2}\right)^{-1} dr^2 + r^2 \big(d\theta^2 + \sin^2\theta d\phi^2\big).
\end{equation}
The spacetime of a spinning black hole is given by the Kerr metric with angular momentum per unit mass $a$. In Boyer-Lindquist coordinates, its line-element is
\begin{multline}
\label{eq:KerrMetric}
ds^2 = - \left[1-\frac{2Mr}{\Sigma}\right]\,dt^2 
    	- \frac{4aMr\sin^2\theta}{\Sigma}\,dt\,d\phi
    	+ \frac{\Sigma}{\Delta}\,dr^2
        \\ 
    	+ \Sigma\,d\theta^2
    	+ \left[\Delta+\frac{2Mr(r^2+a^2)} {\Sigma}\right] \sin^2\theta\, d\phi^2,
\end{multline}
where $\Sigma := r^2+a^2\cos^2\theta$ and $\Delta := r^2-2Mr+a^2 = (r-r_+)(r-r_-)$ with $r_\pm := M \pm \sqrt{M^2-a^2}$ the locations of the inner and outer horizons.
As was the case with Schwarzschild spacetime, the Kerr spacetime may be generalized to allow the black hole to have a charge per unit mass, $Q$, giving the Kerr-Newman solution of the Einstein-Maxwell equations. In Boyer-Lindquist coordinates, the Kerr-Newman metric is:
\begin{multline}
\label{eq:KerrNewmanMetric}
ds^2 = - \left[1-\frac{2Mr-Q^2}{\Sigma} \right]\,dt^2 
	- \frac{2(2Mr - Q^2) a \sin^2\theta}{\Sigma}\,dt\,d\phi
	+ \frac{\Sigma}{\Delta+Q^2}\,dr^2\\
	+ \Sigma  \,d\theta^2
	+  \left[\Delta+Q^2 + \frac{(2Mr-Q^2)(a^2+r^2)}{\Sigma}\right] \sin^2\theta\,d\phi^2.
\end{multline}

In astrophysical scenarios, a charged black hole will quickly be neutralized. For that reason, in later sections we will restrict our attention to the Kerr spacetime. We will also later use $Q$ to denote the Carter constant, associated with the Kerr metric's third, hidden symmetry discussed below. However, we include the charged black hole metrics  here for completeness.

\subsection{Null tetrads}
\label{sec:null-tetrads}

The black hole spacetimes above are all of Petrov type D and thus have two non-degenerate principal null directions. This gives us a natural way to define a complex null tetrad by having two of the tetrad legs aligned with the principal null directions. Choosing $l^\alpha := e^\alpha_{(1)}$ to align with the outward null direction and $n^\alpha := e^\alpha_{(2)}$ to align with the inward null direction, there is still residual freedom in the choice of scaling of each tetrad leg, and also in the relative orientation of the remaining two tetrad legs, $m^\alpha := e^\alpha_{(3)}$ and $\mb^\alpha := e^\alpha_{(4)}$. The two most common choices in Kerr spacetime are Carter's canonical tetrad \cite{Carter:1987hk}\footnote{Carter's original tetrad had interchanged $l^\mu \leftrightarrow n^\mu$ and $m^\mu \leftrightarrow \bar{m}^\mu$. Carter also worked in different coordinates $(\tilde{t} = t - a \phi, r, q = a \cos \theta, \tilde{\phi}= \phi/a)$  which more fully reflect the inherent symmetries of Kerr. We deviate from that here and keep with the convention of having $l^\alpha$ point outwards and working in the more common Boyer-Lindquist coordinates.},
\begin{alignat}{4}
l^\alpha &= \frac{1}{\sqrt{2\Delta \Sigma}}\Big[r^2+a^2,\Delta,0,a\Big], \quad &
n^\alpha &= \frac{1}{\sqrt{2\Delta \Sigma}}\Big[r^2+a^2,-\Delta,0,a\Big], \nonumber \\
m^\alpha &= \frac{1}{\sqrt{2\Sigma}}\Big[i a \sin \theta,0,1,\frac{i}{\sin \theta}\Big], \quad &
\bar{m}^\alpha &= \frac{1}{\sqrt{2\Sigma}}\Big[-i a \sin \theta,0,1,-\frac{i}{\sin \theta}\Big],
\end{alignat}
and the Kinnersley tetrad \cite{Kinnersley:1969zza}, which is related to Carter's canonical tetrad by a simple rescaling:
$l^\alpha   = \sqrt{\frac{\Delta}{2\Sigma}}\,l^\alpha_\mathrm{K}$,
$n^\alpha   = \sqrt{\frac{2\Sigma}{\Delta}}\,n^\alpha_\mathrm{K}$,
$m^\alpha   = \frac{\bar{\zeta}}{\sqrt{\Sigma}} m^\alpha_\mathrm{K}$ and $\mb^\alpha = \frac{\zeta}{\sqrt{\Sigma}} \mb^\alpha_\mathrm{K}$,
where
\begin{equation}
    \zeta := r-i a \cos \theta
\end{equation}
is an important quantity that we will encounter again later (note that $\Sigma = \zeta \bar{\zeta}$). The Carter tetrad transforms as $l \leftrightarrow - n$, $m \leftrightarrow \mb$ under $\{t,\phi\} \to \{-t, -\phi\}$.
Although the Kinnersley tetrad formed a crucial part of Teukolsky's separability result for perturbations of the Weyl tensor \cite{Teukolsky:1972my} it has
two unfortunate features that make it less than ideal for elucidating the symmetric structure of Kerr spacetime: (i) it violates the
$\{t, \phi\} \to \{-t,-\phi\}$ symmetry; and (ii) it destroys a symmetry in $\{r, \theta\}$. 
Carter's canonical tetrad does not suffer from either of these deficiencies and is slightly preferable from that point of view. Note, however, that all of the results that follow can be derived using either tetrad.

\subsection{Symmetries}

Much of the success in studying Kerr spacetime has arisen from the inherent symmetries it possesses. Two of these are associated with the existence of two Killing vectors, $\xi^\alpha$ and $\eta^\alpha$, which satisfy Killing's equation,\footnote{Note that the Killing vector $\delta^\alpha_\phi = \frac{1}{a}\eta^\alpha - a \delta^\alpha_t$ is often used in place of $\eta^\alpha$ when working in Boyer-Lindquist coordinates.}
\begin{equation}
    \xi_{(\alpha;\beta)} = 0 = \eta_{(\alpha;\beta)}.
\end{equation}
In Kerr spacetime these are related to the timelike and axial symmetries,
\begin{equation}
    \xi^\alpha = \delta_t^\alpha, \quad \eta^\alpha = \delta_{\tilde{\phi}}^\alpha = a (\delta_\phi^\alpha + a \delta_t^\alpha).
\end{equation}
The spacetime also admits a conformal Killing-Yano tensor,
\begin{equation}
    f_{\alpha \beta} = (\zeta + \bar{\zeta}) n_{[\alpha} l_{\beta]} - (\zeta - \bar{\zeta}) \mb_{[\alpha} m_{\beta]},
\end{equation}
which satisfies
\begin{equation}
   f_{\alpha(\beta;\gamma)} = g_{\beta\gamma} \xi_\alpha - g_{\alpha (\beta} \xi_{\gamma)}. 
\end{equation}
Here, we have introduced the Killing spinor coefficient, $\zeta$, which we previously encountered as a coordinate expression in Sec.~\ref{sec:null-tetrads}. Its appearance here can be considered more fundamental, and does not depend on any particular coordinate choice.
The divergence of this conformal Killing-Yano tensor is a Killing vector,
\begin{equation}
   \xi_\alpha = \tfrac13 f_{\alpha \beta}{}^{;\beta},
\end{equation}
and its Hodge dual,
\begin{equation}
    {}^\star f_{\alpha \beta} = \tfrac12 \epsilon_{\alpha\beta}{}^{\gamma\delta} f_{\gamma\delta} = i (\zeta - \bar{\zeta}) n_{[\alpha} l_{\beta]} - i (\zeta + \bar{\zeta}) \mb_{[\alpha} m_{\beta]},
\end{equation}
is a Killing-Yano tensor satisfying
\begin{equation}
 {}^\star f_{\alpha(\beta;\gamma)} = 0.
\end{equation}
The products of these Killing-Yano tensors generate two conformal Killing tensors,
\begin{align}
  K_{\alpha \beta} = f_{\alpha \gamma} f_\beta{}^\gamma &= \tfrac12 (\zeta + \bar{\zeta})^2 n_{(\alpha} l_{\beta)} - \tfrac12 (\zeta - \bar{\zeta})^2 \mb_{(\alpha} m_{\beta)}, \\
  \overset{\star}{K}_{\alpha \beta} = f_{\alpha \gamma} \,{}^\star f_\beta{}^\gamma &= \tfrac12 i (\zeta^2 - \bar{\zeta}^2) (n_{(\alpha} l_{\beta)} + \mb_{(\alpha} m_{\beta)}),
\end{align}
which satisfy
\begin{equation}
   K_{(\alpha\beta;\gamma)} = g_{(\alpha\beta} K_{\gamma)},
   \quad \overset{\star}{K}_{(\alpha\beta;\gamma)} = g_{(\alpha\beta} \overset{\star}{K}_{\gamma)},
\end{equation}
where $K_\alpha = \tfrac16 (2 K_{\beta\alpha}{}^{;\beta} + K_{\beta}{}^{\beta}{}_{;\alpha})$ and $\overset{\star}{K}_\alpha = \tfrac16 (2 \overset{\star}{K}_{\beta\alpha}{}^{;\beta} + \overset{\star}{K}_{\beta}{}^{\beta}{}_{;\alpha})$. They also generate a Killing tensor,
\begin{equation}
  \overset{\star\star}{K}_{\alpha \beta} = {}^\star f_{\alpha \gamma} \,{}^\star f_\beta{}^\gamma = - \tfrac12 (\zeta - \bar{\zeta})^2 n_{(\alpha} l_{\beta)} + \tfrac12 (\zeta + \bar{\zeta})^2 \mb_{(\alpha} m_{\beta)},
\end{equation}
satisfying
\begin{equation}
   \overset{\star\star}{K}_{(\alpha\beta;\gamma)} = 0.
\end{equation}
This Killing tensor gives a relationship between the two Killing vectors,
\begin{equation}
    \eta^\alpha = - \overset{\star\star}{K} {}^{\alpha \beta} \xi_\beta.
\end{equation}

\section{Black hole perturbation theory}\label{sec:black hole perturbation theory}

We now consider perturbations of  the isolated black hole spacetimes. We describe, in a unified notation, how to calculate metric perturbations in the most commonly used gauges: radiation gauges, Regge-Wheeler-Zerilli gauges, and the Lorenz gauge. Our focus is particularly on reconstruction methods, in which  most or all of the metric perturbation is reconstructed from decoupled scalar variables.

Since the left-hand sides of the perturbative Einstein equations~\eqref{EFE1} and \eqref{EFE2} are the same at every order, we specialize to the first-order case. We refer the reader to Refs.~\cite{Campanelli:1998jv,Brizuela:2009qd} for general discussions of second-order perturbation theory in Schwarzschild and Kerr spacetimes.

\subsection{The Teukolsky formalism and radiation gauge}
\label{sec:Teukolsky}

Teukolsky \cite{Teukolsky:1972my} showed that the equations
governing perturbations of rotating black hole spacetimes can be recast into a form where
they are given by decoupled equations. These equations further have the
remarkable property of being separable, reducing the problem to the solution of
a set of uncoupled ordinary differential equations. In the case of metric perturbations,
Teukolsky's results yield solutions for the spin-weight $\pm2$ components of the
perturbed Weyl tensor, but do not give a method for obtaining a corresponding metric perturbation. Subsequent
results (and their equivalents for electromagnetic perturbations) 
\cite{Chrzanowski:1975wv,Kegeles:1979an,Wald:1978vm,Whiting:2005hr,Pound:2013faa,Stewart:1978tm,Green:2019nam,Hollands:2020vjg}
derived a method for reconstructing a metric perturbation from a Hertz potential, which in turn can be
obtained from the spin-weight $\pm2$ components of the Weyl tensor.

\subsubsection{Geroch-Held-Penrose formalism}

Our exposition makes use of the formalism of Geroch, Held and Penrose (GHP) \cite{Geroch:1973am},
which is a simplified and more explicitly covariant version of the Newman-Penrose (NP) \cite{Newman:1961qr} 
formalism originally used by Teukolsky. Here we provide a
concise review of the key features of the formalism needed to understand metric perturbations of black hole
spacetimes; see Refs.~\cite{Price:Thesis,Aksteiner:2014zyp,Penrose:1987uia} for more thorough treatments.

The GHP formalism prioritises the concepts of spin- and boost-weights; within the
formalism, everything has a well-defined type $\{p,q\}$, which is related to its spin-weight
$s=(p-q)/2$ and its boost-weight $b=(p+q)/2$. Only objects of the same type can be added together,
providing a useful consistency check on any equations. Multiplication of two quantities yields
a resulting object with type given by the sum of the types of its constituents.

We first introduce a null tetrad $\{e_{(a)}^\alpha\} = \{l^\alpha, n^\alpha, m^\alpha, \mb^\alpha\}$ with normalisation
\begin{equation}
  l^\alpha n_\alpha = -1, \quad m^\alpha \mb_\alpha = 1,
\end{equation}
and with all other inner products vanishing. In terms of the tetrad vectors, the metric may be written as
\begin{equation}
  g_{\alpha\beta} = -2 l_{(\alpha} n_{\beta)} + 2 m_{(\alpha} \mb_{\beta)}.
\end{equation}
There are three discrete transformations that reflect the inherent symmetry in the GHP formalism,
corresponding to simultaneous interchange of the tetrad vectors:
\begin{enumerate}
    \item $'$: $l^\alpha \leftrightarrow n^\alpha$ and $m^\alpha \leftrightarrow \bar{m}^\alpha$, $\{p,q\} \rightarrow \{-p, -q\}$;
    \item $\bar{\phantom{m}}$: $m^\alpha \leftrightarrow \bar{m}^\alpha$, $\{p,q\} \rightarrow \{q, p\}$;
    \item $\ast$: $l^\alpha \rightarrow m^\alpha$, $n^\alpha \rightarrow -\bar{m}^\alpha$, $m^\alpha \rightarrow -l^\alpha$, $\mb^\alpha \rightarrow \bar{n}^\alpha$.
\end{enumerate}
We next introduce the {\it spin coefficients}, defined to be the 12 directional derivatives of the tetrad vectors.
Of these, the 8 with well-defined GHP type are
\begin{alignat}{2}
  \kappa = - l^\mu m^\nu \nabla_\mu l_\nu, & \quad \sigma = - m^\mu m^\nu \nabla_\mu l_\nu, &\quad
  \rho = -\mb^\mu m^\nu \nabla_\mu l_\nu, & \quad \tau = - n^\mu m^\nu \nabla_\mu l_\nu,
\end{alignat}
along with their primed variants, $\kappa'$, $\sigma'$, $\rho'$ and $\tau'$. These have GHP type given by
\begin{alignat*}{6}
  \kappa : \{3,1\}, & \quad \sigma : \{3,-1\}, & \quad \rho : \{1,1\},\quad \tau : \{1,-1\}.
\end{alignat*}
The remaining 4 spin coefficients are used to define the GHP derivative operators (that depend on the GHP
type of the object on which they are acting),
\begin{alignat}{3}
\th   &:= (l^\alpha \nabla_\alpha - p \epsilon - q \bar{\epsilon}), & \quad
\th'  &:= (n^\alpha \nabla_\alpha + p \epsilon' + q \bar{\epsilon}'), \nonumber \\
\edth  &:= (m^\alpha \nabla_\alpha - p \beta + q \bar{\beta}'),& \quad
\edth' &:= (\bar{m}^\alpha \nabla_\alpha + p \beta' - q\bar{\beta}),
\end{alignat}
where
\begin{align}
  \beta &= \frac{1}{2} (m^\mu \mb^\nu \nabla_\mu m_\nu-m^\mu n^\nu \nabla_\mu l_\nu), & \quad
  \epsilon &= \frac{1}{2} (l^\mu \mb^\nu \nabla_\mu m_\nu-l^\mu n^\nu \nabla_\mu l_\nu),
\end{align}
along with their primed variants, $\beta'$ and $\epsilon'$. These spin coefficients have no well-defined GHP type and never appear explicitly in
covariant equations. The action of a GHP derivative causes the type to change by an amount $\{p,q\}\to\{p+r,q+s\}$ where $\{r,s\}$ for each of the operators is given by
\begin{alignat*}{4}
  \th : \{1,1\}, & \quad \th' : \{-1,-1\}, & \quad \edth : \{1,-1\},\quad \edth' : \{-1,1\}.
\end{alignat*}
In this sense we interpret $\th$ and $\th'$ as boost raising and lowering operators, respectively,
while we interpret $\edth$ and $\edth'$ as spin raising and lowering operators, respectively. The adjoints of the GHP operators are given by
\begin{alignat}{3}
\th^\dag   &:= -(\th- \rho - \bar{\rho}), & \quad
\th'^\dag  &:= -(\th'- \rho' - \bar{\rho}'),\nonumber \\
\edth^\dag  &:= -(\edth- \tau -\bar{\tau}'),& \quad
\edth'^\dag &:= -(\edth'- \tau' - \bar{\tau}),
\end{alignat}
or, alternatively,
\begin{equation}
\mathscr{D}^\dag = - (\Psi_2 \bar{\Psi}_2)^{1/3} \mathscr{D} (\Psi_2 \bar{\Psi}_2)^{-1/3}, \quad \mathscr{D}\in\{\th, \th', \edth, \edth'\}.
\end{equation}
In vacuum spacetimes, the only non-zero components of the Riemann tensor are given by the tetrad components of the Weyl tensor, which can be represented by five complex Weyl scalars,
\begin{gather}
  \Psi_0 = C_{lmlm},\quad
  \Psi_1 = C_{lnlm}, \quad
  \Psi_2 = C_{lm\mb n},\quad
  \Psi_3 = C_{ln\mb n},\quad
  \Psi_4 = C_{n\mb n\mb},
\end{gather}
with types inherited from the tetrad vectors that appear in their definition,
\begin{gather*}
  \Psi_0 : \{4,0\}, \quad \Psi_1 : \{2,0\}, \quad \Psi_2 : \{0,0\},\quad
    \Psi_3 : \{-2,0\}, \quad \Psi_4 : \{-4,0\}.
\end{gather*}

Many of the results that follow will be specialised to type-D spacetimes with $l^\mu$ and $n^\mu$ aligned to
the two principal null directions, in which case the Goldberg-Sachs theorem implies that 4 of the of the spin coefficients vanish,
\begin{equation}
  \kappa = \kappa' = \sigma = \sigma' = 0,
\end{equation}
and also that most of the Weyl scalars vanish
\begin{equation}
  \Psi_0 = \Psi_1 = \Psi_3 = \Psi_4 = 0.
\end{equation}

The GHP equations give relations between the Weyl scalars and the directional derivatives of the spin coefficients.
For type-D spacetimes they are given by
\begin{alignat}{4}
  \th \rho &= \rho^2, & \quad \th \tau &= \rho(\tau-\bar{\tau}'), \nonumber \\
  \edth \tau &= \tau^2, & \quad  \edth \rho &= \tau(\rho-\bar{\rho}), \nonumber \\
  \th' \rho &= \rho\bar{\rho}'  - \tau \bar{\tau} &- \Psi_2 &+ \edth' \tau,
\end{alignat}
along with the Bianchi identity,
\begin{equation}
  \th \Psi_2 = 3 \rho \Psi_2, \quad \edth \Psi_2 = 3 \tau \Psi_2,
\end{equation}
and the conjugate, prime, and prime conjugate of these equations. Similarly the commutator of any pair of directional derivatives can be written in terms of a linear combination of
spin coefficients multiplying single directional derivatives. Again for type-D, they are given by
\begin{subequations}
\begin{align}
  [\th, \th'] &= (\bar{\tau} - \tau')\edth + (\tau - \bar{\tau}')\edth'
    - p (\Psi_2 - \tau \tau') - q (\bar{\Psi}_2 - \bar{\tau}\bar{\tau}'), \\
  [\th, \edth] &= \bar{\rho}\edth - \bar{\tau}'\th + q \bar{\rho}\bar{\tau}', \\
  [\edth, \edth'] &= (\bar{\rho}' - \rho')\th + (\rho-\bar{\rho})\th'
    + p (\Psi_2 + \rho \rho') - q (\bar{\rho}\bar{\rho}' + \bar{\Psi}_2),
\end{align}
\end{subequations}
along with the conjugate, prime, and prime conjugate of these. 

If we further restrict to spacetimes that admit a Killing tensor, $\overset{\star\star}{K}_{\alpha \beta}$, 
the associated symmetries lead to additional identities relating the spin coefficients,
\begin{equation}
  \frac{\rho}{\bar{\rho}} = \frac{\rho'}{\bar{\rho}'} =  -\frac{\tau}{\bar{\tau}'}=  -\frac{\tau'}{\bar{\tau}} = \frac{\bar{M}^{1/3}}{M^{1/3}}\frac{\Psi_2^{1/3}}{\bar{\Psi}_2^{1/3}} = \frac{\bar{\zeta}}{\zeta},
\end{equation}
for some complex function $M$ that is annihilated by $\th$.\footnote{In the case of Kerr spacetime, $M$ is the mass of the spacetime as one might anticipate.}
Here, we have used the fact that the Killing spinor coefficient is related to $\Psi_2$ by
\begin{equation}
  \zeta = -M^{1/3} \Psi_2^{-1/3}.
\end{equation}
These identities can be used along with the GHP equations to obtain a complementary set of identities,
\begin{subequations}
\begin{gather}
  \th \tau' = 2\rho \tau' = \edth' \rho, \\
  \th' \rho = \rho \rho' + \tau' (\tau - \bar{\tau}') -\frac12 \Psi_2 - \frac{\bar{\zeta}}{2\zeta} \bar{\Psi}_2, \\
  \edth' \tau = \tau \tau' + \rho (\rho' - \bar{\rho}') + \frac12 \Psi_2 - \frac{\bar{\zeta}}{2\zeta} \bar{\Psi}_2,
\end{gather}
\end{subequations}
along with the conjugate, prime, and prime conjugate of these equations. A consequence
of these additional relations is that there is an operator
\begin{equation}
  \mathcal{\pounds}_\xi = -\zeta \big( - \rho' \th + \rho \th' + \tau' \edth - \tau \edth') - \frac{p}{2} \zeta \Psi_2 - \frac{q}{2} \bar{\zeta} \bar{\Psi}_2,
\end{equation}
associated with the Killing vector
\begin{equation}
  \xi^\alpha = -\zeta(-\rho' l^\alpha + \rho n^\alpha + \tau' m^\alpha - \tau \mb^\alpha).
\end{equation}
There is a second operator
\begin{align}
  \mathcal{\pounds}_\eta &= -\tfrac{\zeta}{4}
    \big[(\zeta-\bar{\zeta})^2(\rho' \th - \rho \th') - (\zeta+\bar{\zeta})^2(\tau' \edth - \tau \edth') \big] 
      + p\, {}_\eta h_1 + q\, {}_\eta \bar{h}_1
\end{align}
where
\begin{align}
  {}_\eta h_1 &= \tfrac18 \zeta(\zeta^2+\bar{\zeta}^2)\Psi_2 - \tfrac14 \zeta\bar{\zeta}^2 \bar{\Psi}_2
     + \tfrac12 \rho \rho' \zeta^2 (\bar{\zeta}-\zeta) + \tfrac12 \tau \tau' \zeta^2 (\bar{\zeta}+\zeta).
\end{align}
This is associated with the second Killing vector
\begin{equation}
  \eta^\alpha = -\tfrac{\zeta}{4}\big[(\zeta-\bar{\zeta})^2 (\rho' l^\alpha - \rho n^\alpha) -(\zeta+\bar{\zeta})^2 (\tau' m^\alpha - \tau \mb^\alpha) \big].
\end{equation}
Both $\mathcal{\pounds}_\xi$ and $\mathcal{\pounds}_\eta$ commute with all of the GHP operators and annihilate all of the spin coefficients and $\Psi_2$.

\subsubsection{Teukolsky equations}
\label{sec:TeukolskyEquations}

We now consider perturbations of vacuum type-D spacetimes. Teukolsky \cite{Teukolsky:1972my} showed
that the perturbations to $\Psi_0$ and $\Psi_4$ (which we will denote by $\psi_0$ and $\psi_4$) are gauge invariant 
and satisfy decoupled and fully separable second order equations. These perturbations may be written in GHP form as
\begin{align}
  \psi_0 = C^{(1)}_{lmlm}[h] = \mathcal{T}_0 h, \quad
  \psi_4 = C^{(1)}_{n\mb n \mb}[h] = \mathcal{T}_4 h,\label{eq:Weyl-scalars-definition}
\end{align}
where the operators $\mathcal{T}_I$ are given by
\begin{subequations}
\begin{align}
  \mathcal{T}_0 h &= - \frac12 \Big[(\edth - \bar{\tau}')(\edth-\bar{\tau}') h_{ll} + (\th-\bar{\rho})(\th-\bar{\rho}) h_{mm} \nonumber \\
     & \qquad \qquad- \big((\th-\bar{\rho})(\edth-2\bar{\tau}')+ (\edth-\bar{\tau}')(\th-2\bar{\rho})\big) h_{(lm)} \Big],\label{eq:T0} \\
  \mathcal{T}_4 h &= - \frac12 \Big[(\edth' - \bar{\tau})(\edth'-\bar{\tau}) h_{nn} + (\th'-\bar{\rho}')(\th'-\bar{\rho}') h_{\mb\mb} \nonumber \\
     & \qquad \qquad - \big((\th'-\bar{\rho}')(\edth'-2\bar{\tau})+ (\edth'-\bar{\tau})(\th'-2\bar{\rho}')\big) h_{(n\mb)} \Big].\label{eq:T4}
\end{align}
\end{subequations}
We will later also need the adjoints of these, which are given by
\begin{subequations}
\begin{align}
  (\mathcal{T}_0^\dag \Psi)_{\alpha \beta} &= - \frac12 \Big[l_\alpha l_\beta (\edth - \tau)(\edth - \tau) + m_\alpha m_\beta (\th-\rho)(\th-\rho) \nonumber \\
     & - l_{(\alpha} m_{\beta)} \big((\edth-\tau+\bar{\tau}')(\th-\rho)+ (\th-\rho+\bar{\rho})(\edth-\tau)\big) \Big]\Psi,\label{eq:T0dag} \\
  (\mathcal{T}_4^\dag \Psi)_{\alpha \beta} &= - \frac12 \Big[n_\alpha n_\beta (\edth' - \tau')(\edth' - \tau') + \mb_\alpha \mb_\beta (\th'-\rho')(\th'- \rho')\nonumber \\
     & - n_{(\alpha} \mb_{\beta)} \big((\edth'-\tau'+\bar{\tau})(\th'-\rho') + (\th'-\rho'+\bar{\rho}')(\edth'-\tau')\big)  \Big]\Psi.\label{eq:T4dag}
\end{align}
\end{subequations}

The scalars $\psi_0$ and $\psi_4$ satisfy the Teukolsky equations,\footnote{Note that $\mathcal{O}' \psi_4= \zeta^{-4} \mathcal{O} \zeta^4 \psi_4$ and $\mathcal{O} \psi_0= \zeta^{-4} \mathcal{O}' \zeta^4 \psi_0$.}
\begin{alignat}{4}
  \mathcal{O} \psi_0 &= 8 \pi \mathcal{S}_0 T,& \qquad \mathcal{O}' \psi_4 &= 8 \pi \mathcal{S}_4 T,
\end{alignat}
where
\begin{align}
  \mathcal{O} &:= \big(\th - 2 \,s\, \rho - \bar{\rho}\big)\big(\th'-\rho'\big) - \big(\edth - 2\, s \,\tau - \bar{\tau}'\big)\big(\edth' - \tau'\big) + \tfrac12 \big[\big(6 s-2\big)-4s^2\big] \Psi_2
\end{align}
is the spin-weight $s$ Teukolsky operator.\footnote{Some authors (e.g. \cite{Wald:1978vm,Green:2019nam}) define $\mathcal{O}$ to be the operator with $s=+2$. Then, the operator for the negative $s$ fields is its adjoint $\mathcal{O}^\dag$.} The decoupling operators
\begin{subequations}
\label{eq:S}
\begin{align}
\mathcal{S}_0 T &= 
  \tfrac12 (\edth-\bar{\tau}'-4\tau) 
    \big[(\th-2\bar{\rho}) T_{(lm)} -(\edth-\bar{\tau}') T_{ll}  \big] \nonumber \\ & \quad
  + \tfrac12(\th-4\rho-\bar{\rho})
    \big[(\edth-2\bar{\tau}') T_{(lm)} - (\th-\bar{\rho}) T_{mm}  \big], \\
\mathcal{S}_4 T &= 
  \tfrac12(\edth'-\bar{\tau}-4\tau') 
    \big[(\th'-2\bar{\rho}') T_{(n\mb)} -(\edth'-\bar{\tau}) T_{nn}  \big] \nonumber \\ & \quad
  + \tfrac12(\th'-4\rho'-\bar{\rho}')
    \big[(\edth'-2\bar{\tau}) T_{(n\mb)} - (\th'-\bar{\rho}') T_{\mb\mb}  \big],
\end{align}
\end{subequations}
allow the sources for the Teukolsky equations to be constructed from the stress-energy tensor. We will
later also need the adjoints of these, which are given by
\begin{subequations}
\label{eq:Sdag}
\begin{align}
\label{eq:Sdag0}
(\mathcal{S}_0^\dag \Psi)_{\alpha \beta} &= - \tfrac12 l_\alpha l_\beta (\edth-\tau)(\edth+3\tau) \Psi - \tfrac12 m_{\alpha}m_\beta (\th-\rho)(\th + 3\rho) \Psi \nonumber \\
  & \quad + \tfrac12 l_{(\alpha}m_{\beta)} \big[(\th-\rho+\bar{\rho})(\edth+3\tau) +(\edth-\tau+\bar{\tau}')(\th+3\rho)] \Psi, \\
\label{eq:Sdag4}
(\mathcal{S}_4^\dag \Psi)_{\alpha \beta} &= - \tfrac12 n_\alpha n_\beta (\edth'-\tau')(\edth'+3\tau') \Psi  - \tfrac12 \mb_{\alpha}\mb_\beta (\th'-\rho')(\th' + 3\rho') \Psi
  \nonumber \\
  & \quad + \tfrac12 n_{(\alpha}\bar{m}_{\beta)} \big[(\th'-\rho'+\bar{\rho}')(\edth'+3\tau') +(\edth'-\tau'+\bar{\tau})(\th'+3\rho')] \Psi.
\end{align}
\end{subequations}

Introducing the index-free linearised Einstein operator
$(\mathcal{E}h)_{\alpha\beta} := G_{\alpha \beta}^{(1)}[h]$, we see that Teukolsky's result for
decoupling the equations are a consequence of the operator identities
\label{eq:SEOT}
\begin{align}
\mathcal{S}_0 \mathcal{E} = \mathcal{O} \mathcal{T}_0, \quad \mathcal{S}_4 \mathcal{E} = \mathcal{O}' \mathcal{T}_4.
\end{align}

In vacuum Kerr-NUT spacetimes, the Teukolsky operator may be written in manifestly separable form by rewriting it in terms of the commuting operators \cite{Aksteiner:2014zyp}
\begin{align}
  \mathscr{R} &:= \zeta \bar{\zeta} (\th - \rho - \bar{\rho})(\th' - 2 b \rho') + \frac{2b-1}{2} (\zeta + \bar{\zeta}) \mathcal{\pounds}_\xi,
\end{align}
and
\begin{align}
  \mathscr{S} &:= \zeta \bar{\zeta} (\edth - \tau - \bar{\tau}')(\edth' - 2 s \tau') + \frac{2s-1}{2} (\zeta - \bar{\zeta}) \mathcal{\pounds}_\xi.
\end{align}
Then, Teukolsky operator is given by
\begin{equation}
   \zeta \bar{\zeta} \mathcal{O} = \mathscr{R} - \mathscr{S}.
\end{equation}
The symmetry operators satisfy the commutation relations $\big[\mathscr{R}, \mathscr{S}\big] = 0$ when acting on a type $\{p,0\}$ object. We will see later that when written as a coordinate expression in Boyer-Lindquist coordinates in Kerr spacetime the operators $\mathscr{R}$ and $\mathscr{S}$ reduce to the radial Teukolsky and spin-weighted spheroidal operators (with a common eigenvalue).

\subsubsection{Teukolsky-Starobinsky identities}

In regions where they satisfy the homogeneous Teukolsky equations, the scalars $\psi_0$ and $\psi_4$ are not independent. Instead, they are related by the Teukolsky-Starobinsky identities, which are given in GHP form by
\begin{subequations}
\begin{align}
  \th^4 \zeta^4 \psi_4 &= \edth'^4 \zeta^4 \psi_0 - 3 M \mathcal{\pounds}_\xi \bar{\psi}_0, \\
  \th'^4 \zeta^4 \psi_0 &= \edth^4 \zeta^4 \psi_4 + 3 M \mathcal{\pounds}_\xi \bar{\psi}_4,
\end{align}
\end{subequations}
where we recall that $M = - \zeta^3 \psi_2$.
From these, we can also derive eighth-order Teukolsky-Starobinsky identities that do not mix the scalars
\begin{subequations}
\begin{align}
  \th^4 \bar{\zeta}^4 \th'^4 \zeta^4 \psi_0 &= \edth^4 \bar{\zeta}^4 \edth'^4 \zeta^4 \psi_0 - 9 M^2 \mathcal{\pounds}_\xi^2 \psi_0, \\
  \th'^4 \bar{\zeta}^4 \th^4 \zeta^4 \psi_4 &= \edth'^4 \bar{\zeta}^4 \edth^4 \zeta^4 \psi_4 - 9 M^2 \mathcal{\pounds}_\xi^2 \psi_4.
\end{align}
\end{subequations}

\subsubsection{Reconstruction of a metric perturbation in radiation gauge}
\label{sec:metric-reconstruction}

Solutions of the Teukolsky equations can be related
back to solutions for the metric perturbation $h_{\alpha \beta}$ by use of a Hertz potential
 \cite{Wald:1978vm, Chrzanowski:1975wv, Kegeles:1979an, Lousto:2002em,
Whiting:2005hr}. In fact, there are two different Hertz potentials: $\hpsi^{\rm IRG}$, which produces a metric perturbation in the ingoing radiation gauge; and $\hpsi^{\rm ORG}$, which produces a metric perturbation in the outgoing radiation gauge.

In the ingoing radiation gauge (IRG), the metric perturbation may be reconstructed by applying a second-order differential operator to a scalar Hertz potential $\hpsi^{\rm IRG}$ of type $\{-4,0\}$  (i.e. the same type as $\psi_4$). 
In terms of this Hertz potential, the IRG metric perturbation is given explicitly by
\begin{equation}
  h_{\alpha\beta}^{\rm IRG} =
2 \Re \big[(\mathcal{S}_0^\dag \hpsi^{\rm IRG})_{\alpha \beta}\big].\label{eq:reconstruction}
\end{equation}
where $\mathcal{S}_0^\dag$ is the operator given in Eq.~\eqref{eq:Sdag0}.
The IRG Hertz
potential satisfies $\mathcal{O} \hpsi^{\rm IRG} = \eta^\text{IRG} $, where $\eta^\text{IRG}$
satisfies $2\Re(\mathcal{T}_0^\dag \eta^\text{IRG})_{\alpha\beta} = 8\pi T_{\alpha\beta}$. In other words,
$\hpsi^{\rm IRG}$ is a solution of the equation satisfied by
$\zeta^{4} \ppsi_4$ (equivalently, the adjoint of the equation satisfied by $\ppsi_0$), but with
a different source.

The IRG Hertz potential manifestly satisfies the gauge conditions $l^\alpha h_{\alpha\beta}=0$ and $h=0$ and it necessarily requires that $(\mathcal{E} h^{\rm IRG})_{ll} = 0 = T_{ll}$. Computing the perturbed Weyl scalars from it, we find
\begin{subequations}
\begin{align}
  \ppsi_0 &= \frac14 \th^4 \overline{\hpsi^\text{IRG}} \label{eq:psi0-IRG}\\
  \ppsi_4 &= \frac14 \edth'^4\overline{\hpsi^\text{IRG}} - \frac34 M \zeta^{-4} \mathcal{\pounds}_\xi \hpsi^\text{IRG},\nonumber \\
  & \quad + \frac14 \Big[ \zeta^{-2}\mathcal{O}\zeta^{2} + 2 \zeta^{-1}\mathcal{\pounds}_\xi - 2(\tau' \tau-\rho' \rho - \bpsi_2)\Big] \eta^\text{IRG}.\label{eq:psi4-IRG}
\end{align}
\end{subequations}
The IRG Hertz potential may therefore be obtained either by solving the sourced (adjoint) Teukolsky equation or by solving either of the fourth-order equations sourced by the perturbed Weyl scalars. The equations involving $\ppsi_0$ and $\ppsi_4$ are often referred to as the ``radial'' and ``angular'' inversion equations, respectively. Acting on the perturbed Weyl scalars with the Teukolsky operator and commuting operators, we find
\begin{subequations}
\begin{align}
  \mathcal{O}\ppsi_0 &= \frac14 (\th-\rho-\bar{\rho})^4 \overline{\eta^\text{IRG}} \label{eq:Opsi0-IRG}\\
  \mathcal{O}'\ppsi_4 &= \frac14 (\edth'-\tau'-\bar{\tau})^4 \overline{\eta^\text{IRG}} - \frac34 M \zeta^{-4} \mathcal{\pounds}_\xi \eta^\text{IRG},\nonumber \\
  & \quad + \frac14 \mathcal{O}'\Big[ \zeta^{-2}\mathcal{O}\zeta^{2} + 2 \zeta^{-1}\mathcal{\pounds}_\xi - 2(\tau' \tau-\rho' \rho - \bpsi_2)\Big] \eta^\text{IRG}.\label{eq:Opsi4-IRG}
\end{align}
\end{subequations}
Thus, in regions where the Hertz potential satisfies the homogenous equation $\mathcal{O}\hpsi^\text{IRG}=0$ the second line of Eq.~\eqref{eq:psi4-IRG} vanishes and the perturbed Weyl scalars satisfy the homogeneous Teukolsky equations.

A similar procedure also works in the outgoing radiation gauge (ORG), the prime of the IRG. There,
we have\footnote{Some authors \cite{vandeMeent:2015lxa} define a slightly different ORG Hertz
potential related to the one here by $\hat{\hpsi}^{\rm ORG} = \zeta^{-4} \hpsi^{\rm ORG}$ and
$(\hat{\mathcal{S}}_4^\dag)_{\alpha\beta} = (\mathcal{S}_4^\dag \zeta^{4})_{\alpha\beta}$. Both
conventions yield the same metric perturbation, $(\hat{\mathcal{S}}_4^\dag\hat{\hpsi}^{\rm ORG})_{\alpha\beta}
 = (\mathcal{S}_4^\dag \hpsi^{\rm ORG})_{\alpha\beta})$.}
\begin{equation}
  h_{\alpha\beta}^{\rm ORG} = 2\Re \big[ (\mathcal{S}_4^\dag \hpsi^{\rm ORG})_{\alpha \beta}\big],
\end{equation}
where the ORG Hertz potential, $\hpsi^{\rm ORG}$, is of type $\{4,0\}$ (i.e. the same as $\ppsi_0$). The ORG Hertz potential satisfies
$\mathcal{O}' \hpsi^\text{ORG} = \eta^{\text{ORG}}$, where $\eta^\text{ORG}$ satisfies $2\Re(\mathcal{T}_4^\dag
\eta^\text{ORG})_{\alpha\beta} = 8\pi T_{\alpha\beta}$. In other words, $\hpsi^{\rm ORG}$ is a solution of the equation satisfied by
$\zeta^4\ppsi_0$ (equivalently, the adjoint of the equation satisfied by $\ppsi_4$), but with a different source.
 
The ORG Hertz potential manifestly satisfies the gauge conditions $n^\alpha h_{\alpha\beta}=0$ and $h=0$, and it necessarily requires that $(\mathcal{E} h^{\rm IRG})_{nn} = 0 = T_{nn}$. Computing the perturbed Weyl scalars from it, we find
\begin{subequations}
\begin{align}
  \ppsi_0 &= \frac14 \edth^4 \overline{\hpsi^{\rm ORG}} + \frac34 M \zeta^{-4} \mathcal{\pounds}_\xi\hpsi^{\rm ORG}  \nonumber \\
  & \quad + \frac14 \Big[ \zeta^{-2}\mathcal{O}'\zeta^{2} - 2 \zeta^{-1}\mathcal{\pounds}_\xi - 2(\tau' \tau-\rho' \rho - \bpsi_2)\Big] \eta^\text{ORG}\label{eq:psi0-ORG}\\
  \ppsi_4 &= \frac14 \th'^4 \overline{\hpsi^{\rm ORG}}\label{eq:psi4-ORG},
\end{align}
\end{subequations}
The ORG Hertz potential may therefore be obtained either by solving the sourced (adjoint) Teukolsky equation or by solving either of the fourth-order equations sourced by the perturbed Weyl scalars. The equations involving $\ppsi_0$ and $\ppsi_4$ are often referred to as the ``angular'' and ``radial'' inversion equations, respectively. Acting on the perturbed Weyl scalars with the Teukolsky operator and commuting operators, we find
\begin{subequations}
\begin{align}
  \mathcal{O}\ppsi_0 &= \frac14 (\edth-\tau-\bar{\tau}')^4 \overline{\eta^{\rm ORG}} + \frac34 M \zeta^{-4} \mathcal{\pounds}_\xi\eta^{\rm ORG}  \nonumber \\
  & \quad + \frac14 \mathcal{O} \Big[ \zeta^{-2}\mathcal{O}'\zeta^{2} - 2 \zeta^{-1}\mathcal{\pounds}_\xi - 2(\tau' \tau-\rho' \rho - \bpsi_2)\Big] \eta^\text{ORG} \label{eq:Opsi0-ORG}\\
  \mathcal{O}'\ppsi_4 &= \frac14 (\th'-\rho'-\bar{\rho}')^4 \overline{\eta^{\rm ORG}}.\label{eq:Opsi4-ORG}
\end{align}
\end{subequations}
Thus, in regions where the Hertz potential satisfies the homogenous equation $\mathcal{O}'\hpsi^\text{ORG}=0$ the second line of Eq.~\eqref{eq:psi0-ORG} vanishes and the perturbed Weyl scalars satisfy the homogeneous Teukolsky equations.

As with the Weyl scalars, the IRG and ORG Hertz potentials are not independent in the homogeneous case. By demanding that they produce the same $\ppsi_0$ and $\ppsi_4$ we obtain Teukolsky-Starobinsky identities relating them,
\begin{align}
  \th^4 \hpsi^{\rm IRG} &= \edth'^4 \hpsi^{\rm ORG} + 3 \overline{M \zeta^{-4} \mathcal{\pounds}_\xi \hpsi^{\rm ORG}} \\
  \th'^4 \hpsi^{\rm ORG} &= \edth^4 \hpsi^{\rm IRG} - 3 \overline{M \zeta^{-4} \mathcal{\pounds}_\xi \hpsi^{\rm IRG}}.
\end{align}

The fact that the Hertz potentials yield solutions of the homogeneous linearised Einstein equations was succinctly summarised by Wald \cite{Wald:1978vm} using the method of adjoints: since the operators satisfy the identity $\mathcal{S} \mathcal{E} = \mathcal{O} \mathcal{T}$, by taking the adjoint and using the fact that $\mathcal{E}$ is self-adjoint we find that $\mathcal{E} \mathcal{S}^\dag = \mathcal{T}^\dag \mathcal{O}^\dag$ so we have a homogeneous solution of the linearised Einstein equations provided the Hertz potential satisfies the (adjoint) homogeneous Teukolsky equation.

Finally, we note that in addition to imposing conditions on the stress-energy, the standard radiation gauge reconstruction procedure fails to reproduce certain ``completion'' portions of the metric
perturbation associated with small shifts in the central mass and angular momentum, and gauge. A more generally valid metric perturbation may be obtained by supplementing the reconstructed piece described here with completion pieces and with a ``corrector'' tensor $x_{\alpha \beta}$ that is designed to eliminated any restrictions on the stress-energy,
\begin{equation}
 h_{\alpha \beta} = 2 \Re (\mathcal{S}^\dag \Psi)_{\alpha \beta} + x_{\alpha \beta} + \dot{g}_{\alpha \beta} + (\mathcal{\pounds}_X g)_{\alpha \beta}.
\end{equation}
The interested reader may refer to
\cite{Wald:1978vm,Kegeles:1979an,Chrzanowski:1975wv} for the original derivations of the reconstruction procedure, to \cite{Barack:2017oir} for an analysis of the sourced equation satisfied by the Hertz potential, to \cite{Merlin:2016boc,vandeMeent:2017fqk} for details of metric completion, and to \cite{Green:2019nam,Hollands:2020vjg} for a thorough explanation of the corrector tensor approach.

\subsubsection{Gravitational waves}

In order to determine gravitational wave strain, we require the metric perturbation far from the source.
If we consider the metric perturbation reconstructed in radiation gauge, then to leading order in a large-distance expansion from the source the components $h_{mm}$ and $h_{\mb\mb}$ dominate, with both falling of as $(\text{distance})^{-1}$. It is common to write these in terms of the two gravitational wave polarizations,
\begin{equation}
\label{eq:Strain}
  h_{mm} = h_+ + i h_\times, \quad
  h_{\mb\mb} = h_+ - i h_\times.
\end{equation}
Furthermore, at large radius the operator $\mathcal{T}_4$ of Eq.~\eqref{eq:T4} reduces to a second derivative along the $l^\mu$ null direction, leading to a simple relationship between $\psi_4$ and the second time derivative of the strain,
\begin{equation}
  \label{eq:Psi4-Strain}
  \psi_4 \sim -\frac12 \ddot{h}_{\mb\mb}.
\end{equation}
This gives us a straightforward way to determine the strain by computing two time integrals of $\psi_4$.
Further mathematical details on the relationship between $\psi_4$ and outgoing gravitational radiation are given in Refs.~\cite{Newman:1961qr,Newman:1962cia,Szekeres:1965ux}, on the equivalent relationship between $\psi_0$ and incoming radiation are given in Ref.~\cite{Walker:1979zk}, and on numerical implementation considerations in Refs.~\cite{Reisswig:2010di,Lehner:2007ip}.

\subsubsection{GHP formalism in Kerr spacetime}

We now give explicit expressions for the various quantities defined in the previous sections specialized to Kerr spacetime. The
spin coefficients are tetrad dependent. When working with the Carter tetrad, the non-zero spin coefficients have a particularly symmetric form given by
\begin{alignat}{4}
 \rho &= -\rho' = -\frac{1}{\zeta} \sqrt{\frac{\Delta}{2 \Sigma}}, \quad &
 \tau &= \tau' = -\frac{i a \sin \theta}{\zeta\sqrt{2 \Sigma}}, \nonumber\\
 \beta  &= \beta' = -\frac{i}{\zeta} \frac{a+i r \cos \theta}{2\sin\theta\sqrt{2\Sigma}},\quad &
 \epsilon &= - \epsilon' = \frac{M r - a^2 - i a (r-M) \cos \theta}{2\zeta\sqrt{2 \Sigma \Delta}},
\end{alignat}
while for the Kinnersley tetrad they are given by
\begin{gather}
 \rho_K = -\frac{1}{\zeta}, \quad  \rho_K' = \frac{\Delta}{2 \zeta^2 \bar{\zeta}}, \quad 
 \tau_K = -\frac{i a \sin \theta}{\sqrt{2}\zeta \bar{\zeta}}, \quad  \tau_K' = -\frac{i a \sin \theta}{\sqrt{2}\zeta^2},\nonumber \\
 \beta_K  = \frac{\cot \theta}{2\sqrt{2} \bar{\zeta}}, \quad  \beta_K' = \frac{\cot \theta}{2\sqrt{2} \zeta} -\frac{i a \sin \theta}{\sqrt{2}\zeta^2} ,\quad 
 \epsilon_K = 0, \quad  \epsilon_K' = \frac{\Delta}{2 \zeta^2 \bar{\zeta}} - \frac{r-M}{2 \zeta \bar{\zeta}}.
\end{gather}
The commuting GHP operators have the same form in both tetrads,
\begin{equation}
  \mathcal{\pounds}_{\xi} = \partial_t,\quad
  \mathcal{\pounds}_{\eta} = a^2 \partial_t + a \partial_\phi = \partial_{\tilde{\phi}}.
\end{equation}

\subsubsection{Mode decomposed equations in Kerr spacetime}
\label{sec:KerrModes}

In additional to decoupling the equations, Teukolsky further showed that the Teukolsky equations are fully separable using a mode ansatz. The specific form of the ansatz depends on the choice of null tetrad. Teukolsky worked with the Kinnersley tetrad \cite{Kinnersley:1969zza}, in which case the Teukolsky equations are separable using the ansatz\footnote{A similar separability result also holds when working with the Carter tetrad by replacing the left hand sides as follows:
\begin{equation*}
  \ppsi_0 \to \frac{2 \zeta^2}{\Delta} \ppsi_0, \quad \zeta^4 \ppsi_4 \to \frac{\zeta^2 \Delta}{2} \ppsi_4, \quad
  (\mathcal{S}_0 T)  \to \frac{2 \zeta^2}{\Delta} (\mathcal{S}_0 T), \quad \zeta^4 (\mathcal{S}_4 T) \to \frac{\zeta^2 \Delta}{2} (\mathcal{S}_4 T).
\end{equation*}
The factors of $\Delta$ here are not required for separability, but are included so that the radial functions are consistent with Teukolsky's original radial functions.}
\begin{align}
  \ppsi_0 &= \int_{-\infty}^\infty \sum_{\ell=2}^\infty \sum_{m=-\ell}^\ell \, {}_2 \ppsi_{\ell \emm \omega}(r) \, {}_2 S_{\ell \emm}(\theta, \phi; a \omega) e^{-i \omega t}\, d\omega , \label{eq:psi0-FD}\\
  \zeta^4 \ppsi_4 &= \int_{-\infty}^\infty \sum_{\ell=2}^\infty \sum_{m=-\ell}^\ell  \,{}_{-2} \ppsi_{\ell \emm \omega}(r) \, {}_{-2} S_{\ell \emm}(\theta, \phi; a \omega) e^{-i \omega t} d\omega, \label{eq:psi4-FD}
\end{align}
with the functions ${}_{s} \ppsi_{\ell \emm \omega}(r)$ and ${}_{s} S_{\ell \emm}(\theta, \phi; a \omega)$ satisfying the spin-weighted spheroidal harmonic and Teukolsky radial equations, respectively,
\begin{equation}
  \label{eq:SWSH}
  \bigg[\dfrac{d}{d\chi} \bigg((1-\chi^2)\dfrac{d}{d\chi} \bigg)
  +a^2 \omega^2 \chi^2 -\frac{(m+s \chi)^2}{1-\chi^2} - 2 a s \omega \chi +s + A\bigg] {}_{s} S_{\ell \emm} = 0,
\end{equation}
and
\begin{equation}
  \label{eq:TeukolskyR}
  \bigg[\Delta^{-s} \dfrac{d}{dr}\bigg( \Delta^{s+1}\dfrac{d }{dr}\bigg)
  +\frac{K^2 - 2 i s (r-M)K}{\Delta} + 4 i s \omega r - {}_s \lambda_{\ell \emm} \bigg]{}_{s} \ppsi_{\ell \emm \omega} = {}_{s} T_{\ell \emm \omega},
\end{equation}
where $\chi := \cos \theta$, $A:= {}_s \lambda_{\ell \emm}+2 a m \omega -a^2 \omega^2$ and $K:=(r^2+a^2)\omega-a m$, and where the eigenvalue ${}_s \lambda_{\ell \emm}$ depends on the value of $a\omega$.

As with the standard spherical harmonics, the dependence of the spin-weighted spheroidal harmonics on the azimuthal coordinate is exclusively through an overall complex exponential factor,
\begin{equation}
  {}_{s} S_{\ell \emm}(\theta, \phi; a \omega) = {}_{s} S_{\ell \emm}(\theta, 0; a \omega) e^{i \emm \phi}.
\end{equation}
With this definition, the spin-weighted spheroidal harmonics are orthonormal,
\begin{equation}
  \int {}_{s} S_{\ell \emm}(\theta, \phi; a \omega) {}_{s} \bar{S}_{\ell' \emm'}(\theta, \phi; a \omega) d\Omega = \delta_{\ell \ell'} \delta_{\emm \emm'},
\end{equation}
where $d \Omega = \sin \theta d\theta d\phi$ is the volume element on the two-sphere. They also satisfy two symmetry identities,
\begin{subequations}
\begin{align}
  {}_{s} S_{\ell \emm}(\theta, \phi; a \omega) &= (-1)^{\ell+\emm}\, {}_{-s} S_{\ell \emm}(\pi - \theta, \phi; a \omega), \\
  {}_{s} S_{\ell \emm}(\theta, \phi; a \omega) &= (-1)^{\ell+s}\, {}_{s} \bar{S}_{\ell -\emm}(\pi - \theta, \phi; -a \omega)
\end{align}
\end{subequations}
which can be combined to obtain the useful identity
\begin{equation}
  {}_{s} S_{\ell \emm}(\theta, \phi; a \omega) = (-1)^{\emm+s} {}_{-s} \bar{S}_{\ell -\emm}(\theta, \phi; -a \omega),
\end{equation}
which relates an $(s, \ell, \emm, a \omega)$ harmonic to the conjugate of an $(-s, \ell, -\emm, -a \omega)$ harmonic. Similarly, the eigenvalue satisfies the identities
\begin{subequations}
\begin{align}
  {}_{s} \lambda_{\ell \emm}(a \omega) &= {}_{-s} \lambda_{\ell \emm}(a \omega) - 2 s, \\
  {}_{s} \lambda_{\ell \emm}(a \omega) &= {}_{s} \lambda_{\ell -\emm}(-a \omega)
\end{align}
\end{subequations}
which can be combined to obtain
\begin{equation}
  {}_{s} \lambda_{\ell \emm}(a \omega) = {}_{-s} \lambda_{\ell -\emm}(-a \omega) - 2 s.
\end{equation}

The sources for the radial Teukolsky equation are defined by
\begin{subequations}
\begin{align}
  8 \pi (\mathcal{S}_0 T) &= -\frac{1}{2\Sigma}\int_{-\infty}^\infty \sum_{\ell=2}^\infty \sum_{\emm=-\ell}^\ell {}_2 T_{\ell \emm \omega}(r) {}_2 S_{\ell \emm}(\theta, \varphi; a \omega) e^{-i \omega t} d\omega, \label{eq:T0-FD}\\
  8 \pi \zeta^4 (\mathcal{S}_4 T) &= -\frac{1}{2 \Sigma} \int_{-\infty}^\infty \sum_{\ell=2}^\infty \sum_{\emm=-\ell}^\ell {}_{-2} T_{\ell \emm \omega}(r) {}_{-2} S_{\ell \emm}(\theta, \varphi; a \omega) e^{-i \omega t} d\omega.\label{eq:T4-FD}
\end{align}
\end{subequations}
Finally, when acting on a single mode of the mode-decomposed Weyl scalars the symmetry operators yield
\begin{alignat}{4}
  \mathscr{S} \psi_0 &= - \frac12 {}_{|2|} \lambda_{\ell \emm} \psi_0,& \quad
  \mathscr{R} \psi_0 &= - \frac12 {}_{|2|} \lambda_{\ell \emm} \psi_0 + \zeta \bar{\zeta} \mathcal{S}_0 T, \nonumber \\
  \mathscr{S}' \psi_4 &= - \frac12 {}_{|-2|} \lambda_{\ell \emm} \psi_4,& \quad
  \mathscr{R}' \psi_4 &= - \frac12 {}_{|2|} \lambda_{\ell \emm} \psi_4 + \zeta \bar{\zeta} \mathcal{S}_ T,
\end{alignat}
where ${}_{|s|} \lambda_{\ell \emm \omega} := {}_{s} \lambda_{\ell \emm \omega} + |s| + s$ is independent of the sign of $s$.\footnote{This is distinct from Chandrasekhar's eigenvalue which is given in Eq.~\eqref{eq:lambdaCh}.}

\begin{figure}[t]
\begin{center}
	\includegraphics[width=11cm]{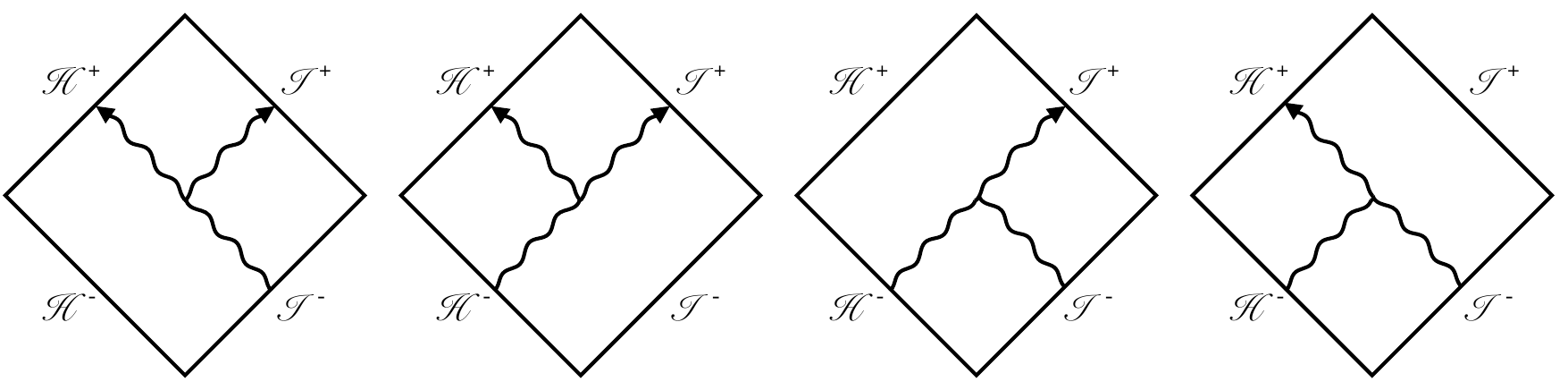}
	\caption{
	Left to right: boundary conditions satisfied by the ``in'', ``up'', ``out'' and ``down'' solutions.}
  \label{fig:BCs}
\end{center}
\end{figure}

Solutions to the radial Teukolsky equation may be written in terms of a pair of homogeneous mode
basis functions chosen according to their asymptotic behavior at the four null boundaries to the
spacetime. For radiative ($\omega \ne 0$) modes, the four common choices are denoted
\begin{itemize}
  \item ``in'': representing waves coming in from $\mathcal{I}^-$ then partially falling into the horizon and partially scattering back out to $\mathcal{I}^+$; these modes are purely ingoing into the horizon;
  \item ``up'': representing waves coming up from $\mathcal{H}^-$ then partially travelling out to $\mathcal{I}^+$ and partially scattering back into $\mathcal{H}^+$; these modes are purely outgoing at infinity;
  \item ``out'': representing waves coming from $\mathcal{I}^-$ and $\mathcal{H}^-$ then travelling out to $\mathcal{I}^+$; these modes are purely outgoing from the horizon;
  \item ``down'': representing waves coming from $\mathcal{I}^-$ and $\mathcal{H}^-$ then travelling down to $\mathcal{H}^+$; these modes are purely incoming at infinity;
\end{itemize}
These have asymptotic behaviour given by
\begin{subequations}
\begin{alignat}{5}
\label{eq:bcRin}
{}_s R^{\text{in}}_{\ell \emm \omega}(r) &\sim
\Big\{&
\begin{array}{c}
 0 \\
 {}_s R^{\text{in,ref}}_{\ell \emm \omega} r^{-1-2s} e^{+i\omega r_*}
\end{array}&
\begin{array}{c}
 + \\ +
\end{array}
\begin{array}{c}
 {}_s R^{\text{in,trans}}_{\ell \emm \omega} \Delta^{-s} e^{-i k r_*}
 \\
 {}_s R^{\text{in,inc}}_{\ell \emm \omega} r^{-1} e^{-i\omega r_*}
\end{array}
&\qquad
\begin{array}{l}
  r \to r_+\\
  r \to \infty
\end{array}
\\
\label{eq:bcRup}
{}_s R^{\text{up}}_{\ell \emm \omega}(r) &\sim
\Big\{&
\begin{array}{c}
 {}_s R^{\text{up,inc}}_{\ell \emm \omega} e^{+i k r_*}
 \\
 {}_s R^{\text{up,trans}}_{\ell \emm \omega} r^{-1-2s} e^{+i\omega r_*}
\end{array}&
\begin{array}{c}
 + \\ +
\end{array}
\begin{array}{c}
 {}_s R^{\text{up,ref}}_{\ell \emm \omega} \Delta^{-s} e^{-i k r_*}
 \\
 0
\end{array}
&\qquad
\begin{array}{l}
  r \to r_+\\
  r \to \infty
\end{array} 
\\
\label{eq:bcRout}
{}_s R^{\text{out}}_{\ell \emm \omega}(r) &\sim
\Big\{&
\begin{array}{c}
 {}_s R^{\text{out,trans}}_{\ell \emm \omega} e^{+i k r_*}
 \\
 {}_s R^{\text{out,inc}}_{\ell \emm \omega} r^{-1-2s} e^{+i \omega r_*}
\end{array}&
\begin{array}{c}
 + \\ +
\end{array}
\begin{array}{c}
 0
 \\
 {}_s R^{\text{out,ref}}_{\ell \emm \omega} r^{-1} e^{-i \omega r_*}
\end{array}
&\qquad
\begin{array}{l}
  r \to r_+\\
  r \to \infty
\end{array} 
\\
\label{eq:bcRdown}
{}_s R^{\text{down}}_{\ell \emm \omega}(r) &\sim
\Big\{&
\begin{array}{c}
 {}_s R^{\text{down,ref}}_{\ell \emm \omega} e^{+i k r_*}
 \\
 0
\end{array}&
\begin{array}{c}
 + \\ +
\end{array}
\begin{array}{c}
 {}_s R^{\text{down,inc}}_{\ell \emm \omega} \Delta^{-s} e^{-i k r_*}
 \\
 {}_s R^{\text{down,trans}}_{\ell \emm \omega} r^{-1} e^{-i \omega r_*}
\end{array}
&
\begin{array}{l}
  r \to r_+\\
  r \to \infty
\end{array} 
\end{alignat}
\end{subequations}
where $k := \omega - m \Omega_+$ with $\Omega_+ := \frac{a}{2 M r_+}$ the angular velocity of the horizon, and where $r_* := r + \frac{1}{2\kappa_+} \ln \frac{r-r_+}{2M} + \frac{1}{2\kappa_-} \ln \frac{r-r_-}{2M}$ with $\kappa_\pm := \frac{r_\pm - r_\mp}{2(r_\pm^2+a^2)}$ the surface gravity on the outer/inner horizon. This behaviour is depicted graphically in Fig.~\ref{fig:BCs}.

Inhomogeneous solutions of the radial Teukolsky equation can then be written in terms of a linear combination of the basis functions,
\begin{align} \label{eq:TeukolskyInhomogeneousModes}
  {}_s \ppsi_{\ell \emm \omega}(r) &= {}_s
    C^{\text{in}}_{\ell \emm \omega}(r) {}_s R^{\text{in}}_{\ell \emm \omega}(r)+{}_s
    C^{\text{up}}_{\ell \emm \omega}(r) {}_s R^{\text{up}}_{\ell \emm \omega}(r),
\end{align}
where the weighting coefficients are determined by variation of parameters,
\begin{subequations}
  \label{eq:weighting-coefficients}
  \begin{align}
  {}_s C_{\ell \emm \omega}^{\text{in}}(r) &= \int^{\infty}_r \frac{{}_s R^{\text{up}}_{\ell \emm \omega}(r')}{W(r')\Delta} {}_s T_{\ell \emm \omega}(r') dr', \\
  {}_s C_{\ell \emm \omega}^{\text{up}}(r) &= \int_{r_+}^r \frac{{}_s R^{\text{in}}_{\ell \emm \omega}(r')}{W(r')\Delta} {}_s T_{\ell \emm \omega}(r') dr',
\end{align}
\end{subequations}
with $W(r) = {}_s R^{\text{in}}_{\ell \emm \omega}(r) \partial_r [{}_s R^{\text{up}}_{\ell \emm \omega}(r)] - {}_s R^{\text{up}}_{\ell \emm \omega}(r) \partial_r [{}_s R^{\text{in}}_{\ell \emm \omega}(r)]$ the Wronskian [in practice, it is convenient to use the fact that $\Delta^{s+1} W(r) = \text{const}$].

If one computes the ``in'' and ``up'' mode functions with normalisation such that transmission coefficients are unity, ${}_s R^{\text{in,trans}}_{\ell \emm \omega} = 1 = {}_s R^{\text{up,trans}}_{\ell \emm \omega}$, then the gravitational wave strain can be determined directly from $\psi_4$ using Eq.~\eqref{eq:Psi4-Strain} to give
\begin{equation}
  \lim_{r\to \infty} r(h_+ - i h_\times) = 2 \int_{-\infty}^\infty \sum_{\ell=2}^\infty \sum_{m=-\ell}^\ell  \, \frac{{}_{-2} C_{\ell \emm \omega}^{\text{up}}}{\omega^2} \, {}_{-2} S_{\ell \emm}(\theta, \phi; a \omega) e^{-i \omega (t-r_*)} d\omega, \label{eq:Teukolsky-waveform}
\end{equation}
where the weighting coefficient ${}_{-2} C_{\ell \emm \omega}^{\text{up}}$ is to be evaluated in the limit $r\to \infty$. Similarly, the time averaged flux of energy carried by gravitational waves\footnote{Strictly speaking, the horizon fluxes given here have been derived from the rates of change of the black hole parameters due to shear of the horizon generators~\cite{Teukolsky:1974yv}. It is generally assumed that these are equivalent to the gravitational wave fluxes, although this has not, to our knowledge, been shown explicitly.} passing through infinity and the horizon can be computed from the ``in'' and ``up'' normalization coefficients \cite{Hughes:1999bq},
\begin{align}
  \fluxEH &= \lim_{r \to r_+} \sum_{\ell\emm\omega} \frac{2\pi \alpha_{\ell \emm \omega}}{\omega^2} |{}_{-2} C^{\text{in}}_{\ell \emm \omega}|^2, \label{EdotH v1} \\
  \fluxEI &= \lim_{r \to \infty} \sum_{\ell\emm\omega} \frac{2\pi}{\omega^2}|\, {}_{-2} C^{\text{up}}_{\ell \emm \omega}|^2,
\end{align}
where $\alpha_{\ell \emm \omega} := \frac{256(2M r_+)^5 k(k^2+4\varepsilon^2)(k^2+16\varepsilon^2)\omega^3}{|\mathcal{C}_{\ell \emm \omega}|^2}$ with 
$\varepsilon := \sqrt{M^2-a^2}/(4 M r_+)$. Similarly, the flux of angular momentum is given by
\begin{align}
  \fluxLzH &= \lim_{r \to r_+} \sum_{\ell\emm\omega} \frac{2\pi \emm \alpha_{\ell \emm \omega}}{\omega^3} |{}_{-2} C^{\text{in}}_{\ell \emm \omega}|^2, \label{LdotH v1} \\
  \fluxLzI &= \lim_{r \to \infty} \sum_{\ell\emm\omega} \frac{2\pi \emm}{\omega^3}|\, {}_{-2} C^{\text{up}}_{\ell \emm \omega}|^2.\label{LdotI v1}
\end{align}
Similar expressions can be obtained in terms of the modes ${}_2 \ppsi_{\ell\emm\omega}$ of $\ppsi_0$ by using the Teukolsky-Starobinsky identities to relate ${}_{-2} C^{\text{up}}_{\ell \emm \omega}$ to ${}_{2} C^{\text{up}}_{\ell \emm \omega}$. The necessary details of how these asymptotic amplitudes are related can be found in Refs.~\cite{Ori:2002uv,vandeMeent:2015lxa}.

When decomposed into modes, each of the Teukolsky-Starobinsky identities separate to yield identities relating the positive spin-weight spheroidal and radial functions to the negative spin-weight ones,
\begin{subequations}
  \label{eq:TS-mode}
\begin{align}
 \mathcal{D}_0^4 ({}_{-2} \psi_{\ell \emm \omega}) &= \tfrac14 \mathcal{C}_{\ell \emm \omega} \, {}_{2} \psi_{\ell \emm \omega}, \\
 \Delta^{2} (\mathcal{D}^\dag_0)^4 (\Delta^{2}\, {}_{2} \psi_{\ell \emm \omega}) &= 4 \bar{\mathcal{C}}_{\ell \emm \omega} \,{}_{-2} \psi_{\ell \emm \omega}, \\
 \mathcal{L}_{-1}\mathcal{L}_{0}\mathcal{L}_{1}\mathcal{L}_{2} ({}_{2} S_{\ell \emm \omega}) &= D \, {}_{-2} S_{\ell \emm \omega}, \\
 \mathcal{L}^\dag_{-1}\mathcal{L}^\dag_{0}\mathcal{L}^\dag_{1}\mathcal{L}^\dag_{2} ( {}_{-2} S_{\ell \emm \omega}) &= D \, {}_{2} S_{\ell \emm \omega},
\end{align}
\end{subequations}
where
\begin{subequations}
  \label{eq:DandL}
\begin{alignat}{4}
  \mathcal{D}_n &:= \partial_r - \frac{i K}{\Delta} + 2 n \frac{r-M}{\Delta}, \quad&
  \mathcal{D}^\dag_n &:= \partial_r + \frac{i K}{\Delta} + 2 n \frac{r-M}{\Delta}, \\
  \mathcal{L}_n &:= \partial_\theta + Q + n \cot \theta, \quad&
  \mathcal{L}^\dag_n &:= \partial_\theta - Q + n \cot \theta,
\end{alignat}
\end{subequations}
(with $K$ defined above and $Q:=-a \omega \sin \theta + m
\csc \theta$) are essentially mode versions of the GHP differential operators.
The constants of proportionality are given by
\begin{subequations}
\begin{align}
  \label{eq:TS_C}
  \mathcal{C}_{\ell \emm \omega} &= D + (-1)^{\ell+m} 12 i M \omega,\\
  \label{eq:TS_D}
  D^2 &= ({}_{s}\lambda^{\rm Ch}_{\ell \emm})^2 ({}_{s}\lambda^{\rm Ch}_{\ell \emm}-2)^2 + 8 a \omega(m-a
\omega)({}_{s}\lambda^{\rm Ch}_{\ell \emm}-2)(5 {}_s \lambda^{\rm Ch}_{\ell \emm}-4) \nonumber \\
& \qquad +48
(a\omega)^2\big[2({}_{s}\lambda^{\rm Ch}_{\ell \emm}-2)+3(m-a\omega)^2\big],
\end{align}
\end{subequations}
where
\begin{equation}
\label{eq:lambdaCh}
  {}_{s} \lambda^{\rm Ch}_{\ell \emm \omega} := {}_{s} \lambda_{\ell \emm \omega} + s^2 + s
\end{equation}
is the eigenvalue used by Chandrasekhar \cite{Chandrasekhar:1985kt}.
This particular choice of $\mathcal{C}_{\ell \emm \omega}$ ensures that the $s=+2$ and $s=-2$ modes represent the same physical perturbation.\footnote{An 
alternative proportionality constant can be derived such that the $s=+2$ and $s=-2$ modes have the same
transmission coefficient; see \cite{Ori:2002uv} for details.}

Finally, when written in terms of modes the homogeneous radiation gauge angular inversion equations can be algebraically inverted to give the modes of the Hertz potentials in terms of the modes of the Weyl scalar,
\begin{align}
\label{eq:angular-inversion-ORG}
\hpsi^{\text{ORG}}_{\ell \emm \omega} & = 
 16\frac{(-1)^m D\;\! {}_2\bar{\ppsi}_{-\omega\ell-m}+12 i M \omega\;\! {}_2\ppsi_{\ell \emm \omega} }{|\mathcal{C}_{\ell \emm \omega}|^2}, \\
\label{eq:angular-inversion-IRG}
 \hpsi^{\text{IRG}}_{\ell \emm \omega} & = 
 16\frac{(-1)^m D\;\! {}_{-2}\bar{\ppsi}_{-\omega\ell-m}-12 i M \omega\;\! {}_{-2}\ppsi_{\ell \emm \omega} }{|\mathcal{C}_{\ell \emm \omega}|^2}.
\end{align}
where the separability ansatz for the Hertz potentials differs by a factor of $\zeta^{-4}$ from that of the Weyl scalars,
\begin{align}
  \zeta^{-4} \hpsi^{\text{ORG}} &= \int_{-\infty}^\infty \sum_{\ell=2}^\infty \sum_{m=-\ell}^\ell \, \hpsi^{\text{ORG}}_{\ell \emm \omega}(r) \, {}_2 S_{\ell \emm}(\theta, \phi; a \omega) e^{-i \omega t}\, d\omega , \\
  \hpsi^{\text{IRG}} &= \int_{-\infty}^\infty \sum_{\ell=2}^\infty \sum_{m=-\ell}^\ell  \, \hpsi^{\text{IRG}}_{\ell \emm \omega}(r) \, {}_{-2} S_{\ell \emm}(\theta, \phi; a \omega) e^{-i \omega t} d\omega.
\end{align}
Alternatively, one can use the radial inversion equations to relate the asymptotic amplitudes of $\hpsi^{\text{IRG}}_{\ell \emm \omega}$ to the asymptotic amplitudes of ${}_{2}\ppsi_{\ell \emm \omega}$ and to relate the asymptotic amplitudes of $\hpsi^{\text{ORG}}_{\ell \emm \omega}$ to the asymptotic amplitudes of ${}_{-2}\ppsi_{\ell \emm \omega}$. Further details are given in \cite{Ori:2002uv} for the IRG case and in \cite{vandeMeent:2015lxa} for the ORG case.

Note that in order to transform back to the time-domain solution, as a final step we must perform an inverse Fourier transform. This poses a challenge in gravitational self-force calculations, where non-smoothness of the solutions in the vicinity of the worldline lead to the Gibbs phenomenon of non-convergence of the inverse Fourier transform. Resolutions to this problem typically rely on avoiding directly transforming the inhomogeneous solution by using the methods of extended homogeneous or extended particular solutions. For further details, see \cite{Hopper:2010uv,Hopper:2012ty}.

\subsubsection{Sasaki-Nakamura transformation}

In numerical implementations, the Teukolsky equation can be problematic to work with due to the presence of a long-ranged potential. One approach to this problem is to transform to an alternative master function that satisfies an equation with a more short-ranged potential. The Sasaki-Nakamura transformation is designed to do exactly this. It introduces a new function of the form
\begin{equation}
  X \sim
    \begin{cases}
      \frac{\bar{\zeta}^{2}}{\zeta^{2}} (r^2+a^2)^{1/2} r^{2} \th' \th' \frac{1}{r^2} \zeta^{4} \ppsi_0 \\
      (r^2+a^2)^{1/2} r^{2} \th \th \frac{1}{r^2} \zeta^{4} \ppsi_4
    \end{cases},
\end{equation}
where the factors of $\zeta$ ensure that these are purely radial operators.\footnote{This expression is appropriate when working with the Kinnersley tetrad; for the Carter tetrad both definitions for $X$ need to be scaled by a common factor of $\frac{\bar{\zeta}}{\zeta}$ to obtain a radial operator.} There is considerable freedom to rescale these expressions by inserting appropriate functions of $r$, for more details see Ref.~\cite{Hughes:2000pf} in which case the $X$ given here corresponds to $\sqrt{r^2+a^2} r^2 J_- J_- \frac{1}{r^2} R$ in the $s=-2$ case and to $\frac14 \sqrt{r^2+a^2} r^2 J_+ J_+ \frac{\Delta^2}{r^2} R$ in the $s=+2$ case.

\subsection{Metric perturbations of Schwarzschild spacetime}
\label{sec:schw-perturbations}

On a Schwarzschild background spacetime, separability is readily achieved without having to rely on the Teukolsky
formalism. Writing the metric perturbation in terms of its null tetrad components, they have GHP type
\begin{alignat*}{5}
    s&=0: &\qquad h_{ln} &: \{0,0\}, &\quad h_{m\mb} &: \{0,0\}, &\quad h_{ll} &: \{2,2\}, &\quad h_{nn} &: \{-2,-2\} \nonumber \\
    s&=\pm1: &\qquad h_{lm} &: \{2,0\}, &\quad h_{l\mb} &: \{0,2\}, &\quad h_{nm} &: \{0,-2\}, &\quad h_{n\mb} &: \{-2,0\} \nonumber \\
    s&=\pm2: &\qquad h_{mm} &: \{2,-2\}, &\quad h_{\mb\mb} &: \{-2,2\}. & & & &
\end{alignat*}
Here we have gathered the components into scalar ($s=0$), vector ($s=\pm1$) and tensor ($s=\pm2$) sectors.

In some instances, it is convenient to work with the trace-reversed metric perturbation, $\bar{h}_{\alpha\beta} = h_{\alpha\beta} - \frac12 h \,g_{\alpha \beta}$. In terms of null tetrad components, the trace is given by $h = -2 (h_{ln} - h_{m\mb})$ so a trace reversal simply corresponds to the interchange $h_{ln} \leftrightarrow h_{m\mb}$: $\bar{h}_{ln} = h_{m\mb}$ and $\bar{h}_{m\mb} = h_{ln}$, with all other components unchanged.

The tetrad components may be decomposed into a basis of spin-weighted spherical harmonics
\begin{align}
  h_{ab} &= \sum_{\ell=|s|}^\infty \sum_{m=-\ell}^\ell \, h_{ab}^{\ell \emm}(t, r) \, {}_s Y_{\ell \emm}(\theta, \phi)
\end{align}
where $s=0$ for $h_{ln}$, $h_{m\mb}$, $h_{ll}$ and $h_{nn}$; $s=+1$ for $h_{lm}$ and $h_{nm}$; $s=-1$ for $h_{l\mb}$ and $h_{n\mb}$; $s=+2$ for $h_{mm}$; and $s=-2$ for $h_{\mb\mb}$. Here, we have introduced the
spin-weighted spherical harmonics ${}_s Y_{\ell \emm}(\theta, \phi) = {}_s S_{\ell \emm \omega}(\theta, \phi; 0)$ with the associated eigenvalue ${}_s \lambda_\ell := {}_s \lambda_{\lm} (a\omega = 0) = \ell(\ell+1) - s(s+1)$.

In the Schwarzschild case the GHP derivative operators split into operators that (up to an overall factor of $\frac{1}{r}$) act only on the two-sphere,
\begin{equation}
  \label{eq:edth-Schw}
  \edth = \tfrac{1}{\sqrt{2}r}( \partial_\theta + i \csc \theta \partial_\phi - s \cot\theta),\quad
  \edth' = \tfrac{1}{\sqrt{2}r}( \partial_\theta - i \csc \theta \partial_\phi + s \cot\theta).
\end{equation}
and operators that act only in the $t-r$ subspace\footnote{These expressions are obtained when working with the Carter tetrad. The equivalent operators for the Kinnersley tetrad are
\begin{equation}
  \label{eq:th-Schw-Kinnersley}
  \th = f^{-1} \partial_t + \partial_r, \quad
  \th' = \tfrac12 (\partial_t -f \partial_r-2 b M/r^2).
\end{equation}
}
\begin{equation}
  \label{eq:th-Schw-Carter}
  \th = \frac{1}{\sqrt{2 f}}\bigg[\partial_t + f \partial_r - \frac{b M}{r^2}\bigg], \quad
  \th' = \frac{1}{\sqrt{2 f}}\bigg[\partial_t - f \partial_r - \frac{b M}{r^2}\bigg].
\end{equation}
The two-sphere operators act as spin-raising and lowering operators to relate spin-weighted spherical harmonics of different spin-weight
\begin{subequations}
  \label{eq:spin-raising-lowering}
\begin{align}
  \sqrt{2} r \,\edth \big[{}_s Y_{\ell \emm}(\theta, \phi)\big] &= -\big[\ell(\ell+1)-s(s+1)\big]^{1/2} \, {}_{s+1} Y_{\ell \emm} (\theta, \phi),\\
  \sqrt{2} r \, \edth' \big[{}_s Y_{\ell \emm}(\theta, \phi)\big] &= \big[\ell(\ell+1)-s(s-1)\big]^{1/2} \, {}_{s-1} Y_{\ell \emm} (\theta, \phi).
  \end{align}
\end{subequations}
In particular, this provides a relationship between the spin-weighted spherical harmonics and the scalar spherical harmonics.

It is convenient to split the six vector and tensor sector components of the metric perturbation into real (even parity) and imaginary (odd parity) parts, representing whether they are even or odd under the transformation $(\theta,\phi)\rightarrow (\pi-\theta, \phi+\pi)$:
\begin{alignat*}{6}
    h_{lm}^{\ell \emm} &= h_{l,\even}^{\ell \emm} + i\, h_{l,\odd}^{\ell \emm},\qquad &
    h_{l\mb}^{\ell \emm} &= -h_{l,\even}^{\ell \emm} + i\, h_{l,\odd}^{\ell \emm},\\
    h_{nm}^{\ell \emm} &= h_{n,\even}^{\ell \emm} + i\, h_{n,\odd}^{\ell \emm},\qquad &
    h_{n\mb}^{\ell \emm} &= -h_{n,\even}^{\ell \emm} + i\, h_{n,\odd}^{\ell \emm}, \\
    h_{mm}^{\ell \emm} &= h_{2,\even}^{\ell \emm} + i\, h_{2,\odd}^{\ell \emm},\qquad &
    h_{\mb\mb}^{\ell \emm} &= h_{2,\even}^{\ell \emm} - i\, h_{2,\odd}^{\ell \emm}.
\end{alignat*}
The four scalar sector components are necessarily even parity, so we therefore have seven fields in the even-parity sector and three in the odd-parity sector. The even and odd parity sectors decouple, meaning that they can be solved for independently. In instances where there is symmetry under reflection about the equatorial plane this decoupling is more explicit in that the even parity sector only contributes for $\ell+\emm$ even and the odd parity sector only contributes for $\ell+\emm$ odd.

Finally, we can also optionally further decompose into the frequency domain,
\begin{align}
  h_{ab}^{\ell \emm}(t,r) &= \int_{-\infty}^\infty h_{ab}^{\ell \emm \omega}(r) e^{-i\omega t}\, d\omega
\end{align}
in order to obtain functions of $r$ only. This has the advantage of reducing the problem of computing the metric perturbation to that of
solving systems of $7+3$ coupled ordinary differential equations, one for each $(\ell,\emm,\omega)$.

\subsubsection{Alternative tensor bases}

There is some freedom in the specific choice of basis into which tensors are decomposed. In particular, the relative scaling of the $l^\mu$ and $n^\mu$ tetrad vectors leads to a slightly different basis if one works with the Kinnersley tetrad rather than the Carter tetrad. It is also possible to work with alternative basis vectors spanning the $t-r$ space. In some instances it is convenient to work with coordinate basis vectors $\delta^\mu_t$ and $\delta^\mu_r$ rather than null vectors. One can also choose to omit the factor of $\frac{1}{r}$ in the definition of $m^\mu$ and $\mb^\mu$. The choice of basis does not have a fundamental impact, but some choices lead to more straightforward or natural interpretations of the resulting equations.

Additionally, as an alternative to a spin-weighted harmonic basis, one could equivalently work with an orthonormal basis of vector and tensor spherical harmonics, which are related to the spin-weighted spherical harmonics by
\begin{subequations}
\begin{align}
  Z_A^{\ell \emm} &:= \big[\ell(\ell+1)]^{-1/2} D_A Y^{\ell \emm}
     = \frac{1}{\sqrt{2}} \Big( {}_{-1}Y_{\ell \emm} m_A - {}_{1}Y_{\ell \emm} \mb_A \Big), \\
  Z_{AB}^{\ell \emm} &:= \bigg[2\frac{(\ell-2)!}{(\ell+2)!}\bigg]^{1/2} \big[D_{A} D_B + \tfrac12 \ell(\ell+1) \Omega_{AB}\big] Y^{\ell \emm} \nonumber \\
     &= \frac{1}{\sqrt{2}} \Big( {}_{-2}Y_{\ell \emm} m_A m_B + {}_{2}Y_{\ell \emm} \mb_A \mb_B \Big),
\end{align}
\end{subequations}
for the even-parity sector and
\begin{subequations}
\begin{align}
  X_A^{\ell \emm} &:= -\big[\ell(\ell+1)]^{-1/2}\epsilon_A{}^B D_B Y^{\ell \emm}
     = -\frac{i}{\sqrt{2}} \Big( {}_{-1}Y_{\ell \emm} m_A +  {}_{1}Y_{\ell \emm} \mb_A \Big), \\
  X_{AB}^{\ell \emm} &:= -\bigg[2\frac{(\ell-2)!}{(\ell+2)!}\bigg]^{1/2} \epsilon_{(A}{}^C D_{B)} D_C Y^{\ell \emm} 
     = -\frac{i}{\sqrt{2}} \Big( {}_{-2}Y_{\ell \emm} m_A m_B -  {}_{2}Y_{\ell \emm} \mb_A \mb_B \Big),
\end{align}
\end{subequations}
for the odd-parity sector. Here, $m_A = \frac{1}{\sqrt{2}}[1, i \sin \theta]$ and $\mb_A$ form a complex orthonormal basis on the two-sphere and are related to the two-sphere components of the tetrad vectors $m_\alpha$ and $\mb_\alpha$ by a factor of $r$. The differential operator $D$ is the covariant derivate on the two-sphere with metric $\Omega_{AB} = {\rm diag}(1,\sin^2 \theta)$. Note that the definitions of the vector and tensor harmonics given here differs from those of Martel and Poisson \cite{Martel:2005ir} in their overall normalisation but are otherwise the same. The choice here ensures that the harmonics are unit-normalised,
\begin{subequations}
\begin{align}
  \int Z^A_{\ell \emm}(\theta, \phi) \bar{Z}_A^{\ell' \emm'}(\theta, \phi) d\Omega = \delta_{\ell \ell'} \delta_{\emm \emm'}, \\
  \int X^A_{\ell \emm}(\theta, \phi) \bar{X}_A^{\ell' \emm'}(\theta, \phi) d\Omega = \delta_{\ell \ell'} \delta_{\emm \emm'}, \\
  \int Z^{AB}_{\ell \emm}(\theta, \phi) \bar{Z}_{AB}^{\ell' \emm'}(\theta, \phi) d\Omega = \delta_{\ell \ell'} \delta_{\emm \emm'}, \\
  \int X^{AB}_{\ell \emm}(\theta, \phi) \bar{X}_{AB}^{\ell' \emm'}(\theta, \phi) d\Omega = \delta_{\ell \ell'} \delta_{\emm \emm'}.
\end{align}
\end{subequations}

For simplicity we opt to work exclusively with a spin-weighted spherical harmonic basis, but point out that equivalent results hold for other choices of basis. In particular, the expressions that follow can be transformed to the commonly-used Barack-Lousto-Sago \cite{Barack:2005nr,Barack:2007tm,Wardell:2015ada} basis, $h^{(i)}_{\ell \emm}$ , the $A$--$K$ basis \cite{Chen:2016plo,Thompson:2018lgb}, the Martel-Poisson basis \cite{Martel:2005ir}, and the Berndtson basis \cite{Berndtson:2009hp} using the relations given in Table \ref{tab:bases}. The table also gives the translation between a Carter null-tetad basis $(l^\alpha,n^\alpha,m^\alpha,\mb^\alpha)$ and a $t$--$r$ coordinate basis $(\delta_t^\alpha,\delta^\alpha,m^\alpha,\mb^\alpha)$. Note that the Barack-Lousto-Sago expressions involve the non-trace-reversed metric. A trace reversal in the Barack-Lousto-Sago basis corresponds to the interchange $h^{(3)}_{\ell\emm} \leftrightarrow h^{(6)}_{\ell\emm}$, consistent with the trace reversal in the null tetrad basis corresponding to the interchange $h_{ln}^{\ell\emm} \leftrightarrow h_{m\mb}^{\ell\emm}$.
\begin{table}
\label{tab:bases}
\centering
\resizebox{\columnwidth}{!}{
\begin{tabular}{|c|c|c|c|c|c|}
  \hline
  Tetrad & Barack-Lousto-Sago & $A$--$K$ & Martel-Poisson & Berndtson & Coord.
  \\ \hline \hline
    $\frac{f}{2} \Big(h_{ll}^{\ell\emm}+h_{nn}^{\ell\emm} + 2 h_{ln}^{\ell\emm}\Big)$
  & $\frac{1}{2 r}\Big(h^{(1)}_{\ell\emm}+f h^{(3)}_{\ell\emm}\Big)$
  & $A^{\ell\emm}$
  & $h^{\ell\emm}_{tt}$
  & $f H_0$
  & $h_{tt}^{\ell\emm}$
  \\ \hline
    $\frac{1}{2f} \Big(h_{ll}^{\ell\emm}+h_{nn}^{\ell\emm} - 2 h_{ln}^{\ell\emm}\Big)$
  & $\frac{1}{2 r f^2}\Big(h^{(1)}_{\ell\emm}- fh^{(3)}_{\ell\emm}\Big)$
  & $K^{\ell\emm}$
  & $h^{\ell\emm}_{rr}$
  & $\frac{1}{f} H_2$
  & $h_{rr}^{\ell\emm}$
  \\ \hline
    $\frac{1}{2} \Big(h_{ll}^{\ell\emm}-h_{nn}^{\ell\emm}\Big)$
  & $\frac{1}{2 r f}h^{(2)}_{\ell\emm}$
  & $-D^{\ell\emm}$
  & $h^{\ell\emm}_{tr}$
  & $H_1$
  & $h_{tr}^{\ell\emm}$
  \\ \hline
    $h_{m\mb}^{\ell\emm}$
  & $\frac{1}{2 r} h^{(6)}_{\ell\emm}$
  & $E^{\ell\emm}$
  & $K^{\ell\emm}$
  & $K$
  & $h_{m\mb}^{\ell\emm}$
  \\ \hline
    $\sqrt{\frac{f}{2}}\Big(h_{l,\rm{even}}^{\ell\emm} + h_{n,\rm{even}}^{\ell\emm}\Big)$
  & $-\frac{h^{(4)}_{\ell\emm}}{2 r \sqrt{2} \sqrt{\ell(\ell+1)}}$
  & $\frac{\sqrt{\ell(\ell+1)}}{\sqrt{2}} B^{\ell\emm}$
  & $-\frac{\sqrt{\ell(\ell+1)}}{r \sqrt{2}} j_t^{\ell\emm}$
  & $-\frac{\sqrt{\ell(\ell+1)}}{r \sqrt{2}} h_0$
  & $h_{t,\rm{even}}^{\ell\emm}$
  \\ \hline
    $\sqrt{\frac{1}{2f}}\Big(h_{l,\rm{even}}^{\ell\emm} - h_{n,\rm{even}}^{\ell\emm}\Big)$
  & $-\frac{h^{(5)}_{\ell\emm}}{2 r f \sqrt{2} \sqrt{\ell(\ell+1)}} $
  & $-\frac{\sqrt{\ell(\ell+1)}}{\sqrt{2}} H^{\ell\emm}$
  & $-\frac{\sqrt{\ell(\ell+1)}}{r \sqrt{2}} j_r^{\ell\emm}$
  & $-\frac{\sqrt{\ell(\ell+1)}}{r \sqrt{2}} h_1$
  & $h_{r,\rm{even}}^{\ell\emm}$
  \\ \hline
    $\sqrt{\frac{f}{2}}\Big(h_{l,\rm{odd}}^{\ell\emm} + h_{n,\rm{odd}}^{\ell\emm}\Big)$
  & $\frac{h^{(8)}_{\ell\emm}}{2 r \sqrt{2} \sqrt{\ell(\ell+1)}} $
  & $\frac{\sqrt{\ell(\ell+1)}}{\sqrt{2}} C^{\ell\emm}$
  & $-\frac{\sqrt{\ell(\ell+1)}}{r \sqrt{2}} h_t^{\ell\emm}$
  & $\frac{\sqrt{\ell(\ell+1)}}{r \sqrt{2}} h_0$
  & $h_{t,\rm{odd}}^{\ell\emm}$
  \\ \hline
    $\sqrt{\frac{1}{2f}}\Big(h_{l,\rm{odd}}^{\ell\emm} - h_{n,\rm{odd}}^{\ell\emm}\Big)$
  & $\frac{h^{(9)}_{\ell\emm}}{2 r f \sqrt{2} \sqrt{\ell(\ell+1)}}$
  & $-\frac{\sqrt{\ell(\ell+1)}}{\sqrt{2}} J^{\ell\emm}$
  & $-\frac{\sqrt{\ell(\ell+1)}}{r \sqrt{2}} h_r^{\ell\emm}$
  & $\frac{\sqrt{\ell(\ell+1)}}{r \sqrt{2}} h_1$
  & $h_{r,\rm{odd}}^{\ell\emm}$
  \\ \hline
    $h_{2,\rm{even}}^{\ell\emm}$
  & $\frac{1}{2r} \sqrt{\frac{(\ell-2)!}{(\ell+2)!}} h^{(7)}_{\ell\emm}$
  & $\frac12 \sqrt{\frac{(\ell+2)!}{(\ell-2)!}} F^{\ell\emm}$
  & $\frac12 \sqrt{\frac{(\ell+2)!}{(\ell-2)!}} G^{\ell\emm}$
  & $\sqrt{\frac{(\ell+2)!}{(\ell-2)!}} G^{\ell\emm}$
  & $h_{2,\rm{even}}^{\ell\emm}$
  \\ \hline
    $h_{2,\rm{odd}}^{\ell\emm}$
  & $-\frac{1}{2r} \sqrt{\frac{(\ell-2)!}{(\ell+2)!}} h^{(10)}_{\ell\emm}$
  & $\frac12 \sqrt{\frac{(\ell+2)!}{(\ell-2)!}} G^{\ell\emm}$
  & $\frac{1}{2r^2} \sqrt{\frac{(\ell+2)!}{(\ell-2)!}} h_2^{\ell\emm}$
  & $\frac{1}{r^2} \sqrt{\frac{(\ell+2)!}{(\ell-2)!}} h_2$
  & $h_{2,\rm{odd}}^{\ell\emm}$
  \\ \hline
\end{tabular}}
\caption{Relationship between choices of basis for perturbations of Schwarzschild spacetime.}
\end{table}

\subsection{Regge-Wheeler formalism and Regge-Wheeler gauge}

The Regge-Wheeler formalism is based on the idea of constructing solutions to the linearised Einstein equations from
solutions to the scalar wave equation with a potential. In the case of the Regge-Wheeler master function,
it is a solution of
\begin{equation}
\label{eq:RW}
  \bigg[\Box + \frac{2Ms^2}{r^3} \bigg] \RW_s = S_s,
\end{equation}
where $s$ is the spin of the field ($s=0$ for scalar fields, $s=1$ for electromagnetic fields and $s=2$ for gravitational fields).

Equation \eqref{eq:RW} is separable in Schwarzschild spacetime using the ansatz
\begin{align}
  \RW_s &= \sum_{\ell=0}^\infty \sum_{m=-\ell}^\ell \, \frac{1}{r} \RW_{s \ell \emm}(t, r) \, {}_0 Y_{\ell \emm}(\theta, \phi),
\end{align}
with $\RW_{s \ell \emm}(t,r)$ satisfying the Regge-Wheeler equation,
\begin{equation}
  \label{eq:RWl}
  \bigg[ \dfrac{\partial}{\partial r} \bigg(f \dfrac{\partial }{\partial r}\bigg) - \frac{1}{f}\frac{\partial }{\partial t^2} - \bigg(\frac{\ell(\ell+1)}{r^2} + 
  \frac{2M(1-s^2)}{r^3} \bigg)\bigg]\RW_{s\ell \emm}  = S^{\rm RW}_{s\ell \emm}.
\end{equation}

In order to study metric perturbations of Schwarzschild spacetime, we consider the $s=2$ case. The Regge-Wheeler master function is then defined in terms of the metric perturbation by
\begin{equation}
  \RW_{2 \ell \emm} := -\frac{f}{r}\bigg[\frac{\sqrt{2} r\, h^{\ell\emm}_{r,\odd}}{\sqrt{\ell(\ell+1)}} + \frac{r^2 \partial_r  h^{\ell\emm}_{2,\odd} }{\sqrt{(\ell-1)\ell(\ell+1)(\ell+2)}} \bigg].
\end{equation}
It satisfies the $s=2$ Regge-Wheeler equation with source derived from the mode-decomposed stress-energy tensor,\footnote{Different conventions
for the source exist in the literature. For example, the source given in Ref.~\cite{Hopper:2010uv} differs from that given here by a factor of $f$; this is a consequence of the left hand side of their Regge-Wheeler equation [Eq.~(2.13) in Ref.~\cite{Hopper:2010uv}] also differing from our Eq.~\eqref{eq:RWl} by a factor of $f$.}
\begin{equation}
  S^{\rm RW}_{2\ell \emm} =  16\pi \bigg[\frac{\sqrt{2} f T^{\ell\emm}_{r,\odd}}{\sqrt{\ell(\ell+1)}} + \frac{r \,\partial_r (f  T^{\ell\emm}_{2,\odd})}{\sqrt{ (\ell-1)\ell(\ell+1)(\ell+2)}} \bigg].
\end{equation}

Rather than working with the Regge-Wheeler master
function itself, it is often preferable to introduce two closely related functions: the Cunningham-Price-Moncrief (CPM)
master function defined by
\begin{equation}
  \CPM_{\ell \emm} := -\frac{\sqrt{2}}{\sqrt{\ell(\ell+1)}} \frac{2r}{(\ell-1)(\ell+2)}\bigg[\partial_r (r h^{\ell\emm}_{t,\odd}) - \partial_t (r h^{\ell\emm}_{r,\odd})  - 2 h^{\ell\emm}_{t,\odd} \bigg],
\end{equation}
and the Zerilli-Moncrief (ZM) master function defined by
\begin{equation}
  \ZM_{\ell \emm} := \frac{2r}{\ell(\ell+1)}\bigg[ \tilde{K}^\lm + \frac{2}{\Lambda} (f^2 \tilde{h}_{rr}^\lm- r f \partial_r \tilde{K}^\lm) \bigg],
\end{equation}
where $\Lambda := (\ell-1)(\ell+2) + \frac{6M}{r}$ and
\begin{subequations}
\begin{align}
    \tilde{K}^\lm &:= h_{m\mb}^{\ell\emm} - \frac{2 f \, h^{\ell\emm}_{r,\even}}{\sqrt{\ell(\ell+1)}} + \frac{\Big[\ell(\ell+1) - 2 r\, f\, \partial_r \Big] h^{\ell\emm}_{2,\even}}{\sqrt{2(\ell-1)\ell(\ell+1)(\ell+2)}}, \\
    \tilde{h}_{rr}^\lm &:= h_{rr}^\lm - \frac{2 \partial_{r} (r\, h^{\ell\emm}_{r,\even})}{\sqrt{\ell(\ell+1)}} + \frac{\partial_{r} \big( r^2 \partial_{r} h_2^\even \big)}{\sqrt{2(\ell-1)\ell(\ell+1)(\ell+2)}}
\end{align}
\end{subequations}
are gauge invariant fields.

The CPM master function satisfies the same $s=2$ Regge-Wheeler equation as the Regge-Wheeler master function, but with a different source given by
\begin{equation}
  S^{\rm CPM}_{\ell \emm} =  16 \pi \frac{\sqrt{2}}{\sqrt{\ell(\ell+1)}} \frac{2r}{(\ell-1)(\ell+2)} \bigg[\partial_r (r\, T^{\ell\emm}_{t,\odd}) - \partial_t (r\, T^{\ell\emm}_{r,\odd}) \bigg].
\end{equation}
The RW and CPM master functions are related by a time derivative (plus source terms),
\begin{equation}
    \RW_{2 \ell \emm} = \frac12 \partial_t \CPM_{\ell \emm} - \frac{16 \pi r^2 f}{(\ell-1)(\ell+2)} \frac{\sqrt{2}}{\sqrt{\ell(\ell+1)}} T^{\ell\emm}_{r,\odd}.
\end{equation}

The ZM master function satisfies the Zerilli equation (the Regge-Wheeler equation with a different potential),
\begin{equation}
  \bigg[ \dfrac{\partial}{\partial r} \bigg(f \dfrac{\partial }{\partial r}\bigg) - \frac{1}{f}\frac{\partial}{\partial t^2} - V^{\rm ZM} \bigg]\ZM_{\ell \emm}  = S^{\rm ZM}_{\ell \emm},
\end{equation}
where
\begin{equation}
    V^{\rm ZM} = \frac{\ell(\ell+1)}{r^2}- \frac{6M}{r^3} + \frac{72 M^2 f}{\Lambda^2 r^4} - \frac{24 M (r-3M)}{\Lambda r^4 },
\end{equation}
and where the ZM source is
\begin{align}
  S^{\rm ZM}_{\ell \emm} &=  \frac{4f}{\Lambda} \frac{16\pi}{\sqrt{\ell(\ell+1)}} T^{\ell\emm}_{r,\even} - \frac{16\pi\sqrt{2}}{\sqrt{(\ell-1)\ell(\ell+1)(\ell+2)}} r \, T^{\ell\emm}_{2,\even} \nonumber \\
  &+ \frac{2}{\ell(\ell+1)\Lambda}\Big\{\Big[\frac{r}{\Lambda}\Big((\ell-1)(\ell+2)(l^2+l-4) + 12 (\ell^2+\ell-5)\frac{M}{r} + 84 \frac{M^2}{r^2}\Big)\nonumber \\
  & \qquad\qquad -2r^2 f \partial_r \Big](f\, T^{\ell\emm}_{rr}-f^{-1}\, T^{\ell\emm}_{tt}) + \frac{24M}{\Lambda} f^2 T^{\ell\emm}_{rr} + 2 r\,f\,T^{\ell\emm}_{m\mb} \Big\}.
\end{align}

\subsubsection{Regge-Wheeler formalism in the frequency domain}

Transforming to the frequency domain, the Regge-Wheeler and Zerilli equations become a set of ordinary differential equations,
one for each $(\ell,\emm,\omega)$ mode. Solutions to these equations may be written in terms of a pair of homogeneous mode
basis functions chosen according to their asymptotic behavior at the four null boundaries to the
spacetime. For radiative ($\omega \ne 0$) modes, the four common choices are denoted ``in'', ``up'', ``out'' and ``down'',
with the same interpretation as described in Sec.~\ref{sec:KerrModes} for the Teukolsky equation. These have asymptotic behaviour given by
\begin{subequations}
\begin{alignat}{5}
\label{eq:bcXin}
{}_s X^{\text{in}}_{\ell \emm \omega}(r) &\sim
\Big\{&
\begin{array}{c}
 0 \\
 {}_s X^{\text{in,ref}}_{\ell \emm \omega} e^{+i\omega r_*}
\end{array}&
\begin{array}{c}
 + \\ +
\end{array}
\begin{array}{c}
 {}_s X^{\text{in,trans}}_{\ell \emm \omega} e^{-i \omega r_*}
 \\
 {}_s X^{\text{in,inc}}_{\ell \emm \omega} e^{-i\omega r_*}
\end{array}
&\qquad
\begin{array}{l}
  r \to 2M\\
  r \to \infty
\end{array}
\\
\label{eq:bcXup}
{}_s X^{\text{up}}_{\ell \emm \omega}(r) &\sim
\Big\{&
\begin{array}{c}
 {}_s X^{\text{up,inc}}_{\ell \emm \omega} e^{+i \omega r_*}
 \\
 {}_s X^{\text{up,trans}}_{\ell \emm \omega} e^{+i\omega r_*}
\end{array}&
\begin{array}{c}
 + \\ +
\end{array}
\begin{array}{c}
 {}_s X^{\text{up,ref}}_{\ell \emm \omega} e^{-i \omega r_*}
 \\
 0
\end{array}
&\qquad
\begin{array}{l}
  r \to 2M\\
  r \to \infty
\end{array} 
\\
\label{eq:bcXout}
{}_s X^{\text{out}}_{\ell \emm \omega}(r) &\sim
\Big\{&
\begin{array}{c}
 {}_s X^{\text{out,trans}}_{\ell \emm \omega} e^{+i \omega r_*}
 \\
 {}_s X^{\text{out,inc}}_{\ell \emm \omega} e^{+i \omega r_*}
\end{array}&
\begin{array}{c}
 + \\ +
\end{array}
\begin{array}{c}
 0
 \\
 {}_s X^{\text{out,ref}}_{\ell \emm \omega} e^{-i \omega r_*}
\end{array}
&\qquad
\begin{array}{l}
  r \to 2M\\
  r \to \infty
\end{array} 
\\
\label{eq:bcXdown}
{}_s X^{\text{down}}_{\ell \emm \omega}(r) &\sim
\Big\{&
\begin{array}{c}
 {}_s X^{\text{down,ref}}_{\ell \emm \omega} e^{+i \omega r_*}
 \\
 0
\end{array}&
\begin{array}{c}
 + \\ +
\end{array}
\begin{array}{c}
 {}_s X^{\text{down,inc}}_{\ell \emm \omega} e^{-i \omega r_*}
 \\
 {}_s X^{\text{down,trans}}_{\ell \emm \omega} e^{-i \omega r_*}
\end{array}
&
\begin{array}{l}
  r \to 2M\\
  r \to \infty
\end{array} 
\end{alignat}
\end{subequations}
where $r_* = r + 2 M \ln (\frac{r}{2M}-1)$ is the Regge-Wheeler tortoise coordinate.

Inhomogeneous solutions of the Regge-Wheeler equation can then be written in terms of a linear combination of the basis functions,
\begin{align} \label{eq:RWInhomogeneousModes}
  {}_s \psi_{\ell \emm \omega}(r) &= {}_s
    C^{\text{in}}_{\ell \emm \omega}(r) {}_s X^{\text{in}}_{\ell \emm \omega}(r)+{}_s
    C^{\text{up}}_{\ell \emm \omega}(r) {}_s X^{\text{up}}_{\ell \emm \omega}(r),
\end{align}
where the weighting coefficients are determined by variation of parameters,
\begin{subequations}
  \label{eq:weighting-coefficients-RW}
  \begin{align}
  {}_s C_{\ell \emm \omega}^{\text{in}}(r) &= \int_{r}^{\infty} \frac{{}_s X^{\text{up}}_{\ell \emm \omega}(r')}{W(r')f} {}_s S_{\ell \emm \omega}(r') dr', \\
  {}_s C_{\ell \emm \omega}^{\text{up}}(r) &= \int^{r}_{2M} \frac{{}_s X^{\text{in}}_{\ell \emm \omega}(r')}{W(r')f} {}_s S_{\ell \emm \omega}(r') dr',
\end{align}
\end{subequations}
with $W(r) = {}_s X^{\text{in}}_{\ell \emm \omega}(r) \partial_r [{}_s X^{\text{up}}_{\ell \emm \omega}(r)] - {}_s X^{\text{up}}_{\ell \emm \omega}(r) \partial_r [{}_s X^{\text{in}}_{\ell \emm \omega}(r)]$ the Wronskian [in practice, it is convenient to use the fact that $f(r) W(r) = \text{const}$].

\subsubsection{Transformation between Regge-Wheeler and Zerilli solutions}

Homogeneous solutions to the Zerilli equation can be obtained from homogeneous solutions to the Regge-Wheeler equation
by applying differential operators,
\begin{subequations}
\begin{align}
X^{\text{ZM,up}}_{\ell\emm\omega}&=\frac{\Big[(\ell-1)\ell(\ell+1)(\ell+2)+\tfrac{72 M^2 f}
{r^2 \Lambda}\Big]{}_2 X^{\text{RW,up}}_{\ell\emm\omega}
+12 M f 
\frac{d{}_2 X^{\text{RW,up}}_{\ell\emm\omega}}{dr}}{(\ell-1)\ell(\ell+1)(\ell+2)+12 i \omega M},
\\
X^{\text{ZM,in}}_{\ell\emm\omega}&=\frac{\Big[(\ell-1)\ell(\ell+1)(\ell+2)+\tfrac{72 M^2 f}
{r^2 \Lambda}\Big]{}_2 X^{\text{RW,in}}_{\ell\emm\omega}
+12 M f 
\frac{d{}_2 X^{\text{RW,in}}_{\ell\emm\omega}}{dr}}{(\ell-1)\ell(\ell+1)(\ell+2)-12 i \omega M}.
\end{align}
\end{subequations}
The constant of proportionality here is such that the transmission coefficients of the two Zerilli solutions is the
same as that of the Regge-Wheeler solution.

\subsubsection{Transformation between Regge-Wheeler and Teukolsky formalism}
The modes of the CPM master function are related to
the modes of the Teukolsky radial function by the Chandrasekhar transformation,
\begin{subequations}
\label{eq:ChandrasekharTransformation}
\begin{align}
  {}_2 \psi_{\ell \emm \omega} &= -\tfrac{i \sqrt{D}}{4 r^2} \mathcal{D} \mathcal{D} \big(r \CPM_{\ell\emm\omega}\big), \\
  {}_{-2} \psi_{\ell \emm \omega} &= -\tfrac{i \sqrt{D}}{16} r^2 f^2 \mathcal{D}^\dag \mathcal{D}^\dag \big(r \CPM_{\ell\emm\omega}\big),
\end{align}
\end{subequations}
where $D = (\ell-1)\ell(\ell+1)(\ell+2)$ is the Schwarzschild limit of the constant that appears in the Teukolsky-Starobinsky identities, Eq.~\eqref{eq:TS_D}.
In the absence of sources this can be inverted to give
\begin{subequations}
\begin{align}
  \CPM_{\ell\emm\omega} &= \tfrac{1}{\sqrt{D} \mathcal{C}_{\ell\emm\omega}} r^3 \mathcal{D} \mathcal{D} \big(\tfrac{1}{r^2} {}_{-2} \psi_{\ell \emm \omega} \big), \\
  \CPM_{\ell\emm\omega} &= \tfrac{1}{4\sqrt{D} \mathcal{C}^\dag_{\ell\emm\omega}} r^2 f^{-1} \mathcal{D}^\dag r^2 f^2 \mathcal{D}^\dag \big(r f {}_{2} \psi_{\ell \emm \omega} \big),
\end{align}
\end{subequations}
where $\mathcal{C}_{\ell\emm\omega}$ is the Schwarzschild limit of the second constant that appears in the Teukolsky-Starobinsky identities, Eq.~\eqref{eq:TS_C}.

\subsubsection{Gravitational waves}

As in the radiation gauge case, the gravitational wave strain can be determined directly from $\ZM$ and $\CPM$. There is a slight subtlety in that the Regge-Wheeler-Zerilli gauge in which the metric is typically reconstructed is not compatible with the transverse-traceless gauge in which gravitational waves are normally defined (it is easy to see this since $h_{mm} = 0 = h_{\mb\mb}$ in the Regge-Wheeler-Zerilli gauge). Instead, we can use the Chandrasekhar transformation in Eq.~\eqref{eq:ChandrasekharTransformation} to first transform to $\psi_4$ and then compute the strain from that as we did in radiation gauge. Doing so we have
\begin{equation}
  r(h_+ - i h_\times) = \sum_{\ell=2}^\infty \sum_{m=-\ell}^\ell  \, \frac{\sqrt{D}}{2} (\ZM_{\ell \emm}-i \CPM_{\ell \emm}) {}_{-2} Y_{\ell \emm}(\theta, \phi), \label{eq:RW-waveform}
\end{equation}
where it is understood that equality holds in the limit $r \to \infty$ (at fixed $u=t-r^*$). If we work in the frequency domain and compute the ``in'' and ``up'' mode functions with normalisation such that transmission coefficients are unity, ${}_s X^{\text{in,trans}}_{\ell \emm \omega} = 1 = {}_s X^{\text{up,trans}}_{\ell \emm \omega}$, then $\ZM_{\ell \emm}$ and $\CPM_{\ell \emm}$ are given by  the ``up'' weighting coefficients $ C_{\ell \emm \omega}^{\text{ZM,up}}$ and $ C_{\ell \emm \omega}^{\text{CPM,up}}$ evaluated in the limit $r\to \infty$. Similarly, the time averaged flux of energy carried by gravitational waves passing through infinity and the horizon can be computed from the ``in'' and ``up'' weighting coefficients,
\begin{subequations}
\begin{align}
  \fluxEH &= \lim_{r \to 2M} \sum_{\lm\omega} \frac{D}{64\pi} \omega^2 \bigg[| C^{\text{ZM,in}}_{\ell \emm \omega}|^2 + | C^{\text{CPM,in}}_{\ell \emm \omega}|^2 \bigg], \label{EdotH v2}\\
  \fluxEI &= \lim_{r \to \infty} \sum_{\lm\omega} \frac{D}{64\pi} \omega^2 \bigg[| C^{\text{ZM,up}}_{\ell \emm \omega}|^2 + | C^{\text{CPM,up}}_{\ell \emm \omega}|^2 \bigg].
\end{align}
\end{subequations}
Similarly, the flux of angular momentum through infinity and the horizon can be computed from the ``in'' and ``up'' normalization coefficients,
\begin{subequations}
\begin{align}
  \fluxLzH &= \lim_{r \to 2M}\sum_{\lm\omega} \frac{D}{64\pi} \emm\omega \bigg[| C^{\text{ZM,in}}_{\ell \emm \omega}|^2 + | C^{\text{CPM,in}}_{\ell \emm \omega}|^2 \bigg], \label{LdotH v2}\\
  \fluxLzI &= \lim_{r \to \infty} \sum_{\lm\omega} \frac{D}{64\pi} \emm\omega \bigg[| C^{\text{ZM,up}}_{\ell \emm \omega}|^2 + | C^{\text{CPM,up}}_{\ell \emm \omega}|^2 \bigg].
\end{align}
\end{subequations}

\subsubsection{Metric reconstruction in Regge-Wheeler gauge}
Much like in the Teukolsky formalism, the CPM master function is gauge invariant and may be used to reconstruct the metric perturbation in a chosen gauge. In the Regge-Wheeler gauge, defined by the choice $h^\lm_{a,\even} = h^\lm_{2,\even} = h^\lm_{2,\odd} = 0$, the odd parity metric perturbation is given by
\begin{subequations}
\begin{align}
  h^{\lm}_{l,\odd} &= \frac{\sqrt{f}\sqrt{\ell(\ell+1)}}{2 \sqrt{2} r}\big(\partial_r + f^{-1} \partial_t\big) \big(r \CPM_{\lm}\big) + \frac{16 \pi r^2}{(\ell-1)(\ell+2)} T^\lm_{l,\odd}, \\
  h^{\lm}_{n,\odd} &= \frac{\sqrt{f}\sqrt{\ell(\ell+1)}}{2 \sqrt{2} r}\big(\partial_r - f^{-1} \partial_t\big)  \big(r \CPM_{\lm}\big) + \frac{16 \pi r^2}{(\ell-1)(\ell+2)} T^\lm_{n,\odd},
\end{align}
\end{subequations}
and the even parity metric perturbation is given by \cite{Hopper:2010uv}
\begin{subequations}
\begin{align}
h_{m\mb}^\lm &=  f \partial_r \ZM_\lm + A \ZM_\lm - \frac{32\pi r^2}{\ell(\ell+1) \Lambda} T^\lm_{tt}, \\
h_{rr}^\lm &= \frac{\Lambda}{2f^2} \left[ \frac{\ell(\ell+1)}{2r} \ZM_\lm - h^\lm_{m\mb} \right] + \frac{r}{f} \partial_r h^\lm_{m\mb}, \\
h_{tr}^\lm &=   r \partial_t \partial_r \ZM_\lm + r B \, \partial_t \ZM_\lm  + \frac{16 \pi r^2}{\ell(\ell+1)} \left[ T^\lm_{tr} - \frac{2r}{f\Lambda} \partial_t T^{\lm}_{tt} \right], \\
h_{tt}^\lm &= f^2 h_{rr} +  \frac{8\pi f }{\sqrt{2(\ell-1)\ell(\ell+1)(\ell+2)}}T^\lm_{2,\even},
\end{align}
\end{subequations}
where
\begin{subequations}
\begin{align}
A(r) &:= \frac{2}{r \Lambda} 
\left[ \frac14 (\ell-1)\ell(\ell+1)(\ell+2) + \frac{3M}{2r} \Big((\ell-1)(\ell+2) + \frac{4M}{r} \Big) \right], 
\\
B(r) &:= \frac{2}{r f \Lambda} 
\left[ \frac{(\ell-1)(\ell+2)}{2} \left( 1 - \frac{3M}{r} \right) - \frac{3M^2}{r^2}  \right].
\end{align}
\end{subequations}

As in the Teukolsky case, in order to transform back to the time-domain solution, as a final step we must perform an inverse Fourier transform. This poses a challenge in gravitational self-force calculations, where non-smoothness of the solutions in the vicinity of the worldline lead to the Gibbs phenomenon of non-convergence of the inverse Fourier transform. Resolutions to this problem typically rely on avoiding directly transforming the inhomogeneous solution by using the methods of extended homogeneous or extended particular solutions. For further details, see \cite{Hopper:2010uv,Hopper:2012ty}.

\subsection{Lorenz gauge}

In the case of perturbations of a Schwarzschild black hole, the equations for the metric perturbation itself are separable. This makes it
practical to work in the Lorenz gauge and to directly solve the Lorenz gauge field equations for the metric perturbation.

Rewriting the Lorenz gauge condition, Eq.~\eqref{eq:LorenzGauge}, in terms of null tetrad components we have four gauge equations,
\begin{subequations}
\begin{align}
    (\th' - 2 \rho') h_{ll} + (\th - 2 \rho) h_{m\mb} - 2 \rho h_{ln} - (\edth h_{l\mb} + \edth' h_{lm}) &= 0, \\
    (\th - 2 \rho) h_{nn} + (\th' - 2 \rho') h_{m\mb} - 2 \rho' h_{ln} - (\edth' h_{nm} + \edth h_{n\mb}) &= 0, \\
    (\th' - 3 \rho') h_{lm} + (\th - 3 \rho) h_{nm} - \edth h_{ln} - \edth' h_{mm} &= 0, \\
    (\th' - 3 \rho') h_{l\mb} + (\th - 3 \rho) h_{n\mb} - \edth' h_{ln} - \edth h_{\mb\mb} &= 0.
\end{align}
\end{subequations}
These decouple into 3 even parity equations (the first two and the real part of either the third or fourth) and 1 odd-parity equation (the imaginary part
of either the third or fourth equation).
Similarly, the Lorenz gauge linearised Einstein equation, Eq.~\eqref{eq:LorenzField}, yields ten field equations (7 even and 3 odd) given by
\begin{subequations}
\begin{align}
    &\hat{\Box} (h_{m\mb} - h_{ln}) = 8 \pi T, \\
    &(\hat{\Box} - 8 \psi_2 + 8 \rho \rho') (h_{ln} + h_{m\mb}) + 4 \rho^2 h_{nn} + 4 \rho'^2 h_{ll} \nonumber \\
    & \qquad + 4 \rho(\edth h_{n\mb} + \edth' h_{nm}) + 4 \rho'(\edth h_{l\mb} + \edth' h_{lm}) = -16 \pi (T_{ln} + T_{m\mb}), \\
    &(\hat{\Box} + 4 \rho \rho') h_{ll} + 4 \rho^2 (h_{ln} + h_{m\mb}) + 4 \rho (\edth h_{l\mb}+ \edth' h_{lm}) = -16 \pi T_{ll}, \\
    &(\hat{\Box}' + 4 \rho \rho') h_{nn} + 4 \rho'^2 (h_{ln} + h_{m\mb}) + 4 \rho' (\edth' h_{nm}+ \edth h_{n\mb}) = -16 \pi T_{nn}, \\
    &(\hat{\Box} - 6 \psi_2 + 4 \rho \rho') h_{lm} + 4 \rho^2 h_{nm} + 2 \rho \edth (h_{ln}+ h_{m\mb}) \nonumber \\
    & \qquad + 2 \rho' \edth h_{ll} + 2 \rho \edth' h_{mm} = -16 \pi T_{lm}, \\
    &(\bar{\hat{\Box}} - 6 \psi_2 + 4 \rho \rho') h_{l\mb} + 4 \rho^2 h_{n\mb} + 2 \rho \edth' (h_{ln}+ h_{m\mb}) \nonumber \\
    & \qquad  + 2 \rho' \edth' h_{ll} + 2 \rho \edth h_{\mb\mb} = -16 \pi T_{l\mb}, \\
    &(\bar{\hat{\Box}}' - 6 \psi_2 + 4 \rho \rho') h_{nm} + 4 \rho'^2 h_{lm} + 2 \rho' \edth (h_{ln}+ h_{m\mb}) \nonumber \\
    & \qquad  + 2 \rho \edth h_{nn} + 2 \rho' \edth' h_{mm} = -16 \pi T_{nm}, \\
    &(\hat{\Box}' - 6 \psi_2 + 4 \rho \rho') h_{n\mb} + 4 \rho'^2 h_{l\mb} + 2 \rho' \edth' (h_{ln}+ h_{m\mb}) \nonumber \\
    & \qquad  + 2 \rho \edth' h_{nn} + 2 \rho' \edth h_{\mb\mb} = -16 \pi T_{n\mb}, \\
    &\hat{\Box} h_{mm} + 4 \rho \edth h_{nm} + 4 \rho' \edth h_{lm} = -16 \pi T_{mm}, \\
    &\bar{\hat{\Box}} h_{\mb\mb} + 4 \rho \edth' h_{n\mb} + 4 \rho' \edth' h_{l\mb} = -16 \pi T_{\mb\mb}
\end{align}
\end{subequations}
where the operators
\begin{align*}
    \hat{\Box} := - 2 \th \th' + 2 \rho' \th + 2 \rho \th' + 2 \edth \edth',\qquad
    \hat{\Box}' := - 2 \th' \th + 2 \rho \th' + 2 \rho' \th + 2 \edth' \edth,\\
    \bar{\hat{\Box}} := - 2 \th \th' + 2 \rho' \th + 2 \rho \th' + 2 \edth' \edth,\qquad
    \bar{\hat{\Box}}' := - 2 \th' \th + 2 \rho \th' + 2 \rho' \th + 2 \edth \edth',
\end{align*}
all coincide with the scalar wave operator when acting on type $\{0,0\}$ objects (but differ when acting on objects of generic GHP type). 
Note that we have chosen here to work with the non-trace-reversed metric perturbation; equivalent equations for the trace-reversed perturbation can be obtained by noting that a trace-reversal corresponds to the interchange $h_{ln} \leftrightarrow h_{m\mb}$.

The Lorenz gauge equations can be decomposed into the same basis of spin-weighted spherical harmonics as for the metric perturbation itself. The mode decomposed
equations follow immediately from the above GHP expressions along with Eqs.~\eqref{eq:spin-raising-lowering} and either \eqref{eq:th-Schw-Kinnersley} or 
\eqref{eq:th-Schw-Carter} for the GHP derivative operators (the specific form for the mode decomposed equations depends on the choice of tetrad).

\subsubsection{Lorenz gauge formalism in the frequency domain}

Following a procedure much like in the Regge-Wheeler and Teukolsky cases, one can construct solutions to the Lorenz gauge equations by working in the frequency domain and solving
ordinary differential equations \cite{Akcay:2013wfa,Osburn:2014hoa,Wardell:2015ada}. The only additional complexity is that for each $(\ell,\emm,\omega)$ mode we must now work with a system of $k$ coupled second order radial equations with $2k$ linearly independent homogeneous solutions.\footnote{There are $k=7$ even parity equations and $k=3$ odd parity equations in general, although these can be reduced to $4+2$ equations using the $3+1$ gauge conditions. The number of equations are also further reduced in certain special cases such as static or low multipole modes.}  As we did in the Regge-Wheeler and Teukolsky cases, it is natural to divide these into $k$ ``in'' solutions and $k$ ``up'' solutions satisfying appropriate boundary conditions at the horizon or radial infinity. Then, using variation of parameters the inhomogeneous solutions are given by
\begin{align}
	h^{(i)}_{\ell\emm\omega}(r) = \mathbf{C}^{\rm in}_{\ell\emm\omega}(r) \cdot \mathbf{h}^{(i),{\rm in}}_{\ell\emm\omega}(r) + \mathbf{C}^{\rm up}_{\ell\emm\omega}(r) \cdot \mathbf{h}^{(i),{\rm up}}_{\ell\emm\omega}(r)
\end{align}
where $i=1,\ldots,k$ represent the $k$ components of the metric perturbation and where $\mathbf{h}^{(i),{\rm in}}_{\ell\emm\omega}(r)$ are vectors of $k$ linearly independent homogeneous solutions for a given $i$. To compute the weighting coefficient vectors $\mathbf{C}^{\rm in/up}_{\ell\emm\omega}(r)$ we define a $2k\times2k$ matrix of homogeneous solutions by
\begin{eqnarray}\label{eq:Phi_matrix}
	\arraycolsep=1.4pt\def\arraystretch{1.5}
	\Phi(r) = \left(\begin{array}{c | c}-\mathbf{h}^{(i),{\rm in}}_{\ell\emm\omega} & \mathbf{h}^{(i),{\rm up}}_{\ell\emm\omega}	\\ \hline -\partial_r \mathbf{h}^{(i),{\rm in}}_{\ell\emm\omega} & \partial_r \mathbf{h}^{(i),{\rm up}}_{\ell\emm\omega} \end{array}\right).
\end{eqnarray}
The vectors of weighting coefficients are then obtained with the standard variation of parameters prescription:
\begin{align}
	\left(\begin{array}{c} \mathbf{C}^{\rm in}(r) \\ \mathbf{C}^{\rm up}(r)\end{array}\right) = \int \Phi^{-1}(r')\left(\begin{array}{c} \mathbf{0} \\ \mathbf{T}(r')\end{array}\right)\,dr',
\end{align}
where $\mathbf{T}(r')$ represents the vector of $k$ sources constructed from the components of the stress energy tensor projected onto the basis and decomposed into modes. The limits on the integral depend upon whether the ``in'' or ``up'' weighting coefficient are being solved for, in the same way as for the Regge-Wheeler and Teukolsky cases.

As in the Regge-Wheeler and Teukolsky cases, in order to transform back to the time-domain solution, as a final step we must perform an inverse Fourier transform. This poses a challenge in gravitational self-force calculations, where non-smoothness of the solutions in the vicinity of the worldline lead to the Gibbs phenomenon of non-convergence of the inverse Fourier transform. Resolutions to this problem typically rely on avoiding directly transforming the inhomogeneous solution by using the methods of extended homogeneous or extended particular solutions. For further details, see \cite{Hopper:2010uv,Hopper:2012ty}.

\subsubsection{Lorenz gauge metric reconstruction from Regge-Wheeler master functions}

As an alternative to directly solving the $7+3$ coupled Lorenz gauge field equations, Berndtson \cite{Berndtson:2009hp} showed
that the solutions could instead be reconstructed from particular solutions to the $s=0$, $1$ and $2$ Regge-Wheeler-Zerilli equations, along with a fourth field obtained by solving the $s=0$ Regge-Wheeler equation sourced by the other $s=0$ field. The explicit expressions are quite unwieldy, particularly when sources are included. Focusing only on the relatively simple odd sector and ignoring the special case such as low multipoles or $\omega=0$ modes, Berndtson's expressions may be written as
\begin{align}
  h_{l,\odd}^{\ell \emm} &= -\frac{\sqrt{f}\sqrt{\ell(\ell+1)}}{2(i \omega)^2 r} \bigg[r^2 \mathcal{D}_0 \bigg(\frac{ \RW_{1\ell\emm}}{r^{2}}\bigg)+ \frac{2\lambda}{3 r}\mathcal{D}_0 \big(r \RW_{2\ell\emm}\big) \bigg] + \frac{8 \pi f}{(i\omega)^2} (T^{\ell\emm}_{l,\rm odd}-T^{\ell\emm}_{n,\rm odd}), \\
  h_{n,\odd}^{\ell \emm} &= \frac{\sqrt{f}\sqrt{\ell(\ell+1)}}{2(i \omega)^2 r} \bigg[r^2 \mathcal{D}^\dag_0 \bigg(\frac{ \RW_{1\ell\emm}}{r^{2}}\bigg)+ \frac{2\lambda}{3 r}\mathcal{D}^\dag_0 \big(r \RW_{2\ell\emm}\big)  \bigg] - \frac{8 \pi f}{(i\omega)^2} (T^{\ell\emm}_{l,\rm odd}-T^{\ell\emm}_{n,\rm odd}), \\
  h_{2,\odd}^{\ell \emm} &= \frac{\sqrt{(\ell-1)\ell(\ell+1)(\ell+2)}}{(i \omega)^2 r^2} \Big[ \RW_{1\ell\emm} + f \partial_r \big( r \RW_{2\ell\emm} \big) + \frac{2\lambda}{3} \RW_{2\ell\emm} \Big]  +  \frac{16 \pi f}{(i\omega)^2} T^{\ell\emm}_{2,\rm odd} ,
\end{align}
where $\mathcal{D}_0$ and $\mathcal{D}_0^\dag$ are the operators defined in Eq.~\eqref{eq:DandL} specialized to the Schwarzschild ($a=0$) case.
Equivalent expressions for the even sector are significantly more complicated and are given in Appendix A of Ref.~\cite{Berndtson:2009hp}, while expressions for low multipoles and $\omega=0$ modes are given elsewhere in the same reference.

\subsubsection{Gravitational waves}
As in the Regge-Wheeler-Zerilli and Teukolsky cases, the flux of gravitational wave energy and angular momentum may be computed from the asymptotic values of the fields. In the Lorenz gauge case where one solves for the metric perturbation directly, the gravitational wave strain is simply given by $h_{mm}$ as in Eq.~\eqref{eq:Strain},
\begin{equation}
  r(h_+ + i h_\times) = r \, h_{mm} = \sum_{\ell=2}^\infty \sum_{\emm=-\ell}^\ell \int_{-\infty}^\infty r\, h_{mm}^{\ell\emm\omega} {}_{2} Y_{\ell\emm}(\theta,\phi) e^{-i \omega (t-r_*)} d \omega,\label{eq:LorenzGauge-waveform}
\end{equation}
where it is understand that the equality holds in the limit $r\to\infty$.
Similarly, the energy fluxes are given explicitly by \cite{Barack:2007tm}
\begin{subequations}
\begin{align}
  \fluxEI &= \lim_{r \to \infty} \sum_{\lm\omega} \frac{\omega^2 r^2}{16\pi} |h_{mm}^{\ell \emm \omega} |^2,\\
  \fluxEH &= \lim_{r \to 2M} \sum_{\lm\omega} \frac{1}{256\pi M^2(1+16 M^2 \omega^2)} \times  \nonumber \\ & \qquad
   \bigg|
    \sqrt{(\ell-1)\ell(\ell+1)(\ell+2)} r f h_{ll}^{\ell \emm \omega} \nonumber \\ & \qquad \,
    - 2 \sqrt{(\ell-1)(\ell+2)} (1+4 i M \omega) r \sqrt{f} h_{lm}^{\ell \emm \omega}  \nonumber \\ & \qquad \,
    + 4 i M \omega (1+4 i M \omega) r h_{mm}^{\ell \emm \omega} \bigg|^2,\label{EdotH v3}
\end{align}
\end{subequations}
and the angular momentum fluxes are given by
\begin{subequations}
\begin{align}
  \fluxLzI &= \lim_{r \to \infty} \sum_{\lm\omega} \frac{\emm\omega r^2}{16\pi} |h_{mm}^{\ell \emm \omega} |^2,\\
  \fluxLzH &= \lim_{r \to 2M} \sum_{\lm\omega} \frac{\emm}{256\pi M^2 \omega (1+16 M^2 \omega^2)} \times  \nonumber \\ & \qquad
   \bigg|
    \sqrt{(\ell-1)\ell(\ell+1)(\ell+2)} r f h_{ll}^{\ell \emm \omega} \nonumber \\ & \qquad \,
    - 2 \sqrt{(\ell-1)(\ell+2)} (1+4 i M \omega) r \sqrt{f} h_{lm}^{\ell \emm \omega}  \nonumber \\ & \qquad \,
    + 4 i M \omega (1+4 i M \omega) r h_{mm}^{\ell \emm \omega} \bigg|^2.\label{LdotH v3}
\end{align}
\end{subequations}

\section{Small objects in General Relativity}

In the previous section we reviewed black hole perturbation theory with a generic source term. In this section, we consider how to formulate the source describing  a small object. This is the {\em local problem} in self-force theory: In a spacetime perturbed by a small body, what are the sources in the field equations~\eqref{EFE1} and \eqref{EFE2}? Moreover, if the body's bulk motion is described by an equation of motion~\eqref{perturbed geodesic equation}, what are the forces on the right-hand side? 

The result  of the analysis is (i) a skeletonization of the small body, in which the body is reduced to a singularity equipped with the body's multipole moments, together with (ii) an equation of motion governing the singularity's trajectory. The setting here is very general: the background can be any vacuum spacetime. Our coverage of the subject is terse, and we refer to Refs.~\cite{Poisson:2011nh,Pound:2015tma} for detailed reviews or to Ref.~\cite{Barack:2018yvs} for a non-expert introduction.

\subsection{Matched asymptotic expansions}

For simplicity, we assume that outside of the small object, the spacetime is vacuum, and that the perturbations are due solely to the object. Over most of the spacetime, the metric is well described by the external background metric $g_{\alpha\beta}$. However, very near the object, in a region comparable to the object's own size, the object's gravity dominates. In this region, which we call the {\em body zone}, the approximation~\eqref{g expansion} breaks down.

This problem is usually overcome in one of two ways: using effective field theory~\cite{Galley:2008ih} (common in post-Newtonian and post-Minkowskian theory~\cite{Porto:2016pyg}) or using the method of matched asymptotic expansions (see, e.g., Refs.~\cite{Eckhaus:79,Kevorkian-Cole:96} for broad introductions, Refs.~\cite{Damour:83,Futamase:2007zz,Poisson:2020vap} for applications in post-Newtonian theory, and Refs.~\cite{DEath:1975jps,Kates:1980zz,Thorne:1984mz,Mino:1996nk,Mino:1997wh,Detweiler:2000gt,Poisson:2003wz,Detweiler:2005kq,Gralla:2008fg,Pound:2009sm,Detweiler:2011tt,Pound:2012nt,Gralla:2012db,Pound:2017psq} and the reviews~\cite{Poisson:2011nh,Pound:2015tma} for the work most relevant here). Here we adopt the latter approach. We let $\e=m/{\cal R}$, where $m$ is the small object's mass and ${\cal R}$ is a characteristic length scale of the external universe; in a small-mass-ratio binary, ${\cal R}$ will be the mass $M$ of the primary, while in a weak-field binary it can be the orbital separation. We then assume Eq.~\eqref{g expansion}, which we dub the {\em outer expansion}, is accurate outside the body zone. Near the object, we assume the metric is well approximated by a second expansion, called an {\em inner expansion}, that effectively zooms in on the body zone. To make this ``zooming in''  precise, we first choose some measure, $\mathscr{r}$, of radial distance from the object, with $\mathscr{r}$ an order-1 function of the external coordinates $x^\alpha$. We then define the scaled distance $\tilde{ \mathscr{r}} := \mathscr{r}/\e$. The body zone corresponds to $\mathscr{r}\sim \e{\cal R}$, but to $\tilde{\mathscr{r}}\sim {\cal R}$. The outer expansion~\eqref{g expansion} is an approximation in the limit $\e\to0$ at fixed coordinate values and therefore at  fixed $\mathscr{r}$. The inner expansion is instead an approximation in the limit $\e\to0$ at fixed $\tilde{\mathscr{r}}$,
\beq
g^{\rm exact}_{\mu\nu}(\tilde{\mathscr{r}},\e) = g^{\rm obj}_{\mu\nu}(\tilde{\mathscr{r}}) + \e H^{(1)}_{\mu\nu}(\tilde{\mathscr{r}}) + \e^2 H^{(2)}_{\mu\nu}(\tilde{\mathscr{r}})+O(\e^3).
\eeq
(We suppress other coordinate dependence.) In the body zone, the coefficients $g^{\rm obj}_{\mu\nu}$ and $H^{(n)}_{\mu\nu}$ are order unity. The background metric $g^{\rm obj}_{\mu\nu}$ represents the metric of the small object's spacetime as if it were isolated, and the perturbations $H^{(n)}_{\mu\nu}$ arise from the tidal fields of the external universe and nonlinear interactions between those tidal fields and the body's own gravity.  

In our construction of the inner expansion, we have assumed that there is only one scale that sets the size of the body zone: the object's mass $m$. This implicitly assumes that the object is compact, such that its typical diameter $d$ is comparable to $m$. That in turn implies that the object's $\ell$th multipole moment scales as
\beq\label{moments scaling}
m d^\ell\sim m^{\ell+1} = \e^{\ell+1}{\cal R}^{\ell+1}.
\eeq
For a noncompact object, we would need to introduce additional perturbation parameters in the outer expansion, and additional scales in the inner expansion. 

Our inner expansion also assumes that while there is a small length scale associated with the object, there is no analogous time scale; in other words, the object is not undergoing changes on its own internal time scale $\sim m$. This is equivalent to assuming the object is in quasi-equilibrium with its surroundings. In practice it corresponds to a spatial derivative near the object dominating over a time derivative by one power of $\mathscr{r}$ (in the outer expansion) or by one power of $\e$ (in the inner expansion).

To date, inner expansions have been calculated for tidally perturbed Schwarz\-schild and Kerr black holes as well as nonrotating or slowly rotating neutron stars; see Refs.~\cite{Damour:2009vw,Binnington:2009bb,Landry:2014jka,Poisson:2014gka,Pani:2015hfa,Pani:2015nua,Landry:2015zfa,Poisson:2018qqd,LeTiec:2020bos,Poisson:2020vap} for recent examples of such work.\footnote{Ref.~\cite{Poisson:2020mdi} alerts readers to a significant error in some of the work on slowly rotating bodies.}  These calculations represent one of the major applications of the methods of black hole perturbation theory reviewed in the previous section, and they form part of an ongoing endeavour to include tidal effects in gravitational-wave templates and to infer properties of neutron stars from observed signals~\cite{Flanagan:2007ix,Yagi:2013awa}. 

However, in self-force applications we require only a minimal amount of information from the inner expansion, often much less than is provided in the above references. The necessary information is extracted from a {\em matching condition}: because the two expansions are expansions of the same metric, they must match one another when appropriately compared. The most pragmatic formulation of this condition is that the inner and outer expansions must commute. If we perform an outer expansion of the inner expansion (or equivalently, re-expand it for $\mathscr{r}\gg\e{\cal R}$), and if we perform an inner expansion of the outer expansion (or equivalently, expand for $\mathscr{r}\ll{\cal R}$), and express the end results as functions of $\mathscr{r}$, then both procedures yield a double series for small $\e$ and small $\mathscr{r}$. We assume that these two double expansions agree with one another, order by order in $\e$ and $\mathscr{r}$. A primary consequence of this matching condition is that near the small object, the metric perturbations in the outer expansions must behave as
\beq\label{hn behavior}
h^{(n)}_{\mu\nu} = \frac{h^{(n,-n)}_{\mu\nu}}{\mathscr{r}^n}+\frac{h^{(n,-n+1)}_{\mu\nu}}{\mathscr{r}^{n-1}}  +\frac{h^{(n,-n+2)}_{\mu\nu}}{\mathscr{r}^{n-2}} + \ldots,
\eeq
growing large at small $\mathscr{r}$. If $h^{(n)}_{\mu\nu}$ grew more rapidly (for example, if $h^{(n)}_{\mu\nu} \sim \frac{1}{\mathscr{r}^{n+1}}$), then the outer expansion could not match an inner expansion. Moreover, the coefficient of $\frac{1}{\mathscr{r}^n}$ matches a term in the $\mathscr{r}\gg\e{\cal R}$ expansion of $g^{\rm obj}_{\mu\nu}$: 
\beq\label{gobj behavior}
g^{\rm obj}_{\mu\nu} = \eta_{\mu\nu} + \frac{\e h^{(1,-1)}_{\mu\nu}}{\mathscr{r}} + \frac{\e^2 h^{(2,-2)}_{\mu\nu}}{\mathscr{r}^2} + \frac{\e^3 h^{(3,-3)}_{\mu\nu}}{\mathscr{r}^3} +\ldots,
\eeq
where $\eta_{\mu\nu}$ is the metric of flat spacetime. The terms in this series are in one-to-one correspondence with the multipole moments of $g^{\rm obj}_{\mu\nu}$, which in turn can be interpreted as the multipole moments of the object itself. This allows us to write $h^{(n,-n)}_{\mu\nu}$ in terms of the object's first $n$ moments; one new moment arises at each new order in $\e$, just as one would expect from the scaling~\eqref{moments scaling}. The moments, together with the general form~\eqref{hn behavior}, are all we require from the inner expansion. After it is obtained, we can effectively ``integrate out'' the body zone from the problem, as described below.

To intuitively understand the meanings of the double expansions, and of expressions such as~\eqref{hn behavior} and \eqref{gobj behavior}, we can interpret them as being valid in the {\em buffer region} $\e{\cal R}\ll \mathscr{r}\ll\cal R$. This region is the large-$\tilde{\mathscr{r}}$ limit of the body zone but the small-$\mathscr{r}$ limit of the external universe.

\subsection{Tools of local analysis}

To determine more than just the general form of the perturbations, we substitute Eq.~\eqref{hn behavior} into the Einstein equations~\eqref{EFE1}--\eqref{EFE2} and then solve order by order in $\e$ and $\mathscr{r}$. These types of local calculations are carried out using two tools: covariant near-coincidence expansions and expansions in local coordinate systems. Ref.~\cite{Poisson:2011nh} contains a thorough, pedagogical introduction to both methods. Here we summarize only the basic ingredients.


Covariant expansions are based on Synge's world function, 
\beq
\sigma(x^\alpha,x^{\alpha'}) = \frac{1}{2}\left(\int_\beta ds\right)^2,
\eeq
which is equal to 1/2 the square of the proper distance $s$ (as measured in $g_{\mu\nu}$) between the points $x^{\alpha'}$ and $x^\alpha$ along the unique geodesic $\beta$ connecting the two points; for a given $x^{\alpha'}$, this is a well-defined function of $x^\alpha$ so long as $x^\alpha$ is within the convex normal neighbourhood of $x^{\alpha'}$. The other necessary tool is the bitensor $g_{\mu}^{\mu'}(x^\alpha,x^{\alpha'})$, which parallel propagates vectors from $x^{\alpha'}$ to $x^\alpha$. A smooth tensor field $A_\mu{}^\nu$ at $x^\alpha$ can be expanded around $x^{\alpha'}$ as
\beq
A_\mu{}^\nu(x^\alpha) = g_\mu^{\mu'}g^\nu_{\nu'}\left[A_{\mu'}{}^{\nu'} - A_{\mu'}{}^{\nu'}{}_{;\alpha'}\sigma^{\alpha'} + \tfrac{1}{2}A_{\mu'}{}^{\nu'}{}_{;\alpha'\beta'}\sigma^{\alpha'}\sigma^{\beta'}+O(\lambda^3)\right],\label{covariant expansion}
\eeq
where we use $\lambda:=1$ to count powers of distance between $x^{\alpha'}$ and $x^\alpha$. The vector $\sigma_{\alpha'}:=\nabla_{\!\alpha'}\sigma$ is tangent to $\beta$ and has a magnitude $\sqrt{2\sigma}$ equal to the proper distance between $x^{\alpha'}$ and $x^\alpha$. The (perhaps unexpected) minus sign in Eq.~\eqref{covariant expansion} arises because $\sigma_{\alpha'}$ points {\em away} from $x^\alpha$ rather than toward it. 

When a derivative, either at $x^\alpha$ or at $x^{\alpha'}$, acts on an expansion like~\eqref{covariant expansion}, it involves derivatives of $g^{\mu}_{\mu'}$ and $\sigma_{\alpha'}$. These can then be re-expanded using, for example, 
\beq
g^{\mu}_{\mu';\nu} = \frac{1}{2}g^\mu_{\rho'}g_{\nu}^{\nu'}R^{\rho'}{}_{\!\!\mu'\nu'\delta'}\sigma^{\delta'}+O(\lambda^2)
\eeq
and 
\beq
\sigma_{;\mu\mu'} = -g_\mu^{\nu'}\left(g_{\mu'\nu'}+\tfrac{1}{6}R_{\mu'\alpha'\nu'\beta'}\sigma^{\alpha'}\sigma^{\beta'}\right)+O(\lambda^3);
\eeq
see Eqs.~(6.7)--(6.11) of Ref.~\cite{Poisson:2011nh}.

To make use of these tools, we install a curve $\gamma$, with coordinates $z^\alpha(\tau)$, in the spacetime of $g_{\mu\nu}$, which will be a representative worldline for the small object. Recall that $\tau$ is proper time as measured in $g_{\mu\nu}$. If the object is a material body, the worldline will be in its physical interior. If the object is a black hole, the worldline will only serve as a reference point for the field {\em outside} the black hole; mathematically, $\gamma$ resides in the manifold on which $g_{\mu\nu}$ lives, not the manifold on which $g^{\rm obj}_{\mu\nu}$ lives. In either case, we only analyze the metric in the object's exterior, never in its interior. 

A suitable measure of distance from $\gamma$ is
\beq
\s(x^\alpha,x^{\alpha'}):=\sqrt{P_{\mu'\nu'}\sigma^{\mu'}\sigma^{\nu'}}, 
\eeq
where $x^{\alpha'}=z^\alpha(\tau')$ is a point on $\gamma$ near $x^\alpha$, and $P_{\mu\nu}:=g_{\mu\nu}+u_\mu u_\nu$ projects orthogonally to $\gamma$. Here we have introduced $\gamma$'s four-velocity $u^\mu=\frac{dz^\mu}{d\tau}$, normalized to $g_{\mu\nu}u^\mu u^\nu=-1$. Note that $\s$ remains positive regardless of whether $x^{\alpha'}$ and $x^{\alpha}$ are connected by a spacelike, timelike, or null geodesic. In terms of these covariant quantities, the expansion~\eqref{hn behavior} can be written more concretely as
\beq\label{hn covariant}
h^{(n)}_{\mu\nu}(x^{\alpha}) = g_\mu^{\mu'}g_\nu^{\nu'}\left[\frac{h^{(n,-n)}_{\mu'\nu'}(x^{\alpha'},\sigma^{\alpha'}/\s)}{\s^n}+\frac{h^{(n,-n+1)}_{\mu'\nu'}(x^{\alpha'},\sigma^{\alpha'}/\s)}{\s^{n-1}}+O(\lambda^{-n+2})\right]\!.
\eeq
$\s$ represents the distance from $x^{\alpha'}$ to $x^{\alpha}$ [playing the role of $\mathscr{r}$ in~\eqref{hn behavior}], and the vector $\sigma^{\alpha'}/\s$ represents the direction of the geodesic connecting $x^{\alpha'}$ to $x^{\alpha}$. Generically, $\log(\s)$ terms also appear~\cite{Pound:2012dk}, but we suppress them for simplicity.

Rather than directly substituting an ansatz of the form~\eqref{hn covariant} into the vacuum field equations and solving for the coefficients $h^{(n,p)}_{\mu'\nu'}$, it is typically more convenient to adopt a local coordinate system centred on $\gamma$ and afterward recover~\eqref{hn covariant}. Here we adopt Fermi-Walker coordinates $(\tau,x^a)$, which are quasi-Cartesian coordinates constructed from a tetrad $(u^\alpha,e^\alpha_a)$ on $\gamma$. The spatial triad $e^\alpha_a$ is Fermi-Walker transported along the worldline according to
\begin{equation}
\frac{De^\alpha_a}{d\tau}=a_au^\alpha,
\end{equation}
where $\frac{D}{d\tau} :=u^\mu \nabla_\mu$. $a_a:= a_\mu e^\mu_a$ is a spatial component of the covariant acceleration $a^\mu:=\frac{Du^\mu}{d\tau}$; this will eventually become the left-hand side of Eq.~\eqref{perturbed geodesic equation}. At each value of proper time $\tau$, we send a space-filling family of geodesics orthogonally outward from $\bar x^{\alpha}=z^{\alpha}(\tau)$, generating a spatial hypersurface $\Sigma_{\tau}$. Each such surface is labelled with a coordinate time $\tau$, and each point on the surface is labeled with spatial coordinates
\begin{equation}\label{xa_def}
x^a=-e^a_{\bar \alpha}\sigma^{\bar \alpha},
\end{equation}
where $\sigma_{\bar\alpha}:=\nabla_{\!\bar\alpha}\sigma$ is tangent to $\Sigma_{\bar\tau}$, satisfying $\sigma_{\bar\alpha} u^{\bar\alpha} = 0$. The magnitude of these coordinates, given by $s := \sqrt{\delta_{ab}x^ax^b} = \sqrt{g_{\bar\alpha\bar\beta}\sigma^{\bar\alpha}\sigma^{\bar\beta}}$, is the proper distance from $\bar x^{\alpha}$ to $x^{\alpha}$. In the special case that $x^{\alpha'}=\bar x^{\alpha}$, $\s$ and $s$ are identical. The analog of Eq.~\eqref{covariant expansion} is the coordinate Taylor series
\beq
A_\mu{}^\nu(\tau,x^a) = A_\mu{}^\nu(\tau,0) +  A_\mu{}^\nu{}_{,a}(\tau,0)x^a +  \tfrac{1}{2}A_\mu{}^\nu{}_{,ab}(\tau,0)x^a x^b + O(s^3).
\eeq

In these coordinates, the four-velocity reduces to $u^\mu=(1,0,0,0)$, and the acceleration to $a^\mu=(0,a^i)$. The external background metric, which is smooth at $x^a=0$, is given by
\begin{subequations}\label{FW metric}
\begin{align}
g_{\tau\tau} &= -1-2a_ix^i-\left(R_{\tau i\tau j}+a_ia_j\right)x^ix^j+O(s^3),\\
g_{\tau a} &= -\tfrac{2}{3}R_{\tau iaj}x^ix^j+O(s^3),\\
g_{ab} &= \delta_{ab}-\tfrac{1}{3}R_{aibj}x^ix^j+O(s^3),
\end{align}
\end{subequations}
reducing to the Minkowski metric on $\gamma$, and the only nonzero Christoffel symbols on $\gamma$ are $\Gamma^a_{\tau\tau}=a^a$ and $\Gamma^\tau_{\tau a}=a_a$. If the worldline is not accelerating, the coordinates become inertial along $\gamma$. 

The Riemann tensor components in Eq.~\eqref{FW metric} are evaluated on the worldline. Higher powers of $x^a$ in the expansion come with higher powers of the acceleration, derivatives of the Riemann tensor, and nonlinear combinations of the Riemann tensor. In a vacuum background, the Riemann tensor on the worldline is commonly decomposed into tidal moments. The quadrupolar moments are defined as
\begin{align}
{\cal E}_{ab} &:= R_{\tau a\tau b}, \label{Eab}\\
{\cal B}_{ab} &:= \tfrac{1}{2}\epsilon^{pq}{}_{(a}R_{b)\tau pq}. \label{Bab}
\end{align}
Higher moments involve derivatives of the Riemann tensor. Equations~(44)--(48) of Ref.~\cite{Pound:2014xva} display the background metric~\eqref{FW metric} through order $s^3$ and the octupolar tidal moments. Ref.~\cite{Poisson:2009qj} presents the background metric in an alternative, lightcone-based coordinate system through order $\lambda^4$ and the hexadecapolar moments.

Given the local Fermi-Walker coordinates, one can adopt a coordinate analog of Eq.~\eqref{hn covariant},
\beq\label{hn FW}
h^{(n)}_{\mu\nu} = \frac{h^{(n,-n)}_{\mu\nu}(\tau,n^a)}{s^{n}} + \frac{h^{(n,-n+1)}_{\mu\nu}(\tau,n^a)}{s^{n-1}} + \frac{h^{(n,-n+1)}_{\mu\nu}(\tau,n^a)}{s^{n-2}} +O(s^{-n+3}).
\eeq
Here $n^a = \frac{x^a}{s}=\delta^{ab}\partial_b s$ is a radial unit vector. To facilitate solving the field equations, we can expand the coefficients in angular harmonics:
\beq
h^{(n,p)}_{\mu\nu}(\tau,n^a) = \sum_{\ell\geq0}h^{(n,p,\ell)}_{\mu\nu L}(\tau)\hat n^L,
\eeq
where $L:=i_1\cdots i_\ell$ is a multi-index, and $\hat n^L:=n^{\langle L\rangle}$, where $n^L:=n^{i_1}\cdots n^{i_\ell}$. The angular brackets denote the symmetric, trace-free (STF) combination of indices, where the trace is defined with $\delta_{ab}$. This is equivalent to expanding the coefficients $h^{(n,p)}_{\mu\nu}$ in scalar spherical harmonics:
\beq
h^{(n,p)}_{\mu\nu}(\tau,n^a) = \sum_{\ell=0}^\infty\sum_{m=-\ell}^\ell h^{(n,p,\ell \emm)}_{\mu\nu}(\tau)Y^{\ell \emm}(\vartheta,\varphi),
\eeq
where the angles $(\vartheta,\varphi)$ are defined in the natural way from 
\beq
n^a=(\sin\vartheta\cos\varphi,\sin\vartheta\sin\varphi,\cos\vartheta).
\eeq
Like spherical harmonics, $\hat n^L$ is an eigenfunction of the flat-space Laplacian, satisfying $\delta^{ab}\partial_a\partial_b \hat n^L = -\frac{\ell(\ell+1)}{s^2}\hat n^L$. One can further decompose $h^{(n,p,\ell)}_{\mu\nu L}$ into irreducible STF pieces that are in one-to-one correspondence with the coefficients in a tensor spherical harmonic decomposition. We refer the reader to Appendix A of Ref.~\cite{Blanchet:1985sp} for a detailed introduction to such expansions and a collection of useful identities.

The general local solution in the buffer region can be found by substituting the expansions~\eqref{FW metric} and~\eqref{hn FW} into the vacuum field equations and working order by order in $\e$ and $s$. Because spatial derivatives increase the power of $1/s$, dominating over $\tau$ derivatives, this process reduces to solving a sequence of stationary field equations. 

An alternative approach is to instead solve for the perturbations $H^{(n)}_{\mu\nu}$ in the inner expansion, starting with a large-$\tilde{\mathscr{r}}$ ansatz complementary to Eq.~\eqref{hn behavior}, and then translate the results into the small-$\mathscr{r}$ expansions for $h^{(n)}_{\mu\nu}$. This approach can draw on existing, high-order inner expansions (e.g., Refs.~\cite{Poisson:2009qj,Poisson:2018qqd,Poisson:2020vap}), though doing so often requires transformations of the coordinates and of the perturbative gauge to arrive at a practical form for the outer expansion (see, e.g., Ref.~\cite{Pound:2017psq}).

\subsection{Local solution: self-field and an effective external metric}

The general solutions for $h^{(1)}_{\mu\nu}$ and $h^{(2)}_{\mu\nu}$ in the buffer region are known to varying orders in $\e$ and $\mathscr{r}$ in a variety of gauges, including classes of ``rest gauges'' (terminology from Ref.~\cite{Pound:2017psq}), ``P smooth'' gauges~\cite{Gralla:2012db}, ``highly regular'' gauges~\cite{Pound:2017psq} (in which no $1/s^2$ term appears in $h^{(2)}_{\mu\nu}$), radiation gauges~\cite{Pound:2013faa}, and the Regge-Wheeler-Zerilli gauge~\cite{Thompson:2018lgb}. (In the last two cases, the gauge choices are restricted to particular classes of external backgrounds.) However, nearly all covariant expressions, and expansions to the highest order in $\mathscr{r}$, are in the Lorenz gauge. Ref.~\cite{Pound:2012dk} provides an algorithm for generating the local solution in the Lorenz gauge, and a large class of similar gauges, to arbitrary order in $\e$.

In all gauges, the general solution is typically divided into two pieces:
\beq\label{S-R split}
h^{(n)}_{\mu\nu} = h^{\S(n)}_{\mu\nu} + h^{\R(n)}_{\mu\nu}.
\eeq
This is akin to the usual split of a general solution into a particular and a homogeneous solution. $h^{\S}_{\mu\nu} = \sum_n\e^n h^{\S(n)}_{\mu\nu}$ is the object's {\em self-field}, encoding all the local information about the object's multipole structure (including the entirety of $g^{\rm obj}_{\mu\nu}$). Although this field is defined only outside the object, it would be singular at $s=0$ if the expansion~\eqref{hn FW} were taken to hold for all ${\cal R}\gg s>0$; it contains all the negative powers of $s$ in~\eqref{hn FW}, as well as all non-negative powers of $s$ with finite differentiability (e.g., all terms proportional to $s^p n^L$ with $p\geq0$ but $p\neq\ell$). For that reason, it is also known as the {\em singular field}.

The second piece of the general solution, $h^\R_{\mu\nu}=\sum_n\e^n h^{\R(n)}_{\mu\nu}$, encodes effectively {\em external} information linked to global boundary conditions. It takes the form of a power series, 
\beq\label{hRn FW}
h^{\R(n)}_{\mu\nu} = \sum_{\ell} c^{(n)}_{\mu\nu L}(\tau)x^L; 
\eeq
unlike $h^\S_{\mu\nu}$, which involves the locally determined multipole moments, every coefficient $c^{(n)}_{\mu\nu L}$ is an unknown that can only be determined when external boundary conditions are imposed. Although, once again, the field is defined {\em outside} the object, we can identify $\sum_{\ell} c^{(n)}_{\mu\nu L}(\tau)x^L$ with a Taylor series, where the coefficients $c^{(n)}_{\mu\nu L}(\tau)=\frac{1}{\ell!}\partial_L h^{\R(n)}_{\mu\nu}(\tau,0)$ define $h^{\R(n)}_{\mu\nu}$ and its derivatives on the worldline. Moreover, $h^{\R}_{\mu\nu}$ can be combined with the external background to form an {\em effective metric}
\beq
\breve{g}_{\mu\nu} := g_{\mu\nu} + h^\R_{\mu\nu}
\eeq
that is a vacuum solution, satisfying
\beq
G_{\mu\nu}[\breve g] = 0
\eeq
even on $\gamma$. $\breve{g}_{\mu\nu}$ characterizes the object's rest frame and local tidal environment. Because $h^{\R}_{\mu\nu}$ is smooth at $x^a=0$, it is also referred to as the {\em regular field}.

This type of division of the local solution into $h^\S_{\mu\nu}$ and $h^{\R}_{\mu\nu}$ was first emphasized by Detweiler and Whiting at first order in $\e$~\cite{Detweiler:2000gt,Detweiler:2002mi}. There is considerable freedom in the specific division, as smooth vacuum perturbations can be interchanged between the two pieces, and multiple distinct choices have been made in practice, particularly beyond linear order~\cite{Rosenthal:2006nh,Harte:2011ku,Pound:2012nt,Gralla:2012db}. However, one can always choose the division such that (i) $\breve{g}_{\mu\nu}$ is a smooth vacuum metric, and (ii) $\breve{g}_{\mu\nu}$ is effectively the ``external'' metric, in the sense that the object moves as a test body in it, as described in the next section. Here for concreteness we adopt the choice introduced in Ref.~\cite{Pound:2012nt} (see also Refs.~\cite{Pound:2012dk,Pound:2014xva,Pound:2015tma}), and we provide the explicit forms of the first- and second-order self-fields in the Lorenz gauge, as presented in Ref.~\cite{Pound:2014xva}.

For the purpose of explicitly displaying factors of the object's multipole moments, from this point forward we take $\e$ to be a formal counting parameter that can be set equal to unity.

At first order, the self-field is determined by the object's mass. It is given in Fermi-Walker coordinates by
\begin{subequations}\label{hS1 FW}
\begin{align}
h^{\S(1)}_{\tau\tau} &= \frac{2m}{s}+3ma_i n^i+\tfrac{5}{3} ms\mathcal{E}_{ab}\hat n^{ab} + O(s^2),\\
h^{\S(1)}_{\tau a} &= 2ms\left(\tfrac{1}{3}\mathcal{B}^{bc} \epsilon_{acd}\hat n_{b}{}^{d}-\dot a_a\right)+O(s^2),\\
h^{\S(1)}_{ab} &= \frac{2m\delta_{ab}}{s}-m\delta_{ab}a_in^i+ms\left(
								\tfrac{4}{3} \mathcal{E}_{(a}{}^{c} \hat{n}_{b)c} 
								- \tfrac{38}{9} \mathcal{E}_{ab} 
								- \mathcal{E}_{cd}\delta_{ab} \hat{n}^{cd}\right)+O(s^2)
\end{align}
\end{subequations}
and in covariant form by
\begin{align}\label{hS1 covariant}
h^{\S(1)}_{\mu\nu} &=\frac{2 m}{\lambda\s} g^{\alpha'}_{\mu} g^{\beta'}_{\nu} \left(g_{\alpha'\beta'} + 2 u_{\alpha'} u_{\beta'}\right) +\frac{m\lambda^0}{\s^3} g^{\alpha'}_\mu g^{\beta'}_{\nu} \left[
														\left(\s^2- \r^2\right) a_{\sigma} (g_{\alpha'\beta'}+2u_{\alpha'} u_{\beta'})\right.\nonumber\\
												&\quad \left.+8\r\s^2 a_{(\alpha'} u_{\beta')}\right] 
 + \lambda\frac{mg^{\alpha'}_\mu g^{\beta'}_{\nu}}{3\s^3} \Big[\!\left(\r^2 - \s^2\right)
			\left(g_{\alpha'\beta'}+2u_{\alpha'} u_{\beta'}\right)R_{u\sigma u\sigma} \nonumber\\
&\quad - 12\s^4 R_{\alpha' u\beta' u}- 12\r\s^2 u_{(\alpha'}R_{\beta')u\sigma u}+12\s^2 (\r^2 + \s^2)\dot{a}_{(\alpha'}u_{\beta')}\nonumber\\
												&\quad
													 +  \r (3\s^2-\r^2)\dot{a}_{\sigma}(g_{\alpha'\beta'}+ 2u_{\alpha'} u_{\beta'})\Big] +O(\lambda^2),
\end{align}
where $x^{\alpha'}$ is an arbitrary point on $\gamma$ near the field point $x^{\alpha}$. In the covariant expressions we have adopted the notation $a_\sigma:=a_{\alpha'}\sigma^{\alpha'}$, $R_{u\sigma u\sigma}:=R_{\mu'\alpha'\nu'\beta'}u^{\mu'}\sigma^{\alpha'}u^{\nu'}\sigma^{\beta'}$, etc. The quantity $\r:=u_{\mu'}\sigma^{\mu'}$ is a measure of the proper time between $x^{\alpha'}$ and $x^\alpha$.  Equations~\eqref{hS1 FW} and \eqref{hS1 covariant} are given in Ref.~\cite{Pound:2014xva} through order $\lambda^2$. Equation~(4.7) of Ref.~\cite{Heffernan:2012su} presents the covariant expansion of $h^{\S(1)}_{\mu\nu}$ through order $\lambda^4$ (omitting acceleration terms).

At second order, the self-field involves both the mass and spin of the object. It can be written as the sum of three pieces,\footnote{Ref.~\cite{Pound:2014xva} further divides $h^{\S\R}_{\mu\nu}$ into two pieces, labeled $h^{\S\R}_{\mu\nu}$ and $h^{\delta m}_{\mu\nu}$.}
\begin{equation}\label{hS2}
h^{\S(2)}_{\mu\nu}=h^{\S\S}_{\mu\nu}+h^{\S\R}_{\mu\nu}+h^{\rm spin}_{\mu\nu}.
\end{equation}
The spin contribution is
\beq
h^{\rm spin}_{\tau a} = \frac{2S_{ai}n^i}{s^2}+O(s^0),
\eeq
where other components are $O(s^0)$, and where $S_{ab}=\epsilon_{abi}S^i$ is the spin tensor and $S^i$ the spin vector. The other two pieces are either quadratic in the mass,
\begin{subequations}\label{hSS FW}%
\allowdisplaybreaks\begin{align}
h^{\S\S}_{\tau\tau} &= -\frac{2m^2}{s^2} - \tfrac{7}{3}m^2\mathcal{E}_{ab} 
							\hat{n}^{ab}+O(s\ln s),\\
h^{\S\S}_{\tau a} &= - \tfrac{10}{3} m^2\mathcal{B}^{bc} \epsilon_{acd} \hat{n}_{b}{}^{d} +O(s\ln s),\\
h^{\S\S}_{ab} &= \frac{\tfrac{8}{3}m^2\delta_{ab} - 7m^2\hat{n}_{ab}}{s^2}
							+ m^2\left(4 \mathcal{E}_{c(a} \hat{n}_{b)}{}^{c} - \tfrac{4}{3} \mathcal{E}_{cd} \delta_{ab} \hat{n}^{cd} + \tfrac{7}{5}\mathcal{E}_{cd} \hat{n}_{ab}{}^{cd}\right)\nonumber\\
						&\quad-\tfrac{16}{15}m^2\mathcal{E}_{ab}\ln s +O(s\ln s),
\end{align}
\end{subequations}
or involve products of the mass with the regular field,
\begin{subequations}\label{hSR FW}%
\begin{align}
h^{\S\R}_{\tau\tau} &= -\frac{m}{s}\left(h^{\R1}_{ab} \hat{n}^{ab}+ \tfrac{1}{3}  h^{\R1}_{ab}\delta^{ab}+ 2 h_{\tau\tau}^{\R1}\right)+O(s^0),\\
h^{\S\R}_{\tau a} &= -\frac{m}{s}\left(h_{\tau b}^{\R1}\hat{n}_a{}^b+ \tfrac{4}{3}h_{\tau a}^{\R1}\right)+O(s^0),\\
h^{\S\R}_{ab} &= \frac{m}{s}\Big[2h^{\R1}_{c(a}\hat{n}_{b)}{}^c -\delta_{ab} h^{\R1}_{cd} \hat{n}^{cd}  - \left(h^{\R1}_{ij}\delta^{ij}+h_{\tau\tau}^{\R1}\right)\hat{n}_{ab}\nonumber\\
						&\quad +\tfrac{2}{3} h^{\R1}_{ab} + \tfrac{1}{3}\delta_{ab} h^{\R1}_{cd}\delta^{cd} + \tfrac{2}{3} \delta_{ab} h_{\tau\tau}^{\R1}\Big] +O(s^0).
\end{align}
\end{subequations}
On the right, the components of $h^{\R1}_{\mu\nu}$ are evaluated at $s=0$. At order $s^0$, $h^{\S\R}_{\mu\nu}$ also depends on first derivatives of $h^{\R1}_{\mu\nu}$ evaluated at $s=0$; at order $s$, it depends on second derivatives of $h^{\R1}_{\mu\nu}$ evaluated at $s=0$; and so on.  $h^{\S(2)}_{\alpha\beta}$ is given in Fermi coordinates through order $s$ in Appendix D of Ref.~\cite{Pound:2012dk}. 

In covariant form, these fields are
\beq
h^{\rm spin}_{\mu\nu} = \frac{4g^{\alpha'}_{\mu} g^{\beta'}_{\nu}u_{(\alpha'}S_{\beta')\gamma'}\sigma^{\gamma'}}{\lambda^2\s^3} + O(\lambda^0),
\eeq
with $S_{\alpha'\beta'}:=e^a_{\alpha'}e^b_{\beta'}S_{ab}$,
\begin{align}\label{hSS covariant}
h^{\S\S}_{\mu\nu}&= \frac{m^2}{\lambda^2\s^4} g^{\alpha'}_{\mu} g^{\beta'}_{\nu} \Bigl\{ 5\s^2g_{\alpha'\beta'}
				-7 \sigma_{\alpha'}\sigma_{\beta'} - 14 \r \sigma_{(\alpha'} u_{\beta')}- (7 \r^2 - 3 \s^2) u_{\alpha'} u_{\beta'}\bigr]\Bigr\} \nonumber\\
&\quad
				- \frac{16}{15} m^2 g^{\alpha'}_{\mu} g^{\beta'}_{\nu} \ln(\lambda\s) R_{\alpha' u\beta' u}  + \frac{m^2\lambda^0}{150 \s^6} g^{\alpha'}_{\mu} g^{\beta'}_{\nu} 
				\bigg\{10\s^2 g_{\alpha'\beta'}\left(25 \r^2+\s^2\right)R_{\sigma u\sigma u}\nonumber\\
&\quad
			+ 20 \r \s^2 \left[35 \r \sigma_{(\alpha'} R_{\beta')u \sigma u} 
			+ \left(35 \r^2 - 31 \s^2\right) u_{(\alpha'} R_{\beta')u\sigma u} 
			- \s^2 R_{\sigma(\alpha' \beta') u}\right] \nonumber\\
&\quad + 10 \s^4 R_{\alpha' \sigma \beta' \sigma} - 350 \r\s^2\sigma_{(\alpha'}R_{\beta')\sigma u\sigma} 
			- 10\s^2\bigl(35 \r^2 - 17 \s^2\bigr) u_{(\alpha'}R_{\beta')\sigma u\sigma} \nonumber\\
&\quad
			+ 2\s^4 \left(5 \r^2 + 26 \s^2\right)R_{\alpha' u\beta' u} - 70 \bigl[\left(10 \r^2 - 3 \s^2\right) \sigma_{\alpha'}\sigma_{\beta'}  \nonumber\\
&\quad
			+ 4 \r \left(5 \r^2 - 4 \s^2\right) u_{(\alpha'}\sigma_{\beta')}\bigr] R_{\sigma u\sigma u}  
			- 20\left(35 \r^4 - 53 \r^2 \s^2 - 6 \s^4\right) u_{\alpha'}u_{\beta'} R_{\sigma u\sigma u} \bigg\}\nonumber\\
&\quad	+O\left(\lambda\ln \lambda\right),
\end{align}
and
\begin{align}\label{hSR covariant}
h^{\S\R}_{\mu\nu} &= \frac{m}{\lambda\s^3} g^{\alpha'}_{\mu} g^{\beta'}_{\nu} \Biggl\{g_{\alpha'\beta'}\left[\frac{2}{3}\s^2h^{\R1}_{\mu'\nu'}g^{\mu'\nu'} 
			- \left(\r^2 -  \s^2\right)h^{\R1}_{uu} - h^{\R1}_{\sigma\sigma} - 2 \r h^{\R1}_{u\sigma}\right]\nonumber\\
			&\quad+\s^2\delta m_{\alpha'\beta'} -\frac{2}{3} h^{\R1}_{\alpha' \beta'} \s^2 
			+ 2h^{\R1}_{\sigma(\alpha'} \sigma_{\beta')} + 2\r h^{\R1}_{\sigma(\alpha'}u_{\beta')} - 2 h^{\R1}_{\sigma\sigma} u_{\alpha'} u_{\beta'} \nonumber\\
			&\quad -  h^{\R1}_{\mu'\nu'} g^{\mu'\nu'}\Bigl[\sigma_{\alpha'}\sigma_{\beta'} + 2\r \sigma_{(\alpha'}u_{\beta')} 
			+ (\r^2 - \s^2) u_{\alpha'}u_{\beta'}\Bigr] + 2\r h^{\R1}_{u(\alpha'}\sigma_{\beta')} 
			 \nonumber\\
&\quad+ 2(\r^2-\s^2)h^{\R1}_{u(\alpha'}u_{\beta')}	+ 4 h^{\R1}_{u\sigma} \sigma_{(\alpha'} u_{\beta')}- 2 h^{\R1}_{uu} \sigma_{\alpha'} \sigma_{\beta'}\Biggr\}  + O\left(\lambda^0\right),
\end{align}
where
\begin{align}\label{dm_SC_cov}
\delta m_{\alpha\beta} &= \frac{1}{3}m\left(2h^{\R1}_{\alpha\beta}+g_{\alpha\beta}g^{\mu\nu}h^{\R1}_{\mu\nu}\right)
				+4mu_{(\alpha}h^{\R1}_{\beta)\mu}u^\mu\nonumber\\
&\quad +m(g_{\alpha\beta}+2u_{\alpha} u_{\beta})u^\mu u^\nu h^{\R1}_{\mu\nu}.
\end{align}
The covariant expressions for $h^{\S\S}_{\mu\nu}$ and $h^{\S\R}_{\mu\nu}$ are known through order $\lambda$~\cite{Pound:2014xva} and are available upon request to the authors. The covariant expansion of $h^{\rm spin}_{\mu\nu}$ appears explicitly here for the first time, but it is known to higher order in $\lambda$~\cite{Mathews:2020}.

In this section, we have stated results from the so-called {\em self-consistent} expansion of the metric~\cite{Pound:2009sm}. In this framework, the metric is not expanded in an ordinary Taylor series in $\e$. Instead, it takes the form
\beq
g^{\rm exact}_{\mu\nu}(x^\alpha,\e) = g_{\mu\nu}(x^\alpha) + \e h^{(1)}_{\mu\nu}(x^\alpha,{\cal P}) + \e^2 h^{(2)}_{\mu\nu}(x^\alpha,{\cal P}) + O(\e^3),\label{self-consistent expansion}
\eeq
where ${\cal P}$ represents a list of system parameters: the worldline $\gamma$ and multipole moments of the small object, along with any evolving external parameters. If the small object is orbiting a black hole that is approximately Kerr, the external parameters will consist of small, slowly evolving corrections to the black hole's mass and spin~\cite{Miller:2020bft}. These parameters all evolve with time in an $\e$-dependent way, meaning that Eq.~\eqref{self-consistent expansion} is not a Taylor series; this allowance for $\e$-dependent coefficients is a hallmark of singular perturbation theory~\cite{Kevorkian-Cole:96}. In the self-force problem, it must be allowed in order to construct a uniformly accurate approximation on large time scales~\cite{Pound:2010pj}. It will lead naturally into the multiscale expansion described in the later sections of this review.

If we use an ordinary Taylor series in place of Eq.~\eqref{self-consistent expansion}, then $z^\mu$ is replaced with the series expansion $z^\mu(\tau,\e) = z^\mu_0(\tau) + \e z^\mu_1 (\tau) + \ldots$ (referred to as a Gralla-Wald expansion after the authors of Ref.~\cite{Gralla:2008fg}). Here $z^\mu_0$ is a geodesic of the external background spacetime, and the local analysis described above is carried out with series expansions in powers of distance from this geodesic. The acceleration $a^\mu$ in this approach is thus set to zero in all the above formulas. The $\e$ dependence of $z^\mu$ then manifests itself in $h^{\S(2)}_{\mu\nu}$ through an additional term,
\begin{align}\label{dipole term}
h^{\rm dipole}_{\mu\nu} &= \frac{2m_i n^i(g_{\mu\nu}+2u_\mu u_\nu)}{s^2}+O(1/s)\\
										&= g^{\alpha'}_{\mu} g^{\beta'}_{\nu}\left[-\frac{2m_{\mu'} \sigma^{\mu'}}{\lambda^2\s^3}(g_{\alpha'\beta'}+2u_{\alpha'} u_{\beta'})+O(1/\lambda)\right],
\end{align}
proportional to a mass dipole moment $m^\alpha = e_a^\alpha m^a = m z^\alpha_1$. $m^\alpha$ describes the position of the object's center of mass relative to $z^\mu_0$. It appears in the second-order metric perturbation in the outer expansion but in the zeroth-order inner metric, $g^{\rm obj}_{\mu\nu}$. By setting $m_i$ to zero in the self-consistent expansion, one defines $\gamma$ to be the center of mass at this order. A correction to $m_i$ generically appears in $h^{(3)}_{\mu\nu}$ and in $g^{\rm obj}_{\mu\nu}+\e H^{(1)}_{\mu\nu}$, and it is likewise set to zero in a self-consistent expansion~\cite{Pound:2012nt,Pound:2017psq}. In a Gralla-Wald expansion, $m_i$ and corrections to it are allowed to be  nonzero; for that case, $h^{\rm dipole}_{\mu\nu}$ is given through order $\lambda^0$ in Fermi coordinates in Sec.~IVC of Ref.~\cite{Pound:2009sm} (where $m_i$ is denoted $M_i$). Explicit expressions through order $\lambda$, in both Fermi-coordinate and covariant form, are known through order $\lambda$~\cite{Pound:2014xva} and are available upon request. 

In the context of a binary, the small object inspirals, eventually moving very far from any initially nearby background geodesic. This causes $z^\mu_1$ and higher corrections to grow large with time, spelling the breakdown of the Gralla-Wald expansion. For this reason, we have focused on the self-consistent formulation in this review. Refs.~\cite{Pound:2009sm,Pound:2015fma,Pound:2015tma} provide detailed explications of the relationship between the two types of expansions.

\subsection{Equations of motion}

Along with the local form of the metric perturbations, the Einstein equations determine the motion of the small object and the evolution of its mass and spin. Specifically, if we let $\gamma$ be the object's center of mass (by setting the mass dipole moment in $h^{(2)}_{\mu\nu}$ to zero), then the vacuum field equations uniquely determine the first-order equations of motion~\cite{Mino:1996nk,Gralla:2008fg,Pound:2009sm}
\beq
\frac{D^2 z^\alpha}{d\tau^2} = -\frac{1}{2}P^{\alpha\delta}\!\left(2h^{\R(1)}_{\delta\beta;\gamma}-h^{\R(1)}_{\beta\gamma;\delta}\right)\!u^\beta u^\gamma -\frac{1}{2m}R^\alpha{}_{\beta\gamma\delta}u^\beta S^{\gamma\delta}+ O(\e^2)\label{EOM spin}
\eeq
and
\beq\label{mass and spin}
\frac{dm}{d\tau} = O(\e^2) \quad \text{and}\quad \frac{DS^{\alpha\beta}}{d\tau} = O(\e^3).
\eeq
The first term on the right of Eq.~\eqref{EOM spin} is referred to as the first-order gravitational self-force (per unit mass) or as the MiSaTaQuWa force (after the authors of Refs.~\cite{Mino:1996nk,Quinn:1996am}); the second term on the right is the Mathisson-Papapetrou spin force~\cite{Mathisson:1937zz,Papapetrou:1951pa}. 
 
Equation~\eqref{EOM spin} represents the leading correction to geodesic motion for a gravitating, extended, compact object.\footnote{For a non-compact object, finite-size effects from higher multipole moments will dominate over self-force effects.} However, these equations are equivalent to those of a test body, not in the background or in the physical spacetime but in the effective metric $\breve{g}_{\mu\nu}$. In particular, Eq.~\eqref{EOM spin} can be rewritten as
\beq
\frac{\breve{D}^2 z^\alpha}{d\breve{\tau}^2} = -\frac{1}{2m}\breve{R}^\alpha{}_{\beta\gamma\delta}\breve{u}^\beta S^{\gamma\delta}+ O(\e^2),
\eeq
where $\breve{\tau}$ is proper time in $\breve g_{\mu\nu}$,  $\frac{\breve{D}}{d\breve\tau}:=\breve{u}^\alpha \breve\nabla_{\!\alpha}$, $\breve\nabla$ is a covariant derivative compatible with $\breve g_{\mu\nu}$, and $\breve u^\mu = \frac{dz^\mu}{d\breve\tau}$. This is the equation of motion of a spinning test particle. Similarly, the evolution equations~\eqref{mass and spin} are the equations of a test mass and spin, which are constant and parallel propagated, respectively.

If we specialize to a spherical, nonspinning object (and set the subleading mass dipole moment to zero), the field equations determine the second-order equation of motion~\cite{Pound:2012nt,Pound:2017psq}
\beq
\frac{D^2 z^\alpha}{d\tau^2} = -\frac{1}{2}P^{\alpha\mu}\left(g_\mu{}^\delta-h^{\R\ \delta}_\mu\right)\!\left(2h^{\R}_{\delta\beta;\gamma}-h^{\R}_{\beta\gamma;\delta}\right)\!u^\beta u^\gamma +O(\e^3)\label{EOM2}
\eeq
and $\frac{dm}{d\tau} = O(\e^3)$. This can be rewritten as the geodesic equation in $\breve{g}_{\mu\nu}$,
\beq\label{EOM2 v2}
\frac{\breve{D}^2 z^\mu}{d\breve{\tau}^2} = O(\e^3).
\eeq
See Sec.~IIIA of Ref.~\cite{Pound:2015fma} for the (simple) steps involved in rewriting Eq.~\eqref{EOM2} as Eq.~\eqref{EOM2 v2}.

For a generic compact object, the spin and quadrupole moments will both appear in Eq.~\eqref{EOM2}. Although the second-order equations of motion have not been derived directly from the field equations in that case, it is known that at least through this order, the motion remains that of a test body in {\em some} effective metric~\cite{Thorne:1984mz}. At least for a material body, this remains true even in the fully nonlinear setting~\cite{Harte:2011ku}. The spin's evolution and its contribution to the acceleration  through second order (but omitting quadratic-in-spin terms), extracted from the nonlinear results for a material body, are given in Eq.~(6) of Ref.~\cite{Mathews:2021rod}; this corrects Eq.~(2.11) of Ref.~\cite{Akcay:2019bvk}.

In this section we have again presented results for the self-consistent expansion. In the Gralla-Wald approach, one instead obtains evolution equations for the mass dipole moment. Such equations are derived at first order in Refs.~\cite{Gralla:2008fg,Pound:2009sm,Gralla:2011zr} and at second order in Ref.~\cite{Gralla:2012db} (see also Ref.~\cite{Pound:2015fma}, which derives such second-order equations in a more compact, parametrization-invariant form).

We stress that the equations in this section follow directly from the vacuum Einstein equations, together with a center-of-mass condition, {\em outside} the small object. There is no assumption about the object's internal composition, nor is there any regularization of singular quantities. We refer to Refs.~\cite{Detweiler:2000gt,Rosenthal:2006iy,Detweiler:2011tt} for variants of the approach described here and to Refs.~\cite{Quinn:1996am,Galley:2008ih,Harte:2011ku} for alternatives to the matched-expansions approach.

\subsection{Skeleton sources: punctures and particles}

After having derived the local form of the metric, and the equations of motion, we can effectively remove the body zone from the problem. We do so by allowing the local forms~\eqref{hRn FW}--\eqref{hSR covariant} to hold all the way down to $\gamma$. This causes the self-field to diverge at $\gamma$, artificially introducing a singular field. However, this does not alter the physics in the buffer region or external universe, and the singularity is more easily handled than the small-scale physics of the small object.

Once the fields have been extended to $\gamma$, one can solve the field equations throughout the spacetime using either a puncture scheme or point-particle methods. The puncture scheme is the more general of the two approaches. We define the {\em puncture field} 
\beq
h^{\P(n)}_{\mu\nu} := h^{\S(n)}_{\mu\nu}{\cal W}
\eeq
as the local expansion of $h^{\S(n)}_{\mu\nu}$ truncated at some order $\lambda^k$, multiplied by a window function ${\cal W}$ that is equal to 1 in a neighbourhood of $z^\alpha$ and transitions to zero at some finite distance from $z^\alpha$.\footnote{Our description may seem (incorrectly) to imply that the puncture field is only defined in a convex normal neighbourhood of the body. For numerical purposes, the puncture is extended over a region of any convenient size. Typically this is done by converting the local, covariant expressions in terms of Synge's world function into expansions in coordinate distance, using, e.g., the Boyer-Lindquist coordinates of the background spacetime. The punctures can then be extended as these coordinate functions. The end result for the combined field $h^{(n)}_{\mu\nu}=h^{\res(n)}_{\mu\nu}+h^{\P(n)}_{\mu\nu}$ is insensitive to the choice of extension.} This implies that $h^{\P(n)}_{\mu\nu}=h^{\S(n)}_{\mu\nu}+O(\lambda^{k+1})$. We then define the {\em residual field} 
\beq
h^{\res(n)}_{\mu\nu}:=h^{(n)}_{\mu\nu}-h^{\P(n)}_{\mu\nu}, 
\eeq
which satisfies $h^{\res(n)}_{\mu\nu}=h^{\R(n)}_{\mu\nu} + O(\lambda^{k+1})$, making $h^{\res(n)}_{\mu\nu}$ a $C^k$ field at $\gamma$. Outside the support of $h^{\P(n)}_{\mu\nu}$, $h^{\res(n)}_{\mu\nu}$ becomes identical to the full field $h^{(n)}_{\mu\nu}$. 

Moving $h^{\P(1)}_{\mu\nu}$ to the right-hand side of the vacuum field equations, we obtain field equations for $h^{\res(n)}_{\mu\nu}$:\footnote{In the self-consistent approach, some care is required in formulating these equations. Specifically, they can only be split into a sequence of equations, one at each order in $\e$, after imposing a gauge condition~\cite{Pound:2009sm}; this is required in order to allow the puncture to move on an accelerated trajectory. We do not belabour this point because we ultimately formulate the equations in a somewhat different, multiscale form tailored to binary inspirals.}
\begin{align}
G^{(1)}_{\mu\nu}[h^{\res(1)}] &= - G^{(1)}_{\mu\nu}[h^{\P(1)}]:=S^{\rm eff(1)}_{\mu\nu},\\
G^{(1)}_{\mu\nu}[h^{\res(2)}] &= - G^{(2)}_{\mu\nu}[h^{(1)},h^{(1)}] - G^{(1)}_{\mu\nu}[h^{\P(2)}]:=S^{\rm eff(2)}_{\mu\nu}.
\end{align}
These equations hold at all points off $\gamma$. The $C^k$ behaviour of the solution is then enforced by defining the effective sources $S^{\rm eff(n)}_{\mu\nu}$ as ordinary integrable functions at $\gamma$, rather than treating $G^{(1)}_{\mu\nu}[h^{\P(n)}]$ in the distributional sense of a linear operator acting on an integrable function; this distinction is important to rule out delta functions in the source, which would create spurious singularities in the residual field. 

If $k\geq1$, then we can replace $h^{\R(n)}_{\mu\nu}$ with $h^{\res(n)}_{\mu\nu}$ in the equations of motion~\eqref{EOM spin} and \eqref{EOM2}. The total field $h^{(n)}_{\mu\nu}=h^{\res(n)}_{\mu\nu}+h^{\P(n)}_{\mu\nu}$ is also guaranteed to satisfy the physical boundary condition in the buffer region (i.e., the matching condition) and at the outer boundaries of the problem. 

An alternative to the puncture scheme is to solve directly for the total fields $h^{(n)}_{\mu\nu}$. Once extended to $\gamma$, they satisfy 
\begin{align}
G^{(1)}_{\mu\nu}[\e h^{(1)} + \e^2h^{(2)}] + \e^2 G^{(2)}_{\mu\nu}[h^{(1)},h^{(1)}] = 8\pi T_{\mu\nu} + O(\e^3),\label{skeleton EFE}
\end{align}
where here we {\em do} interpret each term on the left-hand side in a distributional sense. The stress-energy tensor is then defined by the left-hand side. Through second order, it can be shown to be the stress-energy of a spinning particle in the effective metric~\cite{DEath:1975jps,Gralla:2008fg,Pound:2009sm,Pound:2012dk,Upton:2021}:\footnote{At second order, this is true in a class of highly regular gauges. In other gauges, it requires a direct use of the puncture via a particular distributional definition of the nonlinear quantity $G^{(2)}_{\mu\nu}[h^{(1)},h^{(1)}]$~\cite{Upton:2021}.}
\beq
T_{\mu\nu} = m\int_\gamma \breve{u}_\mu \breve{u}_\nu \breve{\delta}(x,z(\breve\tau))d\breve{\tau} + \int_\gamma \breve u_{(\mu}S_{\nu)}{}^{\alpha}\breve{\nabla}_{\!\alpha}\breve{\delta}(x,z(\breve\tau))d\breve\tau,\label{skeleton Tab}
\eeq
where $\breve{\delta}(x,x')=\frac{\delta^4(x^\alpha-x'^\alpha)}{\sqrt{-\breve{g}}}$ and $\breve{u}_\mu:=\breve{g}_{\mu\nu}\breve{u}^\nu$. We refer to this point-particle stress-energy as the {\em Detweiler stress-energy} after the author of Ref.~\cite{Detweiler:2011tt}. Like the equations of motion, the point-particle approximation is a derived consequence of the vacuum Einstein equations and the matching condition, rather than an input.

In cases where the point-particle method is well defined, it and the puncture scheme yield identical full fields $h^{(n)}_{\mu\nu}$. However, unlike a puncture scheme, a point-particle method does not yield the regular fields $h^{\R(n)}_{\mu\nu}$ as output. The regular fields, and self-forces, must instead be extracted from $h^{(n)}_{\mu\nu}$. This is most often done using the method of {\em mode-sum regularization}~\cite{Barack:2001gx,Barack:2002mh} reviewed in detail in Refs.~\cite{Barack:2009ux,Wardell:2015kea} and sketched in Sec.~\ref{mode decompositions of hS} below. 

We will refer to both the effective sources $S^{\rm eff(n)}_{\mu\nu}$ and the point-particle source $T_{\mu\nu}$ as {\em skeleton sources}. This terminology follows Mathisson's notion~\cite{Mathisson:1937zz,Dixon:2015vxa} of a ``gravitational skeleton'' (see also Refs.~\cite{Dixon:1970zza,Dixon:1970zz,Dixon:74}): an extended body can be represented by a singularity equipped with an infinite set of multipole moments. Punctures provide a generalization of this concept to settings where the singularities are too strong to be represented by distributions. For that reason, although the nomenclature of punctures and effective sources originated from methods of solving the first-order field equations in Refs.~\cite{Barack:2007we,Vega:2007mc}, punctures have a more fundamental role at second and higher orders~\cite{Rosenthal:2006nh,Rosenthal:2006iy,Detweiler:2011tt,Gralla:2012db,Pound:2012dk,Pound:2015tma}. For the same reason, we have presented punctures as a more primitive concept than the point-particle stress-energy.

In either approach, the skeleton sources presented here apply equally for all compact objects, whether black holes or material bodies. The only distinguishing feature of a material body would be a spin that surpasses the Kerr bound (i.e., $|S^i|>m^2$). However, at third order in perturbation theory, the quadrupole moment will appear in the perturbation $h^{(3)}_{\mu\nu}$. Unlike the mass and dipole moments, the quadrupole moment is not governed by the Einstein equation~\cite{Dixon:74,Harte:2011ku,Harte:2014wya}, and its evolution must be determined from the object's equation of state. Hence, at third order the interior composition of the object begins to influence the external metric, and we can begin to distinguish between black holes and material bodies. But note that the quadrupole moments of compact objects differ primarily due to their differing tidal deformability, and this difference is suppressed by an additional five powers of $\e$~\cite{Binnington:2009bb}, suggesting it is almost certainly irrelevant for small-mass-ratio binaries.


\section{Orbital dynamics in Kerr spacetime}

The previous section summarized the local problem in self-force theory: the reduction of an extended body to a skeleton source in the Einstein equations, along with an equation of motion for that source. In the remaining sections, we turn to the {\em global problem}: solving the perturbative Einstein equations, coupled to the equation of motion~\eqref{EOM spin} or \eqref{EOM2}, globally in a specific background metric. 

In the context of a small-mass-ratio binary, the background geometry is the Kerr spacetime of the central black hole. According to the equations of motion, the small body in the binary is only slightly accelerated away from geodesic motion in that background. This section summarizes (i) properties of bound geodesic motion in Kerr spacetime and (ii) how to exploit those properties to analyze accelerated orbits. We emphasize action-angle methods that mesh specifically with our treatment of the Einstein equations in the final section of this review. However, much of our treatment is valid for a more generic acceleration.

We warn the reader that the notation in this section differs in several ways from that of the preceding section. The differences are noted in the first subsubsection below.

\subsection{Geodesic motion}

\subsubsection{Constants of motion, separable geodesic equation, and conventions}

Geodesics in Kerr spacetime are integrable, with three constants of motion associated with the spacetime's three Killing symmetries: (specific) energy $E=-u_\alpha \xi^\alpha$, (specific) azimuthal angular momentum $L_z = u_\alpha \delta^\alpha_\phi$, and the {\em Carter constant} $Q=u_\alpha u_\beta (\overset{\star\star}{K}{}^{\alpha\beta} - \frac{1}{a^2}\eta^\alpha\eta^\beta)$.\footnote{The constant $K=u_\alpha u_\beta \overset{\star\star}{K}{}^{\alpha\beta}$ is also sometimes referred to as Carter's constant.}

Inverting these three equations, together with $g^{\alpha\beta}u_\alpha u_\beta=-1$, for the four-velocity components, we obtain~\cite{Fujita:2009bp}
\begin{align}
\Sigma^2\left(\frac{dr}{d\tau}\right)^2 &= R(r),\label{drdtau}\\
\Sigma^2\left(\frac{dz}{d\tau}\right)^2 &= Z(z),\label{dzdtau}\\
\Sigma\frac{dt}{d\tau} &= T_r(r) + T_z(z) + aL_z:=\mathscr{f}_t,\label{dtdtau}\\
\Sigma\frac{d\phi}{d\tau} &= \Phi_r(r) + \Phi_z(z) - a E:=\mathscr{f}_\phi.\label{dphidtau}
\end{align}
Here $(t,r,z:=\cos\theta,\phi)$ refer to Boyer-Lindquist coordinates,\footnote{Refs.~\cite{Drasco:2003ky,Fujita:2009bp} and many other references instead define $z$ as $\cos^2\theta$, with analogous differences in their definitions of the roots $z_n$ defined below.} and
\begin{align}
R(r) &:= [P(r)]^2-\Delta\left[r^2+(aE-L_z)^2+Q\right],\\
Z(z) &:= Q-\left(Q+a^2\gamma+L_z^2\right)z^2+a^2\gamma\, z^4,\\
T_r(r) &:= \frac{r^2+a^2}{\Delta}P(r),\label{Tr}\\
T_z(z) &:= -a^2E(1-z^2),\label{Tz}\\
\Phi_r(r) &:= \frac{a}{\Delta}P(r),\\
\Phi_z(z) &:= \frac{L_z}{1-z^2},\label{Phiz}
\end{align}
with $P(r):=E(r^2+a^2)-aL_z$ and $\gamma:=1-E^2$. We opt to use $z$ rather than $\theta$ throughout this section.

The equations for $r(\tau)$ and $z(\tau)$ are coupled, but they are immediately decoupled by adopting a new parameter $\lambda$, called {\rm Mino time}~\cite{Mino:2003yg}, that satisfies 
\beq\label{Mino time}
\frac{d\lambda}{d\tau} = \Sigma^{-1}. 
\eeq
(This is not to be confused with the bookkeeping parameter used in the local expansions of the previous section.) The equations also take a hierarchical form: once $r(\lambda)$ and $z(\lambda)$ are known, Eqs.~\eqref{dtdtau} and \eqref{dphidtau} can be straightforwardly integrated to obtain  $t(\lambda)$ and $\phi(\lambda)$. 

Given this hierarchical form, we will focus on the $r$--$z$ dynamics. In Eq.~\eqref{drdtau}, $R(r)$ is a fourth-order polynomial in $r$, meaning it can also be written as $R(r)=-\gamma(r-r_1)(r-r_2)(r-r_3)(r-r_4)$, with $r_1\geq r_2\geq r_3\geq r_4$. Similarly, in Eq.~\eqref{dzdtau}, $Z(z)=a^2\gamma(z^2-z_1^2)(z^2-z_2^2)$, with $|z_1|>|z_2|$. For bound orbits, the radial motion oscillates between the turning points $r_a=r_1$  (apoapsis) and $r_p=r_2$ (periapsis), and the polar motion between $z_{\rm max}=|z_2|$ and $z_{\rm min}=-|z_2|$.\footnote{The other roots ($r_3$, $r_4$, and $z_1$) do not correspond to physical turning points. In particular, $|z_1|>1$.} Hence, the geodesic is confined to a torus-like region $r_p\leq r\leq r_a$, $|z|<z_{\rm max}$. If $Q=0$, the motion is confined to the equatorial plane $z=0$. If $a=0$ (i.e., in Schwarzschild spacetime), the geodesic is likewise confined to a plane, which, due to Schwarzschild's spherical symmetry, can be freely chosen as $z=0$. However, a generic orbit ergodically fills the torus-like region. 

For convenience in the remaining sections, 
we use lowercase Latin indices from the beginning of the alphabet ($a,b,c$) to denote $r$ or $z$ and define $\bm{x} = (r,z)$. However, repeated indices, as in an expression such as $f_a x^a$, are not summed over; instead, such sums will be written as $\bm{f}\cdot \bm{x} := f_r x^r + f_z x^z$. An expression such as $f_a(x^a)$ will denote either one of $f_r(r)$ or $f_z(z)$, while an expression such as $f_a(\bm{x})$ will denote either one of $f_r(r,z)$ or $f_z(r,z)$. $f_\alpha(x^\beta)$ will denote $f_\alpha(t,r,z,\phi)$. 

We use lowercase Latin indices from the middle of the alphabet ($i,j,k$) to label elements of a set of orbital parameters. For example, $P^i=(E,L_z,Q)$. For these indices, unlike $a,b,c$, we use Einstein summation.

We use $f$ throughout this section to denote a generic function, not the specific function $f(r)$ that appears in the Schwarzschild metric~\eqref{eq:SchwMetric}. An overdot will denote a derivative with respect to $\lambda$. 

Finally, we preemptively refer the reader to Refs.~\cite{Schmidt:2002qk,Mino:2003yg,Drasco:2003ky,Drasco:2005kz,Fujita:2009bp,Warburton:2013yj,Stein:2019buj} for additional details about geodesic orbits in Kerr.


\subsubsection{Quasi-Keplerian parametrization}

Unlike Keplerian orbits, generic geodesics in Kerr do not close; the periods of radial, polar, and azimuthal motion are all, generically, incommensurate. Nevertheless, because of their doubly oscillatory nature, it is often useful in applications to express the geodesic trajectories in a quasi-Keplerian form, replacing the constants $\{E,L_z,Q\}$ with an alternative set $\{p,e,z_{\rm max}\}$. In terms of these, $r$ and $z$ can be written in the manifestly periodic form~\cite{Drasco:2003ky}
\begin{align}
r(\psi_r) &= \frac{pM}{1+e\cos\psi_r},\label{r(psi)}\\
z(\psi_z) &= z_{\rm max}\cos\psi_z,\label{z(psi)}
\end{align}
where, for a bound orbit, $0 \leq e < 1$. The phases $(\psi_r,\psi_z)$ are multiples of $2\pi$ at periapsis and at $z=z_{\rm max}$, respectively. Unlike $r$ and $z$, which change direction every half cycle, $\psi_r$ and $\psi_z$ grow monotonically, leading to better numerical behavior at the turning points. 

Because none of the periods are commensurate, $\psi_r$ and $\psi_z$ evolve independently (of each other and of $\phi$). Using $\frac{d\psi_a}{d\lambda} =\frac{dx^a}{d\lambda}/\frac{dx^a}{d\psi_a}$, one finds~\cite{Drasco:2003ky}
\begin{align}
\frac{d\psi_r}{d\lambda} &= \frac{M\sqrt{\gamma[(p-p_3)-e(p+p_3\cos\psi_r)][(p-p_4)+e(p-p_4\cos\psi_r)]}}{1-e^2}\nonumber\\
&:=\mathscr{f}_r,\label{psi_r dot}\\
\frac{d\psi_z}{d\lambda} &= \sqrt{a^2\gamma(z^2_1-z^2_{\rm max}\cos^2\psi_z)}:=\mathscr{f}_z,\label{psi_theta dot}
\end{align}
where $p_3:=r_3(1-e)/M$ and $p_4:=r_4(1+e)/M$. These can be integrated subject to arbitrary choices of initial phase $\psi_a(0)=\psi^0_a$.

The parameters $\{p,e,z_{\rm max}\}$, unlike $\{E,L_z,Q\}$, are related directly to the coordinate shape of the orbit, specifically to its turning points. Equation~\eqref{r(psi)} is the formula for an ellipse, and it implicitly defines $p$ and $e$ to be the semi-latus rectum and eccentricity of that ellipse, related to the periapsis and apoapsis by
\beq\label{rp and ra}
r_p = \frac{pM}{1+e} \quad \text{and}\quad r_a = \frac{pM}{1-e}.
\eeq
As stated above, $z_{\rm max}=z_2$, but to further the analogy with Keplerian orbits, we can also define an inclination angle $\iota$ such that\footnote{Ref.~\cite{Drasco:2003ky} and some other authors use the alternative, inequivalent definition $\cos\iota = \frac{L_z}{\sqrt{L_z^2+Q}}$. This does not describe the maximum coordinate inclination angle but has other useful properties~\cite{Hughes:2002ei}.}
\beq\label{zmax}
z_2 = z_{\rm max} = \sin\iota. 
\eeq

The remaining roots of $R(r)$ and $Z(z)$ are also compactly expressed in terms of these parameters~\cite{Fujita:2009bp}: 
\beq
r_3 = \frac{1}{2}\left(\alpha+\sqrt{\alpha^2-4\beta}\right) \quad \text{and}\quad r_4 = \beta/r_3,\\ 
\eeq
where $\alpha:=2M/\gamma-(r_a+r_p)$ and $\beta:=a^2 Q/(\gamma r_a r_p)$, and 
\beq
z_1 = \sqrt{\frac{Q}{a^2\gamma z_{\rm max}^2}}. 
\eeq
These expressions are in a ``mixed'' form that involves both sets of constants. However, $\{E,L_z,Q\}$ can be written in terms of $\{p,e,\iota\}$ as~\cite{Drasco:2005kz}\footnote{Note that $r_1$ and $r_2$ have the opposite meaning in Ref.~\cite{Drasco:2005kz} than their meaning here. Our notation for the roots $r_n$ follows Ref.~\cite{Fujita:2009bp}.}
\begin{align}
E^2 &= \frac{2|d,g,h|-|d,h,f|-2\chi\sqrt{|d,g,h|^2 +|h,d,g,h,f|+|h,d,h,g,f|}}{|f,h|^2+4|f,g,h|},\label{E(pi)}\\
L_z &= -\frac{g_p ME}{h_p}+M\chi\sqrt{\frac{g_p^2 E^2}{h_p^2}+\frac{f_pE^2-d_p}{h_p}},\label{L(pi)}\\
Q &= z_{\rm max}^2\left(a^2\gamma+\frac{L_z^2}{\cos^2\iota}\right),\label{Q(pi)}
\end{align}
where $\chi:={\rm sgn}(L_z)$ is $+1$ for prograde orbits and $-1$ for retrograde, 
\begin{align}
d(r) &:=\Delta(r^2+z_{\rm max}^2a^2)/M^4,\\
f(r) &:=(r/M)^4+a^2[r(r+2M)+z_{\rm max}^2\Delta]/M^4,\\
g(r) &:=2ar/M^2, \\
h(r) &:=[r(r-2M)+\Delta\tan^2\iota]/M^2,
\end{align}
 and a subscript $a$ or $p$ indicates evaluation at $r_a$ or $r_p$. The quantities $|\cdot|$ appearing in $E^2$ are determinants or products of determinants that we define recursively as $| f,g|:=f_p g_a-f_ag_p$ and $|f,g,\ldots|:=|f,g| |g,\ldots|$.

Given the parametrizations~\eqref{r(psi)} and \eqref{z(psi)} and the equations of motion~\eqref{dtdtau} and \eqref{dphidtau}, $t(\lambda)$ and $\phi(\lambda)$ can be written as
\begin{align}
t(\lambda) &= t_0 + t_r(\psi_r(\lambda)) + t_z(\psi_z(\lambda)) +aL_z\lambda ,\label{t(psi)}\\
\phi(\lambda) &= \phi_0 + \phi_r(\psi_r(\lambda)) + \phi_z(\psi_z(\lambda)) -aE\lambda,\label{phi(psi)}
\end{align}
with 
\beq
t_a(\psi_a) = \int_{\psi^0_{a}}^{\psi_a}\frac{T_a(\psi'_a)}{\mathscr{f}_a(\psi'_a)}d\psi'_a \quad \text{and} \quad  \phi_a(\psi_a) = \int_{\psi^0_{a}}^{\psi_a}\frac{\Phi_a(\psi'_a)}{\mathscr{f}_a(\psi'_a)}d\psi'_a.\label{t_a(psi_a)}
\eeq
Here $t_0$ and $\phi_0$ are integration constants.

This completes the quasi-Keplerian description of geodesic orbital motion. Trajectories are described by the three constants of motion $p^i:=(p,e,\iota)$ and the four secularly growing phase variables $\psi_\alpha:=(t,\psi_r,\psi_z,\phi)$. A given trajectory is uniquely specified by the set of seven constants $\{p,e,\iota, t_0,\psi_r^0,\psi_z^0,\phi_0\}$, called orbital elements. The solution to the geodesic equation can also be put in closed, analytical form~\cite{Fujita:2009bp} by expressing  $\psi_\alpha(\lambda)$ in terms of elliptic integrals and their inverses (the Jacobi elliptic functions). 

\subsubsection{Fundamental Mino frequencies and action angles}\label{geodesic Mino angle variables}

It is often essential to decompose quantities on the worldline into Fourier series, particularly when solving the perturbative Einstein equations in the frequency domain. This procedure is expedited by knowing the orbit's fundamental frequencies. In this section, we summarize the calculation of frequencies and of phase variables (action angles) associated with those frequencies. Unlike the phases $\psi_\alpha$, the angle variables are strictly linear in $\lambda$, facilitating Fourier expansions in that time variable.

In the right-hand sides of Eqs.~\eqref{dtdtau}, \eqref{dphidtau}, \eqref{psi_r dot}, and \eqref{psi_theta dot}, we have defined the ``frequencies'' $\mathscr{f}_\alpha(\bm{\psi})$ as the rates of change of $\psi_\alpha$,
\beq
\frac{d\psi_\alpha}{d\lambda} = \mathscr{f}_\alpha(\bm{\psi}).
\eeq
The true frequencies $\Upsilon_\alpha$ associated with $\lambda$ are the average rates of change of $\psi_\alpha$,
\beq\label{Upsilon = <f>}
\Upsilon_\alpha = \left\langle\mathscr{f}_\alpha\right\rangle_\lambda,
\eeq
and the corresponding action angles are
\beq
q_\alpha = \Upsilon_\alpha \lambda + q^0_\alpha,
\eeq
with arbitrary constants $q^0_\alpha$. For a function $f[r(\lambda),z(\lambda)]$ on the worldline, the average is defined as
\begin{align}
\left\langle f\right\rangle_\lambda := \lim_{\Lambda\to\infty}\frac{1}{2\Lambda}\int_{-\Lambda}^\Lambda f d\lambda. \label{lambda average}
\end{align}
For a generic, nonresonant orbit, this average agrees with the {\em torus average}\footnote{We focus only on functions of $r$ and $z$, which are automatically periodic functions of the intrinsic phases $\bm{\psi}$ and $\bm{q}$. The averaging operation immediately generalizes in the natural way to functions $f[z^\alpha(\lambda)]$ that are periodic in $t$ and $\phi$.}
\beq
\left\langle f\right\rangle_{\bm{q}} = \frac{1}{(2\pi)^2}\oint f d^2q. \label{q average}
\eeq
We use $\oint d^2q$ to denote $\int_0^{2\pi}dq_r\int_0^{2\pi}dq_z$ and $\oint d^2\psi$ for the analogous integral over $\bm{\psi}$. 

To simplify the analysis, we choose our phase space coordinates $\bm{q}$ such that $q_r$ vanishes at some periapsis and $q_z$ vanishes at some $z=z_{\rm max}$. We furthermore choose $q_t$, $q_\phi$, our spacetime coordinates $t$ and $\phi$,  and our parameter $\lambda$ such that they all vanish at some particular passage through periapsis. These choices, which do not represent any loss of generality, imply
\begin{subequations}\label{q and psi ICs}
\begin{align}
q^0_\alpha &= \psi_\alpha^0 = 0 \quad\text{for } \alpha=t,r,\phi, \\
q^0_z &= - \Upsilon_z\lambda^0_z,
\end{align}
\end{subequations}
where $\lambda^0_z$ is the first value of $\lambda$ at which $z=z_{\rm max}$. $\psi^0_z$ can be inferred from $q^0_z$. One can easily do without these specifications if desired.

With our choices, $q_a$ represents the mean growth of $\psi_a$ from the first radial or polar turning point, and we can express it in terms of $\psi_a$ as
\beq\label{q_a(psi_a)}
q_a(\psi_a) = \Upsilon_a \int_{0}^{\psi_a}\frac{d\psi'_a}{\mathscr{f}_a(\psi_a')}.
\eeq
This allows us to straightforwardly write the torus average as as an integral over $\bm{\psi}$,
\beq
\left\langle f\right\rangle_{\bm{q}} = \frac{1}{\Lambda_r\Lambda_z}\displaystyle\oint \frac{ f d^2\psi}{\mathscr{f}_r(\psi_r) \mathscr{f}_z(\psi_z)}, \label{psi average}\\
\eeq
where $\Lambda_a = \int_{0}^{2\pi}\frac{d\psi_a}{\mathscr{f}_a(\psi_a)}$ is the radial or polar period with respect to  $\lambda$. Although they agree generically, $\langle f\rangle_\lambda$ and $\langle f\rangle_{\bm{q}}$ differ in the special case of resonant orbits, discussed in later sections. 

For the $r$ and $z$ motion, the frequencies reduce to $\Upsilon_a = 2\pi/\Lambda_a$, which can be analytically evaluated to~\cite{Fujita:2009bp}
\begin{align}
\Upsilon_r &= \frac{\pi\sqrt{\gamma(r_a-r_3)(r_p-r_4)}}{2{\sf K}(k_r)},\\
\Upsilon_z &= \frac{\pi\sqrt{a^2\gamma}\,z_1}{2{\sf K}(k_z)},
\end{align}
where
\begin{align}
{\sf K}(x)&:=\int_0^{\pi/2}\frac{d\theta}{\sqrt{1-x\sin^2\theta}}
\end{align}
is the complete elliptic integral of the first kind, and its arguments are $k_r:=\frac{r_a-r_p}{r_a-r_3}\frac{r_3-r_4}{r_p-r_4}$ and $k_z:=(z_{\rm max}/z_1)^2$. 

The frequencies of $t$ and $\phi$ motion can also be found analytically. Because of the additive forms of $\frac{dt}{d\lambda}$ and $\frac{d\phi}{d\lambda}$ in ~\eqref{dtdtau} and \eqref{dphidtau}, the averages reduce to a sum of one-dimensional integrals. Evaluating those integrals leads to~\cite{Warburton:2013yj}
\begin{align}
\Upsilon_t &=  \frac{E}{2}\left[r_3(r_a+r_p+r_3) - r_a r_p + (r_a + r_p + r_3 + r_4)F_r +(r_a-r_3)(r_p-r_4)G_r\right]	\nonumber\\
&\quad + \frac{2M}{r_+ - r_-}\left[ \frac{(4M^2E - aL_z)r_+ - 2Ma^2E}{r_3 - r_+}\left(1-\frac{F_+}{r_p-r_+}\right) - (+\leftrightarrow -) \right]\nonumber\\
&\quad +4M^2E + \frac{EQ(1-G_z)}{\gamma\, z_{\rm max}^2}+ 2ME(r_3+F_r) ,\label{Gamma}\\
\Upsilon_\phi &= \frac{a}{r_+-r_-}\left[\frac{2MEr_+-aL_z}{r_3-r_+}\left(1-\frac{F_+}{r_p-r_+}\right)-(+\leftrightarrow-)\right]\nonumber\\
&\quad +\frac{L_z\Pi(z^2_{\rm max},k_z)}{{\sf K}(k_z)},
\end{align}
where we have introduced $G_a := \frac{{\sf E}(k_a)}{{\sf K}(k_a)}$ and $F_A:=(r_p-r_3)\frac{\Pi(h_A,k_r)}{{\sf K}(k_r)}$ for $A=\{r,+,-\}$, with $h_r = \frac{r_a-r_p}{r_a-r_3}$ and $h_\pm:=\frac{(r_a-r_p)(r_3-r_{\pm})}{(r_a-r_3)(r_p-r_{\pm})}$. Here $r_\pm = M \pm \sqrt{M^2-a^2}$ denote the inner and outer horizon radii, and 
\begin{align}
{\sf E}(x)&:=\int_{0}^{\pi/2}d\theta\sqrt{1-x\sin^2\theta},\\
\Pi(x,y)&:=\int_0^{\pi/2}\frac{d\theta}{(1-x\sin^2\theta)\sqrt{1-y\sin^2\theta}} 
\end{align}
are the complete elliptic integrals of the second and third kind, respectively.

In terms of the angle variables, a quantity $f(r,z)$ on the worldline can be expanded in the Fourier series  
\beq\label{Fourier q}
f[r(\lambda),z(\lambda)] = \sum_{\bm{k}}f^{(q)}_{\bm{k}} e^{-iq_{\bm{k}}(\lambda)},
\eeq
where $q_{\bm{k}} := \bm{k}\cdot\bm{q}=  k_r q_r + k_z q_z$, and unless stated otherwise, sums range over $\bm{k}\in\mathbb{Z}^2$. The coefficients are given by
\beq\label{Fourier q coeffs}
f^{(q)}_{\bm{k}} = \frac{1}{(2\pi)^2}\oint f e^{iq_{\bm{k}}}d^2q,
\eeq
which can also be calculated as 
\beq\label{Fourier q coeffs from psi}
f^{(q)}_{\bm{k}} = \frac{\Upsilon_r\Upsilon_z}{(2\pi)^2}\oint \frac{f e^{iq_{\bm{k}}(\bm{\psi})}}{\mathscr{f}_r(\psi_r)\mathscr{f}_z(\psi_z)}  d^2\psi
\eeq
with $q_{\bm{k}}(\bm{\psi}) = k_r q_r(\psi_r)+i k_z q_z(\psi_z)$ given by Eq.~\eqref{q_a(psi_a)}.
The torus average of the function (and infinite $\lambda$ average for nonresonant orbits) is the zero mode in the Fourier series: $\left\langle f\right\rangle_{\bm{q}} =  f^{(q)}_{00}$. 

Using such Fourier expansions, we can invert Eq.~\eqref{q_a(psi_a)} to write the phases $\psi_\alpha$ in terms of the angle variables. 
The transformation $q_\alpha\to \psi_\alpha(q_\beta)$ must satisfy $\frac{\partial\psi_\alpha}{\partial q_\beta}\Upsilon_\beta = \frac{d\psi_\alpha}{d\lambda} = \mathscr{f}_\alpha$ together with our  choice $\psi_\alpha(q_\beta=0)=0$. The solution is the sum of a secular and a purely oscillatory piece:
\beq
\psi_\alpha(q_\beta) = q_\alpha  - \Delta\psi_\alpha(0) + \Delta \psi_\alpha(\bm{q}),\label{psi to q}
\eeq
where
\begin{align}
\Delta\psi_a &= \sum_{k\neq0} \frac{\mathscr{f}^{k}_a e^{-ikq_a}}{-ik\Upsilon_a},\label{Dpsia}\\
\Delta t &=\Delta t_r + \Delta t_z := \sum_{k\neq0}\left(\frac{T^{k}_r e^{-ikq_r}}{-ik\Upsilon_r}+\frac{T^{k}_z e^{-ikq_z}}{-ik\Upsilon_z}\right),\label{Dt}\\
\Delta \phi &=\Delta\phi_r + \Delta\phi_z := \sum_{k\neq0}\left(\frac{\Phi^{k}_r e^{-ikq_r}}{-ik\Upsilon_r}+\frac{\Phi^{k}_z e^{-ikq_z}}{-ik\Upsilon_z}\right).\label{Dphi}
\end{align}
$T_a$ and $\Phi_a$ are given in Eqs.~\eqref{Tr}--\eqref{Phiz}, and we have written them, along with $\mathscr{f}_a(q_a)$, as one-dimensional Fourier series in $q_a$; e.g., $T_a = \sum_k T_a^k e^{-ikq_a}$. We will consistently use $\Delta$ to indicate that a quantity is periodic in $\bm{q}$ with zero average.

We can conveniently write the coordinate trajectory in terms of the action angles as the sum of a secular term and an oscillatory term:
\beq\label{zG(lambda)}
z^\alpha(\lambda) = z^\alpha_{\rm sec}[q_\beta(\lambda)] + \Delta z^\alpha[\bm{q}(\lambda)],
\eeq
where the secular piece is
\begin{align}
z^\alpha_{\rm sec}(q_\beta) &= (q_t,0,0,q_\phi) + \left[-\Delta t(0),r^{(q)}_{0},z^{(q)}_{0},-\Delta\phi(0)\right],
\end{align}
and the purely oscillatory pieces are given by Eqs.~\eqref{Dt}, \eqref{Dphi}, and
\beq
\Delta x^a(q_a) = \sum_{k\neq0}x^{a}_{(q)k}\,e^{-ikq_{a}},
\eeq
with coefficients readily calculated from Eq.~\eqref{Fourier q coeffs from psi}. Ref.~\cite{Fujita:2009bp} gives $x^a(q_a)$ in closed form  in their Eqs.~(26) and (38), $\Delta t$ as the sum of their $t^{(r)}$ and $t^{(\theta)}$ in their Eqs.~(28) and (39), and  $\Delta\phi$ as the sum of their $\phi^{(r)}$ and $\phi^{(\theta)}$ in those same equations. (We caution the reader that the notation in Ref.~\cite{Fujita:2009bp} differs from ours in several ways.)

Our description here has followed the constructive approach of Refs.~\cite{Drasco:2003ky,Drasco:2005kz,Fujita:2009bp}, finding the frequencies and angle variables by directly solving the geodesic equation. There is an alternative, historically prior approach~\cite{Schmidt:2002qk} based on the Hamiltonian description of geodesics, which builds on Carter's original proof~\cite{Carter:1968rr} of integrability using the Hamilton-Jacobi equation. That approach derives action angles and their associated fundamental frequencies from a canonical transformation $(z^\alpha,u_\alpha)\to(q^\alpha,J_\alpha)$, where the actions $J_\alpha$ are the canonical momenta conjugate to the action angles $q^\alpha$. 


\subsubsection{Fundamental Boyer-Lindquist frequencies and action angles}\label{geodesic BL angle variables}

For the purpose of decomposing fields, such as the metric perturbation, into Fourier modes, it is more useful to know the frequencies with respect to coordinate time $t$. These are the frequencies observed at infinity and that appear in the gravitational waveform. They are given by
\beq
\Omega_\alpha = \frac{\Upsilon_\alpha}{\Upsilon_t}.
\eeq
The angle variables associated with them are
\beq\label{varphi geodesic}
\varphi_\alpha = \Omega_\alpha t +\varphi^0_\alpha 
\eeq
with $\Omega_t=1$. We choose the origin of this phase space in analogy with Eq.~\eqref{q and psi ICs}:
\beq
\varphi^0_\alpha = 0 \quad\text{for }\alpha=t,r,\phi, \quad \text{and }\varphi^0_z = -\Omega_z t^0_z,
\eeq
where $t^0_z$ is the first value of $t$ at which $z=z_{\rm max}$.

These new angle variables are related to $q_\alpha$ by a transformation that must satisfy $\frac{\partial\varphi_\alpha}{\partial q_\beta}\Upsilon_\beta = \frac{d\varphi_\alpha}{d\lambda} = \Omega_\alpha\mathscr{f}_t(\bm{q})$. Such a transformation, with our choice of origin $\varphi_\alpha(q_\beta=0)=0$, is
\beq\label{varphi(q)}
\varphi_\alpha(q_\beta) = q_\alpha - \Omega_\alpha \Delta t(0) +  \Omega_\alpha \Delta t(\bm{q})
\eeq
with $\Delta t$ given by Eq.~\eqref{Dt}.

In analogy with Eq.~\eqref{Fourier q}, a function of $r$ and $z$ on the worldline can be expanded in a Fourier series
\beq
f[r(t),z(t)] = \sum_{\bm{k}} f^{(\varphi)}_{\bm{k}}  e^{-i\varphi_{\bm{k}}(t)},
\eeq
with $\varphi_{\bm{k}}:=\bm{k}\cdot\bm{\varphi}=k_r\varphi_r+k_z\varphi_z$ and with coefficients given by the analog of Eq.~\eqref{Fourier q coeffs}. Using the Jacobian ${\rm det}\left(\frac{\partial\varphi_a}{\partial q_b}\right)=\mathscr{f}_t/\Upsilon_t$, we can also write the coefficients as integrals over $\bm{q}$,
\beq\label{Fourier coeff relationship}
f^{(\varphi)}_{\bm{k}} = \frac{e^{-i\Omega_{\bm{k}}\Delta t(0)}}{(2\pi)^2\Upsilon_t}\oint \mathscr{f}_t\,e^{iq_{\bm{k}}+i\Omega_{\bm{k}}\Delta t(\bm{q})}f d^2q,
\eeq
where $\Omega_{\bm{k}}:=k_r\Omega_r+k_z\Omega_z$. Or we can write them as  integrals over $\bm{\psi}$,
\beq\label{Fourier coeff relationship psi}
f^{(\varphi)}_{\bm{k}} = \frac{\Upsilon_r\Upsilon_z }{(2\pi)^2\Upsilon_t}\oint \frac{\mathscr{f}_t(\bm{\psi})\,e^{iq_{\bm{k}}(\bm{\psi})+i\Omega_{\bm{k}}[\delta t_r(\psi_r)+\delta t_z(\psi_z)]}f}{\mathscr{f}_r(\psi_r)\mathscr{f}_z(\psi_z)} d^2\psi,
\eeq
where $q_a(\psi_a)$ is given by Eq.~\eqref{q_a(psi_a)}, and
\beq
\delta t_a(\psi_a) := \Delta t_a[q_a(\psi_a)] - \Delta t_a(0) = \int^{\psi_a}_0 \frac{T_a(\psi'_a)-\left\langle T_a\right\rangle_\lambda}{\mathscr{f}_a(\psi'_a)}d\psi'_a,\label{delta t}
\eeq
with $T_a$ given by Eq.~\eqref{Tr} and \eqref{Tz}. If $f$ is separable [i.e., if it can be written as a sum of products of the form $f_r(r)f_z(z)$], then expressing the integrals in terms of $\bm{q}$ or $\bm{\psi}$ allows one to evaluate the two-dimensional integral as a product of one-dimensional integrals.

We can further define an average over $t$, 
\beq
\left\langle f[r(t),z(t)]\right\rangle_t := \lim_{T\to\infty}\frac{1}{2T}\int_{-\infty}^\infty f dt, 
\eeq
which for nonresonant orbits is equal to the torus average 
\beq
\langle f\rangle_{\bm{\varphi}} := \frac{1}{(2\pi)^2}\oint f d^2\varphi = f^{(\varphi)}_{00}. 
\eeq
Note that the meaning of a time average (and associated torus average) inherently depends on one's choice of time parameter~\cite{Pound:2007ti}, and that $\left\langle f\right\rangle_t$ differs from $\left\langle f\right\rangle_\lambda$: 
\beq
\left\langle f\right\rangle_t = \frac{1}{\Upsilon_t}\left\langle \mathscr{f}_t f \right\rangle_\lambda = \left\langle f\right\rangle_\lambda + \frac{1}{\Upsilon_t}\sum_{\bm{k}\neq0}\mathscr{f}^{(q)}_{t\bm{k}}f^{(q)}_{-\bm{k}} . 
\eeq
The relevance of each average depends on context. 

Using these Fourier expansions, we can express the phases $\psi_\alpha$ in terms of $\varphi_\alpha$. The two are related by a transformation satisfying $\frac{\partial\psi_\alpha}{\partial\varphi_\beta}\Omega_\beta = \frac{d\psi_\alpha}{dt}$. With our choice of origin $\psi_\alpha(\varphi_\beta=0) = 0$, the solution  is
\beq\label{psi(phi)}
\psi_\alpha(\varphi_\beta) = \varphi_\alpha -\Delta_\varphi \psi_\alpha(0) + \Delta_\varphi \psi_\alpha(\bm{\varphi}).
\eeq
The oscillatory terms are $\Delta_\varphi \psi_t = 0$ and
\beq
\Delta_\varphi\psi_\alpha = \sum_{\bm{k}\neq0}\frac{\left(\frac{d\psi_\alpha}{dt}\right)^{\!(\varphi)}_{\!\bm{k}}}{-i\Omega_{\bm{k}}}e^{-i\varphi_{\bm{k}}}\quad \text{for }\alpha=r,z,\phi.\label{Dpsi varphi}
\eeq
Here we use $\Delta_\varphi$ rather than $\Delta$ to indicate that a quantity is  purely oscillatory (i.e., periodic with zero mean) with respect to $\bm{\varphi}$ rather than $\bm{q}$. $(d\psi_\alpha/dt)^{(\varphi)}_{\bm{k}}$ can be calculated using Eq.~\eqref{Fourier coeff relationship psi} with $d\psi_\alpha/dt = \mathscr{f}_\alpha/\mathscr{f}_t$.

Just as we did with $q_\alpha$, we can express the coordinate trajectory in terms of $\varphi_\alpha$ as the sum of a secular and an oscillatory piece,
\beq\label{zG(t)}
z^\alpha(t) = z^\alpha_{(\varphi)\rm sec}[\varphi_\beta(t)] + \Delta_\varphi z^\alpha[\bm{\varphi}(t)],
\eeq
where the secular piece is
\begin{align}
z^\alpha_{(\varphi)\rm sec}(\varphi_\beta) &= \left(\varphi_t,0,0,\varphi_\phi\right) -\left[0,r^{(\varphi)}_{00},z^{(\varphi)}_{00},-\Delta_\varphi\phi(0)\right],
\end{align}
and the oscillatory pieces are $\Delta_\varphi t = 0$, $\Delta_\varphi \phi$ given by Eq.~\eqref{Dpsi varphi} (recalling $\psi_\phi:=\phi$), and
\beq
\Delta_\varphi x^a(\bm{\varphi}) = \sum_{\bm{k}\neq0}x^{a}_{(\varphi)\bm{k}}\,e^{-i\varphi_{\bm{k}}},
\eeq
with coefficients calculated from Eq.~\eqref{Fourier coeff relationship psi}.


\subsubsection{Resonant orbits}

Recall that the radial and polar motions are restricted to a torus-like region $r_p\leq r\leq r_a$ and $|z|\leq z_{\rm max}$ in physical space. If the periods of radial and polar motion are incommensurate, then the orbit ergodically fills this region. The transformation $x^a\to q^a$ maps the $r$--$z$ motion onto the surface of a torus in phase space, which the angles $q^a$ ergodically cover. However, for some values of the orbital parameters, the periods are commensurate, meaning $k^{\rm res}_r\Upsilon_r+k^{\rm res}_z\Upsilon_z=0$ for some nonzero integers $(k^{\rm res}_r,k^{\rm res}_z)$. [Since integer multiples of $(k^{\rm res}_r,k^{\rm res}_z)$ will also satisfy this condition, we take $(k^{\rm res}_r,k^{\rm res}_z)$ to be the smallest two such integers.] In these cases, rather than having two independent frequencies, the $r$--$z$ motion has a single frequency, $\Upsilon=\Upsilon_z/|k^{\rm res}_r|=\Upsilon_r/|k^{\rm res}_z|$, and rather than ergodically covering the torus, the orbit closes on the torus and in the $r$--$z$ plane. The actual shape of this closed orbit is not uniquely specified by its frequencies but depends strongly on the relative initial phase $\psi^0_r-\psi^0_z$. 

Such orbits are referred to as {\em resonant}~\cite{Flanagan:2010cd}. For resonant orbits, the average over the torus no longer represents a meaningful average over the orbit. Rather than having the single stationary mode $f^{(q)}_{00}$, a function $f(r,z)$ on the worldline has an infinite set of stationary modes corresponding to all integer multiples of $\bm{k}^{\rm res}$. The infinite Mino-time average in Eq.~\eqref{lambda average} is then
\beq
\left\langle f[r(\lambda),z(\lambda)]\right\rangle_\lambda = \lim_{\Lambda\to\infty}\frac{1}{2\Lambda}\int_{-\Lambda}^\Lambda f d\lambda = \sum_{N=-\infty}^\infty f^{(q)}_{N\bm{k}^{\rm res}};
\eeq
for a resonant orbit, the infinite $\lambda$ average does not, generically, agree with the torus average  $f^{(q)}_{00}$. 

More broadly, the Fourier series~\eqref{Fourier q} becomes degenerate: $q_{\bm{k}}(\lambda) = q_{\bm{k}+N\bm{k}^{\rm res}}(\lambda)$ for all integers $N$. However, since there is a common period, we can replace the two action angles $q_a$ with a single angle $q(\lambda) =\Upsilon \lambda$ and rewrite the two-dimensional Fourier series~\eqref{Fourier q} as a non-degenerate one-dimensional one,
\beq
f[r(\lambda),z(\lambda)] = \sum_{k\in\mathbb{Z}}f^{(q)}_k e^{-kq(\lambda)}.
\eeq
The coefficients are related to those in Eq.~\eqref{Fourier q} by $f^{(q)}_k = \sum_{\bm{k}} f^{(q)}_{\bm{k}}$, where the sum ranges over all $(k_r,k_z)$ satisfying $k_r |k^{\rm res}_z| +k_z |k^{\rm res}_r|=k$. We then have $\left\langle f\right\rangle_\lambda = f^{(q)}_0$.


The set of resonant orbits is dense in the space of frequencies, though it is of measure zero. A given resonant ratio $\Upsilon_r/\Upsilon_z=|k^{\rm res}_z|/|k^{\rm res}_r|$ describes a surface in the parameter space spanned by $p^i$. We refer to Ref.~\cite{Brink:2015roa} for the characterization of the locations of these surfaces and to Refs.~\cite{Grossman:2011im,Flanagan:2012kg,vandeMeent:2013sza,Brink:2013nna} for further discussion of resonant geodesic orbits.

\subsection{Accelerated motion} 

\subsubsection{Evolution of orbital parameters}

We now consider an accelerated orbit satisfying the equation of motion~\eqref{perturbed geodesic equation}, which we write compactly as
\begin{equation}\label{eq mot}
\frac{D^2 z^\alpha}{d\tau^2} = f^{\alpha}.
\end{equation}
The normalization $u^{\alpha}u_{\alpha} = -1$ implies that $f^\alpha$ is orthogonal to the worldline: $f^{\alpha}u_{\alpha} = 0$.

If we continue to define $E=-u_t$, $L_z=u_\phi$, and $Q=u_\alpha u_\beta \overset{\star\star}{K}{}^{\alpha\beta} - \frac{1}{a^2}(u_{\tilde\phi})^2$  on the accelerated orbit, then 
\begin{align}
\frac{dE}{d\tau} = -f_t, \quad \frac{dL_z}{d\tau} = f_\phi, \quad \frac{dQ}{d\tau} = 2\overset{\star\star}{K}{}^{\alpha\beta}u_\alpha f_\beta - \frac{2}{a^2}u_{\tilde \phi}f_{\tilde\phi},\label{dEdtau}
\end{align}
where $f_{\tilde\phi} = a(f_\phi + a f_t)$ and $u_{\tilde\phi} = a(L_z - aE)$. In other words, the ``constants'' of motion are no longer constant. However, if $f^\alpha$ is small, each parameter will change only slowly or oscillate slightly around a slowly varying mean. 

Our treatment of accelerated orbits mirrors that of geodesics, beginning with quasi-Keplerian methods and then describing the calculation of fundamental frequencies and perturbed angle variables. In the quasi-Keplerian treatment we place no restriction on $f^\alpha$, and in particular we do not assume it is small. In the treatment of perturbed angle variables we restrict the analysis to a small perturbing force, setting $f^\mu_{(0)}=0$ in Eq.~\eqref{perturbed geodesic equation}.

\subsubsection{Method of osculating geodesics}

In celestial mechanics, perturbed Keplerian orbits have historically been described using the method of osculating orbits. The idea in this method is, given an exact solution to the unperturbed problem in terms of a set of orbital elements $p^i=\{p,e,\iota,\ldots\}$, to write the perturbed orbit in precisely the same form but to promote the orbital elements to functions of time. At each instant $t$, the perturbed orbit with elements $\{p(t),e(t),\iota(t),\ldots\}$ is tangent to a Keplerian ellipse (called the osculating orbit) with those same values of orbital elements. 

In general relativity, this idea is referred to as the method of osculating geodesics \cite{Mino:2003yg,Pound:2007th,Gair:2010iv,Warburton:2017sxk}. Our geodesics in Kerr are described by Eqs.~\eqref{r(psi)}, \eqref{z(psi)}, \eqref{t(psi)}, and \eqref{phi(psi)}, which involve the seven orbital elements $I^A =\{p,e,\iota,t_0,\psi_r^0,\psi_z^0,\phi_0\}$. If we let $z^\alpha_G(I^A,\lambda)$ denote a geodesic with these orbital elements, and $\dot z^\alpha_G(I^A,\lambda)=\partial z^\alpha_G/\partial\lambda$ its tangent vector, then the {\em osculation conditions} are 
\begin{align}
z^\alpha(\lambda) =  z^\alpha_G[I^A(\lambda),\lambda]\quad\text{and}\quad \frac{dz^\alpha}{d\lambda}(\lambda)  = \dot z^\alpha_G[I^A(\lambda),\lambda].
\end{align}
These conditions define a one-to-one transformation $(z^\alpha,\dot z^\alpha)\to I^A$. Such a transformation is possible because the number of orbital elements is equal to the number of degrees of freedom on the orbit: the eight degrees of freedom $\{z^\alpha,\dot z^{\alpha}\}$ minus the constraint $\dot z_\alpha f^\alpha=0$. 

The osculation conditions can be used to transform the equation of motion~\eqref{eq mot} into evolution equations for $I^A(\lambda)$. 
Appealing to the chain rule $\frac{dz^\alpha}{d\lambda} = \frac{\partial z_G^\alpha}{\partial I^A}\frac{d I^A}{d\lambda} + \frac{\partial z_G^\alpha}{\partial \lambda}$, to the geodesic equation for $z_G^\alpha$ (in terms of the non-affine parameter $\lambda)$, and to the equation of motion~\eqref{eq mot} for $z^\alpha$ (converted to the non-affine parametrization), we find~\cite{Pound:2007th}
\begin{align}
\frac{\partial z^\alpha_G}{\partial I^A}\frac{d I^A}{d\lambda} & =  0,\label{diffI 1b} \\
\frac{\partial\dot z_G^{\alpha}}{\partial I^A}\frac{d I^A}{d\lambda} & =  f^{\alpha} \left(\frac{d\tau}{d\lambda}\right)^2 + [\kappa(\lambda)-\kappa_G(\lambda)]\dot z^\alpha_G, \label{diffI 2b v0}
\end{align}
where $\kappa = \left(\frac{d\tau}{d\lambda}\right)^{-1}\frac{d^2\tau}{d\lambda^2}$. If we define $\lambda$ to satisfy Eq.~\eqref{Mino time} on both the geodesic and accelerated orbit, then $\kappa=\kappa_G = \Sigma^{-1}\frac{d\Sigma}{d\lambda}$, simplifying Eq.~\eqref{diffI 2b v0} to
\begin{align}
\frac{\partial\dot z_G^{\alpha}}{\partial I^A}\frac{d I^A}{d\lambda} & = \Sigma^2  f^{\alpha} 
\label{diffI 2b}.
\end{align}
These equations are exact, and $f^\alpha$ need not be small.

Equations~\eqref{diffI 1b} and \eqref{diffI 2b} can be straightforwardly inverted to solve for  $\frac{d I^A}{d\lambda}$, providing a system of first-order ordinary differential equations for the orbital elements. However, working with the initial phases $\{t_0,\psi^0_r, \psi^0_z, \phi_0\}$ is cumbersome in practice. In the above evolution equations, the phases $\psi_a$ are given by their geodesic values, meaning the solutions to Eqs.~\eqref{psi_r dot} and \eqref{psi_theta dot} with fixed $I^A$. That is, at each value of $\lambda$, in Eqs.~\eqref{psi_r dot} and \eqref{psi_theta dot} we replace $d\psi_a/d\lambda$ with $d\psi_a/d\lambda'$, then integrate from $\lambda'=0$, with initial values $\psi^0_a(\lambda)$, up to $\lambda'=\lambda$. Similarly, in Eqs.~\eqref{t_a(psi_a)}, the integrals are evaluated with fixed orbital elements in the integrands. The evolution equations also involve derivatives of these integrals with respect to the orbital elements.

Evaluating all these integrals at every time step would be computationally expensive. In applications, it is therefore preferable to work with the variables $\{p,e,\iota,\psi_\alpha\}$ instead of $I^A$. We write a geodesic trajectory and its tangent vector as $z^\alpha_G[p^i,\psi_\beta(\lambda)]$ and $\dot z^\alpha_G[p^i,\psi_\beta(\lambda)]=\dot\psi^G_\beta\frac{\partial z^\alpha_G}{\partial\psi_\beta}$, where $\dot\psi^G_\alpha=\mathscr{f}_\alpha(p^i,\bm{\psi})$ are the geodesic ``frequencies'' given by Eqs.~\eqref{dtdtau}, \eqref{dphidtau}, \eqref{psi_r dot}, and \eqref{psi_theta dot}. The osculation conditions then read 
\begin{align}
z^\alpha(\lambda) = z^\alpha_G[p^i(\lambda),\psi_\beta(\lambda)]\quad\text{and}\quad
\frac{dz^\alpha}{d\lambda}(\lambda) & = \dot z^\alpha_G[p^i(\lambda),\psi_\beta(\lambda)]. 
\end{align}

Appealing to the chain rule $\frac{dz^\alpha}{d\lambda} = \frac{d\psi_\beta}{d\lambda}\frac{\partial z^\alpha_G}{\partial\psi_\beta}+\frac{d p^i}{d\lambda}\frac{\partial z^\alpha_G}{\partial p^i}$, to the geodesic equation for $z_G^\alpha$ (in terms of $\lambda$), and to the equation of motion~\eqref{eq mot} for $z^\alpha$ (in terms of $\lambda$), we find that the osculation conditions imply
\begin{align}
\frac{\partial z^\alpha_G}{\partial p^i}\frac{d p^i}{d\lambda} +\frac{\partial z^\alpha_G}{\partial\bm{\psi}}\cdot \bm{\delta\mathscr{f}} & =  0,\label{diffI 1.1} \\
\frac{\partial\dot z_G^{\alpha}}{\partial p^i}\frac{d p^i}{d\lambda} +\frac{\partial\dot z_G^{\alpha}}{\partial \bm{\psi}}\cdot\bm{\delta\mathscr{f}}  & = \Sigma^2  f^{\alpha} \label{diffI 2.1}.
\end{align}
Here $\frac{\partial z^\alpha_G}{\partial\bm{\psi}}\cdot \bm{\delta\mathscr{f}} = \frac{\partial z^\alpha_G}{\partial\psi_r}\delta\mathscr{f}_r+\frac{\partial z^\alpha_G}{\partial\psi_z}\delta\mathscr{f}_z$, and we have defined
\beq
\delta\mathscr{f}_a := \frac{d\psi_a}{d\lambda} - \dot\psi^G_a.
\eeq

Eq.~\eqref{diffI 1.1} and \eqref{diffI 2.1} provide eight equations for the seven derivatives $\frac{d\psi_\alpha}{d\lambda}$ and $\frac{dp^i}{d\lambda}$; any one of the four equations represented by~\eqref{diffI 2.1} may be eliminated using $f^\alpha u_\alpha = 0$. The $t$ and $\phi$ components of Eq.~\eqref{diffI 1.1} are simply the osculation conditions
\beq\label{tdot and phidot}
\frac{d\psi_\alpha}{d\lambda} = \mathscr{f}_\alpha(\bm{\psi},p^i) \quad \text{for } \alpha=t,\phi.
\eeq
The $r$ and $z$ components of Eq.~\eqref{diffI 1.1} can be rearranged to obtain 
\beq
\delta\mathscr{f}_a = - \frac{\partial z_G^a/\partial p^i}{\partial z^a_G/\partial\psi_a}\frac{d p^i}{d\lambda},\label{Df}
\eeq
where we have used the fact that $r$ is independent of $\psi_z$ and that $z$ is independent of $\psi_r$. Substituting this into Eq.~\eqref{diffI 2.1} yields
\beq
\frac{dp^i}{d\lambda}{\cal L}_i(z^\alpha_G) = \Sigma^2 f^\alpha,\label{pi dot}
\eeq
where
\beq
{\cal L}_i(x) := \frac{\partial \dot{x}}{\partial p^i} - \frac{\partial r/\partial p^i}{\partial r/\partial\psi_r} \frac{\partial\dot{x}}{\partial \psi_r} - \frac{\partial z/\partial p^i}{\partial z/\partial\psi_z} \frac{\partial\dot{x}}{\partial \psi_z}.\label{L_i(x)}
\eeq
One can easily invert Eq.~\eqref{pi dot} to obtain equations for $\frac{dp^i}{d\lambda}$:
\begin{align}
\frac{dp}{d\lambda} &= \frac{\Sigma^2}{D} \left\{ \left[{\cal L}_e(z),{\cal L}_\iota(\phi)\right]f^r+\left[{\cal L}_e(\phi),{\cal L}_\iota(r)\right]f^z  + \left[{\cal L}_e(r),{\cal L}_\iota(z)\right]f^\phi \right\}\!,\label{pdot}\\
\frac{de}{d\lambda} &= \frac{\Sigma^2}{D} \left\{ \left[{\cal L}_\iota(z),{\cal L}_p(\phi)\right]f^r +\left[{\cal L}_\iota(\phi),{\cal L}_p(r)\right]f^z +\left[{\cal L}_\iota(r),{\cal L}_p(z)\right]f^\phi \right\}\!,\label{edot}\\
\frac{d\iota}{d\lambda} &= \frac{\Sigma^2}{D} \left\{ \left[{\cal L}_p(z),{\cal L}_e(\phi)\right]f^r + \left[{\cal L}_p(\phi),{\cal L}_e(r)\right]f^z + \left[{\cal L}_p(r),{\cal L}_e(z)\right]f^\phi \right\}\!,\label{idot}
\end{align}
with $[{\cal L}_i(x),{\cal L}_j(y)]:={\cal L}_i(x){\cal L}_j(y)-{\cal L}_j(x){\cal L}_i(y)$ and
\begin{align}
D := {\cal L}_p(r)[{\cal L}_e(z),{\cal L}_\iota(\phi)]+{\cal L}_e(r)[{\cal L}_\iota(z),{\cal L}_p(\phi)] +{\cal L}_\iota(r)[{\cal L}_p(z),{\cal L}_e(\phi)].
\end{align}

Finally, the evolution equations for $\psi_a$ are obtained by substituting Eqs.~\eqref{pdot}--\eqref{idot} into Eq.~\eqref{Df}, yielding
\beq\label{psidot osc}
\frac{d\psi_a}{d\lambda} = \mathscr{f}_a(p^i,\bm{\psi})+\delta\mathscr{f}_a(p^i,\bm{\psi}),
\eeq
where 
\begin{align}
\delta\mathscr{f}_r &= -\frac{1}{pe\sin\psi_r} \left[(1+e\cos\psi_r) \frac{d p}{d\lambda} - p\frac{d e}{d\lambda}\cos\psi_r\right], \label{Df_r}\\
\delta\mathscr{f}_z &= \frac{d\iota}{d\lambda}\cot\iota\cot\psi_z.\label{dfz}
\end{align}
There are superficial singularities in these formulas when $\psi_a$ is an integer multiple of $\pi$. However, the divergences are analytically eliminated when the formulas are explicitly evaluated.

The full set of evolution equations is given by Eqs.~\eqref{pdot}--\eqref{idot}, \eqref{psidot osc}, and \eqref{tdot and phidot}. In these equations, $x^a(\lambda)=x^a_G[p^i(\lambda),\psi_a(\lambda)]$ is given by Eqs.~\eqref{r(psi)}--\eqref{z(psi)}, $\dot x^a(\lambda) = \dot x^a_G[p^i(\lambda),\psi_a(\lambda)]$ by
\begin{align}
\dot r = \frac{p M e\mathscr{f}_r\sin\psi_r}{(1+e\cos\psi_r)^2} \quad\text{and}\quad
\dot z =  -z_{\rm max}\mathscr{f}_z\sin\psi_z,\label{dot xa}
\end{align}
with Eqs.~\eqref{psi_r dot}--\eqref{psi_theta dot} for $\mathscr{f}_a$, and $(\dot t,\dot\phi) = (\mathscr{f}_t,\mathscr{f}_\phi)$ by Eqs.~\eqref{dtdtau} and \eqref{dphidtau}. Wherever $E$, $L_z$, and $Q$ appear, they are given in terms of $p^i$ by their geodesic expressions~\eqref{E(pi)}--\eqref{Q(pi)}. The quantities $[{\cal L}_i(x),{\cal L}_j(y)]$, when explicitly evaluated, constitute lengthy analytical formulas in terms of $p^i$ and $\bm{\psi}$. However, for several ${\cal L}_i(x)$, the second and third term vanish in Eq.~\eqref{L_i(x)}. Specifically, ${\cal L}_\iota(r) = \frac{\partial \dot r}{\partial \iota} = \frac{\partial r}{\partial \psi_r}\frac{\partial\mathscr{f}_r}{\partial\iota}$, and ${\cal L}_j(z) = \frac{\partial \dot z}{\partial p^j}=\frac{\partial z}{\partial \psi_z}\frac{\partial\mathscr{f}_z}{\partial j}$ for $j=p,e$.

The evolution can be slightly simplified by adopting $\psi_r$ or $\psi_z$ as the parameter along the trajectory. That is easily done by using, e.g., $\frac{dp^i}{d\psi_a} = \frac{1}{d\psi_a/d\lambda} \frac{dp^i}{d\lambda}$. However, for a sufficiently large perturbing force, $\frac{d\psi_a}{d\lambda}$ can vanish at some points in the evolution, making $\psi_a$ an invalid parameter. In that case, we can split $\psi_a$ into $\psi_a = \psi^G_a - \psi^0_a$, where $d\psi^G_a/d\lambda = \mathscr{f}_a$ and $d\psi^0_a/d\lambda = -\delta\mathscr{f}_a$. $\psi^G_a$ is then a convenient, monotonic parameter along the worldline. Alternatively, $t$ can be used, applying, e.g., $\frac{dp^i}{dt}=\mathscr{f}_t^{-1} \frac{d p^i}{d\lambda}$.

The evolution equations simplify more dramatically in the special case of equatorial orbits, for which $z=f^z=0$. In this case, $\iota$ and $\psi_z$ do not appear, and Eqs.~\eqref{pdot}--\eqref{edot} reduce to
\begin{align}
\frac{dp}{d\lambda} &= \frac{r^4\left[{\cal L}_e(\phi)f^r - {\cal L}_e(r)f^\phi\right]}{{\cal L}_p(r){\cal L}_e(\phi)-{\cal L}_e(r){\cal L}_p(\phi)},\label{pdot - equatorial}\\
\frac{de}{d\lambda} &= \frac{r^4\left[{\cal L}_p(r)f^\phi - {\cal L}_p(\phi)f^r\right]}{{\cal L}_p(r){\cal L}_e(\phi)-{\cal L}_e(r){\cal L}_p(\phi)},\label{edot - equatorial}
\end{align}
If $\psi^G_r$ is used as the independent parameter along the orbit, then the other three evolution equations are $d\psi^0_r/d\psi^G_r = -\delta\mathscr{f}_r/\mathscr{f}_r$ and Eq.~\eqref{tdot and phidot} for $t(\psi^G_r)$ and $\phi(\psi^G_r)$.

In our treatment we have left the evolution equations in a highly inexplicit form even in the relatively simple equatorial case. Refs.~\cite{Pound:2007th} and \cite{Warburton:2017sxk} provide explicit formulas in the cases of planar and nonplanar orbits in Schwarzschild spacetime. Ref.~\cite{Gair:2010iv} details the generic case in Kerr spacetime and describes several alternative formulations of the osculating evolution. 

Before proceeding, we note again that the equations in this section are valid for arbitrary forces,\footnote{However, the method has most often been applied~\cite{Warburton:2011fk,Osburn:2015duj,Warburton:2017sxk,vandeMeent:2018rms} in the context of an approximation in which the self-force at each instant is approximated by the value it would take if the particle had spent its entire prior history moving on the osculating geodesic. Since the force is then constructed from the field generated by the osculating geodesic particle, this approximation might more properly be dubbed  the method of osculating sources.} though the orbital elements are most meaningful when the force is small and the orbit is close to a geodesic. In the next section, we restrict to the case of a small perturbing force. 

\subsubsection{Perturbed Mino frequencies and action angles}\label{perturbed Mino frequencies}

In the unperturbed case, the equations of geodesic motion could be written in terms of the orbital parameters and angle variables as
\begin{align}
\frac{d q_\alpha}{d\lambda} &= \Upsilon_\alpha(p^i),\label{qdot geodesic}\\
\frac{d p^i}{d\lambda} &= 0.\label{pidot geodesic}
\end{align}
If the perturbing force is small, with an expansion 
\beq\label{small force1}
f^\alpha = \e f^\alpha_{(1)}(z^\mu,\dot z^\mu)+\e^2 f^\alpha_{(2)}(z^\mu,\dot z^\mu)+O(\e^3),
\eeq
and is periodic in $t$ and $\phi$, then the equations of forced motion can still be written in terms of orbital parameters and angle variables: 
\begin{align}
\frac{d q_\alpha}{d\lambda} &= \Upsilon^{(0)}_\alpha(p_q^j) +\e\Upsilon^{(1)}_\alpha(p_q^j)+ O(\e^2),\label{qdot perturbative}\\
\frac{d p_q^i}{d\lambda} &= \e G_{(1)}^i(p_q^j) + \e^2 G_{(2)}^i(p_q^j) + O(\e^3).\label{pqidot}
\end{align}
Note that the subscript $q$ on the orbital parameters $p_q^i=(p_q,e_q,\iota_q)$ is a label, not an index. The form~\eqref{small force1} is mildly restrictive, and it does not include the Mathisson-Papapetrou spin force, for example; for a spinning body, we must introduce additional parameters and action angles describing the spin's magnitude and direction~\cite{Ruangsri:2015cvg,Witzany:2019nml}. For our purposes we adopt a more restrictive form,
\beq\label{small force}
f^\alpha = \e f^\alpha_{(1)}(\bm{x},\dot z^\mu)+\e^2 f^\alpha_{(2)}(\bm{x},\dot z^\mu)+O(\e^3),
\eeq
which assumes that the force inherits the background spacetime's symmetries. We explain in Sec.~\ref{expansion of source} that the form~\eqref{small force} still needs further, minor alteration to describe the self-force, but it is sufficiently general as a starting point. 

Unlike in the unperturbed case, the orbital parameters $p^i_q$ and frequencies are no longer constant; they evolve slowly with time. However, the variables $(q_\alpha,p_q^i)$ cleanly separate the two scales in the orbit's evolution: the variables $p_q^i$ only change slowly, over the long time scale $\sim1/\e$, while the angle variables $q_\alpha$ change on the orbital time scale $\sim 2\pi/\Upsilon^{(0)}_\alpha$.  In the context of a small-mass-ratio binary, where the inspiral is driven by gravitational-wave emission, the long time scale $\sim 1/\e$ is referred to as the {\em radiation-reaction time}. 

The division of the orbital dynamics into slowly and rapidly varying functions has the same utility as in the geodesic case: it enables convenient Fourier expansions of functions on the worldline, which mesh with a Fourier expansion of the field equations (described in the final section of this chapter). Functions $f(r,z)$ on the accelerated worldline can be expanded in the Fourier series
\beq\label{Fourier q perturbed}
f[r(\lambda),z(\lambda)] = \sum_{\bm{k}}f^{(q)}_{\bm{k}}(p_q^j)e^{-iq_{\bm{k}}(\lambda)},
\eeq
with a clean separation between slowly varying amplitudes and rapidly varying phases. The coefficients remain given by Eq.~\eqref{Fourier q coeffs}. By eliminating oscillatory driving terms in the orbital evolution equations, the transformation to $(q_\alpha,p_q^i)$ also facilitates more rapid numerical evolutions~\cite{vandeMeent:2018rms} and, ultimately, more rapid generation of waveforms~\cite{Chua:2020stf}. In this and the next section, for visual simplicity we shall omit the label ``$(q)$'' from mode coefficients associated with $\bm{q}$. 

Now, to put the equations of motion in the form~\eqref{qdot perturbative}--\eqref{pqidot}, we begin with the evolution equations~\eqref{tdot and phidot}, \eqref{pdot}--\eqref{idot}, and \eqref{psidot osc}. Given the expansion~\eqref{small force}, these equations take the form
\begin{align}
\frac{d \psi_\alpha}{d\lambda} &= \mathscr{f}_\alpha^{(0)}(\bm{\psi},p^j) + \e\mathscr{f}_{\alpha}^{(1)}(\bm{\psi},p^j) + O(\e^2),\label{psidot perturbative}\\
\frac{d p^i}{d\lambda} &= \e g_{(1)}^i(\bm{\psi},p^j) + \e^2 g_{(2)}^i(\bm{\psi},p^j) + O(\e^3).\label{pidot perturbative}
\end{align}
Here $g_{(n)}^i$ is given by Eqs.~\eqref{pdot}--\eqref{idot} with $f^\alpha\to f^\alpha_{(n)}$. We have renamed $\mathscr{f}_\alpha$ to $\mathscr{f}_\alpha^{(0)}$, $\mathscr{f}_a^{(1)}$ is given by $\delta\mathscr{f}_a$ with $f^\alpha\to f^\alpha_{(1)}$, and $\mathscr{f}_\alpha^{(1)}=0$ for $\alpha=t,\phi$. In this form of the equations, every term on the right is a periodic, oscillatory function of the phases. However, one can transform to the new variables $(q_\alpha,p_q^i)$, which have no oscillatory driving terms, using an averaging transformation~\cite{Kevorkian-Cole:96,vandeMeent:2018rms},\footnote{Here we combine a near-identity averaging transformation with a zeroth-order one. }
\begin{align}
\psi_\alpha(q_\beta,p_q^j,\e) &= \psi^{(0)}_\alpha(q_\beta,p_q^j)+\e \psi^{(1)}_\alpha(q_\beta,p_q^j)+O(\e^2),\label{near-identity psi}\\
p^i(q_\beta,p_q^j,\e) & = p_q^i+\e p_{(1)}^i(\bm{q},p_q^j)+\e^2p^i_{(2)}(\bm{q},p_q^j)+O(\e^3),\label{near-identity pi}
\end{align}
where
\beq
\psi^{(0)}_\alpha(q_\beta,p_q^j) = q_\alpha - \Delta \psi^{(0)}_\alpha(0,p_q^i) + \Delta \psi^{(0)}_\alpha(\bm{q},p_q^j)\label{psi0 from q}
\eeq
is the geodesic relationship, and the corrections $\psi^{(n)}_\alpha$ and $p_{(n)}^i$  for $n>0$ are $2\pi$-periodic in each $q_a$ (with a potentially nonzero mean).  In analogy with the geodesic case, we have chosen the origin of phase space such that $\psi^{(0)}_\alpha(q_\beta=0)=0$. This choice will ensure that at fixed $p^i_q$, $\psi^{(0)}_\alpha$ and $q_\alpha$ satisfy all the relationships in Sec.~\ref{geodesic Mino angle variables}. Note that we could replace $\Delta \psi^{(0)}_\alpha(0,p_q^i)$ with any other $q_\alpha$-independent function of $p^i_q$; this would still represent a geodesic relationship between $\psi^{(0)}_\alpha$ and $q_\alpha$, but with different choices of initial phases for different values of $p^i_q$. For convenience in later expressions, we define
\beq
A_\alpha(p^i_q):=- \Delta \psi^{(0)}_\alpha(0,p_q^i).\label{a}
\eeq


By substituting the expansions~\eqref{near-identity psi} and \eqref{near-identity pi} into Eqs.~\eqref{psidot perturbative} and \eqref{pidot perturbative}, appealing to \eqref{qdot perturbative} and~\eqref{pqidot}, and equating coefficients of powers of $\e$, one can solve for the frequencies $\Upsilon_\alpha^{(n)}$ and driving forces $G^i_{(n)}$, as well as for the functions in the averaging transformation. Explicitly, the leading-order terms in Eqs.~\eqref{psidot perturbative} and \eqref{pidot perturbative} are
\begin{align}
\frac{\partial\psi^{(0)}_\alpha}{\partial q_\beta}\Upsilon^{(0)}_\beta = \Upsilon^{(0)}_\alpha + \bm{\Upsilon}^{(0)}\cdot\frac{\partial \Delta\psi^{(0)}_\alpha}{\partial\bm{q}} &= \mathscr{f}^{(0)}_\alpha(\bm{\psi}^{(0)},p_q^j),\label{NITpsi0}\\
G^i_{(1)} + \bm{\Upsilon}^{(0)}\cdot\frac{\partial p^i_{(1)}}{\partial\bm{ q}} &=  g^i_{(1)}(\bm{\psi}^{(0)},p_q^j).\label{NITp1}
\end{align}
Equation~\eqref{NITpsi0} is simply the geodesic relationship between $\psi_\alpha$ and $q_\alpha$. It follows that we can use the geodesic solution~\eqref{q_a(psi_a)} for $q_a(\psi^{(0)}_a,p^i_q)$. 
Concretely, we may write 
\beq
q_a(\psi^{(0)}_a,p^i_q) = \Upsilon^{(0)}_a(p^i_q)\int_{0}^{\psi^{(0)}_a}\frac{d\psi'_a}{\mathscr{f}^{(0)}_a(\psi_a',p^i_q)},\label{qa(psia) perturbed}
\eeq
implying that the Fourier coefficients in Eq.~\eqref{Fourier q perturbed} can  be computed as the integrals over $\bm{\psi}^{(0)}$ in Eq.~\eqref{Fourier q coeffs}, with the replacements $\Upsilon_a\to \Upsilon_a^{(0)}$ and $\psi_a\to\psi_a^{(0)}$. This relies on our particular choice of $A_\alpha$ in Eq.~\eqref{a}; different choices would lead to different $p^i_q$-dependent lower limits of integration in Eq.~\eqref{qa(psia) perturbed}, which in turn would lead to $p^i_q$-dependent phase factors appearing in Eq.~\eqref{Fourier q coeffs}. 

Using either of the forms~\eqref{Fourier q coeffs} or~\eqref{Fourier q coeffs from psi}, we can easily decompose Eqs.~\eqref{NITpsi0} and \eqref{NITp1} into Fourier series, with $\Delta\psi^{(0)}_\alpha = \sum_{\bm{k}\neq0}\Delta\psi^{(0,\bm{k})}_\alpha(p^j_q)e^{-iq_{\bm{k}}}$ and $p^i_{(1)}=\sum_{\bm{k}}p^i_{(1,\bm{k})}(p^j_q)e^{-iq_{\bm{k}}}$. From the $\bm{k}=0$ terms in the equations, we find 
\begin{align}
\Upsilon^{(0)}_\alpha(p_q^j) &= \left\langle\mathscr{f}^{(0)}_\alpha(\bm{\psi}^{(0)},p_q^j)\right\rangle_{\bm{q}},\label{Upsilon0}\\
G_{(1)}^i(p_q^j) &= \left\langle g^i_{(1)}(\bm{\psi}^{(0)},p_q^j)\right\rangle_{\bm{q}},\label{G1}
\end{align}
and from the $\bm{k}\neq0$ terms we find
\begin{align}
\Delta\psi^{(0,\bm{k})}_\alpha(p_q^j) &= -\frac{\mathscr{f}^{(0,\bm{k})}_\alpha(p_q^j)}{i\Upsilon^{(0)}_{\bm{k}}(p_q^j)},\label{Dpsi0}\\
p^i_{(1,\bm{k})}(p_q^j) &= -\frac{g^i_{(1,\bm{k})}(p_q^j)}{i\Upsilon^{(0)}_{\bm{k}}(p_q^j)},\label{Dp1}
\end{align}
where $\Upsilon^{(0)}_{\bm{k}}:=\bm{k}\cdot \bm{\Upsilon}^{(0)} = k_r\Upsilon^{(0)}_r + k_z\Upsilon^{(0)}_z$. Note that these equations leave $p^i_{(1,00)}$ arbitrary. 

As foreshadowed above, Eqs.~\eqref{Upsilon0} and \eqref{Dpsi0} are precisely the same as the geodesic formulas ~\eqref{Upsilon = <f>} and \eqref{psi to q} (with the replacement $p^i\to p^i_q$). The only change is that the orbital parameters $p^i_q$, which determine the frequencies and amplitudes, now adiabatically evolve with time. 

Importantly, Eq.~\eqref{Dp1} requires $\Upsilon^{(0)}_{\bm{k}}\neq0$. This condition fails at resonances, where $\Upsilon^{(0)}_{\bm{k}^{\rm res}}=0$. Therefore, the averaging transformation is impossible when there is a resonance. We discuss this resonant case in the next section. Eq.~\eqref{Dpsi0} also superficially appears to encounter a singularity at resonance, but this is skirted by the particular form of $\mathscr{f}^{(0,\bm{k})}_\alpha$, as we see from the more explicit formula~\eqref{psi to q}.

Moving onto the first subleading order in Eqs.~\eqref{psidot perturbative} and \eqref{pidot perturbative}, we have
\begin{align}
\Upsilon^{(1)}_\alpha + G^j_{(1)}\frac{\partial\psi^{(0)}_\alpha}{\partial p_q^j} +\bm{\Upsilon}^{(1)}\cdot&\frac{\partial\Delta\psi^{(0)}_\alpha}{\partial\bm{ q}}+\bm{\Upsilon}^{(0)}\cdot\frac{\partial\psi^{(1)}_\alpha}{\partial\bm{ q}} \nonumber\\
&= \mathscr{f}^{(1)}_\alpha + p^j_{(1)}\frac{\partial\mathscr{f}^{(0)}_\alpha}{\partial p_q^j} + \bm{\psi}^{(1)}\cdot\frac{\partial\mathscr{f}^{(0)}_\alpha}{\partial\bm{\psi}^{(0)}},\label{psi1 eqn}\\
G^i_{(2)} +G^j_{(1)}\frac{\partial p^i_{(1)}}{\partial p_q^j}+\bm{\Upsilon}^{(1)}\cdot&\frac{\partial  p^i_{(1)}}{\partial \bm{q}}+\bm{\Upsilon}^{(0)}\cdot\frac{\partial p^i_{(2)}}{\partial \bm{q}} \nonumber\\
&= g^i_{(2)} + p^j_{(1)}\frac{\partial g^i_{(1)}}{\partial p_q^j} + \bm{\psi}^{(1)}\cdot\frac{\partial g^i_{(1)}}{\partial \bm{\psi}^{(0)}},\label{p2 eqn}
\end{align}
where all quantities on the left are functions of $(\bm{q},p_q^j)$, and all those on the right are functions of $(\bm{\psi}^{(0)},p_q^j)$. Taking the average of these equations yields 
\begin{align}
\Upsilon^{(1)}_\alpha(p_q^i) &= \left\langle\mathscr{f}^{(1)}_\alpha\right\rangle_{\!\bm{q}}+\left\langle p^i_{(1)}\frac{\partial\mathscr{f}^{(0)}_\alpha}{\partial p_q^i} +\bm{\psi}^{(1)}\cdot\frac{\partial\mathscr{f}^{(0)}_\alpha}{\partial \bm{\psi}^{(0)}}\right\rangle_{\!\bm{q}} -  G^j_{(1)}\frac{\partial A_\alpha}{\partial p^j_q}, \label{Upsilon1}\\
G_{(2)}^i(p_q^j) &= \left\langle g^i_{(2)}\right\rangle_{\bm{q}} + \left\langle  p^j_{(1)}\frac{\partial g^i_{(1)}}{\partial p_q^j} +\bm{\psi}^{(1)}\cdot\frac{\partial g^i_{(1)}}{\partial\bm{ \psi}^{(0)}}\right\rangle_{\!\bm{q}} - G^j_{(1)}\frac{\partial p^i_{(1,00)}}{\partial p^j_q}.\label{G2}
\end{align}

We see from Eq.~\eqref{Upsilon1} that a judicious choice of $p^i_{(1,00)}$ allows us to set
\beq
\Upsilon^{(1)}_\alpha = 0 \quad\text{for }\alpha=r,z,\phi.\label{Upsilon1_rzphi}
\eeq
Such a $p^i_{(1,00)}$ is determined from
\beq
p^i_{(1,00)}\frac{\partial\Upsilon^{(0)}_\alpha}{\partial p_q^i} =  -\left\langle\mathscr{f}^{(1)}_\alpha\right\rangle_{\!\bm{q}} - \sum_{\bm{k}\neq0}p^i_{(1,\bm{k})}\frac{\partial\mathscr{f}^{(0,-\bm{k})}_\alpha}{\partial p_q^i} -\left\langle\bm{\psi}^{(1)}\cdot\frac{\partial\mathscr{f}^{(0)}_\alpha}{\partial \bm{\psi}^{(0)}}\right\rangle_{\!\bm{q}}+ G^j_{(1)}\frac{\partial A_\alpha}{\partial p^j_q}\label{p100}
\eeq
for $\alpha=r,z,\phi$. We could alternatively set $\Upsilon^{(1)}_\alpha = 0$ for a different trio of components, but this choice will be particularly useful in the final section of this review. This freedom is in addition to the freedom discussed above regarding the choice of $A_\alpha$; i.e., the functions $A_\alpha$ and $p^i_{(1,00)}$ in the averaging transformation are degenerate with $\Upsilon^{(1)}_\alpha$. Ref.~\cite{vandeMeent:2018rms} provides a more thorough discussion of the freedom within near-identity averaging transformations.

The averages in  Eqs.~\eqref{Upsilon1}--\eqref{G2} involve  $\psi^{(1)}_a$, which can be obtained from Eq.~\eqref{psi1 eqn}. A $2\pi$-biperiodic solution to that equation is\footnote{This seems to be the unique  $2\pi$-biperiodic solution. Any other solution can only differ by a homogeneous solution to Eq.~\eqref{psi1 eqn}, which must take the form $\exp(\int \mathscr{f}'_a dq_a/\Upsilon^{(0)}_a)f(q_b-q_a\Upsilon^{(0)}_b/\Upsilon^{(0)}_a)$ for some function $f$, with $b\neq a$. It appears that such a function cannot simultaneously be $2\pi$ periodic in both $q_a$ and $q_b$.}
\begin{align}
\psi_a^{(1)}(\bm{q}, p^j_q) = \frac{1}{Y_a(q_a, p^j_q)}\sum_{\bm{k}}\sum_k \frac{S^{\bm{k}}_a(p^j_q)Y^k_a(p^j_q)}{-i\Upsilon^{(0)}_{\bm{k}}-ik\Upsilon^{(0)}_a - \langle \mathscr{f}^{\prime}_a\rangle_{\bm{q}}} e^{-i q_{\bm{k}} - i k q_a},\label{psi1_a soln}
\end{align}
where $Y_a(q_a, p^j_q) := \exp[-F_a(q_a, p^j_q)/\Upsilon^{(0)}_a(p^j_q)] = \sum_k Y^k_a e^{-ikq_a}$, $F_a := \sum_{k\neq0}\frac{\mathscr{f}^{\prime\,k}_a}{-ik} e^{-ikq_a}$ is the  antiderivative of the purely oscillatory part of $\mathscr{f}^\prime_a:=\partial\mathscr{f}^{(0)}_a/\partial\psi^{(0)}_a$, and \beq
S_a(\bm{q},p^i_q) :=  - G^j_{(1)}\frac{\partial\psi^{(0)}_a}{\partial p_q^j}  + \mathscr{f}^{(1)}_a + p^j_{(1)}\frac{\partial\mathscr{f}^{(0)}_a}{\partial p_q^j} = \sum_{\bm{k}}S^{\bm{k}}_a(p^i_q) e^{-i q_{\bm{k}}}.
\eeq

The remaining pieces of Eqs.~\eqref{psi1 eqn} and \eqref{p2 eqn} determine the purely oscillatory parts of $\psi^{(1)}_t$, $\psi^{(1)}_\phi$, and $p^i_{(2)}$. Specifically, 
\begin{align}
\psi^{(1,\bm{k})}_\alpha &= \frac{1}{-i\Upsilon_{\bm{k}}^{(0)}} \left(\mathscr{f}^{(1,\bm{k})}_\alpha + P_\alpha^{\bm{k}} - G^j_{(1)}\frac{\partial\Delta\psi^{(0,\bm{k})}_\alpha}{\partial p_q^j}\right),\label{Dpsi1}\\
p^i_{(2,\bm{k})} &= \frac{1}{-i\Upsilon_{\bm{k}}^{(0)}} \left(g^i_{(2,\bm{k})} + Q^i_{\bm{k}} - G^j_{(1)}\frac{\partial p^i_{(1,\bm{k})}}{\partial p_q^j}\right)
\end{align}
for $\alpha=t,\phi$, where $P_\alpha := p^j_{(1)}\frac{\partial\mathscr{f}^{(0)}_\alpha}{\partial p_q^j} + \bm{\psi}^{(1)}\cdot\frac{\partial\mathscr{f}^{(0)}_\alpha}{\partial\bm{\psi}^{(0)}}$ and $Q^i := p^j_{(1)}\frac{\partial g^i_{(1)}}{\partial p_q^j} + \bm{\psi}^{(1)}\cdot\frac{\partial g^i_{(1)}}{\partial\bm{\psi}^{(0)}}$.  
%

This averaging transformation can be carried to any order. Analogous calculations also apply if we use $P^i = (E,L_z,Q)$ rather than $p^i =(p,e,\iota)$. Ultimately, the coordinate trajectory $z^\alpha$ can be expressed in terms of $(q_\alpha,p^i_q)$ as
\beq
z^\alpha(q_\beta,p^i_q) = z^\alpha_{(0)}(q_\beta,p^i_q) + \e z^\alpha_{(1)}(\bm{q},p^i_q) +O(\e^2).
\eeq
The leading-order trajectory has the same dependence on $q_\alpha$ and $p^i_q$ as a geodesic. That is, if we write a geodesic as $z^\alpha_G(q_\beta,p^i)$, given by Eq.~\eqref{zG(lambda)}, then $z^\alpha_{(0)}(q_\beta,p^i_q)=z^\alpha_G(q_\beta,p^i_q)$. Wherever the geodesic expressions involve $P^i$, they are here evaluated at $P^i_q=(E_q,L_q,Q_q)$, which are related to $p^i_q$ by the geodesic relationships. (We suppress the subscript $z$ on $L_z$.) 
The difference between the geodesic and the accelerated trajectory lies entirely in the evolution of their arguments: rather than evolving according to Eqs.~\eqref{qdot geodesic} and \eqref{pidot geodesic}, $q_\alpha$ and $p^i_q$ now evolve according to Eqs.~\eqref{qdot perturbative} and \eqref{pqidot}. 

In the context of a binary, the small corrections $\e z^\alpha_{(1)}(\bm{q},p^i_q)$ to the trajectory remain uniformly small over the entire inspiral until the transition to plunge~\cite{Ori:2000zn}; because they are periodic functions of $\bm{q}$, they have no large secular terms. $t^{(1)}$ and $ \phi^{(1)}$ are given by Eq.~\eqref{Dpsi1}, with $t^{(1,00)}$ and $\phi^{(1,00)}$ left arbitrary. $r^{(1)}$ and $z^{(1)}$ are given by
\begin{align}
x^a_{(1)} = \bm{\psi}^{(1)}\cdot\frac{\partial x^a_G}{\partial\bm{\psi}^{(0)}} + p^i_{(1)}\frac{\partial x^a_G}{\partial p^i_q},
\end{align}
with $\psi^{(1)}_a$ given by Eq.~\eqref{psi1_a soln}, the oscillatory part of $p^i_{(1)}$ by Eq.~\eqref{Dp1}, and $p^i_{(1,00)}$ by Eq.~\eqref{p100}.

Refs.~\cite{Hinderer:2008dm,vandeMeent:2013sza,Fujita:2016igj,Isoyama:2018sib,vandeMeent:2018rms} contain more detailed action-angle treatments of perturbed orbits. With the exception of Ref.~\cite{vandeMeent:2018rms}, these treatments have not begun with equations of the form~\eqref{psidot perturbative} and \eqref{pidot perturbative}. Instead, they began with approximate angle variables, which we will denote $\hat q_\alpha$ and which satisfy $\hat q_\alpha = q_\alpha + O(\e)$. The equations of motion then take the form
\begin{align}
\frac{d\hat q_\alpha}{d\lambda} &= \Upsilon^{(0)}_\alpha(p^j) + \e U^{(1)}_\alpha(\hat{\bm{q}},p^j)+O(\e^2),\label{q0dot}\\
\frac{dp^i}{d\lambda} &= \e F^i_{(1)}(\hat{\bm{q}},p^j) + \e^2 F^i_{(2)}(\hat{\bm{q}},p^j)+O(\e^3).\label{pdot(q0)}
\end{align}
Ref.~\cite{Hinderer:2008dm} derives concrete equations of this form in the case that proper time $\tau$ is used instead of Mino time and that $\{E,L_z,Q\}$ are used instead of $\{p,e,\iota\}$.\footnote{The notation in Ref.~\cite{Hinderer:2008dm} differs in several significant ways from ours. In particular, Ref.~\cite{Hinderer:2008dm} uses $\lambda$ to denote a rescaled $\tau$, $q_\alpha$ to denote an analogue of our $\hat q_\alpha$ (and associated with $\tau$ rather than Mino time), and $\psi_\alpha$ to denote an analogue of our $q_\alpha$ (again associated with $\tau$).} [The driving forces $F^i_{(n)}$ are then given by Eq.~\eqref{dEdtau}.] The averaging transformation  $(\hat q_\alpha,p^i)\to (q_\alpha,p^i_q)$ can be found as we did above, with substantial simplifications arising from the fact that $\frac{d\hat q_\alpha}{d\lambda}$ is constant at leading order; the transformation is given by Eqs.~\eqref{q0 to q} and \eqref{p to pq} below (without the restriction $\bm{k}\neq N\bm{k}_{\rm res}$ in the nonresonant case). 

Our particular construction in this section and the next is instead designed to link the action-angle description with the quasi-Keplerian one. It appears here for the first time. However, Ref.~\cite{vandeMeent:2018rms} considers more general sets of coupled differential equations that involve variables analogous to our $\psi_\alpha$ as well as variables analogous to $\hat q_\alpha$, though without providing a solution analogous to our~\eqref{psi1_a soln}.

\subsubsection{Perturbed Boyer-Lindquist frequencies and action angles}\label{perturbed BL frequencies}

Because we solve field equations and extract waveforms using Boyer-Lindquist time $t$, it is once again useful to construct variables $(\varphi_\alpha,p^i_{\varphi})$ associated with $t$, where $\varphi_t=\psi_t=t$ and $p^i_\varphi=(p_\varphi,e_\varphi,\iota_\varphi)$. The construction of the variables $(\varphi_\alpha,p^i_{\varphi})$ (and of their evolution equations) is analogous to the construction based on $\lambda$: the osculating-geodesic equations for $d\psi_\alpha/dt$ and $dp^i/dt$ have the same form as Eqs.~\eqref{psidot perturbative} and \eqref{pidot perturbative}, simply with $\mathscr{f}^{(n)}_\alpha\to \mathscr{f}^{(n)}_\alpha/\mathscr{f}^{(0)}_t$ and $g^i_{(n)}\to g^i_{(n)}/\mathscr{f}^{(0)}_t$, and after a near-identity averaging transformation we arrive at the equations of motion
\begin{align}
\frac{d\varphi_\alpha}{dt} &= \Omega^{(0)}_\alpha(p^j_{\varphi}),\label{dphidt}\\
\frac{dp^i_{\varphi}}{dt} &= \e \Gamma_{(1)}^i(p^j_{\varphi}) + \e^2 \Gamma_{(2)}^i(p^j_{\varphi}) + O(\e^3).\label{dpphidt}
\end{align}
$\Omega^{(0)}_\alpha$ are the geodesic frequencies, and $\Omega^{(0)}_t=1$. The corrections $\Omega^{(n)}_\alpha$ for $\alpha=r,z,\phi$ and $n>0$ are eliminated just as in the previous section, while $\Omega^{(n)}_t=0$ trivially for $n>0$ because $d\varphi_t/dt=1$. 


The two sets of variables $(\varphi_\alpha,p^i_\varphi)$ and $(q_\alpha,p^i_q)$ are related by a transformation
\begin{align}
\varphi_\alpha(q_\beta,p_q^j,\e) &= \varphi^{(0)}_\alpha(q_\alpha,p^i_q) + \e  \Phi_\alpha^{(1)}(\bm{q},p_q^j) +O(\e^2),\label{q to phi}\\
p^i_{\varphi}(q_\beta, p_q^j,\e) &= p_q^i +  \e \pi^i_{(1)}(\bm{q},p_q^j)+O(\e^2),\label{pq to pphi}
\end{align}
where the leading term in $\varphi_\alpha$ is given by the geodesic relationship~\eqref{varphi(q)}, which we restate as
\beq
\varphi^{(0)}(q_\alpha,p^i_q) := q_\alpha+B_\alpha(p^i_q)+\Omega_\alpha^{(0)}(p^i_q)\Delta t^{(0)}(\bm{q},p_q^j),\label{phi0}
\eeq
defining
\beq
B_\alpha(p^i_q):=- \Omega_\alpha^{(0)}(p^i_q)\Delta t^{(0)}(0,p_q^j).
\eeq
Like in the geodesic case, this value for $B_\alpha$ imposes that $\varphi_\alpha$ and $q_\alpha$ (and $\psi^{(0)}_\alpha$) have the same origin in phase space. As discussed around Eqs.~\eqref{a} and \eqref{qa(psia) perturbed}, this means that we can immediately utilize all the relationships from Sec.~\ref{geodesic BL angle variables}. The corrections $\Phi^{(1)}_\alpha$ and $\pi^i_{(1)}$ are $2\pi$-biperiodic in $\bm{q}$. 


The terms in this transformation, as well as the driving forces $\Gamma^i_{(n)}$, can be derived by substituting Eqs.~\eqref{q to phi} and \eqref{pq to pphi} into Eqs.~\eqref{dphidt} and \eqref{dpphidt}. Taking the average of the resulting equations and appealing to Eqs.~\eqref{qdot perturbative} and \eqref{pqidot}, we obtain
\begin{align}
\Omega^{(0)}_\alpha = \frac{\Upsilon^{(0)}_\alpha}{\Upsilon^{(0)}_t} \quad \text{and} \quad
\Gamma_{(1)}^i = \frac{G^i_{(1)}}{\Upsilon^{(0)}_t}
\end{align}
at leading order and
\begin{align}
\Omega^{(1)}_\alpha =0 &= -\frac{1}{\Upsilon_t^{(0)}}\left(G^j_{(1)}\partial_{p^j}B_\alpha+\langle R^i\rangle_{\bm{q}} \partial_{p^i}\Omega^{(0)}_\alpha + \langle P_t\rangle_{\bm{q}} \Omega^{(0)}_\alpha\right),\label{Omega1}\\
\Gamma_{(2)}^i &= \frac{1}{\Upsilon^{(0)}_t}\left(G^i_{(2)} +G^j_{(1)}\partial_{p^j}\langle \pi^i_{(1)}\rangle_{\bm{q}} - \langle R^j\rangle_{\bm{q}}\partial_{p^j}\Gamma^i_{(1)} - \langle P_t\rangle_{\bm{q}}\Gamma^i_{(1)} \right)\label{Gamma2}
\end{align}
at the first subleading order, where $R^i:=\mathscr{f}^{(0)}_t \pi^i_{(1)}$ and $P_t=\bm{\psi}^{(1)}\cdot\partial_{\bm{\psi}}\mathscr{f}^{(0)}_t + p^i_{(1)}\partial_{p^i}\mathscr{f}^{(0)}_t$. The average $\langle \pi^i_{(1)}\rangle_{\bm{q}}$ is chosen to enforce $\Omega^{(1)}_\alpha=0$ in Eq.~\eqref{Omega1}. The oscillatory parts of the equations yield
\begin{align}
\Delta t^{(0,\bm{k})} &= \frac{\mathscr{f}^{(0,\bm{k})}_t}{-i\Upsilon^{(0)}_{\bm{k}}},\\
\pi^i_{(1,\bm{k})} &= \frac{\mathscr{f}^{(0,\bm{k})}_t}{-i\Upsilon^{(0)}_{\bm{k}}}\Gamma^i_{(1)}
\end{align}
at leading order and
\begin{align}
\Phi^{(1,\bm{k})}_\alpha &= \frac{1}{-i\Upsilon^{(0)}_{\bm{k}}}\left(R^i_{\bm{k}}\partial_{p^i}\Omega^{(0)}_\alpha+P^{\bm{k}}_t\Omega^{(0)}_\alpha  -G^i_{(1)}\partial_i \Delta\varphi^{(0,\bm{k})}_\alpha\right)
\end{align}
at the first subleading order. In all of the above expressions, $\bm{k}$ refers to a Fourier decomposition into $e^{-iq_{\bm{k}}}$ modes. All functions of $p^i$ are evaluated at $p^i_q$, and inside the integrals~\eqref{Fourier q coeffs}, all functions of $\bm{\psi}$ are evaluated at $\bm{\psi}^{(0)}(\bm{q},p^i_q)$.


When solving the field equations, we shall require Fourier decompositions with respect to  $\bm{\varphi}$,
\beq
f[r(t),z(t)] = \sum_{\bm{k}} f^{(\varphi)}_{\bm{k}}(p^j_\varphi) e^{-i\varphi_{\bm{k}}}.
\eeq
We can calculate the coefficients as integrals over $\bm{q}$ using the transformation~\eqref{q to phi}. However, it is simpler to use the geodesic change of variables defined by the leading term in the transformation. The coefficients are then given by Eq.~\eqref{Fourier coeff relationship} with the replacements $p^i\to p^i_\varphi$, $\Upsilon_\alpha\to\Upsilon^{(0)}_\alpha$, $\mathscr{f}_\alpha\to\mathscr{f}^{(0)}_\alpha$, or by Eq.~\eqref{Fourier coeff relationship psi} with the additional replacement $\bm{\psi}\to\bm{\psi}^{(0)}$.

We will also require the transformation from $(\psi_\alpha,p^i)$ to $(\varphi_\alpha,p^i_\varphi)$:
\begin{align}
\psi_\alpha(\varphi_\beta,p^i_\varphi,\e) &= \psi^{(0)}_\alpha(\varphi_\beta,p^i_\varphi) + \e\psi^{(\varphi,1)}_\alpha(\bm{\varphi},p^i_\varphi) + O(\e^2),\label{psi from phi perturbative}\\
p^i(\varphi_\alpha,p^j_\varphi,\e) &= p^i_\varphi + \e p^i_{(\varphi,1)}(\bm{\varphi},p^j_\varphi)+O(\e^2).\label{p from p_phi perturbative}
\end{align}
Following the same steps as in the previous section, at leading order we recover the geodesic frequencies and find $\psi^{(0)}_\alpha(\varphi_\beta,p^i_\varphi)$ is given by the geodesic relationship~\eqref{psi(phi)}. Solving the subleading-order equations is made difficult because the analogue of Eq.~\eqref{psi1 eqn} has the form
\beq
\bm{\Omega}^{(0)}\cdot\frac{\partial\psi^{(\varphi,1)}_\alpha}{\partial\bm{\varphi}} - \bm{\psi}^{(\varphi,1)}\cdot\frac{\partial}{\partial\bm{\psi}^{(0)}}\left(\frac{\mathscr{f}^{(0)}_\alpha}{\mathscr{f}^{(0)}_t}\right) = \ldots
\eeq
The $\alpha=r,z$ components of this equation, unlike those of Eq.~\eqref{psi1 eqn}, are coupled partial differential equations for $\bm{\psi}^{(\varphi,1)}$, which do not have a solution of the form~\eqref{psi1_a soln}. However, we can find the subleading terms in Eqs.~\eqref{psi from phi perturbative} and \eqref{p from p_phi perturbative} by combining our knowledge of $(\varphi_\alpha,p^i_\varphi)$ and $(\psi_\alpha,p^i)$ as functions of $(q_\alpha,p^i_q)$. Substituting the expansions~\eqref{q to phi} and \eqref{pq to pphi} into the right-hand sides of Eqs.~\eqref{psi from phi perturbative} and \eqref{p from p_phi perturbative} and equating the results with Eqs.~\eqref{near-identity psi} and \eqref{near-identity pi}, we find
\begin{align}
\psi^{(\varphi,1)}_\alpha(\varphi^{(0)}_\beta,p^i_\varphi) &= \psi^{(1)}_\alpha(q_\beta,p^i_\varphi) - \varphi^{(1)}_\beta\frac{\partial\psi^{(0)}_\alpha}{\partial\varphi_\beta}(\varphi^{(0)}_\gamma,p^i_\varphi) -\pi^i_{(1)}\frac{\partial\psi^{(0)}_\alpha}{\partial p^i_\varphi}(\varphi^{(0)}_\gamma,p^i_\varphi),\label{psi(varphi,1)}\\
p^i_{(\varphi,1)}(\bm{\varphi}^{(0)},p^j_\varphi) &= p^i_{(1)}(\bm{q},p^i_\varphi) - \pi^{i}_{(1)}(\bm{\varphi}^{(0)},p^j_\varphi),\label{p(varphi,1)}
\end{align}
where $\varphi^{(0)}_\alpha$ is given by Eq.~\eqref{phi0} with $p^i_q\to p^i_\varphi$. The inverse transformation, which we will also need, is
\begin{align}
\varphi_\alpha(\psi_\beta,p^i,\e) &= \varphi^{(0)}_\alpha(\psi_\beta,p^i) +\e\Phi^{(1)}_\alpha(\bm{q}^{(0)},p^i)-\e\psi^{(1)}_\beta\frac{\partial \varphi^{(0)}_\alpha}{\partial\psi_\beta}-\e p^i_{(1)}\frac{\partial \varphi^{(0)}_\alpha}{\partial p^i},\label{varphi(psi) perturbative}\\
p^i_\varphi(\psi_\beta,p^i,\e) &= p^i +\e\pi^i_{(1)}(\bm{\varphi}^{(0)},p^j) - \e p^i_{(1)}(\bm{q}^{(0)},p^j),\label{pvarphi(psi) perturbative}
\end{align}
where $\varphi^{(0)}_\alpha$ and $q^{(0)}_\alpha$ are the geodesic functions of $\psi_\alpha$ and $p^i$.

Finally, we can reconstruct the coordinate trajectory $z^\alpha$ in the form
\beq
z^\alpha(\varphi_\beta,p^i_\varphi) = z^\alpha_{(0)}(\varphi_\beta,p^i_{\varphi}) + \e z^\alpha_{(\varphi,1)}(\bm{\varphi},p^i_\varphi) + O(\e^2).\label{z expansion - varphi}
\eeq
The zeroth-order trajectory has the same functional dependence as a geodesic; that is, $z^\alpha_{(0)}(\varphi_\beta,p^i_\varphi) = z^\alpha_G(\varphi_\beta,p^i_\varphi)$, where $z^\alpha_G(\varphi_\beta,p^i)$ is given in Eq.~\eqref{zG(t)}. In analogy with the Mino-time solution, wherever the geodesic expressions involve $P^i$, they are here evaluated at $P^i_\varphi=(E_\varphi,L_\varphi,Q_\varphi)$, which are related to $p^i_\varphi$ by the geodesic relationships. The first-order corrections $z^\alpha_{(\varphi,1)}$ are $t_{(\varphi,1)} = 0$,  $\phi_{(\varphi,1)}=\psi^{(\varphi,1)}_\phi$ given by Eq.~\eqref{psi(varphi,1)}, and
\begin{align}
x^a_{(\varphi,1)} &=\bm{\psi}^{(\varphi,1)}\cdot\frac{\partial x^a_G}{\partial\bm{\psi}^{(0)}}+ p^i_{(\varphi,1)}\frac{\partial x^a_G}{\partial p^i_\varphi},\label{xa(varphi,1)}
\end{align}
with $\bm{\psi}^{(\varphi,1)}$ and $p^i_{(\varphi,1)}$ given by Eqs.~\eqref{psi(varphi,1)} and \eqref{p(varphi,1)}.

\subsubsection{Multiscale expansions, adiabatic approximation, and post-adiabatic approximations}\label{multiscale expansion of orbit}

Self-accelerated orbits are often described with a multiscale (or two-timescale) expansion~\cite{Pound:2007ti,Hinderer:2008dm,Mino:2008rr,Pound:2010wa,Pound:2015wva,Hughes:2016xwf,Bonga:2019ycj,Miller:2020bft}. This is essentially equivalent to the averaging transformation described above.

To illustrate the method, we return to Eqs.~\eqref{psidot perturbative} and \eqref{pidot perturbative}. We introduce a {\em slow time} variable $\tilde \lambda := \e\lambda$; this changes by an amount $\sim \e^0$ on the time scale $\sim 1/\e$. In place of the transformations~\eqref{near-identity psi} and \eqref{near-identity pi}, we adopt expansions
\begin{align}
\psi_\alpha(q_\beta,\tilde\lambda,\e) &= q_\alpha + \tilde A_\alpha(\tilde \lambda) + \Delta\tilde\psi^{(0)}_\alpha(\bm{q},\tilde \lambda) + \e\tilde\psi^{(1)}_\alpha(\bm{q},\tilde\lambda) + O(\e^2),\label{multiscale psi}\\
p^i(q_\alpha,\tilde\lambda,\e) &= \tilde p^i_{(0)}(\tilde \lambda) + \e \tilde p^i_{(1)}(\bm{q},\tilde\lambda) + O(\e^2),\label{multiscale pi}
\end{align}
where $q_\alpha$ satisfies 
\beq
\frac{dq_\alpha}{d\lambda} = \tilde\Upsilon^{(0)}_\alpha(\e \lambda) + \e \tilde \Upsilon^{(1)}_\alpha(\e \lambda) + O(\e^2):=\tilde\Upsilon_\alpha(\e \lambda,\e).
\eeq
We then substitute these expansions into Eqs.~\eqref{psidot perturbative} and \eqref{pidot perturbative}, applying the chain rule
\beq
\frac{d}{d\lambda} = \tilde\Upsilon_\alpha \frac{\partial}{\partial q_\alpha} +\e\frac{\partial}{\partial\tilde\lambda}.
\eeq
$q_\alpha$ and $\tilde\lambda$ are then treated as independent variables, making Eqs.~\eqref{psidot perturbative} and \eqref{pidot perturbative} into a sequence of equations, one set at each order in $\e$. These equations are essentially equivalent to \eqref{NITpsi0} and \eqref{NITp1} at leading order and to Eqs.~\eqref{psi1 eqn} and \eqref{p2 eqn} at first subleading order, with tildes placed over all quantities and the following replacements: $p^i_q\to \tilde p^i_{(0)}$, $G^i_{(n)}\to d\tilde p^i_{(n-1)}/d\tilde\lambda$,  $G^i_{(1)}\frac{\partial\psi^{(0)}_\alpha}{\partial p^j_q}\to d(\tilde A_\alpha+\Delta\tilde\psi^{(0)}_\alpha)/d\tilde\lambda$, and $G^i_{(1)}\frac{\partial p^i_{(1)}}{\partial p^j_q}\to 0$. These equations can be solved just as we solved Eqs.~\eqref{NITpsi0}, \eqref{NITp1}, \eqref{psi1 eqn}, and \eqref{p2 eqn}. 

The only difference between this expansion and Eqs.~\eqref{near-identity psi} and \eqref{near-identity pi} is how each parameterizes the orbit's slow evolution, whether with slowly evolving parameters $p^i_q$ or with slow time $\tilde\lambda$. Indeed, the solutions are easily related. The solutions to Eqs.~\eqref{qdot perturbative} and \eqref{pqidot} can be expanded as 
\begin{align}
q_\alpha(\tilde \lambda,\e) &= \frac{1}{\e}\left[\tilde q^{(0)}_\alpha(\tilde \lambda) + \e \tilde q^{(1)}_\alpha(\tilde \lambda) + O(\e^2)\right],\label{multiscale q}\\
p^i_q(\tilde \lambda,\e) &= \tilde p^i_{q(0)}(\tilde\lambda) + \e  \tilde p^i_{q(1)}(\tilde \lambda) + O(\e^2),\label{multiscale pq}
\end{align}
where $\tilde q^{(n)}_\alpha(\tilde\lambda) = \int_0^{\tilde\lambda}\tilde \Upsilon^{(n)}_\alpha(\tilde\lambda^\prime) d\tilde\lambda^\prime + \tilde q^{(n)}_\alpha(0)$ with
\begin{align}
\tilde\Upsilon^{(0)}_\alpha(\tilde \lambda) &= \Upsilon^{(0)}_\alpha,\\ \tilde\Upsilon^{(1)}_\alpha(\tilde \lambda) &= \Upsilon^{(1)}_\alpha + \tilde p^i_{q(1)}(\tilde\lambda)\partial_{p^i}\Upsilon^{(0)}_\alpha.
\end{align}
On the right, $\Upsilon^{(n)}_\alpha$ and its derivatives are evaluated at $\tilde p^i_{q(0)}(\tilde\lambda)$. Substituting these expansions into Eqs.~\eqref{near-identity psi} and \eqref{near-identity pi} and comparing to Eqs.~\eqref{multiscale psi} and \eqref{multiscale pi}, we read off
\begin{align}
\Delta\tilde\psi^{(0)}_\alpha(\bm{q},\tilde\lambda) &= \Delta\psi^{(0)}_\alpha,\\
\tilde p^i_{(0)}(\tilde\lambda) &= \tilde p^i_{q(0)}(\tilde\lambda),\\
\tilde\psi^{(1)}_\alpha(\bm{q},\tilde\lambda) &= \psi^{(1)}_\alpha + \tilde p^j_{q(1)}(\tilde\lambda)\partial_{p^j}\psi^{(0)}_\alpha,\\
\tilde p^i_{(1)}(\bm{q},\tilde\lambda) &= p^i_{(1)}+\tilde p^i_{q(1)}(\tilde\lambda).
\end{align}
Here $\psi^{(0)}_\alpha$, $p^i_{(1)}$, $\psi^{(1)}_\alpha$, and their derivatives are evaluated at $[\bm{q},\tilde p^j_{q(0)}(\tilde\lambda)]$. These particular relationships rely on choosing $\tilde A_\alpha(\tilde \lambda)=A_\alpha[\tilde p^i_{q(0)}(\tilde\lambda)]$. Just as the averaging transformation did, the multiscale expansion has considerable degeneracy between $\tilde A_\alpha$, $\tilde\Upsilon^{(1)}_\alpha$, and $\langle\tilde p^i_{(1)}\rangle$. If different choices are made, then we cannot identify $q_\alpha$ between the two methods. However, regardless of choices, both methods will ultimately output identical solutions of the form $\psi_\alpha(\tilde \lambda,\e)$ and $p^i(\tilde\lambda,\e)$ (assuming identical initial conditions), and when written in that form they can be unambiguously related.

All the same relationships apply if we instead use $t$-based variables with a slow time $\tilde t :=\e t$. When considering the multiscale expansion of the Einstein equation, it will be useful to have at hand the expansions
\begin{align}
\varphi_\alpha(\tilde t,\e) &= \frac{1}{\e}\left[\tilde\varphi^{(0)}_\alpha(\tilde t) + \e\tilde\varphi^{(1)}_\alpha(\tilde t) + O(\e^2)\right],\label{multiscale varphi}\\
p^i_\varphi(\tilde t,\e) &= \tilde p^i_{\varphi(0)}(\tilde t) + \e\tilde p^i_{\varphi(1)}(\tilde t) + O(\e^2).\label{multiscale pivarphi}
\end{align}
It follows from Eqs.~\eqref{dphidt} and \eqref{dpphidt} that the coefficients in these expansions satisfy
\begin{align}
\frac{d\tilde\varphi^{(0)}_\alpha}{d\tilde t} &= \Omega^{(0)}_\alpha(\tilde p^j_{\varphi(0)}),\label{0PA varphi}\\
\frac{d\tilde p^i_{\varphi(0)}}{d\tilde t} &= \Gamma^i_{(1)}(\tilde p^j_{\varphi(0)}),\label{0PA pivarphi}\\
\frac{d\tilde\varphi^{(1)}_\alpha}{d\tilde t} &= \tilde p^j_{\varphi(1)}\partial_j \Omega^{(0)}_\alpha(\tilde p^j_{\varphi(0)}),\label{1PA varphi}\\
\frac{d\tilde p^i_{\varphi(1)}}{d\tilde t} &= \Gamma^i_{(2)}(\tilde p^j_{\varphi(0)}) +\tilde p^j_{\varphi(1)}\partial_j \Gamma^i_{(1)}(\tilde p^j_{\varphi(0)}).\label{1PA pivarphi}
\end{align}
We can also write Eq.~\eqref{multiscale varphi} as 
\beq
\varphi_\alpha(\tilde t,\e) = \frac{1}{\e}\int_0^{\tilde t}\Omega_\alpha(\tilde t',\e)d\tilde t' + \varphi_\alpha(0,\e),
\eeq
with
\beq\label{Omega tilde expansion}
\Omega_\alpha(\tilde t,\e) = \tilde\Omega^{(0)}_\alpha(\tilde t) +\e \tilde\Omega^{(1)}_\alpha(\tilde t) +O(\e^2),
\eeq
where $\tilde\Omega^{(0)}_\alpha(\tilde t) = \Omega^{(0)}_\alpha[\tilde p^i_{\varphi(0)}(\tilde t)]$ and $\tilde\Omega^{(1)}_\alpha(\tilde t) = \tilde p^j_{\varphi(1)}(\tilde t)\partial_j \Omega^{(0)}_\alpha[\tilde p^j_{\varphi(0)}(\tilde t)]$. 

There is a tradeoff in solving Eqs.~\eqref{0PA varphi}--\eqref{1PA pivarphi} rather than Eqs.~\eqref{dphidt} and \eqref{dpphidt}: Eqs.~\eqref{0PA varphi}--\eqref{1PA pivarphi} double the number of numerical variables, but they are independent of $\e$, meaning they can be solved for all values of $\e$ simultaneously. Eqs.~\eqref{dphidt} and \eqref{dpphidt} have half as many variables, but they cannot be solved without first specifying a value of $\e$.

Since the waveform phase in a binary is directly related to the orbital phase, the expansion~\eqref{multiscale varphi} provides a simple means of assessing the level of accuracy of a given approximation. The approximation that includes only the first term, $\tilde\varphi_\alpha^{(0)}$, is called the {\em adiabatic approximation} (denoted 0PA); it consists of the coupled equations~\eqref{0PA varphi} and \eqref{0PA pivarphi}, which describe a slow evolution of the geodesic frequencies. An approximation that includes all terms through $\tilde\varphi_\alpha^{(n)}$ is called an {\em $n$th post-adiabatic approximation} (denoted $n$PA); it consists of the coupled equations~\eqref{0PA varphi}--\eqref{1PA pivarphi}. We return to the efficacy of 0PA and 1PA approximations in the final section of this review.

Ref.~\cite{Hinderer:2008dm} determined what inputs are required for a 0PA or 1PA approximation. To describe these inputs, we define the time-reversal $\psi_\alpha\to-\psi_\alpha$, $f^\alpha(\bm{\psi})\to \varepsilon^\alpha f^\alpha(-\bm{\psi})$, where $f^\alpha$ is the accelerating force, $\epsilon^\alpha := (-1,1,1,-1)$, and there is no summation over $\alpha$. We then define the dissipative and conservative pieces of the force:
\begin{align}
f^\alpha_{\rm diss} &= \frac{1}{2}f^\alpha(\bm{\psi}) - \frac{1}{2}\varepsilon^\alpha f^\alpha(-\bm{\psi}),\label{fdiss}\\
f^\alpha_{\rm con} &= \frac{1}{2}f^\alpha(\bm{\psi}) + \frac{1}{2}\varepsilon^\alpha f^\alpha(-\bm{\psi}).\label{fcon}
\end{align}
These definitions imply that under time reversal, $f^\alpha_{\rm diss}\to - f^\alpha_{\rm diss}$ and $f^\alpha_{\rm con}\to +f^\alpha_{\rm con}$. It is straightforward to see from Eqs.~\eqref{pdot}--\eqref{dfz} and the definition of ${\cal L}_i$ that $dp^i/dt$ only receives a direct contribution from $f^\alpha_{\rm diss}$, while $d\psi_a/dt$ only receives a direct contribution from $f^\alpha_{\rm con}$. At 0PA order, $f^\alpha$ enters the evolution through Eq.~\eqref{0PA pivarphi}, in the quantity
\beq
\Gamma^i_{(1)}=\left\langle\left.\frac{dp^i}{dt}\right|_{f^\alpha\to f^\alpha_{(1)}}\right\rangle_{\bm{\varphi}} = \frac{1}{\Upsilon^{(0)}_t}\left\langle\left.\frac{dp^i}{d\lambda}\right|_{f^\alpha\to f^\alpha_{(1)}}\right\rangle_{\bm{q}}.
\eeq
Hence, the 0PA approximation only requires $f^\alpha_{(1)\rm diss}$. At 1PA, $f^\alpha$ enters the evolution through both $\Gamma^i_{(1)}$ and $\Gamma^i_{(2)}$ in Eq.~\eqref{1PA pivarphi}, where $\Gamma^i_{(2)}$ is given by Eq.~\eqref{Gamma2} with \eqref{G2}, \eqref{Dp1}, \eqref{psi1_a soln}, and with $p^i_{(1,00)}$ chosen such that $\Upsilon^{(1)}_\alpha=0$. These quantities involve $f^\alpha_{(2)\rm diss}$ [via $\langle g^i_{(2)}\rangle_{\bm{q}}=\langle\frac{dp^i}{d\lambda}|_{f^\alpha\to f^\alpha_{(2)}}\rangle_{\bm{q}}$ in Eq.~\eqref{G2}],  $f^\alpha_{(1)\rm con}$ (via $\psi^{(1)}_\alpha$ and $p^i_{(1,00)}$, which are both partially determined by $\mathscr{f}^{(1)}_a = \delta\mathscr{f}_a|_{f^\alpha\to f^\alpha_{(1)}}$), and $f^\alpha_{(1)\rm diss}$. Hence, the 1PA approximation requires the entirety of $f^\alpha_{(1)}$ as well as$f^\alpha_{(2)\rm diss}$.

The fact that dissipative effects dominate over conservative ones on the long time scale of an inspiral is important in practical simulations of binaries. At least at first order, the dissipative self-force is substantially easier to compute than the conservative self-force. We discuss this in the final section of this review.

We refer to Ref.~\cite{Kevorkian-Cole:96} for a pedagogical introduction to multiscale expansions in more general contexts. Ref.~\cite{Hinderer:2008dm} contains a detailed discussion of the multiscale approximation for self-accelerated orbits in Kerr spacetime. Refs.~\cite{Pound:2007ti,Pound:2010wa} present variants of the method in simpler binary scenarios.

\subsubsection{Transient resonances}

Given that the orbital frequencies slowly evolve, they will occasionally encounter a resonance. Typically~\cite{vandeMeent:2013sza}, the frequencies will continue to evolve, transitioning out of the resonance. These transient resonances have significant impact on orbital evolution.

The near-identity averaging transformation $(\psi_\alpha,p^i)\to (q_\alpha,p_q^i)$ becomes singular at a resonance, as described below Eq.~\eqref{Dp1}. Specifically, it becomes singular for the mode numbers $N\bm{k}_{\rm res}$ for which $\Upsilon_{N\bm{k}_{\rm res}}=0$. To assess the effect of a resonance, we start from the equations of motion in the form~\eqref{q0dot}--\eqref{pdot(q0)}.


The driving forces $F^i_{(n)}$ and ``frequencies'' $U^{(1)}_\alpha$ can be expanded in Fourier series such as $F^i_{(n)} = \sum_{\bm{k}}F^i_{(n,\bm{k})}(p^j)e^{-i \hat q_{\bm{k}}}$. However, near a resonance, a set of apparently oscillatory terms becomes approximately stationary. Specifically, near a resonance where $\Upsilon^{(0)}_{\rm res}:=k^{\rm res}_r\Upsilon_r^{(0)}+ k^{\rm res}_z\Upsilon_z^{(0)}=0$, the phase $q_{\rm res} = \bm{k}_{\rm res}\cdot\hat{\bm{q}}$, and all integer multiples of it, ceases to evolve on the orbital time scale. To see this, suppose the resonance occurs at a time $\lambda_{\rm res}$. Near that time, $q_{\rm res}$ can be expanded in a Taylor series
\beq
q_{\rm res}(\lambda) = q_{\rm res}(\lambda_{\rm res}) + \dot q_{\rm res}(\lambda_{\rm res})(\lambda-\lambda_{\rm res}) + \frac{1}{2}\ddot q_{\rm res}(\lambda_{\rm res})(\lambda-\lambda_{\rm res})^2 + \ldots 
\eeq
Since  $\dot q_{\rm res}(\lambda_{\rm res}) \approx \Upsilon^{(0)}_{\rm res}(\lambda_{\rm res})=0$ and $\ddot q_{\rm res} \approx d\Upsilon^{(0)}_{\rm res}/d\lambda$, we see that $q_{\rm res}$ changes on the time scale
\beq
\delta \lambda = \sqrt{\frac{1}{d\Upsilon^{(0)}_{\rm res}/d\lambda}}\sim\frac{1}{\sqrt{\e}},\label{resonant time scale}
\eeq
which is much longer than the orbital time scale (but much shorter than the radiation-reaction time).

During the passage through a resonance, these additional quasistationary driving forces cause secular changes to the orbital parameters. We isolate these effects by performing a partial near-identity averaging transformation that eliminates all oscillations from the evolution equations {\em except} those depending on the resonant angle variable $q_{\rm res}$. An appropriate transformation, through 1PA order, is given by
\begin{align}
\hat q_\alpha(\bm{q},p^j_q,\e) &= q_\alpha +\e B_\alpha{}^\beta(p^i_q)q_\beta+\e\sum_{\bm{k}\neq N\bm{k}_{\rm res}}\frac{U^{(1,\bm{k})}_\alpha-\frac{F^j_{(1,\bm{k})}}{i\Upsilon^{(0)}_{\bm{k}}}\frac{\partial\Upsilon^{(0)}_\alpha}{\partial p^j_q}}{-i\Upsilon^{(0)}_{\rm res}(p^j_q)}e^{-iq_{\bm{k}}},\label{q0 to q}\\
p^i(\bm{q},p^j_q,\e) &= p^i_q + \e C^i(p^j_q) +\e\sum_{\bm{k}\neq N\bm{k}_{\rm res}}\frac{F^i_{(1,\bm{k})}}{-i\Upsilon^{(0)}_{\rm res}}e^{-iq_{\bm{k}}},\label{p to pq}
\end{align}
where  functions of $p^j$ inside the sums are evaluated at $p^j_q$, and $B_\alpha{}^\beta$ and $C^i$ are any functions satisfying
\beq
B_\alpha{}^\beta \Upsilon^{(0)}_\beta = C^j\partial_{p^j_q}\Upsilon^{(0)}_\alpha + \langle U^{(1)}_\alpha\rangle_{\bm{q}}.
\eeq
These transformations satisfy the analogs of Eqs.~\eqref{NITp1} and \eqref{psi1 eqn} but with resonant modes excluded, with $B_\alpha{}^\beta$ and $C^i$ chosen to eliminate the frequency corrections $\Upsilon_\alpha^{(1)}$, and with the simplifications that $\mathscr{f}^{(0)}_a$ is replaced by $\Upsilon_a^{(0)}$ and $\Delta\psi^{(0)}_a$ by 0 (consequences of $q^{(0)}_\alpha$ already being a leading-order action angle). Together, the transformations~\eqref{q0 to q} and \eqref{p to pq} bring the  equations of motion to the form
\begin{align}
\frac{dq_\alpha}{d\lambda} &= \Upsilon^{(0)}_\alpha(p^j_q) + O(\e),\\
\frac{dp^i_q}{d\lambda} &= \e G^i_{(1)}(p_q^j) + \e \sum_{N\neq0}F^i_{(1,N\bm{k}_{\rm res})}(p_q^j)e^{-iNq_{\rm res}}+ O(\e^2),\label{dot pqi with resonance}
\end{align}
with $G^i_{(1)} = \langle F^i_{(1)}\rangle_{\bm{q}}$. For simplicity, we suppress 1PA terms.

How much does the second term  in Eq.~\eqref{dot pqi with resonance} contribute to the evolution of $p^i_q$? Far from resonance, the additional term averages to zero. If we denote the term as $\e\delta G^i$, then across resonance, it contributes an amount $\delta p^i_q = \e \int \delta G^i d\lambda $. Applying the stationary phase approximation to the integral, we find
\beq
\delta p^i = \sum_{N\neq0}F^i_{(1,N\bm{k}_{\rm res})}\sqrt{\frac{2\pi \e}{|N\dot\Upsilon^{(0)}_{{\rm res}}|}}\exp\left[{\rm sgn}\left(N\dot\Upsilon^{(0)}_{{\rm res}}\right)\frac{i\pi}{4}+iNq_{\rm res}(\lambda_{\rm res})\right]+o(\sqrt{\e}),\label{resonant shifts}
\eeq
where we use a dot to denote a derivative with respect to $\tilde\lambda$, $\dot\Upsilon^{(0)}_{{\rm res}}:=\frac{d\Upsilon^{(0)}_{\rm res}}{d\tilde\lambda}=G^i_{(1)}\frac{\partial\Upsilon^{(0)}_{\rm res}}{\partial p^i_q}$, and both $\dot\Upsilon^{(0)}_{{\rm res}}$ and $F^i_{(1,N\bm{k}_{\rm res})}$ are evaluated at $p^j_q(\lambda_{\rm res})$. The magnitude of $\delta p^i$ is $\sim \sqrt{\e}$; intuitively, this corresponds to a quasistationary driving force of size $\sim \e$ multiplied by the resonance-crossing time $\delta \lambda\sim 1/\sqrt{\e}$. But $\delta p^i$ is not a simple product of the two; each quasistationary driving force is weighted by a phase factor, such that $\delta p^i$ depends sensitively on the value of the resonant phase at resonance, $q_{\rm res}(\lambda_{\rm res})$. This implies that in order to determine $\delta p^i$ at leading order, one must know the 1PA phase evolution prior to resonance.

A proper accounting of the passage through resonance requires matching a near-resonance expansion to an off-resonance, multiscale expansion; see Sec. III of Ref.~\cite{vandeMeent:2013sza} or Appendix B of Ref.~\cite{Berry:2016bit} for demonstrations of this matching procedure. Because a resonance shifts the orbital parameters by an amount $\sim\sqrt\e$, and the shifted parameters subsequently evolve over the long time scale $\sim 1/\e$, the resonance introduces half-integer powers into the multiscale expansion. For example, after a resonance, Eqs.~\eqref{multiscale varphi}--\eqref{multiscale pivarphi} become
\begin{align}
\varphi_\alpha(\tilde t,\e) &= \frac{1}{\e}\left[\tilde\varphi^{(0)}_\alpha(\tilde t) +\e^{1/2}\tilde\varphi_\alpha^{(1/2)}(\tilde t) + \e\tilde\varphi^{(1)}_\alpha(\tilde t) + O(\e^{3/2})\right],\\
p^i_\varphi(\tilde t,\e) &= \tilde p^i_{\varphi(0)}(\tilde t) + \e^{1/2}\tilde p^i_{\varphi(1/2)}(\tilde t)+ \e\tilde p^i_{\varphi(1)}(\tilde t) + O(\e^{3/2}).
\end{align}
The effect of a single resonance therefore dominates over all other post-adiabatic effects. However, determining the resonant corrections~$\tilde\varphi^{(1/2)}_\alpha$ and $p^i_{\varphi(1/2)}$ requires the shifts~\eqref{resonant shifts}, which in turn require the resonant phase $q_{\rm res}$ through 1PA order. This means that the 1/2-post-adiabatic-order corrections can be thought of as outsize 1PA corrections.

Further discussions of transient resonances can be found in Refs.~\cite{Flanagan:2010cd,Gair:2011mr,Flanagan:2012kg,Ruangsri:2013hra,vandeMeent:2013sza,Isoyama:2013yor,Berry:2016bit,Mihaylov:2017qwn,Isoyama:2018sib}. Because resonances are dense in the parameter space, an inspiraling body will pass through an infinite number of them. However, because the forcing coefficients $F^i_{(1,\bm{k})}$ decay with increasing $\bm{k}$, the only resonances with significant impact are ``low-order'' resonances, such as $\Upsilon_r/\Upsilon_z=1/2$. A large fraction of inspiraling orbits will encounter such a resonance in the late inspiral~\cite{Ruangsri:2013hra,Berry:2016bit}, but neglecting the effect of resonance in EMRIs may lead to only a small loss of detectable signals~\cite{Berry:2016bit}.

In addition to the intrinsic $r$--$z$ orbital resonances discussed here, resonances can also occur due to a variety of other effects. There can be extrinsic resonances in which $k^{\rm res}_r\Omega_r+k^{\rm res}_z\Omega_z +k^{\rm res}_\phi\Omega_\phi =0$ for some triplet $(k^{\rm res}_r,k^{\rm res}_z,k^{\rm res}_\phi)$; these lead to non-isotropic emission of gravitational waves, causing possibly observable kicks to the system's center of mass~\cite{Hirata:2010xn,vandeMeent:2014raa}, but their effects are subdominant relative to $r$--$z$ resonances. If the secondary is spinning, its spin can also create resonances~\cite{Zelenka:2019nyp}, as can the presence of external matter source such as a third body~\cite{Bonga:2019ycj,Yang:2019iqa}. 

\section{Solving the Einstein equations with a skeleton source}

In this section we describe how to combine the methods of the previous sections to model small-mass-ratio binaries. This consists of solving the global problem in a Kerr background: the perturbative Einstein equations with a skeleton source (i.e., a point particle or effective source) moving on a trajectory governed by Eq.~\eqref{EOM spin} or ~\eqref{EOM2}. The first part of the section summarizes a multiscale expansion of the field equations, building directly on our treatment of orbital dynamics. At adiabatic order, waveforms can be generated by solving the linearized Einstein or Teukolsky equation with a point particle source and calculating the dissipative first-order self-force. At 1PA order, one must solve the second-order Einstein equation and compute the first-order conservative self-force and second-order dissipative self-force. These 1PA calculations require, as a central ingredient, a mode decomposition of the singular field; this is the subject of the second part of the section.

For simplicity, we assume the small object is spherical and nonspinning and that it does not encounter any significant orbital resonances.

\subsection{Multiscale expansion}

\subsubsection{Structure of the expansion}

Like the orbital dynamics, the metric in a binary has two distinct time scales: the orbital periods $T_\alpha = 2\pi/\Omega_\alpha$ and the long radiation-reaction time $\sim 1/(\e T_\alpha)$. The evolution on the orbital time scale is characterized by periodic dependence on the orbital action angles $\varphi_\alpha$, which satisfy Eq.~\eqref{dphidt}. The evolution on the radiation-reaction time is characterized by a slow change of the orbital parameters $p^i_\varphi$, governed by Eqs.~\eqref{dpphidt}, and of the central black hole parameters $(M_{BH},J_{BH})$, which evolve due to absorption of energy and angular momentum according to Eqs.~\eqref{EdotH v1} and \eqref{LdotH v1}. If we did not neglect the small object's spin and higher moments, they would come with additional parameters and phases~\cite{Ruangsri:2015cvg,Witzany:2019nml}.

The black hole parameters change at a rate $\flux^{\cal H}\propto |h|^2\sim \e^2$. Over the radiation-reaction time, this accumulates to a change $\sim \e$, allowing us to write the evolving parameters as $M_{BH}=M + \e \delta M $ and $J_{BH} = J + \e \delta J $, where $M$ and $J$ are constant and $ M_A:=(\delta M,\delta J)$ evolve on the radiation-reaction time. We then work on the fixed Kerr background with parameters $M$ and $a=J/M$, with a set of slowly evolving system parameters ${\cal P}^\alpha = \{p_\varphi^i, M_A\}$.

We will use the split into action angles $\varphi_\alpha$ and system parameters  ${\cal P}^\alpha$ to expand the metric perturbation and stress-energy as
\begin{align}
h_{\mu\nu} &= \sum_{n=1}^2\sum_{\emm,\bm{k}}\e^n h^{(n\emm\bm{k})}_{\mu\nu}({\cal P}^\alpha,\bm{x})e^{i\emm\phi -i\varphi_{\emm\bm{k}}}+O(\e^3),\label{h multiscale} \\
T_{\mu\nu} &= \sum_{n=1}^2\sum_{\emm,\bm{k}}\e^n T_{\mu\nu}^{(n\emm\bm{k})}({\cal P}^\alpha,\bm{x}) e^{i\emm \phi -i\varphi_{\emm\bm{k}}}+O(\e^3),\label{T multiscale}
\end{align}
where $\emm,k_r,k_z$ all run from $-\infty$ to $\infty$, $\bm{x} = (r,z)$, and $\varphi_{\emm\bm{k}} := \emm\varphi_{\phi} + k_r\varphi_r + k_z\varphi_z$. Here $\varphi_\alpha$ and ${\cal P}^\alpha$ are functions of $t$ and $\e$ governed by Eqs.~\eqref{dphidt}, \eqref{dpphidt}, and
\beq
\frac{dM_A}{dt} = \e\flux^{(1)}_A(p^i_\varphi) +O(\e^2),\label{dMAdt}
\eeq
where $\flux_A^{(1)} = (\flux_E^{\cal H}/\e^2,\flux_{L_z}^{\cal H}/\e^2)$ is given by any of Eqs.~\eqref{EdotH v1} and \eqref{LdotH v1}, Eqs.~\eqref{EdotH v2} and \eqref{LdotH v2}, or Eqs.~\eqref{EdotH v3} and \eqref{LdotH v3}; the reason this depends only on $p^i_\varphi$ at leading order is explained in Sec.~\ref{adiabatic approximation} below. The decomposition into azimuthal modes $e^{i\emm\phi}$ is not strictly necessary here, but it simplifies the analysis of the stress-energy in the next subsection, and it dovetails with the decompositions into  angular harmonics in Sec.~4, as all the bases of harmonics involve $\phi$ only through the factor $e^{i\emm\phi}$.

The expansions~\eqref{h multiscale} and \eqref{T multiscale} differ from the ``self-consistent expansion''~\eqref{self-consistent expansion} in that the parameters ${\cal P}$ in the self-consistent expansion include the complete trajectory $z^\mu$ and its derivatives. We can therefore move from Eqs.~\eqref{self-consistent expansion} and \eqref{skeleton Tab} to Eqs.~\eqref{h multiscale} and \eqref{T multiscale} by substituting the expansion of $z^\alpha(t)$ from Eq.~\eqref{z expansion - varphi}. To fully motivate our multiscale expansion, we work through this expansion of $T_{\mu\nu}$ in the next subsection. 

But first, we focus on the overall structure and efficacy of the multiscale expansion. Given Eqs.~\eqref{h multiscale} and \eqref{T multiscale}, the perturbative field equations become equations for the Fourier coefficients $h^{(n\emm\bm{k})}_{\mu\nu}$. These are identical, at leading order, to the usual frequency-domain field equations of black hole perturbation theory, with discrete frequencies
\beq
\frac{d\varphi_{\emm\bm{k}}}{dt} = \omega_{\emm\bm{k}}(p^i_\varphi) := k_r\Omega^{(0)}_r(p^i_\varphi) + k_z\Omega^{(0)}_z(p^i_\varphi) + \emm\Omega^{(0)}_\phi(p^i_\varphi).
\eeq
More concretely, if we substitute the expansions~\eqref{h multiscale} and \eqref{T multiscale} into the Einstein equations, then $t$ derivatives act as
\begin{align}
\frac{\partial}{\partial t} &= \Omega^{(0)}_\alpha(p^j_\varphi)\frac{\partial}{\partial\varphi_\alpha} + \frac{d{\cal P}^\alpha}{dt} \frac{\partial}{\partial{\cal P}^\alpha}\label{ddt multiscale}\\
&\to -i\omega_{\emm\bm{k}}(p^j_\varphi)+\e\left[\Gamma^i_{(1)}(p^j_\varphi)\frac{\partial}{\partial p^i_\varphi}  + {\cal F}^{(1)}_A(p^j_\varphi)\frac{\partial}{\partial  M_A}\right] +O(\e^2).\label{ddt multiscale modes}
\end{align}
Using this, we can write covariant derivatives as
\beq
\nabla_\alpha \to \tilde\nabla^{0\emm\bm{k}}_{\alpha} + \e\delta_\alpha^t  \tilde\partial_t^{1\emm\bm{k}} + O(\e^2),\label{nabla multiscale}
\eeq
where $\tilde\nabla^{0\emm\bm{k}}_{\alpha}$ is an ordinary covariant derivative with $\partial_\phi\to i\emm$ and $\partial_t \to -i\omega_{\emm\bm{k}}$, and $\tilde\partial_t^{1\emm\bm{k}}$ is the operator in square brackets in Eq.~\eqref{ddt multiscale modes}. If we then treat $\varphi_\alpha$ and ${\cal P}^\alpha$ as independent variables, we can split the field equations into coefficients of  $e^{i\emm\phi -i\varphi_{\emm\bm{k}}}$ and of explicit powers of $\e$. This results in a sequence of differential equations in $(r,z)$ for the coefficients $h^{(n\emm\bm{k})}_{\mu\nu}$:
\begin{align}
G^{(1\emm\bm{k})}_{\mu\nu}[h^{(1\emm\bm{k})}] &= 8\pi T^{(1\emm\bm{k})}_{\mu\nu},\label{multiscale EFE1}\\
G^{(1\emm\bm{k})}_{\mu\nu}[h^{(2\emm\bm{k})}] &= 8\pi T^{(2\emm\bm{k})}_{\mu\nu} - \sum_{\emm'\emm''}\sum_{\bm{k}'\bm{k}''}G^{(2\emm\bm{k})}_{\mu\nu}[h^{(1\emm'\bm{k}')},h^{(1\emm''\bm{k}'')}] \nonumber\\
&\quad - \Gamma^i_{(1)}\dot G^{(1\emm\bm{k})}_{\mu\nu}[\partial_{p^i_\varphi}h^{(1\emm\bm{k})}] - {\cal F}^{(1)}_A\dot G^{(1\emm\bm{k})}_{\mu\nu}[\partial_{ M_A}h^{(1\emm\bm{k})}].\label{multiscale EFE2}
\end{align}
Here $G^{(1\emm\bm{k})}_{\mu\nu}$ and $G^{(2\emm\bm{k})}_{\mu\nu}$ are the linearized and quadratic Einstein tensors~\eqref{Einstein1} and \eqref{Einstein2} with the replacement $\nabla_\alpha \to \tilde\nabla^{0\emm\bm{k}}_{\alpha}$. $\dot G^{(1\emm\bm{k})}_{\mu\nu}$ is the piece of $G^{(1)}_{\mu\nu}$ that, after applying the rule~\eqref{nabla multiscale}, is linear in $\tilde\partial_t^{1\emm\bm{k}}$. Explicit expressions for these quantities can be found in Sec. VC of Ref.~\cite{Miller:2020bft} in a Schwarzschild background in the Lorenz gauge.\footnote{The field equations in Ref.~\cite{Miller:2020bft} are further specialized to quasicircular orbits, with frequencies $\omega_\emm = \emm \Omega^{(0)}_\phi$, but they remain valid under the replacement $\omega_\emm\to \omega_{\emm\bm{k}}$. In Sec.~VC of Ref.~\cite{Miller:2020bft} they also include frequency corrections $\Omega^{(1)}_\phi$, which we have eliminated here with our choice of averaged variables $(\varphi_\alpha,p^i_\varphi)$; the analogue of our choice is described in their Appendix A. Beyond these minor differences, they more substantially differ by allowing the phases and system variables to depend on $r$ in addition to $t$. We discuss the reason for this in Sec.~\ref{snapshots}.}

The left-hand side of the field equations~\eqref{multiscale EFE1} and \eqref{multiscale EFE2} is identical to what it would be if we expanded $h_{\mu\nu}$ in Fourier modes $e^{i\emm\phi-i\omega_{\emm\bm{k}}t}$. Such a Fourier expansion is what has  been implemented historically in first-order frequency-domain calculations with geodesic sources (e.g.,~\cite{Detweiler:1978ge,Hughes:1999bq,Hughes:2005qb,Drasco:2005kz,Detweiler:2008ft,Fujita:2009us,Akcay:2010dx,Hopper:2010uv,Keidl:2010pm,Shah:2010bi,Flanagan:2012kg,Hopper:2012ty,Akcay:2013wfa,Osburn:2014hoa,Hopper:2015jxa,vandeMeent:2015lxa,vandeMeent:2017bcc}), and we can now immediately re-interpret those computations as leading-order implementations of the expansion~\eqref{h multiscale}.\footnote{First-order implementations in the time domain~\cite{Barack:2005nr,Barack:2007tm,Sundararajan:2008zm,Barack:2010tm,Dolan:2012jg,Harms:2014dqa,Barack:2017oir} do not mesh quite so readily with a multiscale expansion. 
We discuss their utility within a multiscale expansion in later subsections.} 
This is a principal advantage of using the variables $(\varphi_\alpha,p^i_\varphi)$ instead of $(q_\alpha,p^i_q)$.

Importantly, Eqs.~\eqref{multiscale EFE1} and \eqref{multiscale EFE2} can be solved for any values of the parameters ${\cal P}^\alpha$, without having to simulate complete inspirals. At each point in the parameter space, the solution, comprising the set of amplitudes $h^{(n\emm\bm{k})}_{\mu\nu}$, can loosely be thought of as a ``snapshot'' of the spacetime in the frequency domain. These solutions can be used to calculate the driving forces in the evolution equations~\eqref{dpphidt} and \eqref{dMAdt} for $d{\cal P}^\alpha/dt$. After populating the space of snapshots, one can then use these evolution equations, together with the phase evolution equation~\eqref{dphidt}, to evolve any particular binary spacetime through the space. Note that even though each snapshot is determined by an ``instantaneous'' value ${\cal P}^\alpha$, each snapshot fully accounts for dissipation and for the nongeodesic past history of the binary: because the evolution is slow compared to the orbital time scale, these effects are suppressed by a power of $\e$ and are incorporated through the $\dot G^{(1\emm\bm{k})}_{\mu\nu}$ source terms in Eq.~\eqref{multiscale EFE2}.

What would go wrong if, rather than using this multiscale expansion, we were to actually use $h^{(1)}_{\mu\nu}=\sum_{\emm\bm{k}}h^{(1\emm\bm{k})}_{\mu\nu}(r,z)e^{i\emm\phi-i\omega_{\emm\bm{k}}t}$ as our first-order metric perturbation? This would be approximating the trajectory of the companion as a geodesic of the background black hole spacetime. As explained in the discussion around Eq.~\eqref{dipole term}, such an approximation would accumulate large errors with time: the ``small'' corrections to the trajectory would grow large as the object spirals inward. The growing correction, represented by $z^\alpha_1=m^\alpha/m$ in Eq.~\eqref{dipole term}, would manifest itself as a dipole term in $h^{(2)}_{\mu\nu}$ that would grow until $h^{(2)}_{\mu\nu}$ became larger than $h^{(1)}_{\mu\nu}$, spelling the breakdown of regular perturbation theory. 

We can now understand this behavior directly from the orbital phases. If we were to use the geodesic phases $\omega_{\emm\bm{k}}t$, we would be implicitly expanding the phase
\beq\label{varphi-soln}
\varphi_\alpha(t,\e) = \int_0^{ t}\Omega^{(0)}_\alpha[p^j_\varphi(\e t',\e)]d t' + \varphi_\alpha^0
\eeq
in powers of $\e$, as
\beq
\varphi_\alpha(t,\e) = \tilde\Omega^{(0)}_\alpha(0) t + \e\left[\frac{1}{2} \frac{d\tilde\Omega^{(0)}_\alpha}{d\tilde t}(0)t^2 + \tilde\Omega^{(1)}_\alpha(0)t\right]+ O(\e^2) +\varphi_\alpha^0,
\eeq
where we have used Eq.~\eqref{Omega tilde expansion}. Such an expansion would be accurate on the orbital timescale but would accumulate large errors on the dephasing time $\sim 1/\sqrt{\e}$, which is much shorter than the radiation reaction time. Moreover, the order-$\e$ terms in this expansion would appear as non-oscillatory, linear- and quadratic-in-$t$ terms in $h^{(2)}_{\mu\nu}$, implying that $h^{(2)}_{\mu\nu}$ would not admit a discrete Fourier expansion or correctly describe the system's approximate triperiodicity. The multiscale expansion avoids these errors and maintains uniform accuracy prior to the transition to plunge (and excluding resonances).

The basic idea of this multiscale expansion of the field equations was first put forward in Ref.~\cite{Hinderer:2008dm}. It is described in detail in Ref.~\cite{Miller:2020bft} for the special case of quasicircular orbits in Schwarzschild spacetime. Our presentation here, building on our particular treatment of orbital motion in the preceding section, is the most complete description to date of the generic case. We provide additional details below. A thorough description, focused on the mathematical foundations of the expansion, is in preparation~\cite{two-timescale-0}.

\subsubsection{Multiscale expansion of source terms and driving forces}\label{expansion of source}

We illustrate, and further motivate, the multiscale expansion by examining the multiscale form of the source terms in the coupled equations~\eqref{skeleton EFE} and \eqref{EOM2}: the Detweiler stress-energy and the self-force. 

We start with the stress-energy~\eqref{skeleton Tab}. Writing the trajectory as 
\beq
z^\alpha(t) = [t,r_o(t),z_o(t),\phi_o(t)]
\eeq
 (where the subscript stands for ``object's orbit''), setting the spin to zero, using the $\delta$ function to evaluate the integral, and expanding the factors of $\sqrt{-\breve g}$ and $\frac{d\tau}{d\breve{\tau}}$, we express $T_{\mu\nu}$ as
\begin{align}
T_{\mu\nu} &= \frac{m\breve{g}_{\mu\alpha} \breve{g}_{\nu\beta}u^\alpha u^\beta}{u^t\Sigma}\left[1+\frac{\e}{2}\left(u^\gamma u^\delta - g^{\gamma\delta}\right)h^{\R(1)}_{\gamma\delta}\right]\nonumber\\
&\quad\times\delta^2[\bm{x}-\bm{x}_o(t)]\delta[\phi-\phi_o(t)] + O(\e^3).
\end{align}
We now take as a given our multiscale expansion~\eqref{z expansion - varphi} of $z^\alpha(t)$; this assumed the form~\eqref{small force} for the force, which we return to below. Substituting~\eqref{z expansion - varphi} and using $u^\mu = \dot z^\mu/\Sigma$, we obtain the coefficients in the expansion
\beq
T_{\mu\nu} = \e T^{(1)}_{\mu\nu}(\varphi_\alpha,p^i_\varphi) + \e^2 T^{(2)}_{\mu\nu}(\varphi_\alpha,p^i_\varphi) +O(\e^3).
\eeq
The leading term is
\beq
T^{(1)}_{\mu\nu}(\varphi_\alpha,p^i_\varphi) = \frac{m \dot z^{(0)}_\mu \dot z^{(0)}_\nu}{\mathscr{f}_t^{(0)}\Sigma^2_{(0)}}\delta^2[\bm{x}-\bm{x}_{(0)}(\bm{\varphi},p^i_\varphi)]\delta[\phi-\phi_{(0)}(\varphi_\alpha,p^i_\varphi)],\label{T1}
\eeq
where $\Sigma^2_{(0)} := r^2_{(0)}+a^2 z^2_{(0)}$, $\dot z^{(0)}_\mu := g_{\mu\nu}(\bm{x}_{(0)})\dot z^\nu_{(0)}$,  $\dot z^\mu_{(0)}=\mathscr{f}^{(0)}_\mu(\bm{\psi}^{(0)},p^i_\varphi)$ for $\mu=t,\phi$, and $\dot{\bm{x}}_{(0)}$ is given by Eq.~\eqref{dot xa} with $p^i\to p^i_\varphi$ and $\psi_a\to\psi_a^{(0)}$. The second-order term is 
\begin{align}
T^{(2)}_{\mu\nu}(\varphi_\alpha,p^i_\varphi) &= \frac{m}{\mathscr{f}_t^{(0)}\Sigma^2_{(0)}}\left[2\dot z^{(0)}_{(\mu} h^{\R(1)}_{\nu)\beta} \dot z_{(0)}^\beta+\frac{1}{2}\dot z^{(0)}_\mu \dot z^{(0)}_\nu\left(u^\gamma_{(0)} u^\delta_{(0)} - g^{\gamma\delta}_{(0)}\right)h^{\R(1)}_{\gamma\delta}\right]\nonumber\\
&\quad \times\delta^2[\bm{x}-\bm{x}_{(0)}(\bm{\varphi},p^i_\varphi)]\delta[\phi-\phi_{(0)}(\varphi_\alpha,p^i_\varphi)] + z^\alpha_{(\varphi,1)}\frac{\partial T^{(1)}_{\mu\nu}}{\partial x^\alpha_{(0)}},\label{T2}
\end{align}
where $u^\mu_{(0)} = \dot z^\mu_{(0)}/\Sigma_{(0)}$, $g^{\gamma\delta}_{(0)}:=g^{\gamma\delta}(\bm{x}_{(0)})$, and $h^{\R(1)}_{\gamma\delta}$ is evaluated at $z^\alpha_{(0)}$. The last term in Eq.~\eqref{T2} involves the action of $z^\alpha_{(\varphi,1)}\frac{\partial}{\partial x^\alpha_{(0)}}$ on $\dot z^\mu_{(0)}(\bm{\psi}^{(0)},p^i_\varphi)$; this can be evaluated using 
\beq
\bm{x}_{(\varphi,1)}\cdot\frac{\partial}{\partial \bm{x}_{(0)}} = \bm{\psi}^{(\varphi,1)}\cdot\frac{\partial}{\partial \bm{\psi}^{(0)}} + p^i_{(\varphi,1)}\frac{\partial}{\partial p^i_{\varphi}},
\eeq
with $\bm{\psi}^{(\varphi,1)}$ and $p^i_{(\varphi,1)}$ given by Eqs.~\eqref{psi(varphi,1)} and \eqref{p(varphi,1)}.

Next, we consider the mode decomposition of the expanded stress-energy. We first define the mode coefficients
\beq
T_{\mu\nu}^{(n\emm\emm'\bm{k})} := \frac{1}{(2\pi)^4}\oint T^{(n)}_{\mu\nu}e^{i\varphi_{\emm\bm{k}}-i\emm'\phi}d^2\varphi d\varphi_\phi d\phi,\label{Tnmm'k}
\eeq
which assume no relationship between the dependence on $\phi$ and $\varphi_\phi$. Substituting $T^{(1)}_{\mu\nu}$ from Eq.~\eqref{T1}, using the azimuthal $\delta$ function to evaluate the integral over $\phi$, inserting Eq.~\eqref{psi(phi)} for $\phi_{(0)}$, and using $\oint e^{i(\emm-\emm')\varphi_\phi}d\varphi_\phi=2\pi \delta_{\emm\emm'}$, we obtain
\begin{align}
T_{\mu\nu}^{(1\emm\emm'\bm{k})} 
&=\frac{\delta_{\emm \emm'}}{(2\pi)^3}\oint \frac{m \dot z^{(0)}_\mu \dot z^{(0)}_\nu}{\mathscr{f}_t^{(0)}\Sigma^2_{(0)}}\delta^2(\bm{x}-\bm{x}_{(0)})e^{i\varphi_{\bm{k}} -i\emm\Delta_\varphi \phi^{(0)}}d^2\varphi.\label{T1mm'k}
\end{align}
This enforces $\emm'=\emm$, establishing that the stress-energy only depends on $\phi$ and $\varphi_\phi$ in the combination $e^{i\emm(\phi-\varphi_\phi)}$. We can now do away with the $\emm'$ label and evaluate the integral in Eq.~\eqref{T1mm'k} in the form~\eqref{Fourier coeff relationship} or \eqref{Fourier coeff relationship psi}. The result is
\begin{align}
T_{\mu\nu}^{(1\emm\bm{k})} &= \frac{m\Upsilon^{(0)}_r\Upsilon^{(0)}_z}{(2\pi)^3\Upsilon^{(0)}_t}\sum_{\substack{\sigma_r=\pm\\\sigma_z=\pm}}\frac{\dot z^{(0)}_\mu(\bm{\psi}^{\sigma}) \dot z^{(0)}_\nu(\bm{\psi}^\sigma)}{\Sigma^2(\bm{x})\dot r_{(0)}(\psi^{\sigma_r}_{r})\dot z_{(0)}(\psi^{\sigma_z}_{z})} e^{i\varphi_{\bm{k}}(\bm{\psi}^{\sigma}) -i\emm\Delta_\varphi \phi^{(0)}(\bm{\psi}^{\sigma})}\nonumber\\
&\quad\times[\theta(r-r_p) - \theta(r-r_a)]\theta(z_{\rm max}-|z|),\label{T1 modes}
\end{align}
where the various quantities have been defined as functions of the field point $\bm{x}=(r,z)$, and $\sigma_a=\pm$ refers to a portion of the orbit in which $x^a$ is increasing ($\sigma_a=+$) or decreasing ($\sigma_a=-$). $\psi_r^{\pm}(r)$ is the value of $\psi_r$ satisfying Eq.~\eqref{r(psi)} (with $p^i\to p^i_\varphi$) on an outgoing ($+$) or ingoing ($-$) leg of the radial motion; $\psi_z^{\pm}(z)$ is defined analogously from Eq.~\eqref{z(psi)}. $\varphi_{\bm{k}}(\bm{\psi}^\sigma)$ is given by 
\beq
\varphi_{\bm{k}}(\bm{\psi}^\sigma) = q_{\bm{k}}(\bm{\psi}^{\sigma})+\Omega_{\bm{k}}^{(0)}\cdot[\delta t_r(\psi^{\sigma_r}_{r})+\delta t_z(\psi^{\sigma_z}_{z})],
\eeq
with $q_a(\psi_a)$ by Eq.~\eqref{q_a(psi_a)}, and $\delta t_a(\psi_a)$ by Eq.~\eqref{delta t}. $\Delta_\varphi \phi^{(0)}(\bm{\psi}^\sigma)$ is given by Eq.~\eqref{Dpsi varphi}. The Mino-time velocities are $\dot z^{(0)}_\mu(\bm{\psi}^\sigma) = g_{\mu\nu}(\bm{x}) \dot z^\nu_{(0)}(\bm{\psi}^\sigma)$ with $\dot x^a_{(0)}(\bm{\psi})$ given by Eq.~\eqref{dot xa} [or \eqref{drdtau} and \eqref{dzdtau}] and $\dot t$ and $\dot\phi$  by Eqs.~\eqref{dtdtau} and \eqref{dphidtau}. We can also use $\dot z^{(0)}_t = - E_\varphi \Sigma$ and $\dot z^{(0)}_\phi = L_\varphi\Sigma$; recall that we  suppress the subscript $z$ on $L_z$ in $P^i_\varphi=(E_\varphi,L_\varphi,Q_\varphi)$.

This calculation demonstrates how $T^{(1)}_{\mu\nu}$ inherits the form~\eqref{T multiscale} from the trajectory $z^\mu$. Given this form of $T^{(1)}_{\mu\nu}$, the linearized Einstein equation preserves it (in an appropriate class of gauges), justifying our ansatz for $h^{(1)}_{\mu\nu}$. Given that form of $h^{(1)}_{\mu\nu}$, the second-order stress-energy~\eqref{T2} inherits the same form, as do the other sources in Eq.~\eqref{multiscale EFE2}, and so, finally, does $h^{(2)}_{\mu\nu}$. 

All of this relies on the presumed form~\eqref{small force} for the force, from which we derived the form~\eqref{z expansion - varphi} for $z^\mu$. Our force on the right-hand side of Eq.~\eqref{EOM2} is not quite of that form. To derive its form, first note that, assuming Eq.~\eqref{z expansion - varphi}, the puncture field $h^{\P}_{\mu\nu}$ has a form analogous to Eq.~\eqref{h multiscale}, and therefore $h^{\R}_{\mu\nu}$ does as well. If we write this as $h^{\R}_{\mu\nu}({\cal P}^\alpha,\bm{x},\varphi_\phi-\phi,\bm{\varphi},\e)$, apply a covariant derivative using~\eqref{nabla multiscale}, and evaluate the result on the trajectory $z^\mu(t)$, then the right-hand side of Eq.~\eqref{EOM2} takes the form
\beq
f^\mu({\cal P}^\alpha,\bm{x}_o,\dot z^\alpha,\bm{\varphi},\e),
\eeq
where we have used $\varphi_\phi-\phi_o = - \Delta_{\varphi}\phi^{(0)}(\bm{\varphi}) - \e\phi^{(\varphi,1)}(\bm{\varphi})$ to eliminate dependence on $\varphi_\phi$. This differs from Eq.~\eqref{small force} in two ways: it depends explicitly on $(\bm{\varphi},p^i_\varphi)$, and it depends on the additional parameters $M_A$. 

With respect to the first difference, we can use Eqs.~\eqref{varphi(psi) perturbative} and \eqref{pvarphi(psi) perturbative} to write the force in the form
\beq
f^\mu = \e f^\mu_{(1)}(\bm{\psi},p^i,M_A) + \e^2 f^\mu_{(2)}(\bm{\psi},p^i,M_A)+O(\e^3).
\eeq
The system of equations~\eqref{psidot perturbative} and \eqref{pidot perturbative} thus becomes
\begin{align}
\frac{d \psi_\alpha}{d\lambda} &= \mathscr{f}_\alpha^{(0)}(\bm{\psi},p^j) + \e\mathscr{f}_{\alpha}^{(1)}(\bm{\psi},p^j,M_A) + O(\e^2),\label{psidot amended}\\
\frac{dp^i}{d\lambda} &= \e g_{(1)}^i(\bm{\psi},p^j,M_A) + \e^2 g_{(2)}^i(\bm{\psi},p^j,M_A) + O(\e^3),\\
\frac{dM_A}{d\lambda} &= \e\mathscr{f}^{(0)}_t(\bm{\psi},p^i)\flux^{(1)}_A(p^i)+O(\e^2).\label{Mdot amended}
\end{align}
The analysis of these equations then follows essentially without change as in Secs.~\ref{perturbed Mino frequencies}--\ref{multiscale expansion of orbit}. To see why the use of Eqs.~\eqref{psidot perturbative} and \eqref{pidot perturbative} does not lead to vicious circularity, note that their subleading terms only affect $f^\mu_{(2)}$, which only enters into the dynamics in Eq.~\eqref{G2}. The nongeodesic functions appearing in Eqs.~\eqref{psidot perturbative} and \eqref{pidot perturbative} are therefore determined from lower-order equations prior to requiring $f^\mu_{(2)}$.

Finally, how does $M_A$ influence the orbital dynamics? It enters into the driving forces $g^i_{(n)}$ and $\mathscr{f}^{(1)}_\alpha$. However, it does not enter into $\langle g^{(1)}_i\rangle$. This follows from the fact that $f^\mu_{(1)\rm con}$ depends on $M_A$ but $f^\mu_{(1)\rm diss}$ does not, as explained  in Sec.~\ref{adiabatic approximation} below. $M_A$ therefore contributes to the action-angle dynamics at 1PA order via Eq.~\eqref{G2}, as well as to the coordinate trajectory correction $z^{\mu}_{(\varphi,1)}$ at 1PA order through $\psi_{\alpha}^{(\varphi,1)}$ and $p^i_{(\varphi,1)}$. This is the only material change to our treatment of the orbital dynamics in Secs.~\ref{perturbed Mino frequencies}--\ref{multiscale expansion of orbit}.

Together, the analyses of this section establish the consistency of our multiscale treatments of the field equations and orbital motion. In the following sections, we describe more concretely how to utilize these treatments.

\subsubsection{Snapshot solutions and evolving waveforms}\label{snapshots}

Snapshot solutions, consisting of the mode amplitudes $h^{n\emm\bm{k}}_{\mu\nu}({\cal P}^\alpha,\bm{x})$, can be computed using any of the frequency-domain methods reviewed in Sec.~\ref{sec:black hole perturbation theory}. As an example, in this section we sketch how this is done at first order using the method of metric reconstruction in the radiation gauge, starting from the Teukolsky equation. This summarizes work from Ref.~\cite{vandeMeent:2017bcc}, which provided the first calculation of the full first-order self-force for generic bound orbits in Kerr spacetime. Our summary also appeals to methods and results from Refs.~\cite{Ori:2002uv,Drasco:2005kz,Pound:2013faa,Merlin:2016boc,vandeMeent:2017fqk}.

We first define leading-order Weyl scalars $\psi_0$ and $\psi_4$ related to the $h^{(1)}_{\mu\nu}$ of Eq.~\eqref{h multiscale} by Eqs.~\eqref{eq:Weyl-scalars-definition}--\eqref{eq:T4} with the replacements $\partial_t\to-i\omega_{\emm\bm{k}}$ and $\partial_\phi\to i\emm$. For concreteness, we use $\psi_0$. In analogy with Eqs.~\eqref{eq:psi0-FD} and \eqref{eq:psi4-FD}, it can be written as
\beq
\psi_0 = \sum_{\ell=2}^\infty\sum_{\emm=-\ell}^\ell\sum_{\bm{k}}{}_2\psi_{\ell\emm\bm{k}}(p^i_\varphi,r){}_2S_{\ell\emm}(\theta,\phi;a\omega_{\emm\bm{k}})e^{-i\varphi_{\emm\bm{k}}}.
\eeq
Note that the radial coefficients depend on $p^i_\varphi$ but not on $M_A$; this is because the linearized $\psi_0$ and $\psi_4$ are insensitive to linear perturbations of the central black hole's mass or spin~\cite{Wald:1973}.

The coefficients ${}_{2}\psi_{\ell\emm\bm{k}}(p^i_\varphi,r)$ satisfy the radial Teukolsky equation~\eqref{eq:TeukolskyR} with $\omega\to\omega_{\emm\bm{k}}$ and ${}_s\psi_{\ell\emm\omega}\to {}_s\psi_{\ell\emm\bm{k}}$. The source in that equation is constructed from the stress-energy~\eqref{T1} or its modes~\eqref{T1 modes} using the analog of Eq.~\eqref{eq:T0-FD},
\beq
{}_2T_{\ell\emm\bm{k}} = -32\pi^2\Sigma\int_{-z_{\rm max}}^{z_{\rm max}} (\tilde{\cal S}^{\emm\bm{k}}_0T) (r,z){}_2S_{\ell\emm}(\theta,0;a\omega_{\emm\bm{k}}) dz, \label{T0-lmk}
\eeq
where the integral ranges over the support of $T^{(1\emm\bm{k})}_{\mu\nu}$, $\theta$ is related to $z$ by $z=\cos\theta$, and we have suppressed the dependence on $p^i_\varphi$. The source $\tilde{\cal S}^{\emm\bm{k}}_0T$ in the integrand is given by Eq.~\eqref{eq:S} with $T_{ll}\to T^{(1\emm\bm{k})}_{ll}$ (and the same for other tetrad components), $\partial_t\to-i\omega_{\emm\bm{k}}$, and $\partial_\phi\to i\emm$. What may appear to be an extra factor of $2\pi$ in Eq.~\eqref{T0-lmk} accounts for the factor of $1/(2\pi)$ introduced in the integration over $\phi$ in Eq.~\eqref{Tnmm'k}. 

The retarded solution to the Teukolsky equation, as given in the variation-of-parameters form ~\eqref{eq:TeukolskyInhomogeneousModes}, is
\begin{align}\label{psi0 nonvacuum}
  {}_2 \ppsi_{\ell \emm \bm{k}}(r) &= {}_2
    C^{\text{in}}_{\ell \emm \bm{k}}(r) {}_2 R^{\text{in}}_{\ell \emm \bm{k}}(r)+{}_2
    C^{\text{up}}_{\ell \emm \bm{k}}(r) {}_2 R^{\text{up}}_{\ell \emm \bm{k}}(r), 
\end{align}
where we have defined the homogeneous solutions ${}_2 R^{\text{in/up}}_{\ell \emm \bm{k}}(r):={}_2 R^{\text{in/up}}_{\ell \emm \omega_{\emm\bm{k}}}(r)$. The weighting coefficients are given by Eq.~\eqref{eq:weighting-coefficients}, which we restate here as
\begin{subequations}
  \begin{align}
  {}_2 C_{\ell \emm \bm{k}}^{\text{in}}(r) &:= \int^{r_a}_r \frac{{}_2 R^{\text{up}}_{\ell \emm \bm{k}}(r')}{W(r')\Delta} {}_2 T_{\ell \emm \bm{k}}(r') dr', \\
  {}_2 C_{\ell \emm \bm{k}}^{\text{up}}(r) &:= \int_{r_p}^r \frac{{}_2 R^{\text{in}}_{\ell \emm \bm{k}}(r')}{W(r')\Delta} {}_2 T_{\ell \emm \bm{k}}(r') dr'.
\end{align}
\end{subequations}
In the vacuum regions $r>r_a$ and $r<r_p$, outside the support of ${}_2 T_{\ell \emm \bm{k}}$, the weighting coefficients become constants,
\begin{subequations}\label{mode amplitudes}
\begin{align}
{}_2 \hat C_{\ell \emm \bm{k}}^{\text{in}} &= \int^{r_a}_{r_p} \frac{{}_2 R^{\text{up}}_{\ell \emm \bm{k}}(r')}{W(r')\Delta} {}_2 T_{\ell \emm \bm{k}}(r') dr',\\
 {}_2 \hat C_{\ell \emm \bm{k}}^{\text{up}} &= \int_{r_p}^{r_a} \frac{{}_2 R^{\text{in}}_{\ell \emm \omega_{\emm\bm{k}}}(r')}{W(r')\Delta} {}_2 T_{\ell \emm \bm{k}}(r') dr',
\end{align}
\end{subequations}
and in those regions the solution becomes
\begin{align}\label{psi0 vacuum region}
  {}_2 \ppsi_{\ell \emm \bm{k}}(r) &= \begin{cases}
  {}_2\hat C^{\text{in}}_{\ell \emm \bm{k}}\, {}_2 R^{\text{in}}_{\ell \emm \bm{k}}(r) & \text{for } r<r_p\\
  {}_2\hat C^{\text{up}}_{\ell \emm \bm{k}}\, {}_2 R^{\text{up}}_{\ell \emm \bm{k}}(r) & \text{for } r>r_a.
  \end{cases}
\end{align}

We can evaluate the $r$ and $z$ integrals in ${}_2 C_{\ell \emm \bm{k}}^{\text{in/up}}$ as integrals over $\psi_r$ and $\psi_z$ by using appropriate changes of variables for each value of $\sigma_a$ in Eq.~\eqref{T1 modes}. For $\sigma_r=+$ and a generic function $f(r)$, the radial integrals are $\int^r_{r_p}f dr' = \int_0^{\psi^+_r(r)}f \frac{dr}{d\psi_r}d\psi_r$ and $\int_r^{r_a}f dr' = \int^\pi_{\psi^+_r(r)}f \frac{dr}{d\psi_r}d\psi_r$; for $\sigma_r=-$, they are $\int^r_{r_p}f dr' = -\int^{2\pi}_{\psi^-_r(r)}f \frac{dr}{d\psi_r}d\psi_r$ and $\int_r^{r_a}f dr' = -\int_\pi^{\psi^-_r(r)}f \frac{dr}{d\psi_r}d\psi_r$. The transformations for $\sigma_z=\pm$ are analogous. We can also write the $r$ and $z$ derivatives in $\tilde{\cal S}^{\emm\bm{k}}_0T$ as $\frac{\partial}{\partial x^a} = \frac{\partial\psi_a}{\partial x^a}\frac{\partial}{\partial\psi_a}$ (with no sum over $a$). For more explicit formulas for the integrands, see Sec. 3B of Ref.~\cite{Drasco:2005kz}. See also, e.g., Refs.~\cite{Fujita:2009us,Hopper:2015jxa} for discussion of practical methods of numerically evaluating such integrals.

The modes of $\psi_0$ (or $\psi_4$) are by themselves sufficient to calculate many quantities, such as gravitational-wave fluxes. But for other purposes, such as the calculation of the self-force and the needed input for the second-order field equations, one must compute the entire metric perturbation. Starting from the modes of $\psi_0$ or $\psi_4$, this can be done using the method of metric reconstruction reviewed in Sec.~\ref{sec:metric-reconstruction}. In the presence of a source, metric reconstruction typically yields a metric perturbation that has a gauge singularity extending in a ``shadow'' from the matter source to the black hole horizon or from the matter to infinity~\cite{Ori:2002uv,Green:2019nam}. In the case of a point particle, this shadow becomes a string singularity. However, we can more usefully reconstruct the metric perturbation in a ``no-string'' radiation gauge~\cite{Pound:2013faa}, in which it has no string but does have a jump discontinuity and radial $\delta$ function on a sphere of varying radius $r=r_{(0)}(t)$. 

To construct the no-string solution in practice, we first find a Hertz potential $\psi^{IRG}$ satisfying Eq.~\eqref{eq:psi0-IRG} (at fixed $p^i_\varphi$) in the disjoint vacuum regions $r<r_p$ and $r>r_a$, subject to regularity at infinity and the horizon. The appropriate solution in each region is given by Eq.~(15)  of Ref.~\cite{Ori:2002uv}. In the libration region $r_p<r<r_a$, the radial source ${}_2 T_{\ell \emm \bm{k}}$ is nonzero, as the Fourier decomposition smears the point particle source over the entire toroidal region $\{r_p<r<r_a,|z|<z_{\rm max}\}$. The  solution~(15) of Ref.~\cite{Ori:2002uv} therefore cannot be used in the libration region. However, it can be analytically extended into that region, using Eq.~\eqref{psi0 vacuum region} in place of Eq.~\eqref{psi0 nonvacuum}. Because the time-domain solution is analytic everywhere except on the sphere at $r_{(0)}(t)$, the sum over $\emm\bm{k}$ of the analytically continued functions from $r<r_p$ yields the correct result for all $r\leq r_{(0)}(t)$, and the sum over $\emm\bm{k}$ of the analytically continued functions from $r>r_a$ yields the correct result for all $r\geq r_{(0)}(t)$~\cite{vandeMeent:2015lxa,vandeMeent:2017bcc}; this is the method of extended homogeneous solutions~\cite{Barack:2008ms,Hopper:2010uv}.\footnote{See also Ref.~\cite{Hopper:2012ty} for a generalization of this method to problems with sources that are nowhere vanishing.} As alluded to in Sec.~\ref{sec:KerrModes}, this method was originally devised to alleviate another problem that arises in frequency-domain calculations for eccentric orbits: the sum over $\bm{k}$ modes of the inhomogeneous solution converges slowly within the libration region. In the context of metric reconstruction, the method allows one to avoid the complexities of nonvacuum reconstruction.

From the extended modes of the Hertz potential, we can reconstruct modes of an incomplete metric perturbation, $h^{(1\emm\bm{k})\rm rec}_{\mu\nu}$, using Eq.~\eqref{eq:reconstruction} (as ever, with $\partial_t\to-i\omega_{\emm\bm{k}}$ and $\partial_\phi\to i\emm$). To complete this perturbation, in the region $r>r_{(0)}(t)$ we add mass and spin perturbations, $E_\varphi \frac{\partial g_{\mu\nu}}{\partial M}$ and $L_\varphi \frac{\partial g_{\mu\nu}}{\partial J}$, where the $M$ derivative is taken at fixed $J=Ma$, and the $J$ derivative at fixed $M$; these account for the mass and spin that the particle contributes to the spacetime~\cite{Merlin:2016boc,vandeMeent:2017fqk}. In general we must also add  mass and spin perturbations $\delta M \frac{\partial g_{\mu\nu}}{\partial M}$ and $\delta J \frac{\partial g_{\mu\nu}}{\partial J}$ throughout the spacetime (at fixed $p^i_\varphi$); these account for the slowly evolving corrections to the central black hole's mass and spin.\footnote{These corrections proportional to $M_A$ have not been added historically because for any specific snapshot with parameters ${\cal P}^\alpha$, call them ${\cal P}^\alpha_0$, they can be absorbed with a redefinition $M\to M+\e\delta M_0$ and $J\to J+\e\delta J_0$, setting $M^0_A=0$. However, in the context of an evolution, which moves through the space of ${\cal P}^\alpha$ values, they must always be included at 1PA order. Even at a single value of ${\cal P}^\alpha$ where $M^0_A=0$, their time derivatives must be included in Eq.~\eqref{multiscale EFE2}. See Ref.~\cite{Miller:2020bft} for a discussion.} Finally, in the region $r<r_{(0)}(t)$, we must add gauge perturbations that ensure the coordinates $t$ and $\phi$, and therefore the frequencies $\Omega^{(0)}_\alpha$, have the same meaning in the two regions $r<r_{(0)}(t)$ and $r>r_{(0)}(t)$~\cite{Shah:2015nva} (see also~\cite{Bini:2019xwn}).

With the completed modes $h^{(1\emm\bm{k})}_{\mu\nu}({\cal P}^\alpha,\bm{x})$ in hand, one can calculate any quantity of interest on the orbital timescale with fixed ${\cal P}^\alpha$. In particular, one can calculate the first-order self-force and its dynamical effects using the mode-sum regularization formula derived in Ref.~\cite{Pound:2013faa}; the formula in the no-string gauge is given by Eq.~(125) in that reference. 


To date, Ref.~\cite{vandeMeent:2017bcc} is the only work to carry out the entire calculation we have just described for generic bound orbits in Kerr spacetime. However, for orbits in Schwarzschild spacetime and for equatorial orbits in Kerr, snapshot frequency-domain calculations of the complete $h^{(1)}_{\mu\nu}$ and $f^\mu_{(1)}$ are now routine, whether in the Lorenz gauge, Regge-Wheeler-Zerilli gauge, or no-string radiation gauge \cite{Barack:2007tm,Detweiler:2008ft,Barack:2010tm,Akcay:2010dx,Shah:2010bi,Dolan:2012jg,Akcay:2013wfa,Osburn:2014hoa,Wardell:2015ada,vandeMeent:2015lxa,Thompson:2018lgb}.  Numerical implementations at second order, which are necessary for post-adiabatic accuracy, are still in an early stage but have computed some physical quantities for quasicircular orbits in Schwarzschild spacetime~\cite{Pound:2019lzj}.

We can use the output of these snapshot calculations to obtain the true, evolving gravitational waveforms. Once the snapshot mode amplitudes are calculated, from them we can calculate the inputs for the evolution equations~\eqref{dphidt}, \eqref{dpphidt}, and \eqref{dMAdt}. In analogy with Eqs.~\eqref{eq:Teukolsky-waveform}, \eqref{eq:RW-waveform}, and \eqref{eq:LorenzGauge-waveform}, the waveforms are then given by any of
\begin{subequations}\label{multiscale waveform}
\begin{align}
h_+ - i h_\times &= 2 \sum_{\ell\emm\bm{k}} \, \frac{{}_{-2} \hat C_{1\ell \emm \bm{k}}^{\text{up}}[p^i_\varphi(\tilde u,\e)]}{\omega_{\emm\bm{k}}^2} \, {}_{-2} S_{\ell \emm}(\theta, \phi; a \omega_{\emm\bm{k}}) e^{-i\varphi_{\emm\bm{k}}(\tilde u,\e)}+O(\e^2), \\
&= \sum_{\ell\emm\bm{k}} \, \frac{D}{2} \left(\hat C_{1\ell \emm \bm{k}}^{\text{ZM,up}}[p^i_\varphi(\tilde u,\e)]-i\,\hat C_{1\ell \emm \bm{k}}^{\text{CPM,up}}[p^i_\varphi(\tilde u,\e)]\right) \nonumber\\
&\quad\qquad \times{}_{-2} Y_{\ell \emm}(\theta, \phi)e^{-i\varphi_{\emm\bm{k}}(\tilde u,\e)} +O(\e^2),\\
&= \sum_{\ell\emm\bm{k}} \hat C_{mm}^{1\ell\emm\bm{k}}[p^i_\varphi(\tilde u,\e)] {}_{2} Y_{\ell\emm}(\theta,\phi) e^{-i\varphi_{\emm\bm{k}}(\tilde u,\e)} +O(\e^2), 
\end{align}
\end{subequations}
where $D=\sqrt{(\ell-1)\ell(\ell+1)(\ell+2)}$, and $\omega_{\emm\bm{k}}=\omega_{\emm\bm{k}}[p^i_\varphi(\tilde u,\e)]$. Here we have written the waveform in terms of  $\tilde u:=\e(t-r^*)$; we return to this point below. We also note that we have given the waveform in terms of modes of $\psi_4$ rather than the less natural (for this purpose) $\psi_0$. In analogy with Eq.~\eqref{mode amplitudes}, we have defined the amplitudes $\hat C_{1\ell \emm \bm{k}}^{\text{ZM,up}}$ and $\hat C_{1\ell \emm \bm{k}}^{\text{CPM,up}}$ as the relevant weighting coefficients for $r>r_p$, and we have defined the Lorenz-gauge amplitudes $\hat C_{mm}^{1\ell\emm\bm{k}}:=\lim_{r\to\infty}(r e^{i\omega_{\emm\bm{k}}r^*} h_{mm}^{1\ell\emm\bm{k}})$. We have also intentionally inserted a label ``1'' onto the mode amplitudes and omitted $O(\e^2)$ amplitudes. This is because even if we determine the phase $\varphi_{\emm\bm{k}}(\tilde u,\e)$ through 1PA order, the second-order amplitudes do not increase the waveform's order of accuracy; an order-$\e^2$ amplitude in the waveform is indistinguishable from a 2PA (order-$\e$) correction to the phase.

The waveform \eqref{multiscale waveform} is in the time domain, but it is almost trivially related to the frequency-domain waveform.  Defining $h(\omega) := \frac{1}{2\pi}\int_{-\infty}^\infty (h_+ - i h_\times )e^{i\omega u}du$ and applying the stationary-phase approximation, we obtain, e.g.,
\begin{align}
h(\omega) &= \frac{1}{2\pi}\sum_{\ell\emm\bm{k}}\sqrt{\frac{2\pi \e}{|d\omega_{\emm\bm{k}}/d\tilde t|}}\hat C_{mm}^{1\ell\emm\bm{k}}[\tilde t_{\emm\bm{k}}(\omega)] {}_{2} Y_{\ell\emm}\nonumber\\
&\quad\times\exp\left\lbrace i[\omega\, \tilde t_{\emm\bm{k}}(\omega)-\varphi_{\emm\bm{k}}(\omega)]+{\rm sgn}\left(d\omega_{\emm\bm{k}}/d\tilde t\right)\frac{i\pi}{4}\right\rbrace+o(\sqrt{\e}).
\end{align}
Here $\tilde t_{\emm\bm{k}}(\omega)$ is the solution to $\omega=\omega_{\emm\bm{k}}(\tilde t)$, and the phase as a function of $\omega$ is $\varphi_{\emm\bm{k}}(\omega)=\varphi_{\emm\bm{k}}[\tilde t_{\emm\bm{k}}(\omega)]$.

Before proceeding, we return to the dependence on $\tilde u$ rather than $\tilde t$. In Eq.~\eqref{multiscale waveform}, all functions of $\tilde u$ are the functions obtained by solving~\eqref{dphidt}, \eqref{dpphidt}, and \eqref{dMAdt}, simply evaluated as a function of $\tilde u$. For example, from Eq.~\eqref{varphi-soln},
\beq
\varphi_\alpha(\tilde u,\e) = \frac{1}{\e}\int_0^{\tilde u}\Omega^{(0)}_\alpha[p^j_\varphi(\tilde t',\e)]d \tilde t' + \varphi_\alpha^0.
\eeq
This dependence on $u$ is not a trivial consequence of the multiscale expansion~\eqref{h multiscale}. To justify it, one must adopt a hyperboloidal choice of time that asymptotes to $u$ at ${\cal I}^+$ or perform a matched-expansions calculation, matching the solution~\eqref{h multiscale} to an outgoing-wave solution near ${\cal I}^+$. Ref.~\cite{Miller:2020bft} discusses these points along with several additional advantages of using a hyperboloidal slicing. To see why the replacement $t\to u$ is intuitively sensible, note that with it, Eq.~\eqref{multiscale waveform} correctly reduces to a snapshot waveform on the orbital timescale if we fix $p^i_\varphi$ and replace $\varphi_{\emm\bm{k}}(\tilde u,\e)$ with its geodesic approximation $\Omega^{(0)}_{\emm\bm{k}}u$ (with fixed $\Omega^{(0)}_{\emm\bm{k}}$); without the replacement, the multiscale waveform would not correctly reduce in this way.

In the next two subsections, we outline the steps required to generate multiscale waveforms at adiabatic (0PA) and 1PA order, whether in the time or frequency domain.

\subsubsection{Adiabatic approximation}\label{adiabatic approximation}

At this stage we consider the evolution equations in the form~\eqref{0PA varphi}--\eqref{1PA pivarphi}. There is no difference between that form and Eqs.~\eqref{dphidt}--\eqref{dpphidt} at adiabatic order, but we adopt the notation of Eqs.~\eqref{0PA varphi}--\eqref{1PA pivarphi} here for consistency with our discussion of the 1PA approximation in the next subsection. 

For convenience, we transcribe the adiabatic evolution equations~\eqref{0PA varphi} and \eqref{0PA pivarphi}:
\begin{align}
\frac{d\tilde\varphi^{(0)}_\alpha}{d\tilde t} &= \Omega^{(0)}_\alpha(\tilde p^j_{\varphi(0)}),\label{0PA varphi repeat}\\
\frac{d\tilde p^i_{\varphi(0)}}{d\tilde t} &= \Gamma^i_{(1)}(\tilde p^j_{\varphi(0)}).\label{0PA pivarphi repeat}
\end{align}

An adiabatic waveform-generation scheme consists of the following steps:
\begin{enumerate}
\item Solve the field equation~\eqref{multiscale EFE1} or the associated Teukolsky equation or Regge-Wheeler-Zerilli equations, on a grid of $p^i_\varphi$ values. At each grid point in parameter space, compute and store two things: the driving forces $\Gamma^i_{(1)}(p^i_\varphi)$ in Eq.~\eqref{0PA pivarphi} and the asymptotic mode amplitudes at ${\cal I}^+$ [e.g., ${}_{\pm2}\hat C^{\rm up}_{1\ell\emm\bm{k}}(p^i_\varphi)$ in the Teukolsky case].
\item Using the stored values of $\Gamma^i_{(1)}$, evolve through the parameter space by solving the coupled equations~\eqref{0PA varphi repeat} and \eqref{0PA pivarphi repeat} to obtain the adiabatic parameters $\tilde p^i_{\varphi(0)}$ and phases $\tilde\varphi^{(0)}_r$, $\tilde\varphi^{(0)}_z$, $\tilde\varphi^{(0)}_\phi$ as functions of $\tilde t=\e t$.
\item Construct the adiabatic waveform using, e.g.,
\beq
h_+ - i h_\times = 2 \sum_{\ell\emm\bm{k}} \, \frac{{}_{-2} \hat C_{1\ell \emm \bm{k}}^{\text{up}}[\tilde p^i_{\varphi(0)}(\tilde u)]}{\omega_{\emm\bm{k}}^2} \, {}_{-2} S_{\ell \emm}(\theta, \phi; a \omega_{\emm\bm{k}})e^{-i\varphi^{(0)}_{\emm\bm{k}}(\tilde u)/\e},
\eeq
where $\omega_{\emm\bm{k}}=\omega_{\emm\bm{k}}[\tilde p^i_{\varphi(0)}(\tilde u)]$.
\end{enumerate}

Starting from seminal work in Refs.~\cite{Galtsov:1982hwm,Mino:2003yg}, two groups of authors have developed practical implementations of this scheme~\cite{Hughes:1999bq,Sago:2005gd,Sago:2005fn,Hughes:2005qb,Drasco:2005is,Ganz:2007rf,Hughes:2016xwf,Isoyama:2018sib}. 
 
One of the convenient aspects of the adiabatic approximation is that it can be implemented entirely in terms of the Teukolsky equation with a point-particle source, with no requirement to calculate a reconstructed and completed metric or to extract the regular fields $h^{\R(n)}_{\mu\nu}$. The reason is that, as explained around Eqs.~\eqref{fdiss} and \eqref{fcon}, only the first-order dissipative force $f^\mu_{(1)\rm diss}$ is needed to calculate the driving force $\Gamma^i_{(1)}$. This force is entirely due to the half-retarded minus half-advanced piece of $h^{(1)}_{\mu\nu}$~\cite{Mino:2003yg}, 
\beq
h^{(1)\rm rad}_{\mu\nu} = \frac{1}{2}h^{(1)\rm ret}_{\mu\nu} - \frac{1}{2}h^{(1)\rm adv}_{\mu\nu}. 
\eeq
Because $h^{(1)\rm rad}_{\mu\nu}$ is a vacuum solution to the linearized Einstein equation, it can be reconstructed from the half-retarded minus half-advanced piece of $\psi_0$ or $\psi_4$, using the radiation-gauge reconstruction method reviewed in Sec.~\ref{sec:metric-reconstruction} (as translated to the multiscale expansion in the previous section). Again because it is a vacuum solution, it is smooth at the particle, and it is equal there to the relevant part of $h^{\R(1)}_{\mu\nu}$ that creates $f^\mu_{(1)\rm diss}$. Furthermore, $h^{(1)\rm rad}_{\mu\nu}$ can contain no stationary perturbations, implying it cannot contain any contribution from the mass and spin perturbations $M_A$, so it needs no completion. Hence, one can evolve the orbit and generate the waveform entirely from mode amplitudes of $\psi_0$ or $\psi_4$.

Concrete formulas for adiabatic driving forces in terms of Teukolsky amplitudes were first derived in Ref.~\cite{Galtsov:1982hwm}, which showed that the average rates of change of $E$ and $L_z$ due to $f^\alpha_{(1)\rm diss}$ satisfy a balance law:
\begin{align}
\frac{d\tilde E^{(0)}_\varphi}{dt} &= -\flux_E^{\cal H} - \flux_E^{\cal I},\label{E balance}\\
\frac{d\tilde L^{(0)}_\varphi}{dt} &= -\flux_{L_z}^{\cal H} - \flux_{L_z}^{\cal I},\label{L balance}
\end{align}
where $\tilde P^i_{\varphi(0)} = (\tilde E^{(0)}_\varphi,\tilde L^{(0)}_\varphi, \tilde Q^{(0)}_\varphi)$ are related to $\tilde p^i_{\varphi(0)}$ by the geodesic relationships~\eqref{E(pi)}--\eqref{Q(pi)} between $P^i$ and $p^i$. The fluxes are those due to the retarded field, which we can translate from Eqs.~\eqref{EdotH v1}--\eqref{LdotI v1} as
\begin{align}
  \fluxEH &=  \sum_{\ell\emm\bm{k}} \frac{2\pi \alpha_{\ell \emm \omega_{\emm\bm{k}}}}{\omega_{\emm\bm{k}}^2} |{}_{-2} \hat C^{\text{in}}_{1\ell \emm \bm{k}}|^2 := \sum_{\ell\emm\bm{k}}\flux_E^{{\cal H}\ell\emm\bm{k}}, \\
  \fluxEI &= \sum_{\ell\emm\bm{k}} \frac{2\pi}{\omega_{\emm\bm{k}}^2}|\, {}_{-2} \hat C^{\text{up}}_{1\ell \emm \bm{k}}|^2:= \sum_{\ell\emm\bm{k}}\flux_E^{{\cal I}\ell\emm\bm{k}},
\end{align}
and similarly for $\fluxLzH$ and $\fluxLzI$.

Equation~\eqref{E balance} states that the change in the particle's orbital energy is equal at leading order to the sum total of energy carried out of the system (into the black hole and out to infinity). Equation~\eqref{L balance} states the analog about the particle's angular momentum. 

Some time later, Ref.~\cite{Sago:2005fn} derived a similar formula for the average rate of change of the Carter constant due to $f^\alpha_{(1)\rm diss}$:
\begin{align}\label{adiabatic Qdot}
\frac{d\tilde Q^{(0)}_\varphi}{d t} = - \left(\frac{dQ}{dt}\right)^{\cal H} - \left(\frac{dQ}{dt}\right)^{\cal I},\\
\end{align}
where\footnote{Note that Ref.~\cite{Sago:2005fn} uses $C$ to denote our $Q$ and $Q$ to denote our $K$. We give here the expression for $\left(\frac{dQ}{dt}\right)^{\star}$ as presented in Ref.~\cite{Flanagan:2012kg}. In all cases in the literature, expressions such as these are written in terms of averages $\langle\cdot\rangle$, which we can omit because we work with already averaged orbital variables.}
\beq
\left(\frac{dQ}{dt}\right)^{\star} = 2\sum_{\ell\emm\bm{k}}\frac{L_{\emm\bm{k}}+k_z\tilde\Upsilon^{(0)}_z}{\omega_{\emm\bm{k}}}\flux_E^{\star\,\ell\emm\bm{k}}
\eeq
with
\beq
L_{\emm\bm{k}} = \emm\langle \cot^2\theta_{(0)}\rangle_\lambda L_\varphi^{(0)} - a^2\omega_{\emm\bm{k}}\langle\cos^2\theta_{(0)}\rangle_\lambda E^{(0)}_\varphi.
\eeq
While this evolution equation for $Q$ superficially resembles those for $E$ and $L_z$, it is of fundamentally different character. The quantities $\flux_E^{\star}$ and $\flux_{L_z}^{\star}$ are true physical fluxes across the horizon and out to infinity; they are defined entirely in terms of the metric on the surfaces ${\cal H}^+$ and ${\cal I}^+$. The quantities $\left(\frac{dQ}{dt}\right)^{\star}$, on the other hand, directly involve orbital parameters; they are not locally measurable fluxes. Thus, although Eq.~\eqref{adiabatic Qdot} is sometimes referred to as a flux-balance law, there is no known sense in which it can be meaningfully described as such. However, the evolution equations for $E$, $L_z$, and $Q$ all share the same practical advantage: they can be evaluated directly from the retarded solution to the Teukolsky equation with a point-particle source, with no need to reconstruct the complete metric perturbation or to extract the regular field.

Combining Eqs.~\eqref{E balance}, \eqref{L balance}, and \eqref{adiabatic Qdot}, we can compute the adiabatic driving forces
\beq
\Gamma^i_{(1)}(\tilde p^i_{\varphi(0)}) = \frac{\partial \tilde p^i_{\varphi(0)}}{\partial \tilde P^j_{\varphi(0)}}\frac{d\tilde P^j_{\varphi(0)}}{dt}
\eeq
from the Teukolsky amplitudes ${}_{-2} \hat C^{\text{in/up}}_{1\ell \emm \bm{k}}$ given by Eq.~\eqref{mode amplitudes}. We can then follow the prescription outlined at the beginning of the section. Alternatively, we can express the geodesic frequencies in terms of $\tilde P^i_{\varphi(0)} = (\tilde E^{(0)}_\varphi,\tilde L^{(0)}_\varphi, \tilde Q^{(0)}_\varphi)$ and  work directly with those variables, treating $\tilde p^i_{\varphi(0)}$ as a function of $\tilde P^i_{\varphi(0)}$ by inverting the relationships~\eqref{E(pi)}--\eqref{Q(pi)}.

The adiabatic approximation has been used to evolve equatorial orbits in Kerr spacetime~\cite{Fujita:2020zxe} and to generate waveforms in Schwarzschild spacetime~\cite{Chua:2020stf}. Yet, despite the approximation's efficient formulation, to date no adiabatic waveforms have been generated for orbits in Kerr spacetime, nor have orbital evolutions been performed for generic (eccentric and inclined) orbits. There are two main obstacles. One is generating sufficiently dense data on the $p^i_\varphi$ space to perform accurate interpolation or fitting. The second is the very large ($\sim 10^4$) number of mode amplitudes that are required to achieve an accurate waveform. Both obstacles are expected to be soon overcome~\cite{Fujita:2020zxe,Chua:2020stf}, but as of this writing, the gold standard for generic orbits remains snapshot waveforms~\cite{Drasco:2005is} that use geodesic phases.

\subsubsection{First post-adiabatic approximation}\label{1PA approximation}

The 1PA evolution equations~\eqref{0PA varphi}--\eqref{1PA pivarphi}, as extended following the discussion around Eqs.~\eqref{psidot amended}--\eqref{Mdot amended}, are
\begin{align}
\frac{d\tilde\varphi^{(0)}_\alpha}{d\tilde t} &= \Omega^{(0)}_\alpha(\tilde p^j_{\varphi(0)}),\label{0PA varphi 3}\\
\frac{d\tilde p^i_{\varphi(0)}}{d\tilde t} &= \Gamma^i_{(1)}(\tilde p^j_{\varphi(0)}),\label{0PA pivarphi 3}\\
\frac{d\tilde\varphi^{(1)}_\alpha}{d\tilde t} &= \tilde p^j_{\varphi(1)}\partial_j \Omega^{(0)}_\alpha(\tilde p^j_{\varphi(0)}),\label{1PA varphi 2}\\
\frac{d\tilde p^i_{\varphi(1)}}{d\tilde t} &= \Gamma^i_{(2)}(\tilde p^j_{\varphi(0)},\tilde M^{(1)}_A) +\tilde p^j_{\varphi(1)}\partial_j \Gamma^i_{(1)}(\tilde p^j_{\varphi(0)}),\label{1PA pivarphi 2}\\
\frac{d\tilde M^{(1)}_A}{d\tilde t} &= \flux^{(1)}_A(\tilde p^j_{\varphi(0)}).\label{1PA dMdt}
\end{align}
Here we have assumed $M_A= \tilde M^{(1)}_A(\tilde t)+O(\e)$. Because (i) $M_A$ only contributes stationary modes to $h^{(1\emm\bm{k})}_{\mu\nu}$, (ii) any source term for $h^{(2\emm\bm{k})}_{\mu\nu}$ that is quadratic in these modes will also be stationary, and (iii) a stationary mode of $h^{(2\emm\bm{k})}_{\mu\nu}$ will not contribute to $\Gamma^i_{(2)}$, it follows that $\Gamma^i_{(2)}$ is linear in $M_A$, implying we can write it in the form
\beq
\Gamma^i_{(2)}(\tilde p^j_{\varphi(0)},\tilde M^{(1)}_A) = \Gamma^i_{(2)}(\tilde p^j_{\varphi(0)},0) + \tilde M^{(1)}_A\gamma^i_A(\tilde p^j_{\varphi(0)}),
\eeq
where $A$ is summed over. $\gamma^i_A(\tilde p^j_{\varphi(0)})$ here is defined as the coefficient of $\tilde M^{(1)}_A$ in $\Gamma^i_{(2)}(\tilde p^j_{\varphi(0)},\tilde M^{(1)}_A)$.

A 1PA waveform-generation scheme then consists of the following steps:
\begin{enumerate}
\item Solve the field equations~\eqref{multiscale EFE1} and \eqref{multiscale EFE2} on a grid of $p^i_\varphi$ values. At each grid point, compute and store the following: (i) the driving forces $\Gamma^i_{(1)}(p^i_\varphi)$, $\Gamma^i_{(2)}(p^i_\varphi,0)$, and $\gamma^i_A(p^i_\varphi)$, (ii) the asymptotic first-order mode amplitudes at ${\cal I}^+$ [e.g., ${}_{-2}\hat C^{\rm up}_{1\ell\emm\bm{k}}(p^i_\varphi)$].
\item Using the stored values of the driving forces, evolve through the parameter space by solving the coupled equations~\eqref{0PA varphi 3}--\eqref{1PA dMdt} to obtain $\tilde p^i_{\varphi(0)}$ and the phases $\tilde\varphi^{(0)}_\alpha$ and $\tilde\varphi^{(1)}_\alpha$ as functions of $\tilde t=\e t$.
\item Construct the 1PA waveform 
\beq
h_+ - i h_\times = 2 \sum_{\ell\emm\bm{k}} \, \frac{{}_{-2} \hat C_{1\ell \emm \bm{k}}^{\text{up}}[\tilde p^i_{\varphi(0)}(\tilde u)]}{\omega_{\emm\bm{k}}^2} \, {}_{-2} S_{\ell \emm}(\theta, \phi; a \omega_{\emm\bm{k}})e^{-i\left[\varphi^{(0)}_{\emm\bm{k}}(\tilde u)+\e\varphi^{(1)}_{\emm\bm{k}}(\tilde u)\right]/\e},
\eeq
where $\omega_{\emm\bm{k}}=\omega_{\emm\bm{k}}[\tilde p^i_{\varphi(0)}(\tilde u)]$.
\end{enumerate}

We make two potentially clarifying remarks about these steps. First, even though the 1PA dynamics depend on the black hole parameters $M_A$, we need not include these parameters in our storage grid. This is because the 1PA effect of $M_A$ is linear in $M_A$, allowing us to only store its coefficient. However, note that the background spin parameter $a$ must be included in the grid (the background parameter $M$ need not be, as we can measure all lengths in units of $M$). Our second remark is that though $p^i_\varphi\neq\tilde  p^i_{\varphi(0)}$ at a given value of $\tilde t$ and $\e$, we can still freely solve~\eqref{multiscale EFE1} and \eqref{multiscale EFE2}, working with $p^i_\varphi$, in order to determine the driving forces as functions; it is precisely those functions, simply with $p^i_\varphi\to\tilde  p^i_{\varphi(0)}$, that appear in Eqs.~\eqref{0PA varphi 3}--\eqref{1PA dMdt}. 

A scheme of this sort was first sketched in Ref.~\cite{Pound:2019lzj} and detailed in Ref.~\cite{Miller:2020bft} for the special case of quasicircular orbits into a Schwarzschild black hole. Fig.~3 of Ref.~\cite{Miller:2020bft} gives a more thorough breakdown, though the structure of the multiscale expansion differs slightly from our formulation here. The general case for generic bound orbits in Kerr appears here for the first time.

At its core, the scheme requires three key ingredients for each set of orbital parameter values: the full first-order self-force, the asymptotic mode amplitudes of the first-order waveform, and the second-order dissipative self-force. As we summarized in Sec.~\ref{expansion of source}, the first two ingredients have been calculated for generic bound orbits in Kerr spacetime~\cite{vandeMeent:2017bcc} and are routinely calculated for less generic configurations. The main obstacle to including these ingredients in an evolution scheme is the computational cost and runtime of sufficiently covering the parameter space. The third ingredient has not yet been calculated in even the simplest configurations, though there is ongoing development of a practical implementation~\cite{Pound:2015wva,Miller:2016hjv,Miller:2020bft}, which led to the recent calculation of a second-order conservative effect~\cite{Pound:2019lzj}.

\subsection{Mode decompositions of the singular field}\label{mode decompositions of hS}

In our description so far, we have largely glossed over what is the pivotal step in almost all self-force calculations beyond the adiabatic approximation: the calculation of $h^{\R(n)}_{\mu\nu}$ and its derivatives, which are required for the conservative first-order self-force, the second-order self-force, as inputs for the second-order sources (whether the Detweiler stress-energy, the effective source, or the second-order Einsten tensor~\cite{Miller:2016hjv}), and as the essential ingredient in most dynamical quantities of interest.

In order to compute $h^{\R(n)}_{\mu\nu}$ (either using a puncture, or the point-particle method with regularization) in a mode-decomposed calculation,
a crucial component is a mode-decomposed form for the puncture field. This can be obtained by expanding the puncture field into the same basis
as is used in the calculation of the retarded field and can typically be done analytically, or at least semi-analytically. The specific details depend on the context
(e.g. choice of gauge, whether the mode decomposition needs to be exact or if it can be an approximation, whether the harmonics are spheroidal or spherical and scalar, vector, tensor or spin-weighted). However, the essential ingredients in the method are common to all cases:
\begin{enumerate}
  \item Introduce a rotated angular coordinate system $(\theta',\phi')$ such that the worldline is instantaneously at the pole, $\theta' = 0$. This makes the mode decomposition integrals analytically tractable and in some instances reduces the number of spherical harmonic $\emm$ modes that need to be considered.
  \item Expand the relevant quantity in a coordinate series about the worldline. In doing so, it is important to ensure that the series approximation is well behaved away from the worldline, in particular at $\theta' = \pi/2$. This can be achieved by multiplying by an appropriate window function in the $\theta'$ direction \cite{Wardell:2015ada}.\\
        The resulting expansion can always\footnote{In some instances (e.g. eccentric orbits) obtaining an expression for $\rho$ in this form requires the definition of the rotated coordinates to include a dependence not just on the unrotated angular coordinates, $(\theta,\phi)$, but also on the other coordinates (e.g. $\Delta r$ for the eccentric case).} be algebraically manipulated into the form of a power series (including $\log$ terms in some cases) in 
  \begin{equation}
      \rho := k_1 \chi^{1/2} \Big[\delta^2 + 1 -  \cos \theta'\big]^{1/2}.
  \end{equation}
  Here, $\delta ^2 = \frac{k_2 \Delta r^2}{\chi}$, $\Delta r := r - r_\p$ and $\chi := 1 - k_3^2 \sin^2 \phi'$, and $k_1$, $k_2$ and $k_1$ depend on the
  orbital parameters and can be treated as constants in the mode decomposition. The coefficients in the power series contain powers of $\Delta r$ and $\chi$ and also depend on the orbital parameters. Apart from that, the dependence on
  $\phi'$ is only via one of four possibilities: a. independent of $\phi'$; b. $\cos \phi' \sin \phi'$; c. $\cos \phi'$; d. $\sin \phi'$.
  The resulting dependence on $\phi'$ will then combine in the next step with the $e^{-i \emm' \phi'}$ factor from the harmonic to produce a dependence on $\phi'$ only via powers
  of $\chi$.\\
  When decomposing tensors, certain tensor components may also include an overall factor of $\sin \theta'$, but only ever in such a way that it cancels a singularity in the harmonic at $\theta' = 0$ so that the final integrand is non-singular away from $\Delta r = 0$. 
  \item Integrate against (the conjugate of) the relevant harmonic to obtain a mode decomposition in $(\ell,\emm')$ modes with respect to the rotated coordinate system. In the case of spin-weighted or vector and tensor harmonics, we must also be careful to account for the rotation, $\mathcal{R}$, of the frame, either by including the appropriate factor of $e^{i s \gamma'(\theta', \phi', \mathcal{R})}$ in the spin-weighted case \cite{Boyle:2016tjj}, or by including the appropriate tensor transformation in the case of vector and tensor harmonics \cite{Wardell:2015ada}.\\
  In performing the integrals, we can
  exploit the fact that only certain integrals over $\phi'$ are non-vanishing. In particular for the four possibilities listed in the previous step:
  \begin{enumerate}
      \item \label{poss1} only contributes for $\emm'$ even and only for the real part of $e^{i \emm' \phi'}$;
      \item only contributes for $\emm'$ even and only for the imaginary part of $e^{i \emm' \phi'}$;
      \item only contributes for $\emm'$ odd and only for the real part of $e^{i \emm' \phi'}$;
      \item only contributes for $\emm'$ odd and only for the imaginary part of $e^{i \emm' \phi'}$.
  \end{enumerate}
  The integrals over $\theta'$ can all be done analytically and result in expressions of the form
  \begin{alignat}{3}
      \delta^{n+2}(\delta^2 + 2)^{(n+2)/2} \sum_{i=0}^{\ell-\frac{n+4}{2}} a_i \delta^{2i}\, +\,  &\log \Big(\frac{\delta^2+2}{\delta^2}\Big) \sum_{i=0}^{\ell+\frac{n+2}{2}} b_i \delta^{2i} & \quad \text{$n$ even}
\\
      (\delta^2 + 2)^{(n+2)/2} \sum_{i=0}^{\ell} c_i \delta^{2i}\, +\, &|\delta| \delta^{n+1} \sum_{i=\emm'}^{\ell} d_i \delta^{2i} & \quad \text{$n$ odd}
  \end{alignat} 
  where $n$ is the power of $\rho$ and where the coefficients $a_i$, $b_i$, $c_i$ and $d_i$ are $\ell$-dependent rational numbers. The specific limits on the sums given here is for the $\emm'=0$ scalar harmonic case. Structurally similar expressions also appear for $\log \rho$ terms, for $\emm'\ne 0$ and for spin-weighted harmonics, but with the sums running over different ranges of $i$.\\
  The integrals over $\phi'$ can also be done analytically and result in power series (for integer powers of $\chi$), elliptic integrals (for half-integer powers), or the derivative of a hypergeometric function with respect to its argument (for $\log$ terms). In all three cases, they are functions of $k_3$ and potentially also $\Delta r$.
  \item Transform back to the $(\ell,\emm)$ modes with respect to the unrotated $(\theta,\phi)$ coordinate system using the Wigner-D matrix, $D^\ell_{\emm\emm'}(\mathcal{R})$. With a moving worldline the rotation is time dependent, but this complication is not relevant in many cases, the notable exception being in the effective source method where it is necessary to take time derivatives when computing the source from the puncture \cite{Miller:2016hjv,Heffernan:2017cad}.
\end{enumerate}

In many practical applications, an exact mode decomposition is not
necessary and an approximation is sufficient. For example, in the mode-sum regularization scheme one is only interested in the modes of the puncture (or its radial derivative in the
case of the self-force) evaluated in the limit $\Delta r \to 0$. Similarly, in the effective source scheme a series expansion to some power in $\Delta r$ suffices. Then, the exact expression for the mode-decomposed puncture field has a series expansion in $\Delta r$ of the form
\begin{equation}
    \sum_{\emm',ijk} c_{1,i} \Delta r^i + c_{2,j} \Delta r^j |\Delta r|  + c_{3,k} \Delta_r^k \log \Delta r
\end{equation}
where the coefficients $c_{1,i}$, $c_{2,j}$ and $c_{3,k}$ depend on the orbital parameters. In those cases, the mode
decomposition procedure simplifies significantly and one need only compute up to some maximum value for $i$, $j$ and $k$. Similarly, another simplification arises from the fact that one may only be interested in the decomposition of the puncture accurate to some order in distance from the worldline in the angular directions. This is is reflected in the number of $\emm'$ modes that must be included: for a puncture accurate to $n$ derivatives one must include up to $|\emm'| = |s| \pm n$ for the spin-weighted case (the vector and tensor cases similarly follow from their relation to the spin-weighed harmonics: $|s|=1$ for the vector case and $|s|=2$ for the tensor case).

One particularly important special case is that of mode-sum regularization, where one is only interested in the result for a given quantity summed over $\emm$ and with $\Delta r=0$. This leads to so-called mode-sum regularization formulas. For example, in the case of the first-order gravitational self-force this is given by
\begin{equation}
    F^\alpha = \sum_{\ell} (F^\alpha_{\rm ret} - A_{\pm} (2\ell+1) - B ) + D
\end{equation}
where $A_\pm$, $B$ and $D$ are ``regularization parameters'' that depend on the orbital parameters. Here, the value of $A_\pm$ depends on whether the limit $\Delta r \to 0$ is taken from above or below; this is because is comes from the derivative of the $|\Delta r|$ piece of the puncture. The parameters $B$ does not have this property as it comes from the piece of the puncture that does not $|\Delta r|$ (in particular, for the self-force it comes from the derivative of the $\Delta r^1$ piece of the puncture). The parameter $D$ accounts for the possibility that the subtraction does not exactly capture the behaviour of the contribution from the singular field (and only the singular field) and can often be set to $0$ by appropriately defining the subtraction \cite{Wardell:2015ada,Pound:2013faa}.

\subsubsection{Example: leading order puncture for circular orbits in Schwarzschild spacetime}
As a simple representative example, consider the problem of decomposing the leading-order piece of the Lorenz-gauge puncture [i.e. the first term in Eq.~\eqref{hS1 covariant}] into the spin-weighted spherical harmonic basis introduced in Sec.~\ref{sec:schw-perturbations}. For concreteness, we consider a circular geodesic orbit of radius $\ro$ with four-velocity $u^\alpha = u^t[1, 0, 0, \Omega]$, where $\Omega = \sqrt{\frac{M}{\ro^3}}$ and $u^t = \sqrt{\frac{\ro}{\ro-3M}}$.

As a first step, we expand the covariant expression in a coordinate series. Keeping only the leading term in the coordinate expansion, we have $g^{\alpha'}_{\mu} = \delta^{\alpha}_{\mu} + \mathcal{O}(\Delta x)$ and $\s = \rho + \mathcal{O}(\Delta x^2)$ where $\rho^2 := B^2 (\delta^2 + 1 - \cos \theta')$, $\delta^2 := \frac{\ro \Delta r^2}{B^2(\ro-2M)}$, $\chi := 1 - \frac{M}{\ro-2M} \sin^2 \phi'$, $B^2 := \frac{2 \ro^2 (\ro - 2 M) \chi}{(\ro - 3M)}$ and $\Delta r = r-\ro$. Here, we have made the standard choice of identifying a point on the worldline with the point where the puncture is evaluated by setting $\Delta t = t-\torbit = 0$.

Then, working with the Carter tetrad, the tetrad components of the puncture are
\begin{align}
  h_{ll} = h_{nn} &= \frac{2}{\rho}\frac{\ro-2M}{\ro - 3M},\nonumber \\
  h_{ln} = \frac{M}{\ro-2M} h_{m\mb} &= \frac{1}{\rho} \frac{2M}{\ro-3M},\nonumber \\
  h_{lm} = h_{nm} = -h_{l\mb} = -h_{n\mb} &= - \frac{\cos^2 \big(\tfrac{\theta'}{2}\big)}{\rho} \frac{2i (\ro-2M) \ro \Omega}{\sqrt{\fo}(\ro-3M)},\nonumber \\
  h_{mm} = h_{\mb\mb} &= -\frac{\cos^4 \big(\tfrac{\theta'}{2}\big)}{\rho}\frac{2 M}{\ro - 3M}.
\end{align}
Note that we have included factors of $\cos^2 \big(\tfrac{\theta'}{2}\big)$ and $\cos^4 \big(\tfrac{\theta'}{2}\big)$ to ensure that the puncture is sufficiently regular at $\theta'=\pi$ while not altering its leading-order behaviour near the worldline at $\theta'=0$.

We now integrate these against the appropriate spin-weighted spherical harmonic to obtain mode decomposed versions. In doing so, we must take account of the fact that our integration is with respect to a rotated coordinate system by including a factor of $e^{i s \gamma'(\theta', \phi', \mathcal{R})} \approx i^s e^{i s \phi'} + \mathcal{O}(\theta'^2)$.

Since we are only interested in the leading-order behaviour near the worldline we will only consider the modes $\emm' + s = 0$ series expanded through $\mathcal{O}(\Delta r^1)$. Then, we encounter the following integrals over $\theta'$
\begin{subequations}
\begin{equation}
 \int_0^\pi  \frac{1}{\rho} {}_0 \bar{Y}_{\ell0} (\theta', 0) \sin\theta' d\theta' \approx \frac{1}{B} \frac{1}{\sqrt{2\pi(2\ell+1)}}\bigg[2 - \sqrt{2}(2\ell+1)|\delta|\bigg],
\end{equation}
\begin{gather}
 \int_0^\pi  \frac{\cos^2 \big(\tfrac{\theta'}{2}\big)}{\rho} {}_1 \bar{Y}_{\ell,-1} (\theta', 0) \sin\theta' d\theta'
 = \int_0^\pi  \frac{\cos^2 \big(\tfrac{\theta'}{2}\big)}{\rho} {}_{-1} \bar{Y}_{\ell1} (\theta', 0) \sin\theta' d\theta'
 \nonumber \\
 \approx -\frac{1}{B} \frac{1}{\sqrt{2\pi(2\ell+1)}}\bigg[8 \Lambda_1 - \sqrt{2} (2\ell+1) |\delta|\bigg],
\end{gather}
\begin{gather}
 \int_0^\pi  \frac{\cos^2 \big(\tfrac{\theta'}{2}\big)}{\rho} {}_1 \bar{Y}_{\ell1} (\theta', 0) \sin\theta' d\theta'
 = \int_0^\pi  \frac{\cos^2 \big(\tfrac{\theta'}{2}\big)}{\rho} {}_{-1} \bar{Y}_{\ell,-1} (\theta', 0) \sin\theta' d\theta'
 \nonumber \\
 \approx -\frac{1}{B} \frac{1}{\sqrt{2\pi(2\ell+1)}}\bigg[8 \Lambda_1 - \frac{12}{(2\ell-1)(2\ell+3)}\bigg],
\end{gather}
\begin{gather}
 \int_0^\pi  \frac{\cos^4 \big(\tfrac{\theta'}{2}\big)}{\rho} {}_2 \bar{Y}_{\ell,-2} (\theta', 0) \sin\theta' d\theta'
 = \int_0^\pi  \frac{\cos^4 \big(\tfrac{\theta'}{2}\big)}{\rho} {}_{-2} \bar{Y}_{\ell2} (\theta', 0) \sin\theta' d\theta'
 \nonumber \\
 \qquad \approx \frac{1}{B} \frac{1}{\sqrt{2\pi(2\ell+1)}}\bigg[32 \Lambda_2 - \sqrt{2} (2\ell+1)  |\delta|\bigg],
\end{gather}
\begin{gather}
 \int_0^\pi  \frac{\cos^4 \big(\tfrac{\theta'}{2}\big)}{\rho} {}_2 \bar{Y}_{\ell2} (\theta', 0) \sin\theta' d\theta'
 = \int_0^\pi  \frac{\cos^4 \big(\tfrac{\theta'}{2}\big)}{\rho} {}_{-2} \bar{Y}_{\ell,-2} (\theta', 0) \sin\theta' d\theta'
 \nonumber \\
 \qquad \approx \frac{1}{B} \frac{1}{\sqrt{2\pi(2\ell+1)}}\bigg[32 \Lambda_2 - \frac{80}{(2\ell-1)(2\ell+3)} \bigg],
\end{gather}
\end{subequations}
where $\Lambda_1 := \frac{\ell(\ell+1)}{(2\ell-1)(2\ell+3)}$ and $\Lambda_2 := \frac{(\ell-1)\ell(\ell+1)(\ell+2)}{(2\ell-3)(2\ell-1)(2\ell+3)(2\ell+5)}$ .
Next, performing the integrals over $\phi'$ the integrands all involve integer (for the $|\delta|$ terms) and half-integer (for the $\delta^0$ terms) powers of $\chi$, producing elliptic integrals or polynomial functions of $\frac{M}{\ro - 2M}$, respectively. Putting everything together, transforming to the frequency domain (which in this case amounts to simply dividing by $2\pi$) and transforming back to the unrotated frame using the Wigner-D matrix, we then obtain expressions for the mode-decomposed punctures,
\begin{subequations}
\begin{equation}
  h_{ll}^{\ell\emm\omega} = \frac{D^\ell_{\emm,0}(\mathcal{R})}{\sqrt{(2\ell+1)\pi}}\bigg[ \frac{4 \mathcal{K}}{\pi \ro}\sqrt{\frac{\ro-2M}{\ro-3M}} - \frac{(2\ell+1)}{\ro^{3/2}\sqrt{\ro-3M}}|\Delta r| \bigg],
\end{equation}
\begin{equation}
    h_{ln}^{\ell\emm\omega} = \frac{D^\ell_{\emm,0}(\mathcal{R})}{\sqrt{(2\ell+1)\pi}}\bigg[ \frac{4 M \mathcal{K}}{\pi \ro \sqrt{\ro-2M}\sqrt{\ro-3M}} - \frac{(2\ell+1)M}{\ro^{3/2}\sqrt{\ro-3M}(\ro-2M)}|\Delta r| \bigg],
\end{equation}
\begin{align}
    h_{lm}^{\ell\emm\omega} &= \frac{D^\ell_{\emm,-1}(\mathcal{R})}{\sqrt{(2\ell+1)\pi}}\bigg[ -\frac{16 \Lambda_1 \Omega \mathcal{K}}{\pi} \sqrt{\frac{\ro}{\ro-3M}} + \frac{(2\ell+1)\Omega}{\sqrt{\ro-3M}\sqrt{\ro-2M}} |\Delta r| \bigg] \nonumber \\
     & + \frac{D^\ell_{\emm,1}(\mathcal{R})}{\sqrt{(2\ell+1)\pi}}\bigg[4\Lambda_1 - \tfrac{6}{(2\ell-1)(2\ell+3)}\bigg]\frac{4[(2\ro-5M) \mathcal{K}-2(\ro-2M)\mathcal{E}]}{M^{1/2} \ro \pi\sqrt{\ro-3M} } ,
\end{align}
\begin{align}
    h_{mm}^{\ell\emm\omega} &= \frac{D^\ell_{\emm,-2}(\mathcal{R})}{\sqrt{(2\ell+1)\pi}}\bigg[\frac{64 M \Lambda_2 \mathcal{K}}{\pi \ro \sqrt{\ro-2M}\sqrt{\ro-3M}} - \frac{(2\ell+1)M}{\ro^{3/2}\sqrt{\ro-3M}(\ro-2M)} |\Delta r| \bigg] \nonumber \\
     & + \frac{D^\ell_{\emm,2}(\mathcal{R})}{\sqrt{(2\ell+1)\pi}}\bigg[16\Lambda_2 - \tfrac{40}{(2\ell-1)(2\ell+3)}\bigg] \times \nonumber \\
     & \qquad \frac{4[(4\ro-9M)(4\ro-11M) \mathcal{K}-8(\ro-2M)(2\ro-5M)\mathcal{E}]}{3 M\pi \ro \sqrt{\ro-3M}\sqrt{\ro-2M} } ,
\end{align}
\end{subequations}
with the other components given either by $h_{ll}^{\ell\emm\omega} = h_{nn}^{\ell\emm\omega}$, $h_{m\mb}^{\ell\emm\omega} = \frac{\ro-2M}{M} h_{ln}^{\ell\emm\omega}$, $h_{nm}^{\ell\emm\omega} = h_{lm}^{\ell\emm\omega}$, or by complex conjugation. Here,
\begin{subequations}
\begin{align}
  \mathcal{K} := \frac14 \int_0^{2\pi} \bigg(1-\frac{M}{\ro-2M} \sin^2 \phi'\bigg)^{-1/2} \, d\phi', \\
  \mathcal{E} := \frac14 \int_0^{2\pi} \bigg(1-\frac{M}{\ro-2M} \sin^2 \phi'\bigg)^{1/2} \, d\phi',
\end{align}
\end{subequations}
are complete elliptic integrals of the first and second kind, respectively.

Higher order circular-orbit punctures including the contribution at $\mathcal{O}(\lambda^0)$ are available in Ref.~\cite{Wardell:2015ada}. Yet higher
orders and punctures for more generic cases are available upon request to the authors.

\section{Conclusion}

We stated in the introduction to this review that we aimed to summarize the key methods of black hole perturbation theory and self-force theory rather than summarizing the status of the field, leaving that task to existing reviews. However, it is worth putting this review in the context of the field's current state, and it is worth mentioning key topics that we did {\em not} cover.

Regarding topics we neglected, we first state the obvious: we did not cover any applications of black hole perturbation theory other than small-mass-ratio binaries. Although the bulk of the review is intended to  provide general treatments of black hole perturbation theory, orbital dynamics in black hole spacetimes, and self-force theory in generic spacetimes, without specializing to binaries, it is undoubtedly slanted toward our application of interest.

For that reason, we will also focus exclusively on the state of small-mass-ratio binary modelling. Hinderer and Flanagan's two-timescale analysis of orbital motion~\cite{Hinderer:2008dm} long ago made clear that 1PA waveforms are almost certainly required to perform high-precision measurements of these binaries. Such measurements will require phase errors much smaller than 1 radian, while the errors in 0PA waveforms will have errors of $O(\e^0)$ [or $O(1/\sqrt{\e})$ in the case of a resonance], which could be 1 or many more radians. Ref.~\cite{vandeMeent:2020xgc} has recently provided strong numerical evidence that a 0PA waveform will have significant errors for all mass ratios. Conversely, the same reference shows that a 1PA waveform should be not only highly accurate for EMRIs and IMRIs, but reasonably accurate even for comparable-mass binaries. This bolsters a long line of evidence that perturbative self-force theory is surprisingly accurate well outside its expected domain of validity; see Ref.~\cite{Rifat:2019ltp} for other recent evidence, as well as the reviews~\cite{Tiec:2014lba,Barack:2018yvs}.

Our presentation here provides the first framework for 1PA waveform generation. There are two main hurdles to overcome on the way to implementing this framework. One is the difficulty of efficiently covering the parameter space. Once a region is well covered by snapshots, recent advances make it possible to generate long, accurate waveforms extremely rapidly in that region, with generation times of a few tens of milliseconds for eccentric orbits in Schwarzschild spacetime~\cite{Chua:2020stf}. However, covering the parameter space of generic orbits in Kerr is highly expensive even for adiabatic codes, let alone calculations of the first-order self-force.

The second main hurdle is calculating the necessary second-order inputs for the 1PA evolution. There has  been steady progress in developing practical methods of computing these inputs, but only recently have results begun to materialize~\cite{Pound:2019lzj}. To date, these calculations have been restricted to quasicircular orbits in Schwarzschild spacetime; they must be extended to Kerr and to generic orbits.

In lieu of accurate evolving waveforms, the development of data analysis methods has so far been based on ``kludge'' waveforms constructed using a host of additional approximations (primarily, post-Newtonian approximations for the fluxes)~\cite{Glampedakis:2002cb,Barack:2003fp,Gair:2005ih,Babak:2006uv,Sopuerta:2011te,Chua:2017ujo}. These kludges will be very far from accurate enough to enable precise parameter estimation, but they are sufficiently similar to accurate waveforms to serve as testbeds for analysis methods. They may also be sufficiently accurate for detection of loud signals. There is also ongoing work to improve the accuracy of post-Newtonian 0PA approximations to enable them to accurately fill out the weak-field region of the small-mass-ratio parameter space~\cite{Sago:2015rpa}.

Our summary of multiscale evolution has also omitted some important ingredients in an accurate model. We must correctly account for passages through resonance, and we may need to include the transition to plunge  for mass ratios $\sim 1:50$. We must also account for the secondary's spin, which enters into the 1PA dynamics in three ways: (i) through the Mathisson-Papapetrou spin force~\eqref{EOM spin}, which contributes to $f^\mu_{(1)\rm con}$, (ii) through the spin's contribution to $T^{(2)}_{\mu\nu}$ in Eq.~\eqref{skeleton Tab}, which generates a perturbation that contributes to $f^\mu_{(2)\rm diss}$, and (iii) through a coupling between $h^{\R(1)}_{\mu\nu}$ and the spin, which again contributes a second-order dissipative effect. We refer to Refs.~\cite{Warburton:2017sxk,Akcay:2019bvk,Witzany:2019nml,Zelenka:2019nyp} for a sample of recent work on calculating these effects and incorporating them into waveform-generation schemes.
Specifically, Ref.~\cite{Warburton:2017sxk} generated waveforms from inspirals into a Schwarzschild black hole including first-order conservative (but not second-order dissipative) spin effects; Ref.~\cite{Akcay:2019bvk} derived balance laws incorporating spin; Ref.~\cite{Witzany:2019nml} derived the spin correction to the fundamental frequencies; and Ref.~\cite{Zelenka:2019nyp} computed the spin's contribution to fluxes from spinning particles on generic orbits in Schwarzschild spacetime.

We also note that while we have focused on a multiscale expansion built on frequency-domain methods, there has  been considerable development of time-domain snapshot calculations of $h^{(1)}_{\mu\nu}$ and $f^\mu_{(1)}$ using fixed geodesic sources~\cite{Barack:2005nr,Barack:2007tm,Barack:2010tm,Dolan:2012jg,Barack:2017oir}. The quantities $h^{(1)}_{\mu\nu}$ and $f^\mu_{(1)}$ output from such computations cannot be directly fed into the second-order field equations~\eqref{multiscale EFE2} or into the multiscale evolution scheme. However, if we decompose the outputs into Fourier modes, as in $h^{(1)}_{\mu\nu} = \sum_{\emm\bm{k}}h^{(1\emm\bm{k})}_{\mu\nu}e^{im\phi - \omega_{\emm\bm{k}}t}$, then the coefficients $h^{(1\emm\bm{k})}_{\mu\nu}$ are identical to those in a multiscale expansion, and these can be used as inputs for the multiscale scheme. Moreover, any first-order quantity that depends only on ${\cal P}^\alpha$ will be identical in the time domain with a geodesic source as in the multiscale expansion; this includes any quantity constructed as an average over the orbit, which includes most physical quantities of interest~\cite{Barack:2018yvs}. Because time-domain methods are typically more efficient than frequency-domain ones for highly eccentric orbits, certain dynamical quantities entering into the evolution may be more usefully computed in the time domain. 

Time-domain calculations also offer an alternative framework for waveform generation: rather than using Eq.~\eqref{multiscale waveform}, one can perform a multiscale evolution of ${\cal P}^\alpha$ to generate a self-accelerated trajectory and then solve the Teukolsky equation in the time domain with an accelerated point-particle source~\cite{Sundararajan:2008zm,Harms:2014dqa}. This may seem redundant, given that in the process of generating the multiscale evolution one must already compute all the inputs for Eq.~\eqref{multiscale waveform}. However, it offers significant flexibility, in that it can take as input trajectories generated with any method, such as inspirals which have been produced that include the full first-order self-force but omit second-order dissipative effects~\cite{Warburton:2011fk,Osburn:2015duj,vandeMeent:2018rms}. This gives it the additional advantage of being able to easily evolve through different dynamical regimes, such as the  evolution from the adiabatic inspiral to the transition to plunge~\cite{Taracchini:2014zpa,Rifat:2019ltp}. 

Beyond these alternative methods of wave generation, we have also passed over what has been the main application of self-force calculations. Although such calculations were originally motivated by modelling EMRI waveforms (and more recently, the prospect of using them to model IMRIs), they have also enabled the calculation of numerous physical effects in binaries. These, in turn, have facilitated a rich interaction with other binary models: post-Newtonian and post-Minkowskian theory, effective one body theory, and fully nonlinear numerical relativity~\cite{Tiec:2014lba}. Sections 7 and 8 of Ref.~\cite{Barack:2018yvs} provide a summary of the physical effects that have been computed and the synergies with other models. We highlight Refs.~\cite{Bini:2019nra, Damour:2019lcq,Bini:2020wpo} for more recent discussions of the power of such synergies and of the potential future impact of self-force calculations.

\section*{Cross-References}
\begin{itemize}
    \item \emph{Introduction to gravitational wave astronomy}, N.~Bishop
    \item \emph{Space-based laser interferometers}, J.~Gair, M.~Hewitson, A.~Petiteau
    \item \emph{The gravitational capture of compact objects by massive black holes}, P.~Amaro Seoane
    \item \emph{Post-Newtonian templates for gravitational waves from compact binary inspiral}, S.~Isoyama, R.~Sturani, H.~Nakano
    \item \emph{Non-linear effects in EMRI dynamics and the imprints on gravitational waves}, G.~Lukes-Gerakopoulos, V.~Witzany, O.~Semer\'ak
\end{itemize}

\section*{Acknowledgements}

AP is grateful to Jordan Moxon and Eanna Flanagan for numerous helpful discussions. BW thanks Andrew Spiers and Sam Dolan for independently checking several equations. AP also acknowledges the support of a Royal Society University Research Fellowship.

\bibliographystyle{spbasic}
\bibliography{pt-sf}

\end{document}